\nonstopmode

\newcommand{\ie}{i.e.{}}
\newcommand{\eg}{e.g.{}}

\newcommand{\mul}{\cdot}

\newcommand{\Cal}[1]{{\cal #1}}
\newcommand{\etal}{\textit{et al.}}
\newcommand{\U}[1]{\,{\mathrm{#1}}}

\newcommand{\X}[1]{_{\mathrm{#1}}}
\newcommand{\mat}[1]{\hbox{\boldmath{$#1$}\unboldmath}}
\newcommand{\bfvec}[1]{\hbox{\boldmath{$#1$}\unboldmath}}
\newcommand{\imag}{\mathrm i}
\newcommand{\euler}{\mathrm e}
\newcommand{\Sum}{\sum\limits}
\newcommand{\Prod}{\prod\limits}
\newcommand{\Int}{\int\limits}
\newcommand{\Lim}{\lim\limits}
\newcommand{\unitop}{\hat{\mathbbm{1}}}
\newcommand{\unitmatrix}{\mat{\mathbbm{1}}}
\newcommand{\differential}{\>\mathrm d}
\newcommand{\transpose}{{}^{\textrm{\scriptsize T}}}
\newcommand{\bra}[1]{\left<\right.\!#1\!\left.\right|}
\newcommand{\ket}[1]{\left|\right.\!#1\!\left.\right>}
\newcommand{\bracket}[2]{\left<\right.#1\>\,|\,\>#2\left.\right>}
\newcommand{\diag}{\textbf{diag}}
\newcommand{\cleb}[2]{C(#1 ; #2)}
\newcommand{\threeY}[2]{\Cal Y(#1 ; #2)}
\newcommand{\E}[1]{\times 10^{#1}}
\newcommand{\VUV}{\textsc{vuv}}
\newcommand{\XUV}{\textsc{xuv}}
\newcommand{\NIR}{\textsc{nir}}
\newcommand{\atopa}[2]{\genfrac{}{}{0pt}{}{#1}{#2}}

\newcommand{\sinc}{\textrm{sinc}\,}
\newcommand{\nbh}{\hbox{-}}
\newcommand{\xray}{x\nbh{}ray}
\newcommand{\Xray}{X\nbh{}ray}
\newcommand{\onlinecite}[1]{\cite{#1}}

\documentclass[a4paper,pra,aps,10pt,twocolumn,longbibliography,showpacs,showkeys,nopreprintnumbers]{revtex4-1}
\usepackage{amsmath,bbm,amssymb,subeqnarray,datetime,fancyhdr,color,CJK,graphicx}
\usepackage[nativepdf,bookmarks,bookmarksopen,bookmarksnumbered,raiselinks,%
pdftitle={High-order harmonic generation with resonant core excitation by ultraintense x rays},%
pdfauthor={Christian Buth},%
pdfsubject={Atomic Physics, Optics},%
pdfkeywords={x rays, photoabsorption, core-hole decay, high-order harmonic generation, HHG, x-ray free electron laser, FEL, Rabi flopping, strong-field physics, optical laser}]{hyperref}

\begin{document}
\title{High-order harmonic generation with resonant core excitation
by ultraintense x~rays}
\thanks{The final publication is available at Springer via
\href{http://dx.doi.org/10.1140/epjd/e2015-60357-3}
{dx.doi.org/10.1140/epjd/e2015-60357-3}.}
\author{Christian Buth}
\thanks{World Wide Web: \href{http://www.christianbuth.name}
{www.christianbuth.name}, electronic mail}
\email{christian.buth@web.de}
\affiliation{Max-Planck-Institut f\"ur Kernphysik, Saupfercheckweg~1,
69117~Heidelberg, Germany}
\affiliation{Theoretische Chemie, Physikalisch-Chemisches Institut,
Ruprecht-Karls-Universit\"at Heidelberg, Im Neuenheimer Feld~229,
69120~Heidelberg, Germany}
\date{15 Oct 2015}

\begin{abstract}
High-order harmonic generation~(HHG) is combined with resonant
\xray~excitation of a core electron into the transient
valence vacancy that is created in the course of the HHG~process.
To describe this setting, I develop a two-active-electron quantum
theory for a single atom assuming no Coulomb interaction among the electrons;
one electron performs a typical HHG three-step process
whereas another electron is excited (or even Rabi flops)
by intense x~rays from the core shell into the valence hole
after the first electron has left the atom.
Depending on the amplitude to find a vacancy in the valence
and the core, the returning continuum electron recombines with
the valence and the core, respectively, emitting high-order harmonic~(HH)
radiation that is characteristic of the combined process.
After presenting the theory of \xray~boosted HHG for continuous-wave
light fields, I develop a description for \xray~pulses with a
time-varying amplitude and phase.
My prediction offers novel prospects for nonlinear \xray~physics,
attosecond x~rays, and HHG-based time-dependent chemical imaging
involving core orbitals.￼
\end{abstract}

%
% Physics and Astronomy Classification Scheme (PACS) of 2010
%
% 32.80.Aa     Inner-shell excitation and ionization
%
% 32.30.Rj     X-ray spectra
%
% 41.60.Cr     Free-electron lasers (see also 52.59.Rz Free-electron
%              devices in plasma physics)
%
% 42.65.Ky     Frequency conversion; harmonic generation, including
%              higher-order harmonic generation (see also 42.79.Nv Optical
%              frequency converters)
%



\preprint{arXiv:1303.1332}
\maketitle

\section{Introduction}

Since the discovery of high-order harmonic generation\linebreak
(HHG) by atoms in intense optical laser
fields~\cite{McPherson:MP-87,Ferray:MH-88},
HHG has spawned the field of attoscience, is used for spectroscopy,
and serves as a light source in many optical
laboratories~\cite{Agostini:PA-04,Scrinzi:AP-06,Bucksbaum:FA-07,%
Krausz:AP-09,Kohler:FA-12}.
Initially, there has been a period of learning about the fundamental
physics of HHG~\cite{Kuchiev:AA-87,Krause:HO-92} which lead to a
phenomenological description by a single atom in terms of a three-step
model~\cite{Schafer:AT-93,Corkum:PP-93} and quantum
theories~\cite{Lewenstein:HH-94,Becker:MH-94,%
Kuchiev:QT-99,Milosevic:SI-01,Becker:AT-02,Schafer:NM-09}.
Soon it was realized that high-order harmonic~(HH)~light from
a dense gas is not well-described by the HH~spectrum of a
single atom and one needs to account for
propagation effects of the optical laser and the HH~radiation in the macroscopic
medium~\cite{LHuillier:TA-91,Balcou:QP-99,Priori:NT-00,%
Gaarde:MA-08,Popmintchev:BC-12}.
Nowadays, more and more applications based on~HHG have come into focus
[see, \eg, Refs.~\onlinecite{Agostini:PA-04,Scrinzi:AP-06,Bucksbaum:FA-07,%
Krausz:AP-09,Kohler:FA-12}].
To name a few, in atoms, there is the phase measurement of resonant two-photon
ionization in helium~\cite{Swoboda:PM-10} and intense
\xray~generation~\cite{Popmintchev:BC-12};
for molecules, there is tomographic imaging of
orbitals and related methods for structure
determination~\cite{Itatani:TI-04,Santra:IM-06,Morishita:AR-08,%
Lin:UD-12,Santra:TS-06,Patchkovskii:OT-07}, and the control of
electron localization in photodissociation~\cite{Kling:CE-06}.
Finally, for condensed matter, an attosecond spectroscopy has been
developed~\cite{Cavalieri:AS-07}.

Present-day canonical theory of HHG~\cite{Agostini:PA-04,Scrinzi:AP-06,%
Bucksbaum:FA-07,Krausz:AP-09,Kohler:FA-12}
gravitates around HHG from valence electrons and the
single-active electron~(SAE) approximation~\cite{Kulander:EP-91,Kulander:SI-93}.
Within the SAE approximation and the three-step model of
HHG~\cite{Schafer:AT-93,Corkum:PP-93}, the optical laser tunnel
ionizes a valence electron and accelerates it in the continuum.
When the optical laser field changes direction, the liberated electron
is driven back to rescatter with the parent ion.
This may cause the electron to recombine with the ion whereby the
excess energy due to the atomic potential and due to the
energy gained from the optical laser field is released in terms of
a photon with an energy which is a high-order harmonic of the
optical frequency.
In HHG each rescattering generates attosecond bursts of
radiation~\cite{Pukhov:TS-03}.
HHG from an ultrashort optical laser pulse leads to an
attosecond pulse train~\cite{Paul:TA-01}
in which the individual attosecond pulses are separated by a half
cycle period of the optical laser from which a single attosecond burst can be
isolated by filtering out the harmonics close to the HH~cutoff
(maximum photon energy for a given optical laser
intensity)~\cite{Goulielmakis:SC-08}.
By optimizing the quantum path of the continuum
electron, the HH~cutoff can be increased by a factor
of~$2.5$~\cite{Chipperfield:IW-09}.

Explorations beyond the canonical theory of~HHG~\cite{Agostini:PA-04,%
Scrinzi:AP-06,Bucksbaum:FA-07,Krausz:AP-09,Kohler:FA-12} have been
pursued in only a few works.
In Refs.~\onlinecite{Watson:CS-96,Sanpera:HG-96,Milosevic:TA-06,%
Milosevic:HE-07}, HHG from a coherent superposition of the
ground state and an excited state of an atom was investigated.
A shift of the HH~spectrum was found amounting to the
energy difference between the two states involved with
the unshifted and the shifted plateaux having a similar height.
In Ref.~\onlinecite{Strelkov:AR-10} HHG for a parent ion that is
described by a potential that has an autoionizing state is
reported for HHG from laser-ablation of solids;
a four-level model is proposed in which the recombination from
an electron trapped in the autoionizing state to the ground state occurs.
The autoionizing state is populated by a radiationless transition of
the returning continuum electron.
A one- to two-order of magnitude enhancement of HH intensity for a
harmonic that is resonant with the transition between the autoionizing
state and the ground state is observed.
A somewhat different modeling based on a modification of the
model of Lewenstein~\etal~\cite{Lewenstein:HH-94} with a coherent
superposition of states with a thorough numerical investigation
of the intensity and phase of resonant high harmonics is presented
in Ref.~\onlinecite{Milosevic:RH-10}.
Recently, it was both experimentally observed and theoretically explained
in Refs.~\onlinecite{McFarland:HH-08,Smirnova:HI-09} that molecular
orbitals below the highest occupied molecular orbital have a significant
impact on HH~emission;
the authors of Ref.~\onlinecite{Figueira:MH-10} investigate in strong-field
approximation~\cite{Lewenstein:HH-94}
the influence of quantum interference on HH~spectra of
molecules with two optically active electrons from different orbitals.
In Ref.~\onlinecite{Kraus:HP-13} a coherent superposition of states
in dynamically aligned molecules was exploited to measure the
time-evolution of the molecular dynamics with high sensitivity.
Multiple cutoffs from plasmon-like excitations in ions
were predicted in Ref.~\onlinecite{Zanghellini:PS-06}.
The role of multielectron effects regarding the recombination
probability of the recolliding electron was discussed in
Refs.~\onlinecite{Santra:TS-06,Gordon:RM-06}.
Further a two-electron scheme was considered that uses sequential double
ionization by an optical laser with a subsequent nonsequential
double recombination;
in helium it leads to a second plateau with about 12~orders of magnitude lower
yield than the primary HH~plateau~\cite{Koval:ND-07}.

Frequently, the SAE~approximation for a valence electron is also applied
to two-color HHG where an optical laser is combined with
\VUV/\XUV~light;
the high-frequency radiation assists, thereby, in the ionization process leading
to an overall increased yield~\cite{Ishikawa:PE-03,Takahashi:DE-07,%
Ishikawa:WD-09,Popruzhenko:HO-10}.
Two-color HHG is evolved further by using attosecond \XUV{}~pulses
to manipulate the HHG~process which increases the
HH~yield for a certain frequency range by enhancing the
contribution from specific quantum orbits~\cite{Schafer:SF-04,%
Gaarde:LE-05,Figueira:CH-06,Heinrich:EV-06,Biegert:CH-06,Figueira:HO-07}.
However, there are only few exceptions, \eg,
Refs.~\onlinecite{Santra:TS-06,Gordon:RM-06,Fleischer:AH-08,Fleischer:GH-08},
in which many-electron effects are treated for two-color HHG.
In Refs.~\onlinecite{Fleischer:AH-08,Fleischer:GH-08} the influence of
other electrons is included implicitly by using a frequency-dependent
polarizability for the atoms:
the \XUV~light is found to cause new plateaux to emerge at higher
energies, however, with a much lower HH~yield.

Alternatively to HHG-based light sources for \XUV{}~light and,
particularly, x~rays, there are synchrotrons and---most relevant
for this work---the newly constructed free electron lasers~(FELs)
that provide exciting opportunities for strong-field physics.
For example, the Free Electron Laser in
Hamburg~(FLASH)~\cite{Altarelli:TDR-06} offers \XUV{} to
soft \xray~radiation that is ideal for studying the valence and shallow
core of atoms or the Linac Coherent Light
Source~(LCLS)~\cite{LCLS:CDR-02,Emma:FL-10}
which delivers soft and hard x~rays suitable for \xray~diffraction with atomic
resolution and examining core electrons of the elements in the periodic table.
Existing high-frequency FELs produce \XUV{}- and \xray~light of unprecedented
intensity which can be used---among many other applications---to
manipulate optical strong-field processes.
They operate according to the self-amplification of spontaneous
emission~(SASE) principle which produces chaotic
light~\cite{Kondratenko:GC-79,Bonifacio:CI-84,Saldin:PF-00}
that can be modeled by the partial coherence
method~(PCM)~\cite{Pfeifer:PC-10,Jiang:TC-10,Cavaletto:RF-12}.

The combination of an optical laser and \XUV/\xray~light from synchrotrons
or FELs has proven to be very beneficial in the past, \eg, to control
the interaction of the x~rays with optical light in various ways
involving atoms and molecules~\cite{Wuilleumier:PP-06,Santra:SF-07,%
Santra:SF-08,Adams:QO-13}.
Recently, \xray{} and optical wave mixing---specifically sum-frequency
generation~(SFG) of optical light and x~rays---was demonstrated
at LCLS by Glover~\etal~\cite{Glover:XO-12} with a conversion
efficiency of~$3 \E{-7}$.
Such \xray{} and optical wave mixing was proposed
theoretically~\cite{Freund:OM-70,Freund:OM-71,Eisenberger:XO-71}
many years ago in the 1970s but only recently it has been
become feasible to observe SFG experimentally by to the
construction of FELs with unprecedented \xray~intensity~\cite{Glover:XO-12}.
The experiment of Glover~\etal{} is an encouraging
motivation of theoretical research on \xray~boosted
HHG in which HHG is modified by x~rays~\cite{Buth:NL-11,Kohler:EC-12,%
Buth:KE-13} because SFG is the simplest way to boost x~rays by
optical light.
Within the SAE, HHG with core electrons of neon that are ionized by
x~rays were studied theoretically~\cite{Buth:KE-13}.
A single-atom efficiency comparable with conventional HHG was found
and attosecond pulses were isolated from the \xray-boosted HH~spectra.
The theory of Ref.~\cite{Buth:KE-13} is a form of laser-assisted \xray-atom
scattering extensively examined for hydrogen atoms in
Refs.~\onlinecite{Milosevic:CH-00,Milosevic:SR-03}
and Refs.~therein generalized to multielectron atoms.

Since the publication of the articles~\cite{Buth:NL-11,Kohler:EC-12,Buth:KE-13}
and a preprint of this work~\cite{Buth:HO-13} a few related or inspired
articles have been published.
Using the resonant interaction of \XUV{}~light with an atomic transition
in hydrogen-like atoms that are strongly perturbed by optical light
Antonov \etal~\cite{Antonov:FS-13} predict attosecond pulse generation
based on resonant interaction
in a macroscopic medium for \VUV{} or \XUV{}~light.
My prediction of a splitting of HH~peaks by Rabi oscillations is
demonstrated in Ref.~\onlinecite{Tudorovskaya:HH-14} where an optical laser
and intense \XUV{}~light is used jointly to predict HHG from a model
potential where two cases are explored:
first, a system with two bound states;
second, a system with a shape resonance

\begin{figure}
  \begin{center}
    \includegraphics[clip,width=\hsize]{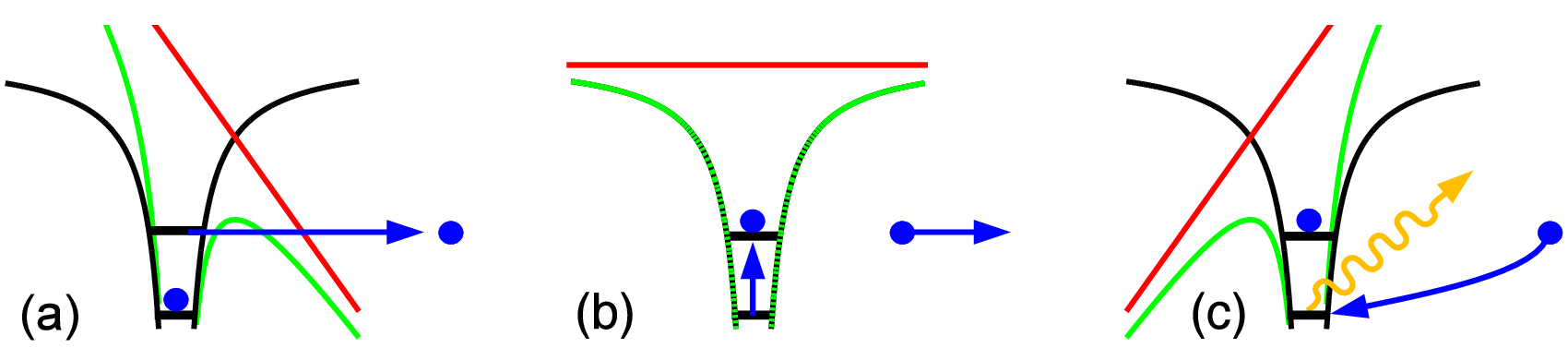}
    \caption{(Color online) Schematic of the three-step model of
             HHG~\cite{Schafer:AT-93,Corkum:PP-93} augmented by resonant
             \xray~excitation of core electrons:
             (a)~tunnel ionization of a valence electron in the optical
             laser field;
             (b)~while the first electron propagates freely
             in the continuum, a second electron from a core orbital
             is excited into the transient valence hole;
             (c)~eventually, the continuum electron is driven back to
             the residual ion by a reversal of the direction of the
             optical laser field and recombines with the newly
             formed core vacancy emitting \xray-boosted HH~light.}
    \label{fig:3stepXray}
  \end{center}
\end{figure}

In this work, I devise a detailed theory of HHG by an optical laser
with resonant excitation (or even Rabi flopping~\cite{Rabi:OP-36,%
Rabi:SQ-37,Rabi:MN-38,Cohen:QM-77,Scully:QO-97,Meystre:QO-99})
of a core electron in intense \XUV/\xray~light.
Specifically, I will focus on the case of continuous-wave~(cw) light, first,
constant-amplitude light with finite pulse duration, second,
and a constant-amplitude optical laser in combination with
arbitrary \xray~pulses, third.
In the parlance of the three-step model~\cite{Schafer:AT-93,Corkum:PP-93},
\xray-boosted HHG [Fig.~\ref{fig:3stepXray}] proceeds as follows:
first, the atomic valence is tunnel ionized;
second, the liberated electron propagates freely in the electric
field of the optical laser;
third, the direction of the optical laser field is reversed and the continuum
electron is driven back to the parent ion and rescatters with it which may
cause it to recombine with the vacancy emitting the excess energy
in terms of a HH~photon.
The excursion time of the liberated electron from the ion
is~${\sim}1 \U{fs}$ for optical laser light.
During this time, one may manipulate the cation such that the returning
electron sees an altered ion as depicted in Fig.~\ref{fig:3stepXray}.
Then, the emitted HH~radiation bears the signature of the change.
I use resonant \xray~excitation of an inner-shell electron into the transient
valence vacancy that is formed by tunnel ionization.
The recombination of the returning electron with the core hole leads
to a large increase of the energy of the emitted HH~light as the energy
of the \xray~photons is added to the unmodified HH~spectrum.
A prerequisite for this to work certainly is that the core hole is
not too short lived~\cite{Als-Nielsen:EM-01,Campbell:WA-01}, \ie, it
should not decay before the electron returns.

Choosing a geometry of parallel, linear polarization vectors for the optical
laser and the x~rays, the equations of \xray-boosted HHG simplify substantially.
Namely, as tunnel ionization of valence electrons takes
place predominantly along the linear optical laser polarization
axis~\cite{Perelomov:I1-66,Perelomov:I2-66,Perelomov:I3-67,Popov:I4-67,%
Ammosov:TI-86,Delone:MP-00,Yudin:NA-01}, mostly the valence
states along the polarization axis are depleted which have a
magnetic quantum number~$m = 0$ which then couple via x~rays to
core electrons with~$m = 0$.
In Ref.~\onlinecite{Buth:NL-11}, we examined HHG in
krypton atoms where tunnel ionization leads to $4p$~vacancies.
There the \XUV{}~light is tuned to the $3d \to 4p$~resonance in the cation
which leads to a second high-yield HH~plateau that is shifted to
higher energies by the \XUV~photon energy.
Similarly, we studied \xray-boosted HHG for neon~$2p$~vacancies where
x~rays are tuned to the $1s \to 2p$~resonance in the
cation~\cite{Kohler:EC-12}.
Both Refs.~\onlinecite{Buth:NL-11,Kohler:EC-12} focus only on $m = 0$~valence
and core states and drop the magnetic quantum number in the notation.
If the x~rays are very intense, even Rabi flopping~\cite{Rabi:OP-36,%
Rabi:SQ-37,Rabi:MN-38,Cohen:QM-77,Scully:QO-97,Meystre:QO-99}
of the core electron is possible.
Such Rabi oscillations have been investigated in neon atoms without optical
laser~\cite{Rohringer:RA-08,Rohringer:PN-08,Rohringer:SD-12,Kanter:MA-11,%
Cavaletto:RF-12,Adams:QO-13}.

My scheme offers new prospects for HHG involving core electrons,
however, if one is only interested in an extension of the HHG
cutoff into the kiloelectronvolt regime,
one may use conventional HHG that is laboratory size and
valence-electron based.
In this case, the optical laser intensity needs to be
high~$\sim${}$10^{15} \U{W/cm^2}$ and its wavelength should
be in the midinfrared instead of the \NIR{} in order to reduce
ionization~\cite{Perelomov:I1-66,Perelomov:I2-66,Perelomov:I3-67,Popov:I4-67,%
Ammosov:TI-86,Delone:MP-00,Yudin:NA-01}
to avoid a large electron background that causes phase mismatching.
The longer wavelength, however, reduces the efficiency of HHG
due to longer continuum propagation and thus enlarged spreading
of the wave packet.
A good HH~yield can, nonetheless, be obtained by a judicious choice of the gas
pressure~\cite{Popmintchev:PM-09,Arpin:EH-09,Chen:BC-10,Popmintchev:BC-12}.

The article is structured as follows.
The theory of HHG in the presence of a cw or constant-amplitude
optical laser and cw or constant-amplitude x~rays, respectively,
is developed in Sect.~\ref{sec:theoryCW}:
I discuss the wave-function ansatz [Sect.~\ref{sec:wavefunction}],
the Hamiltonian [Sect.~\ref{sec:hamiltonian}],
the equations of motion~(EOMs) [Sect.~\ref{sec:EOMs}], the time-dependent
dipole transition matrix element [Sect.~\ref{sec:HHGtrama}], and
the HH~spectrum [Sect.~\ref{sec:HHGspectrum}].
A formalism for a constant-amplitude optical laser and arbitrary
\xray~pulses is derived in Sect.~\ref{sec:theoryGen}:
I specify the Hamiltonian, the wave-function ansatz, and
the EOMs [Sect.~\ref{sec:HamWavEOM}],
decouple and integrate the EOMs [Sect.~\ref{sec:integrationEOM}],
and calculated the time-dependent electric dipole moment and the
HH~spectrum [Sect.~\ref{sec:TDdipHHG}].
Conclusions are drawn in Sect.~\ref{sec:conclusion}.
In the Appendices, I discuss the atomic electronic structure,
Sect.~\ref{sec:elstructmodel}, the saddle-point approximation,
Sect.~\ref{sec:saddlepoint}, and the quantum electrodynamic description
of HH~emission, Sect.~\ref{sec:HPNS}.
Accompanying Electronic Supplementary Material~\cite{SuppData} is available
for this article containing a \textit{Mathematica}~\cite{Mathematica:pgm-V10.1}
Notebook with detailed calculations of a few of the expressions
from this article.
Atomic units are used throughout~\cite{Hartree:WM-28,Szabo:MQC-89}.

\section{High-order harmonic generation with continuous-wave x~rays}
\label{sec:theoryCW}

I consider HHG for an atom in the two-color light of an optical laser
and intense x~rays which are both linearly polarized along the $z$~axis
and copropagate along the $x$~axis.
The optical laser-only problem was treated by
Lewenstein~\etal~\cite{Lewenstein:HH-94} for a valence electron in
SAE~\cite{Kulander:EP-91,Kulander:SI-93}.
Here I consider also core electrons in addition to
valence electrons [Fig.~\ref{fig:3stepXray}].
The combined treatment of HHG and x~rays, which may induce Rabi
flopping~\cite{Rabi:OP-36,Rabi:SQ-37,Rabi:MN-38,Cohen:QM-77,Scully:QO-97,%
Meystre:QO-99,Rohringer:RA-08,Rohringer:PN-08,Kanter:MA-11,Rohringer:SD-12,%
Cavaletto:RF-12,Adams:QO-13}
in the residual ion, represents a genuine two-active-electron problem.

\subsection{Symmetry and basis states}
\label{sec:wavefunction}

The theory is devised for a closed-shell atom with a spin-singlet
ground state.
I set out from the nonrelativistic independent-electron approximation for
the atomic electronic structure of Hartree-Fock-Slater
type~\cite{Slater:AS-51,Slater:XA-72} with spherical averaging
of the one-electron potential leading to a central potential as implemented in
the program by Herman and Skillman~\cite{Herman:AS-63,Buth:TX-07}.
Thus, for an isolated atom without light fields, atomic orbitals
with the same principal and angular quantum numbers are degenerate
with respect to the magnetic quantum number~\cite{Merzbacher:QM-98}.
As no spin-dependent terms are in the Hamiltonian of the atom in
the optical and \xray~light, all quantities can be expressed
in terms of spatial atomic orbitals.
Consequently, I restrict myself to spatial orbitals occupied by a
single electron in the following.
Observables, however, need to be multiplied by a factor of two
in order to account for the two electrons with opposite spin
per spatial orbital.

The essential states necessary to describe HHG in two-color light
are deduced as follows.
Let the polarization of the optical laser~$\vec e\X{L}$
and the x~rays~$\vec e\X{X}$ be linear along the $z$~axis,
\ie, $\vec e\X{L} = \vec e\X{X} = \vec e_z$~holds.
As spin-orbit coupling is not treated which, in any
case, only influences strong-field ionization of heavier
atoms~\cite{Young:XR-06,Santra:SO-06,Rohringer:MC-09,Loh:QS-07}, the
magnetic quantum number is conserved in the interactions
[Appendix~\ref{sec:elstructmodel}].
Hence, one-electron states of relevance to the problem are the valence
states~$\ket{a ; m_a}$ with magnetic quantum numbers~$m_a \in \mathbb M_a$
for principal~$n_a$ and angular momentum~$l_a$ quantum numbers
with~$\mathbb M_a = \{-l_a, \ldots, l_a\}$.
The core states with~$n_c$ and $l_c$ are~$\ket{c ; m_c}$ for~$m_c \in
\mathbb M_c$ with~$\mathbb M_c = \{-l_c, \ldots, l_c\}$.
Continuum states are approximated as free-electron states, \ie,
plane waves, and are denoted by~$\ket{\vec k}$
for~$\vec k \in \mathbb R^3$.
Specifically, I use momentum-normalized free-electron wave functions
\begin{equation}
  \label{eq:freeelmom}
  \bracket{\vec r}{\vec k} = \dfrac{1}{(2 \pi)^{3/2}} \; \euler^{\imag
    \, \vec k \mul \vec r} \; ,
\end{equation}
with the position operator~$\vec r$~\cite{Merzbacher:QM-98}.
The $\ket{a ; m_a}$, $\ket{c ; m_c}$, and $\ket{\vec k}$ form three classes
of one-electron basis states.
They are taken to be mutually orthogonal where discrete states
are normalized and continuum states are Dirac
$\delta$-distribution~\cite{Arfken:MM-05}
normalized, \ie, $\bracket{\vec k}{\vec k^{\,\prime}} = \delta^3(\vec k
- \vec k^{\,\prime})$.
Note that there is a small overlap between continuum states and discrete
states which, however, is neglected throughout.

Taking~$\ket{\vec k}$ for continuum electrons implies that
I make the strong-field approximation~(SFA)~\cite{%Kulander:TD-88,%
Lewenstein:HH-94},%,Paulus:PA-94},
\ie, I neglect the impact of the
Coulomb potential of the parent ion on continuum electrons.
Only electrons with~$m \in \mathbb M_2$ for~$\mathbb M_2 = \mathbb M_a \cap
\mathbb M_c$ are amenable to a core excitation by the x~rays
with~$\vec e\X{X} = \vec e_z$ in electric dipole approximation
[Appendix~\ref{sec:elstructmodel}] and HHG is modified only for these electrons.
For states~$\ket{a ; m_a}$ with~$m_a \in \mathbb M_1$
and $\mathbb M_1 = \mathbb M_a \setminus \mathbb M_2$, HHG is
described within the SAE~approximation~\cite{Kulander:EP-91,Kulander:SI-93}
following Lewenstein~\etal~\cite{Lewenstein:HH-94} because no core electrons
couple to these valence vacancies.
Hence there are $N/2 = \# \mathbb M_a + \# \mathbb M_2$ relevant spatial
orbitals in the system.

To describe the quantum dynamics of resonant \xray~excitation
in the course of HHG, one needs to consider two active electrons:
one in the valence or continuum and one in the core or valence.
The spatial one-electron states are occupied by two electrons each
with opposite spin.
To facilitate a core excitation, the spin of the valence electron
liberated by tunnel ionization and the spin of the core electron
need to be the same as the interaction with the x~rays is spin independent.
Whether the valence electron with spin up or spin down
is ionized does not change the result and thus
it is sufficient to consider only one case~\cite{Szabo:MQC-89}.
In other words, I need to consider an effective two-electron system formed
by one electron from the valence and one electron from the core
with the same spin, \ie, the system has triplet spin.
The two-electron wave function can be factored into a spin part
and a spatial part.
Due to the Pauli exclusion principle for indistinguishable particles
and the fact that the triplet spin part is symmetric with respect to electron
exchange, the spatial part needs to be
antisymmetric~\cite{Szabo:MQC-89,Merzbacher:QM-98}.

I use three different classes of two-electron basis states.
First, the ground state of the two-electron system for~$m \in \mathbb M_2$
is given by the Slater determinant
\begin{equation}
  \label{eq:twoground}
  \ket{a \, c ; m} = \dfrac{1}{\sqrt{2}} \; \bigl[ \ket{a ; m} \otimes
    \ket{c ; m} - \ket{c ; m} \otimes \ket{a ; m} \bigr] \; .
\end{equation}
This is a normalized linear combination of tensorial products of the
two respective one-electron states which is antisymmetric with respect
to electron exchange;
it incorporates probabilistic correlations between
electrons~\cite{Cohen:QM-77,Szabo:MQC-89}.
As core states are energetically separated from valence states,
electron correlations due to Coulomb interaction between core and valence
electrons beyond the Hartree-Fock-Slater
approximation~\cite{Slater:AS-51,Slater:XA-72}
are small and neglecting them has a minor influence on excitation
and ionization energies which is referred to as core-valence
separation~\cite{Cederbaum:CH-80,Angonoa:KS-87}.

Second, the valence-excited states with one electron in the continuum
for~$\vec k \in \mathbb R^3$ and one electron in the core
with~$m \in \mathbb M_2$ are
\begin{equation}
  \label{eq:twovalence}
  \ket{\vec k \, c ; m} = \dfrac{1}{\sqrt{2}} \; \bigl[ \ket{\vec k} \otimes
    \ket{c ; m} - \ket{c ; m} \otimes \ket{\vec k} \bigr] \; .
\end{equation}

Third, the core-excited states with one electron in the continuum,
$\vec k \in \mathbb R^3$, and one electron in the valence state,
$m \in \mathbb M_2$, are
\begin{equation}
  \label{eq:twocore}
  \ket{\vec k \, a ; m} = \dfrac{1}{\sqrt{2}} \; \bigl[ \ket{\vec k} \otimes
    \ket{a ; m} - \ket{a ; m} \otimes \ket{\vec k} \bigr] \; .
\end{equation}

Note that in Refs.~\onlinecite{Buth:NL-11,Kohler:EC-12}, Hartree products
are used for the three classes of two-electron basis
states~(\ref{eq:twoground}), (\ref{eq:twovalence}), and (\ref{eq:twocore})
instead of Slater determinants.
This does not cause any difficulties as long as the two-electron
dipole operator is defined suitably---Eqs.~(\ref{eq:Nhalfdipole}),
(\ref{eq:diptransmat}) restricted to two electrons---such that it
leads to Eq.~(5) of Ref.~\onlinecite{Buth:NL-11}.

From the one- and two-electron basis states of the previous two
paragraphs, I construct the ground state~$\ket{\Phi_0}$ and
the valence-excited~$\ket{\Phi_{\vec k \, c}^{(m)}}$
and core-excited~$\ket{\Phi_{\vec k \, a}^{(m)}}$ states of the
$N/2$~electron system for~$m \in \mathbb M_2$.
If $\mathbb M_1 \neq \emptyset$, then I need to include those
one-electron valence-excited states which do not couple via x~rays to
core electrons~$\ket{\Phi_{\vec k}^{(m)}}$ for~$m \in \mathbb M_1$.
For~$l_a = l_c + 1$, this gives rise to a
left~$\ket{\Phi\X{L}} = \ket{a ; -l_a}$,
and a right~$\ket{\Phi\X{R}} = \ket{a ; l_a}$ valence state.
Then, the $N/2$~electron ground state of the atom is given
for~$l_a = l_c + 1$ by
\begin{equation}
  \label{eq:fullground}
  \ket{\Phi_0} = \dfrac{1}{\sqrt{\# \mathbb M_a}} \; \ket{\Phi\X{L}} \otimes
    \Bigl[ \bigotimes_{m = \min \mathbb M_2}^{\max \mathbb M_2}
    \ket{a \, c ; m} \Bigr] \otimes \ket{\Phi\X{R}} \; .
\end{equation}
For~$l_a = l_c - 1$, I have~$\mathbb M_1 = \emptyset$ and I omit the
factors~$\ket{\Phi\X{L}}$ and $\ket{\Phi\X{R}}$.
Here $\# \mathbb M_a$~stands for the number of elements in the
set~$\mathbb M_a$.
From the ground state~(\ref{eq:fullground}), I form valence-excited
states~$\ket{\Phi_{\vec k \, c}^{(m)}}$ for~$m \in \mathbb M_2$ by
replacing~$\ket{a \, c ; m}$ [Eq.~(\ref{eq:twoground})]
with~$\ket{\vec k \, c ; m}$ [Eq.~(\ref{eq:twovalence})]
for~$\vec k \in \mathbb R^3$.
For $l_a = l_c + 1$, I additionally have the valence-excited
states~$\ket{\Phi_{\vec k}^{(-l_a)}}$ and $\ket{\Phi_{\vec k}^{(l_a)}}$
by substituting~$\ket{\Phi\X{L}}$ or $\ket{\Phi\X{R}}$ in
Eq.~(\ref{eq:fullground}) by~$\ket{\vec k}$, respectively;
otherwise, for~$l_a = l_c - 1$, the extra factors are left out.
Finally, core-excited states~$\ket{\Phi_{\vec k \, a}^{(m)}}$
for~$m \in \mathbb M_2$ are constructed from Eq.~(\ref{eq:fullground})
by replacing~$\ket{a \, c ; m}$ [Eq.~(\ref{eq:twoground})]
with~$\ket{\vec k \, a ; m}$ [Eq.~(\ref{eq:twocore})]
for~$\vec k \in \mathbb R^3$.

In summary, I apply the three assumptions of Lewenstein~\etal{}
from page~2119 of Ref.~\onlinecite{Lewenstein:HH-94} in a somewhat modified
way where I need to account for the interaction
with x~rays in the parent ion by adding clauses:
\begin{enumerate}
 \item[(a)] The bound one-electron states of the system are the
   valence states~$\ket{a ; m_a}$ with~$m_a \in \mathbb M_a$ and
   the core states~$\ket{c ; m_c}$ with~$m_c \in \mathbb M_2$.
 \item[(b)] Continuum electrons are described as free
   electrons~$\ket{\vec k}$ for~$\vec k \in \mathbb R^3$
   [Eq.~(\ref{eq:freeelmom})] and, consequently, the influence of the
   Coulomb potential of the residual ion is omitted.
 \item[(c)] The system is completely described by the three classes of
   $N/2$~electron states: the atomic ground state~$\ket{\Phi_0}$,
   the valence-excited states~$\ket{\Phi_{\vec k}^{(m_1)}}$
   for~$m_1 \in \mathbb M_1$, $\ket{\Phi_{\vec k \, c}^{(m_2)}}$
   for~$m_2 \in \mathbb M_2$, and the core-excited
   states~$\ket{\Phi_{\vec k \, a}^{(m_2)}}$, which are formed using
   the one-electron states from~(a) and (b), and $\vec k \in \mathbb R^3$.
  \item[(d)] Continuum-continuum transitions are only considered for
    the interaction of a continuum electron with the optical laser
    and do not play a role anywhere else.
  \item[(e)] The coherent interaction of x~rays with the atom---where
    the photon energy is tuned in resonance with the core-valence level
    spacing---is confined to the two-level systems formed
    by~$\ket{\Phi_{\vec k \, c}^{(m)}}$ and $\ket{\Phi_{\vec k \, a}^{(m)}}$
    for~$m \in \mathbb M_2$ and $\vec k \in \mathbb R^3$.
  \item[(f)] The depopulation of the states~$\ket{\Phi_0}$,
    $\ket{\Phi_{\vec k}^{(m_1)}}$ for~$m_1\in \mathbb M_1$,
    $\ket{\Phi_{\vec k \, c}^{(m_2)}}$ and $\ket{\Phi_{\vec k \, a}^{(m_2)}}$
    for~$m_2\in \mathbb M_2$ with~$\vec k \in \mathbb R^3$ by
    optical laser-induced and \xray-induced ionization
    is represented by phenomenological destruction
    rates~$\Gamma^{\,\prime}_i(t)$ for~$i \in \{0, a, c\}$ where
    $\Gamma^{\,\prime}_c(t)$ also accounts for core-hole decay.
    The destruction rates~$\Gamma^{\,\prime}_i(t)$ are assumed to be independent
    of~$m_1$, $m_2$, and $\vec k$.
  \item[(g)] A single HH~photon is emitted in~HHG whereby
    HH~emission due to transitions to the ground state occurs.
    For a complete description, I need the vacuum photon state and
    the single HH~photon number states.
    Propagation of HH~light is along the $x$~axis.
\end{enumerate}
Point~(e) is justified by the fact that x~rays oscillate very rapidly and
cause only a small ponderomotive potential~\cite{Madsen:SF-05,Buth:TA-09}.
Ionization of~$\ket{\Phi_{\vec k}^{(m_1)}}$, $\ket{\Phi_{\vec k \, c}^{(m_2)}}$,
and $\ket{\Phi_{\vec k \, a}^{(m_2)}}$ by the optical laser is much smaller
in point~(f) than for the neutral atom as the ionization potential of the
cation is significantly larger than the ionization potential of the
neutral atom and the tunneling rate depends exponentially on the
ionization potential~\cite{Perelomov:I1-66,Perelomov:I2-66,%
Perelomov:I3-67,Popov:I4-67,Ammosov:TI-86,Delone:MP-00,Yudin:NA-01}.
In point~(f), ionization by the x~rays and the optical laser is
treated incoherently via phenomenological destruction rates in contrast
to the coherent formulation of resonant \xray~absorption in point~(e).

The phenomenological destruction rates of point~(f) are determined
following Ref.~\onlinecite{Buth:KE-13}.
I assume approximately monochromatic x~rays, \ie, the photoabsorption
cross sections~$\sigma_0$, $\sigma_a$, and $\sigma_c$ for one-\xray-photon
absorption by the atom in the ground state, valence-excited and
core-excited states, respectively, can be obtained with
Refs.~\onlinecite{Cowan:TA-81,LANL:AP-00} and are approximately
constant over the bandwidth of the \xray~pulse that has a central
angular frequency~$\omega\X{X}$.
Further, the \xray~intensity is represented by~$I\X{X}(t)$ implying
an \xray~photon flux of~$J\X{X}(t) \approx \tfrac{I\X{X}(t)}{\omega\X{X}}$.
With the tunneling rates~$\Gamma_{\mathrm L, 0}(t)$, $\Gamma_{\mathrm L, a}(t)$,
and $\Gamma_{\mathrm L, c}(t)$ induced by the optical
laser~\cite{Perelomov:I1-66,Perelomov:I2-66,Perelomov:I3-67,Popov:I4-67,%
Ammosov:TI-86,Delone:MP-00,Yudin:NA-01},
the instantaneous phenomenological destruction rates of the atom in the
three classes of basis states are given by
\begin{equation}
  \label{eq:phendec}
  \Gamma^{\,\prime}_i(t) = \Gamma_{\mathrm L, i}(t) + \sigma_i \, J\X{X}(t)
    + \delta_{i\,c} \, \gamma_c \; ,
\end{equation}
for~$i \in \{0, a, c\}$~\cite{Buth:KE-13} with the intrinsic decay
width~$\gamma_c$ from Auger and radiative decay of the core-excited
state~\cite{Als-Nielsen:EM-01,Campbell:WA-01} and $\delta_{i\,c}$~is
the Kronecker-$\delta$~\cite{Arfken:MM-05}.
\textit{Nota bene}, that I omitted the dependence of~$\Gamma_{\mathrm L, i}(t)$
for~$i \in \{a, c\}$ on the magnetic quantum
number~\cite{Perelomov:I1-66,Perelomov:I2-66,Perelomov:I3-67,Popov:I4-67,%
Ammosov:TI-86,Delone:MP-00,Yudin:NA-01}.
This is justified as there is only a single ground state
with~$\Gamma_{\mathrm L, 0}(t)$ and destruction by the optical laser
is strongly suppressed for the parent ion.
For cw~light, I determine time-independent destruction rates by
averaging~$\Gamma^{\,\prime}_i(t)$ from Eq.~(\ref{eq:phendec}) over an optical
laser cycle
\begin{equation}
  \label{eq:phendecave}
  \Gamma_i = \dfrac{1}{T\X{L}} \Int_0^{T\X{L}} \Gamma^{\,\prime}_i(t)
    \differential t^{\,\prime} \; .
\end{equation}

\subsection{Hamiltonian}
\label{sec:hamiltonian}

The total $N/2$-electron Hamiltonian of the atom in two-color light
(optical laser and x~rays) reads
\begin{equation}
  \label{eq:hamiltonian}
  \hat H = \hat H\X{A} + \hat H\X{L} + \hat H\X{X} \; ;
\end{equation}
it consist of three parts: the atomic electronic structure~$\hat H\X{A}$,
the interaction with the optical laser~$\hat H\X{L}$, and the interaction
with the x~rays~$\hat H\X{X}$.

The one-electron Hamiltonian for the atomic electronic structure is
\begin{eqnarray}
  \label{eq:oneatstruc}
  \hat h\X{A} &=& \Sum_{m \in \mathbb M_a} \ket{a ; m} \varepsilon_a \bra{a ; m}
    + \Sum_{m \in \mathbb M_2} \ket{c ; m} \varepsilon_c \bra{c ; m}
    \nonumber \\
  &&{} + \Int_{\mathbb R^3} \ket{\vec k} \dfrac{\vec k^{\,2}}{2}
    \bra{\vec k} \differential^3 k \; ,
\end{eqnarray}
with the energies~$\varepsilon_a$, $\varepsilon_c$,
and $\tfrac{\vec k^{\,2}}{2}$ of the one-electron states.
The $N/2$-electron atomic electronic structure Hamiltonian follows to
\begin{eqnarray}
  \label{eq:twoatstruc}
  \hat H\X{A} &=& \Sum_{i=1}^{N/2} \unitop\X{el}^{i-1} \otimes \hat h\X{A}
    \otimes \unitop\X{el}^{N/2-i} - \imag \, \ket{\Phi_0} \dfrac{\Gamma_0}{2}
    \bra{\Phi_0} \nonumber \\
  &&{} - \imag \, \Sum_{m \in \mathbb M_1} \  \Int_{\mathbb R^3}
    \ket{\Phi_{\vec k}^{(m)}} \dfrac{\Gamma_a}{2}
    \bra{\Phi_{\vec k}^{(m)}} \differential^3 k \\
  &&{} - \imag \, \Sum_{m \in \mathbb M_2} \  \Int_{\mathbb R^3} \Bigl[
    \ket{\Phi_{\vec k \, c}^{(m)}} \dfrac{\Gamma_a}{2}
    \bra{\Phi_{\vec k \, c}^{(m)}} \nonumber \\
  &&{} \quad\qquad  + \ket{\Phi_{\vec k \, a}^{(m)}} \dfrac{\Gamma_c}{2}
    \bra{\Phi_{\vec k \, a}^{(m)}} \Bigr] \differential^3 k \; , \nonumber
\end{eqnarray}
with the unity operator in the one-electron Hilbert space~$\unitop\X{el}$
and the time-independent destruction rates~(\ref{eq:phendecave}).

The interaction of a single electron with the optical laser field is described
in electric dipole approximation in length form~\cite{Scully:QO-97} by
\begin{eqnarray}
  \label{eq:honeL}
  \hat h\X{L} &=& E\X{L}(t) \Sum_{m \in \mathbb M_a} \  \Int_{\mathbb R^3}
      \bigl[ \ket{\vec k} \bra{\vec k} \vec e\X{L} \mul \vec r
      \ket{a ; m} \nonumber \\
    &&\qquad\qquad\qquad{} \times \bra{a ; m} + \textrm{h.c.} \bigr]
      \differential^3 k \\
    &&{} + E\X{L}(t) \Int_{\mathbb R^3} \Int_{\mathbb R^3} \ket{\vec k}
      \bra{\vec k} \vec e\X{L} \mul \vec r \ket{\vec k^{\,\prime}}
      \bra{\vec k^{\,\prime}} \differential^3 k \differential^3 k^{\,\prime}
      \; . \nonumber
\end{eqnarray}
With ``$\textrm{h.c.}$'' I denote the Hermitian conjugate of the preceding
summand and I use the electric field of the optical laser
\begin{equation}
  \label{eq:lasfield}
  E\X{L}(t) = E\X{0L} \, \cos (\omega\X{L} \, t) \; ,
\end{equation}
with the peak amplitude~$E\X{0L}$ and the angular
frequency~$\omega\X{L}$~\cite{Scully:QO-97,Meystre:QO-99,Buth:TX-07}.
The $N/2$-electron Hamiltonian for the interaction with the optical laser reads
\begin{equation}
  \label{eq:twoHL}
  \hat H\X{L} = \Sum_{i=1}^{N/2} \unitop\X{el}^{i-1} \otimes \hat h\X{L}
    \otimes \unitop\X{el}^{N/2-i} \; .
\end{equation}
The optical laser couples only valence electrons to the continuum;
the coupling of core electrons to the continuum by the optical laser
is negligible.
The influence of the optical laser on continuum electrons is represented
by the last term on the right hand side of Eq.~(\ref{eq:honeL}).

The interaction of a single electron with the \xray~radiation is described
in electric dipole approximation in length form~\cite{Scully:QO-97} via
\begin{equation}
  \label{eq:honeX}
  \hat h\X{X} = E\X{X}(t) \Sum_{m \in \mathbb M_2} \bigl[ \ket{c ; m}
    \bra{c ; m} \vec e\X{X} \mul \vec r \ket{a ; m} \bra{a ; m} + \textrm{h.c.}
    \bigr] \; ,
\end{equation}
for the \xray~electric field
\begin{equation}
  \label{eq:xrayfield}
  E\X{X}(t) = E\X{0X} \, \cos (\omega\X{X} \, t) = \dfrac{E\X{0X}}{2} \,
    [ \euler^{\imag \, \omega\X{X} \, t} + \euler^{-\imag \, \omega\X{X} \, t}
    ] \; ,
\end{equation}
with the peak amplitude~$E\X{0X}$ and the angular
frequency~$\omega\X{X}$~\cite{Scully:QO-97,Meystre:QO-99,Buth:TX-07}.
The $N/2$-electron Hamiltonian for the \xray~interaction follows
similarly as in Eq.~(\ref{eq:twoHL}) to
\begin{equation}
  \label{eq:htwoX}
  \hat H\X{X} = \Sum_{i=1}^{N/2} \unitop\X{el}^{i-1} \otimes \hat h\X{X}
    \otimes \unitop\X{el}^{N/2-i} \; .
\end{equation}

\subsection{Equations of motion}
\label{sec:EOMs}

I make the following ansatz for the $N/2$-electron wave packet
\begin{eqnarray}
  \label{eq:wavepacket}
  \ket{\Psi, t} &=& a(t) \, \euler^{-\imag \, E_0 \, t} \ket{\Phi_0}
    \nonumber \\
  &&{} + \Sum_{m \in \mathbb M_1} \  \Int_{\mathbb R^3}
    b^{(m)}_a(\vec k,t) \> \euler^{-\imag \, E_0 \, t}
    \ket{\Phi_{\vec k}^{(m)}} \differential^3 k \\
  &&{} + \Sum_{m \in \mathbb M_2} \  \Int_{\mathbb R^3} \bigl[
    b^{(m)}_a(\vec k,t) \> \euler^{-\imag \, E_0 \, t}
    \ket{\Phi_{\vec k \, c}^{(m)}} \nonumber \\
  &&\quad\qquad{} + b^{(m)}_c(\vec k,t) \> \euler^{-\imag \,
    (E_0 + \omega\X{X}) \, t} \ket{\Phi_{\vec k \, a}^{(m)}} \bigr]
    \differential^3 k \; . \nonumber
\end{eqnarray}
The subscripts ``$a$'' and ``$c$'' on the amplitudes~$b^{(m)}_a(\vec k, t)$
and $b^{(m)}_c(\vec k, t)$, respectively, indicate which one-electron state
contains the hole.
The ground-state energy of the atom is~$E_0 = \bra{\Phi_0} \hat H\X{A}
\ket{\Phi_0} + \imag \, \tfrac{\Gamma_0}{2} = \# \mathbb M_a \, \varepsilon_a
+ \# \mathbb M_2 \, \varepsilon_c$.
The phase factors in the ansatz~(\ref{eq:wavepacket}) are chosen such
that the equations of motion~(EOMs), which are derived in the following,
are simplified the most, that fast oscillations are separated, and thus
the expansion coefficients~$a(t)$, $b^{(m)}_a(\vec k, t)$, and
$b^{(m)}_c(\vec k, t)$ vary comparatively slowly.
I insert~$\ket{\Psi, t}$ from Eq.~(\ref{eq:wavepacket}) into the
time-dependent Schr\"odinger equation~\cite{Merzbacher:QM-98} yielding
\begin{equation}
  \label{eq:tdschroedi}
  \imag \> \dfrac{\partial}{\partial t} \ket{\Psi, t} = \hat H \ket{\Psi, t}
    \; ,
\end{equation}
with the Hamiltonian from Eq.~(\ref{eq:hamiltonian}).
I project onto the three classes of basis states from
Sect.~\ref{sec:wavefunction} which yields EOMs for the
expansion coefficients in Eq.~(\ref{eq:wavepacket}).

First, I obtain the EOM for the ground-state amplitude by projecting
onto~$\bra{\Phi_0}$ giving
\begin{eqnarray}
  \label{eq:groundepl}
  \dfrac{\differential}{\differential t} \, a(t) &=& -\dfrac{\Gamma_0}{2}
    \, a(t) \\
  &&{} - \imag \, \Sum_{m \in \mathbb M_a} \  \Int_{\mathbb R^3}
    \bra{a ; m} \hat h\X{L} \ket{\vec k} \, b_a^{(m)}(\vec k, t)
    \differential^3 k \; . \nonumber
\end{eqnarray}
The second term on the right-hand side is the rate of change of the
ground-state amplitude due to tunnel ionization of a valence electron
induced by the optical laser and recombination of a continuum electron
with the valence vacancy.
As the efficiency of HHG is low, tunnel ionization is the only noticeable
contribution;
it can be neglected entirely for low ground-state depopulation by the
optical laser~\cite{Lewenstein:HH-94}.
For high depopulation, the influence of the second term on the right-hand
side of Eq.~(\ref{eq:groundepl}) is included in~$\Gamma_0$ as
in Eqs.~(\ref{eq:phendec}) and (\ref{eq:phendecave}).
Consequently, the term is dropped in
Eq.~(\ref{eq:groundepl})~\cite{Lewenstein:HH-94,Buth:KE-13}.

The Eq.~(\ref{eq:groundepl}) becomes identical to Eq.~(49) of
Ref.~\onlinecite{Lewenstein:HH-94} upon
dropping the first term on the right-hand side, setting~$M_a = \{0\}$
and replacing~$\bra{a ; m} \hat h\X{L}
\ket{\vec k} \to -E \cos(t) \, d^{\,*}_x(\bfvec v)$,
$b_a^{(m)}(\vec k, t) \to b(\bfvec v, t)$, and
$\vec k \to \bfvec v$.

Second, projecting onto~$\bra{\Phi_{\vec k \, c}^{(m)}}$
for~$\vec k \in \mathbb R^3$ and $m \in \mathbb M_2$
[Eq.~(\ref{eq:twovalence})] gives the EOMs for a continuum electron and
a valence hole where the valence hole is coupled to core electrons:
\begin{eqnarray}
  \label{eq:revalence}
  \dfrac{\partial}{\partial t} \, b_a^{(m)}(\vec k, t) &=&
    - \frac{\imag}{2} \, (\vec k^{\,2} - \delta + 2 \, \Cal I_p
    - \imag \, \Gamma_a) \, b_a^{(m)}(\vec k, t) \nonumber \\
  &&{} - \imag \, \Int_{\mathbb R^3} \bra{\vec k} \hat h\X{L}
    \ket{\vec k^{\,\prime}} \, b_a^{(m)}(\vec k^{\,\prime}, t)
    \differential^3 k^{\,\prime} \nonumber \\
  &&{} - \imag \bra{c ; m} \hat h\X{X,RWA} \ket{a ; m} \, b_c^{(m)}(\vec k, t)
    \nonumber \\
  &&{} - \imag \, \bra{\vec k} \hat h\X{L} \ket{a ; m} \, a(t) \; .
\end{eqnarray}
The first term on the right-hand side of Eq.~(\ref{eq:revalence})
specifies the phase of~$b_a^{(m)}(\vec k, t)$.
Following Refs.~\onlinecite{Buth:NL-11,Kohler:EC-12}, I introduce the energy
\begin{equation}
  \label{eq:quasiIP}
  \Cal I\X{P} = -\dfrac{1}{2} \> (\varepsilon_a + \varepsilon_c + \omega_x)
    = -\varepsilon_a + \dfrac{\delta}{2}
\end{equation}
and the detuning~\cite{Meystre:QO-99} of the \xray~photon energy from
the energy difference between the core and valence levels
\begin{equation}
  \delta = \varepsilon_a - \varepsilon_c - \omega\X{X} \; .
\end{equation}
For zero detuning, $\Cal I\X{P}$~corresponds to the valence ionization
potential of the atom, but $\Cal I\X{P}$~is shifted to higher (lower) values,
if $\omega\X{X}$ is smaller (larger) than the level spacing.
The second term on the right-hand side of Eq.~(\ref{eq:revalence})
describes optical-laser-mediated continuum-continuum transitions;
the third term stands for resonant excitation by the x~rays,
where I make the rotating wave approximation~$\hat h\X{X} \,
\euler^{-\imag \, \omega\X{X} \, t} \approx
\hat h\X{X,RWA}$~\cite{Scully:QO-97,Meystre:QO-99};
and the fourth term induces tunnel ionization.

Similarly, projecting onto~$\bra{\Phi_{\vec k}^{(m)}}$
for~$\vec k \in \mathbb R^3$ and $m \in \mathbb M_1$
gives the EOMs for the recombination of the continuum electron
when there is no coupling to core electrons.
The resulting EOMs are the same as Eq.~(\ref{eq:revalence})
apart from the term~$- \imag \, b_c^{(m)}(\vec k, t) \bra{c ; m}
\hat h\X{X,RWA} \ket{a ; m}$ which is missing in this case.
These EOMs contribute only to optical-laser HHG with
valence electrons.

Third, projecting onto~$\bra{\Phi_{\vec k \, a}^{(m)}}$
for~$\vec k \in \mathbb R^3$ and $m \in \mathbb M_2$
provides the EOMs for a continuum electron and a core hole
\begin{eqnarray}
  \label{eq:recore}
  \dfrac{\partial}{\partial t} \, b_c^{(m)}(\vec k, t) &=&
    - \frac{\imag}{2} \, (\vec k^{\,2} + \delta
    + 2 \, \Cal I_p - \imag \, \Gamma_c) \, b_c^{(m)}(\vec k,t) \nonumber \\
  &&{} - \imag \, \Int_{\mathbb R^3} \bra{\vec k} \hat h\X{L}
    \ket{\vec k^{\,\prime}} \, b_c^{(m)}(\vec k^{\,\prime}, t)
    \differential^3 k^{\,\prime} \\
  &&{} - \imag \bra{a ; m} \hat h\X{X,RWA} \ket{c ; m} \,
    b_a^{(m)}(\vec k, t) \; . \nonumber
\end{eqnarray}
The tunnel ionization rate of core electrons is vanishingly small
such that the last term on the right-hand side of Eq.~(\ref{eq:revalence})
is missing here.

In order to evaluate the matrix elements in the EOMs that involve
continuum electrons~(\ref{eq:revalence}), (\ref{eq:recore}), I use
the spatial part of the free-electron wave function of
Eq.~(\ref{eq:freeelmom}).
Specifically, I obtain from Eqs.~(\ref{eq:honeL}) and (\ref{eq:lasfield})
the matrix element that is responsible for tunnel ionization of the valence
state
\begin{equation}
  \label{eq:tunnelmatel}
  \bra{\vec k} \hat h\X{L} \ket{a ; m} = E\X{L}(t) \,
    \wp^{(m)}_{\mathrm L, \vec k \, a} \; ,
\end{equation}
with~$\wp^{(m)}_{\mathrm L, \vec k \, a} = \bra{\vec k} \vec e\X{L}
\mul \vec r \ket{a ; m}$ [Appendix~\ref{sec:BCtrans}].
Optical laser-induced continuum-continuum transitions~\cite{Lewenstein:HH-94}
are described by the matrix element [Appendix~\ref{sec:CCtrans}]:
\begin{equation}
  \label{eq:CCtrans}
  \bra{\vec k} \hat h\X{L} \ket{\vec k^{\,\prime}} = E\X{L}(t) \, (-\imag)
    \, \dfrac{\partial}{\partial k^{\,\prime}_z} \,
    \delta^3(\vec k^{\,\prime} - \vec k) \; .
\end{equation}
Further, \xray~transitions in Eqs.~(\ref{eq:revalence}) and (\ref{eq:recore})
are described by
\begin{equation}
  \label{eq:xrayexcite}
  \bra{c ; m} \hat h\X{X, RWA} \ket{a ; m} = \dfrac{E\X{0X}}{2}
    \, \wp^{(m)}_{\mathrm X, c \, a} = \dfrac{R^{(m)}\X{0X}}{2}
\end{equation}
using Eqs.~(\ref{eq:honeX}) and (\ref{eq:xrayfield})
with~$\wp^{(m)}_{\mathrm X, c \, a} = \bra{c ; m} \vec e\X{X} \mul \vec r
\ket{a ; m}$ [Appendix~\ref{sec:BBtrans}] and defining the real Rabi
frequency~$R^{(m)}\X{0X}$~\cite{Scully:QO-97,Meystre:QO-99}.
I choose the spatial matrix element~$\wp^{(m)}_{\mathrm X, c \, a}$
and thus~$R^{(m)}\X{0X}$ to be real which is feasible as the atomic potential
is approximated by a central potential~\cite{Herman:AS-63,Buth:TX-07}.

I rewrite Eqs.~(\ref{eq:revalence}) and (\ref{eq:recore}) as a vector
equation for~$\vec k \in \mathbb R^3$ and~$m \in \mathbb M_2$ defining
the continuum-electron amplitudes
\begin{equation}
  \label{eq:contelamp}
  \vec b^{\,(m)}(\vec k, t) \equiv \bigl( b^{(m)}_a(\vec k, t),
    b^{(m)}_c(\vec k, t) \bigr)\transpose
\end{equation}
and the complex symmetric Rabi matrix
\begin{equation}
  \label{eq:RabiMat}
  \mat R^{(m)} = \left( \begin{matrix}
    -\delta - \imag \, \Gamma_a & R^{(m)}\X{0X} \\
    R^{(m)}\X{0X}               & \delta - \imag \, \Gamma_c
  \end{matrix} \right) \; .
\end{equation}
This yields, with Eqs.~(\ref{eq:CCtrans}) and (\ref{eq:xrayexcite}), the
combined EOMs for a continuum electron with either a valence hole or
a core hole
\begin{eqnarray}
  \label{eq:coupledAmps}
  \dfrac{\partial}{\partial t} \, \vec b^{\,(m)}(\vec k, t)
    &=& -\dfrac{\imag}{2} \, \bigl( \mat R^{(m)} + (\vec k^{\,2}
    + 2 \, \Cal I\X{P}) \, \unitmatrix \bigr) \, \vec b^{\,(m)}(\vec k, t)
    \nonumber \\
  &&{} + E\X{L}(t) \, \dfrac{\partial}{\partial k_z}
    \, \vec b^{\,(m)}(\vec k, t) \\
  &&{} - \imag \, E\X{L}(t) \, \wp^{(m)}_{\mathrm L,
    \vec k \, a} \, \binom{1}{0} \, a(t) \; . \nonumber
\end{eqnarray}

The matrix of eigenvalues of~$\mat R^{(m)}$ is
\begin{equation}
  \label{eq:mateigvals}
  \mat \Lambda^{(m)} = \diag(\lambda^{(m)}_+, \lambda^{(m)}_-)
\end{equation}
with the eigenvalues~\cite{SuppData}:
\begin{equation}
  \label{eq:RabiEigval}
  \lambda^{(m)}_{\pm} = -\dfrac{\imag}{2} \, (\Gamma_a + \Gamma_c)
    \pm \mu^{(m)} \; ,
\end{equation}
and the complex Rabi frequency~\cite{Meystre:QO-99} given by
\begin{equation}
  \mu^{(m)} = \sqrt{\bigl[ \delta + \frac{\imag}{2} \, (\Gamma_a - \Gamma_c)
    \bigr]^2 + \bigl( R^{(m)}\X{0X} \bigr)^2} \; .
\end{equation}
The matrix of eigenvectors of~$\mat R^{(m)}$ is denoted by~$\mat U^{(m)}$.
With the eigenvectors~$\mat U^{(m)}$ and eigenvalues~$\mat \Lambda^{(m)}$,
I transform to the eigenbasis of~$\mat R^{(m)}$, the \xray-dressed
states~\cite{Meystre:QO-99}.
With Eq.~(\ref{eq:contelamp}), I find the new amplitudes
\begin{equation}
  \label{eq:decoupledAmps}
  \vec {\mathfrak b}^{(m)}(\vec k, t)
    \equiv \binom{{\mathfrak b}^{(m)}_+(\vec k, t)}
    {{\mathfrak b}^{(m)}_-(\vec k, t)}
    = \bigl( \mat U^{(m)} \bigr)^{-1} \, \vec b^{\,(m)}(\vec k, t)
\end{equation}
and define the valence ionization fraction by
\begin{equation}
  \label{eq:valionfrac}
  \vec w^{(m)} \equiv \binom{w^{(m)}_+}{w^{(m)}_-}
    = \bigl( \mat U^{(m)} \bigr)^{-1} \, \binom{1}{0} \; .
\end{equation}
Next, I recast Eq.~(\ref{eq:coupledAmps}) in terms of \xray-dressed states
with Eqs.~(\ref{eq:mateigvals}), (\ref{eq:decoupledAmps}),
(\ref{eq:valionfrac}), leading to
\begin{eqnarray}
  \label{eq:decoupledEOM}
  \dfrac{\partial}{\partial t} \, \vec {\mathfrak b}^{(m)}(\vec k, t)
    &=& -\dfrac{\imag}{2} \, \bigl( \mat \Lambda^{(m)} + (\vec k^{\,2} + 2
    \, \Cal I\X{P})\, \unitmatrix \bigr) \, \vec {\mathfrak b}^{(m)}(\vec k, t)
    \nonumber \\
  &&{} + E\X{L}(t) \, \dfrac{\partial}{\partial k_z} \,
    \vec {\mathfrak b}^{(m)}(\vec k, t) \\
  &&{} - \imag \, E\X{L}(t) \, \wp^{(m)}_{\mathrm L, \vec k \, a}
    \, \vec w^{(m)} \, a(t) \; , \nonumber
\end{eqnarray}
where I exploit that $\mat U^{(m)}$~neither depends on~$\vec k$ nor on~$t$ and
thus commutes with derivatives with respect to both variables.

Equation~(\ref{eq:decoupledEOM}) is a system of two decoupled partial
differential equations.
To transform it into two ordinary differential equations, I
introduce the vector potential~\cite{Madsen:SF-05,Buth:TA-09}
of the optical laser
\begin{eqnarray}
  \label{eq:lasvector}
  \vec A\X{L}(t) &=& -\Lim_{\eta \to 0^+} \Int_{-\infty}^t E\X{L}(t^{\,\prime})
    \, \vec e\X{L} \, \euler^{- \eta \, |t^{\,\prime}|} \differential
    t^{\,\prime} \\
  &=&{} - \dfrac{E\X{0L}}{\omega\X{L}} \, \sin (\omega\X{L} \, t)
      \> \vec e\X{L} \; , \nonumber
\end{eqnarray}
which is determined from its electric field~(\ref{eq:lasfield}).
The factor~$\euler^{-\imag \, \eta \, |t^{\,\prime}|}$ with~$\eta > 0$, thereby,
ensures convergence of the integral~\footnote{%
Equation~(\ref{eq:lasvector}) ist the only expression of this article
where the factor~$\euler^{- \eta \, |t^{\,\prime}|}$ is required.
Thus I took the limit~$\eta \to 0^+$ in all other equations
and, specifically, the factor is omitted from Eq.~(\ref{eq:lasfield}).}.
As the kinetic momentum of the continuum electron~$\vec k$ at time~$t$
in Eq.~(\ref{eq:decoupledEOM}) can be expressed at time~$t^{\,\prime}$
by~$\vec k^{\,\prime} = \vec k - \vec A\X{L}(t) +
\vec A\X{L}(t^{\,\prime})$~\cite{Lewenstein:HH-94}, the sum of the left hand
side of Eq.~(\ref{eq:decoupledEOM}) and the negative of the summand on the right
hand side that contains the derivative with respect to~$k^{\,\prime}_z$
represents a total time derivative of~$\vec {\mathfrak b}(\vec k^{\,\prime},
t^{\,\prime})$ with respect to~$t^{\,\prime}$ and I have
\begin{eqnarray}
  \label{eq:betasubstitution}
  \dfrac{\differential}{\differential t^{\,\prime}} \, \vec{\mathfrak
    b}^{(m)}(\vec k^{\,\prime}, t^{\,\prime}) &=& -\dfrac{\imag}{2} \, \bigl[
    \mat \Lambda^{(m)} + 2 \, \Cal I\X{P} \, \unitmatrix \nonumber \\
  &&\qquad{} + \bigl(\vec k - \vec A\X{L}(t) + \vec A\X{L}(t^{\,\prime})
    \bigr)^2 \, \unitmatrix \bigr] \nonumber \\
  &&\qquad{} \times \vec{\mathfrak b}^{(m)} (\vec k^{\,\prime}, t^{\,\prime}) \\
  &&{} - \imag \, E\X{L}(t^{\,\prime}) \, \wp^{(m)}_{\mathrm L, \vec k
    - \vec A\X{L}(t) + \vec A\X{L}(t^{\,\prime}) \, a} \nonumber \\
  &&\qquad{} \times \vec w^{(m)} \, a(t^{\,\prime}) \; . \nonumber
\end{eqnarray}
This is a system of two ordinary first-order differential equations which
can be integrated exactly~\cite{Arfken:MM-05} yielding
\begin{eqnarray}
  \label{eq:betasolution}
  {\mathfrak b}^{(m)}_{\pm}(\vec k, t) &=& -\imag \, w^{(m)}_{\pm}
    \Int_{0}^t E\X{L}(t^{\,\prime}) \> \wp^{(m)}_{\mathrm L, \vec k
    - \vec A\X{L}(t) + \vec A\X{L}(t^{\,\prime}) \, a} \nonumber \\
  &&{} \times \euler^{-\tfrac{\imag}{2} \, \Int_{t^{\,\prime}}^t (\vec k
    - \vec A\X{L}(t) + \vec A\X{L}(t^{\,\prime\prime}))^2 \differential
    t^{\,\prime\prime}} \\
  &&{} \times \euler^{- \imag\, {\textstyle(}
    \frac{\lambda^{(m)}_{\pm}}{2} + \Cal I\X{P} {\textstyle)} \,
    (t-t^{\,\prime})} \> a(t^{\,\prime}) \differential t^{\,\prime}
    \; , \nonumber
\end{eqnarray}
by letting~$t^{\,\prime} = t$ in the solution of Eq.~(\ref{eq:betasubstitution})
because then $\vec k^{\,\prime} = \vec k$ and thus~${\mathfrak b}_{\pm}
(\vec k, t) = {\mathfrak b}_{\pm}(\vec k^{\,\prime}, t^{\,\prime})$.
Next, I transform back to the amplitudes of the bare
states~(\ref{eq:decoupledAmps}) via~$\vec b^{\,(m)}(\vec k, t)
= \mat U^{(m)} \, \vec {\mathfrak b}^{(m)}(\vec  k, t)$
providing the solution of Eq.~(\ref{eq:coupledAmps}).

The EOMs for~$m \in \mathbb M_1$ follow from Eq.~(\ref{eq:betasolution})
by the replacements~${\mathfrak b}^{(m)}_{\pm}(\vec k, t) \to
b^{(m)}_a(\vec k,t)$, $w^{(m)}_{\pm} \to 1$, and $\lambda^{(m)}_{\pm} \to
- \delta - \imag \, \Gamma_a$.

To transform Eq.~(\ref{eq:betasolution}) into Eq.~(5) of
Lewenstein~\etal~\cite{Lewenstein:HH-94}, I make the
replacements~${\mathfrak b}^{(m)}_{\pm}(\vec k, t) \to b(\bfvec v,t)$,
$-\imag \, w^{(m)}_{\pm} \to \imag$, $\wp^{(m)}_{\mathrm L, \vec k
- \vec A\X{L}(t) + \vec A\X{L}(t^{\,\prime}) \, a} \to
d_x(\bfvec v + \bfvec A(t) - \bfvec A(t^{\,\prime}))$,
$\vec A\X{L} \to - \mat A$, $E\X{L}(t^{\,\prime}) \to E \cos t^{\,\prime}$,
and $\tfrac{\lambda^{(m)}_{\pm}}{2} + \Cal I\X{P} \to I\X{P}$.

\subsection{Electric dipole transition matrix element}
\label{sec:HHGtrama}

The emission of HH~light in electric dipole approximation is governed
by the position operator~$\vec r$ projected onto
the polarization vector of the emitted HH~light~$\vec e\X{H}$, \ie,
the one-electron dipole operator is~$\hat d_1 = \vec e\X{H} \mul
\vec r$~\cite{Scully:QO-97,Meystre:QO-99}.
Furthermore, the two-electron dipole operator is~$\hat D_2 = \hat d_1 \otimes
\unitop\X{el} + \unitop\X{el} \otimes \hat d_1$.
With this, the $N/2$-electron dipole operator,
in terms of the basis states from Sect.~\ref{sec:wavefunction},
follows to
\begin{eqnarray}
  \label{eq:Nhalfdipole}
  \hat D &=& \Sum_{m \in \mathbb M_1} \  \Int_{\mathbb R^3} \ket{\Phi_0}
    \bra{a ; m} \hat d_1 \ket{\vec k} \bra{\Phi_{\vec k}^{(m)}}
    \differential^3 k + \mathrm{h.c.} \nonumber \\
  &&{} + \Sum_{m \in \mathbb M_2} \  \Int_{\mathbb R^3} \bigl[
    \ket{\Phi_0} \bra{a \, c ; m} \hat D_2 \ket{\vec k \, c ; m}
    \bra{\Phi_{\vec k \, c}^{(m)}} \\
  &&{} + \ket{\Phi_0} \bra{a \, c ; m} \hat D_2 \ket{\vec k \, a ; m}
    \bra{\Phi_{\vec k \, a}^{(m)}} \bigr] \differential^3 k + \mathrm{h.c.}
    \; . \nonumber
\end{eqnarray}
The operator~$\hat D$ in Eq.~(\ref{eq:Nhalfdipole}) is defined such that
it describes the recombination of a continuum electron with valence and
core holes~\footnote{%
I omit terms for continuum-continuum transitions, \ie, terms
involving one-electron matrix elements of the type~$\bra{\vec k} \hat d_1
\ket{\vec k^{\,\prime}}$ for~$\vec k, \vec k^{\,\prime} \in \mathbb R^3$
[Appendix~\ref{sec:CCtrans}] because such terms do not play a role for
the description of HH~emission in the electric dipole transition matrix
element in Eq.~(\ref{eq:Nhalfdipole})~\cite{Lewenstein:HH-94,Kuchiev:QT-99}.}.
For~$\vec k \in \mathbb R^3$, the one-electron dipole matrix elements
are~$\wp^{(m)}_{\mathrm H, a \, \vec k} = \bra{a ; m} \hat d_1 \ket{\vec k}$
with~$m \in \mathbb M_1$ whereas the two-electron dipole matrix elements
are~$\wp^{(m)}_{\mathrm H, a \, \vec k} = \bra{a \, c ; m}
\hat D_2 \ket{\vec k \, c ; m}$ and $\wp^{(m)}_{\mathrm H, c \, \vec k}
= -\bra{a \, c ; m} \hat D_2 \ket{\vec k \, a ; m}$ with~$m \in \mathbb M_2$.

High-order harmonic emission is determined by the time-depended dipole
transition matrix element between the atomic ground state and the
valence-excited and core-excited states, respectively.
Namely, it is expressed by $\Cal D(t) = \bra{\Psi_0, t} \hat D
\ket{\Psi\X{c}, t}$ where $\ket{\Psi_0, t}$~is the ground-state part of the
wave packet~(\ref{eq:wavepacket}) at time~$t$ and
$\ket{\Psi\X{c}, t}$~is the continuum part~\cite{Kuchiev:QT-99}.
Then, the dipole transition matrix element
follows from Eqs.~(\ref{eq:wavepacket}) and (\ref{eq:Nhalfdipole}) to
\begin{eqnarray}
  \label{eq:diptransmat}
  \Cal D(t) &=& a_0^*(t) \, \Int_{\mathbb R^3} \Bigl[ \Sum_{m \in \mathbb M_a}
       \wp^{(m)}_{\mathrm H, a \, \vec k} \; b^{(m)}_a(\vec k,t) \\
     &&\qquad{} - \Sum_{m \in \mathbb M_2} \wp^{(m)}_{\mathrm H, c \, \vec k} \;
       \euler^{-\imag \, \omega\X{X} \, t} \;  b^{(m)}_c(\vec k, t) \Bigr]
       \differential^3 k \nonumber \\
     &=& \Sum_{m \in \mathbb M_1} \mathfrak d^{(m)}(t) +
       \Sum_{m \in \mathbb M_2} \Sum_{\atopa{\scriptstyle i \in \{a, c\}}
       {\scriptstyle j \in \{+, -\}}} \mathfrak D^{(m)}_{ij}(t)
       \; . \nonumber
\end{eqnarray}
The dipole components, with~$m \in \mathbb M_2$, for the valence- and
the core-excited states, $i \in \{ a, c\}$, and the \xray-dressed
states, $j \in \{+, -\}$, are introduced by
\begin{eqnarray}
  \label{eq:dipcomp}
  \mathfrak D^{(m)}_{ij}(t) &=& (-1)^{\delta_{i\,c}} \, U^{(m)}_{ij} \>
    \euler^{-\imag \, \delta_{i\,c} \, \omega\X{X} \, t} \, a_0^*(t)
    \nonumber \\
  &&{} \times \Int_{\mathbb R^3} \wp^{(m)}_{\mathrm H, i \, \vec k} \;
    {\mathfrak b}^{(m)}_j(\vec k,t) \differential^3 k \\
  &=& (-1)^{\delta_{i\,c}} \, (-\imag) \> U^{(m)}_{ij} \, w^{(m)}_{j}
    \> \euler^{-\imag \, \delta_{i\,c} \, \omega\X{X} \, t} \, a^*(t)
    \nonumber \\
  &&{} \times \Int_0^t a(t^{\,\prime}) \, E\X{L}(t^{\,\prime}) \,
    \Int_{\mathbb R^3} \wp^{(m) \> *}_{\mathrm H, \vec p + \vec A\X{L}(t) \, i}
    \> \wp^{(m)}_{\mathrm L, \vec p + \vec A\X{L}(t^{\,\prime}) \, a}
    \nonumber \\
  &&{} \times \euler^{-\imag \, S^{(m)}_j(\vec p, t, t^{\,\prime})}
    \differential^3 p \, \differential t^{\,\prime} \; . \nonumber
\end{eqnarray}
Here I inserted the amplitudes of Eq.~(\ref{eq:betasolution})
transformed from the \xray-dressed-state basis back to the
bare-state basis via~$U^{(m)}_{ij}$.
The Kronecker~$\delta_{i\,c}$~\cite{Arfken:MM-05} is used to insert
the phase factor~$\euler^{-\imag \, \omega\X{X} \, t}$ for the terms
involving core-excited states.
I also introduced the canonical momentum~\footnote{%
As the electron propagates freely in SFA in the continuum with kinetic
momentum~$\vec k$, the canonical momentum~$\vec p$ at time~$t$ and
at time~$t^{\,\prime}$ are the same~\cite{Lewenstein:HH-94}.}%
~$\vec p = \vec k - \vec A\X{L}(t)$ and the quasiclassical action
\begin{equation}
  \label{eq:action}
  \textstyle S^{(m)}_j(\vec p, t, t^{\,\prime}) = \tfrac{1}{2}
    \Int_{t^{\,\prime}}^t \bigl( \vec p + \vec A\X{L}(t^{\,\prime\prime})
    \bigr)^2 \differential t^{\,\prime\prime} + \bigl(
    \tfrac{\lambda^{(m)}_{j}}{2} + \Cal I\X{P} \bigr) \, (t-t^{\,\prime}) \; .
\end{equation}
Equation~(\ref{eq:dipcomp}) can be understood term by
term~\cite{Lewenstein:HH-94}.
Namely, $w^{(m)}_j \> \wp^{(m)}_{\mathrm L, \vec p + \vec A\X{L}(t^{\,\prime})
\, a} \, E\X{L}(t^{\,\prime})$~determines valence ionization at
time~$t^{\,\prime}$ via tunneling in the optical laser field.
During propagation of the free electron from time~$t^{\,\prime}$ to~$t$
in the optical laser field, the electron acquires the
phase~$\euler^{-\imag \, S^{(m)}_j(\vec p, t, t^{\,\prime})}$.
The emission of HH~light at time~$t$ due to recombination of the free
electron with the state~$i \in \{ a, c\}$ is governed
by~$\wp^{(m) \, *}_{\mathrm H, \vec p + \vec A\X{L}(t) \, i}$;
the shift of the HH~spectrum by~$\omega\X{X}$ for a recombination
with a core hole is given by~$\euler^{-\imag \, \delta_{i\,c}
\, \omega\X{X} \, t}$.
The quasiclassical action~(\ref{eq:action}) differs from the optical laser-only
case~\cite{Lewenstein:HH-94} by the summand~$\bigl( \tfrac{\lambda^{(m)}_j}{2}
+ \Cal I\X{P} - I\X{P} \bigr) \, (t-t^{\,\prime})$ which represents the
energy splitting of the core and valence states due to \xray~dressing.

Equation~(\ref{eq:action}) is the same as Eq.~(9) in
Ref.~\onlinecite{Lewenstein:HH-94} after
replacing~$\vec p + \vec A\X{L}(t^{\,\prime\prime}) \to \bfvec p
- \bfvec A(t^{\,\prime\prime})$ and
$\tfrac{\lambda^{(m)}_{j}}{2} + \Cal I\X{P} \to I\X{P}$.

To calculate~$\Cal D(t)$ [Eq.~(\ref{eq:diptransmat})], I make the
saddle-point approximation [Appendix~\ref{sec:saddlepoint}] for
the integration over the canonical momentum~$\vec p$ in Eq.~(\ref{eq:dipcomp}).
The stationary points~\footnote{%
I use only the classical part of~$S^{(m)}_j(\vec p, t, t^{\,\prime})$ for
the saddle-point approximation, \ie, I let~$\Cal S(\vec p)
= \tfrac{1}{2} \Int_{t^{\,\prime}}^t \bigl(\vec p + \vec
A\X{L}(t^{\,\prime\prime}) \bigr)^2 \differential t^{\,\prime\prime}$
in Eq.~(\ref{eq:integralgS}) and include the remaining factors in~$f(\vec p)$.
For~$S^{(m)}_j(\vec p, t, t^{\,\prime})$, the series~(\ref{eq:TaylorS})
actually terminates after the second-order term and thus is convergent
in the entire~$\mathbb R^3$.}
of~$S^{(m)}_j(\vec p, t, t^{\,\prime})$ [Eq.~(\ref{eq:action})] are
determined from~$\vec \nabla_p \> S^{(m)}_j(\vec p, t, t^{\,\prime}) = \vec 0
= \vec s(t) - \vec s(t^{\,\prime})$ which is the difference between the
positions of the electron at times~$t$ and~$t^{\,\prime}$ from its classical
trajectory~$\vec s(t) = \vec p \> t + \tfrac{E\X{0L}}{\omega\X{L}^2}
\, \vec e\X{L} \cos(\omega\X{L} \, t) + \vec s_0$ with
Eq.~(\ref{eq:lasfield}), \ie, the electron is liberated
at~$t^{\,\prime}$ and returns to the parent ion at~$t$ where~$\vec s_0$
is the initial position close to the parent ion in the
origin~\cite{Lewenstein:HH-94}.
This finding suggests to introduce the excursion time~$\tau = t - t^{\,\prime}$
of the electron in the continuum.
The matrix elements of the Hessian [Appendix~\ref{sec:saddlepoint}]
are~$\tfrac{\partial^2 S^{(m)}_j(\vec p, t, t - \tau)}
{\partial p_{i'} \, \partial p_{j'}} = \tau \, \delta_{i'\,j'}$ and its
determinant is~$\tau^3$.
At the stationary points of~$S^{(m)}_j(\vec p, t, t^{\,\prime})$, the canonical
momentum satisfies
\begin{equation}
  \label{eq:pstat}
  \vec p\X{st}(t, \tau) = -\dfrac{E\X{0L}}{\omega\X{L} \, \tau} \,
    \vec e\X{L} \, \bigl[ \cos (\omega\X{L} \, t) - \cos \bigl(\omega\X{L} \,
    (t - \tau) \bigr) \bigr] \; ,
\end{equation}
and the quasiclassical action~(\ref{eq:action})~\cite{SuppData} is
\begin{eqnarray}
  \label{eq:actionstat}
  S^{(m)}_{\mathrm{st}, j}(t, \tau) &=& \Bigl(
    \dfrac{\lambda^{(m)}_j}{2} + \Cal I\X{P} + U\X{P} \Bigr) \, \tau
    \nonumber \\
  &&{} - 2 \, \dfrac{U\X{P}}{\omega\X{L}^2 \tau} \, \bigl(1 -
    \cos (\omega\X{L} \, \tau) \bigr) \\
  &&{} - \dfrac{U\X{P}}{\omega\X{L}} \, C(\tau) \,
    \cos \bigl(\omega\X{L} \, (2 \, t - \tau) \bigr) \; , \nonumber
\end{eqnarray}
with the ponderomotive potential~\cite{Madsen:SF-05,Buth:TA-09} of the
optical laser~$U\X{P} = \tfrac{E\X{0L}^2}{4 \, \omega\X{L}}$ and
\begin{equation}
  \label{eq:funC}
  C(\tau) = \sin(\omega\X{L} \, \tau) - \dfrac{4}{\omega\X{L} \, \tau}
    \, \sin^2 \bigl(\omega\X{L} \, \dfrac{\tau}{2} \bigr) \; .
\end{equation}

Equation~(\ref{eq:pstat}) becomes Eq.~(14) in
Ref.~\onlinecite{Lewenstein:HH-94} by dropping the
factors~$\vec e\X{L}$, $\omega\X{L}$ and making the
replacement~$-E\X{0L} \to E$.
I find that Eq.~(\ref{eq:actionstat}) is transformed into Eq.~(15) of
Ref.~\onlinecite{Lewenstein:HH-94} by omitting~$\omega\X{L}$
and replacing~$\tfrac{\lambda^{(m)}_j}{2} + \Cal I\X{P} \to I\X{P}$.
Apart from the extra factors~$\omega\X{L}$ in
Eq.~(\ref{eq:funC}), it is the same as Eq.~(16) in
Ref.~\onlinecite{Lewenstein:HH-94}.

The dipole components~(\ref{eq:dipcomp}) for~$i \in \{a, c\}$,
$j \in \{+, -\}$, and $m \in \mathbb M_2$ in saddle-point approximation
are simplified as the quadruple integral over the canonical
momentum~$\vec p$ and the time~$t^{\,\prime}$ is replaced by a single integral
over the excursion time~$\tau$ leading to
\begin{eqnarray}
  \label{eq:dipcompst}
  \mathfrak D^{(m)}_{ij}(t) &=& -\imag \, (-1)^{\delta_{i\,c}}
    \> U^{(m)}_{ij} \, w^{(m)}_j \> \euler^{-\imag \, \delta_{i\,c} \,
    \omega\X{X} \, t} \, a^*(t) \nonumber \\
  &&{} \times \Int_0^{\infty} \sqrt{\tfrac{(-2 \pi \imag)^3}{\tau^3}} \;
    a(t-\tau) \> \mathfrak A^{(m)}_i(t, \tau) \nonumber \\
  &&\qquad{} \times \euler^{-\imag \, S^{(m)}_{\mathrm{st}, j}(t, \tau)}
    \differential \tau \; ,
\end{eqnarray}
upon extending the integration over~$\tau$ to infinity and introducing
the field-dipole product with~$i \in \{ a, c\}$ at the stationary points of the
quasiclassical action by
\begin{eqnarray}
  \label{eq:atomicTerms}
  \mathfrak A^{(m)}_i(t, \tau) &=& E\X{L}(t - \tau) \,
    \wp^{(m) \, *}_{\mathrm H, \vec p\X{st}(t, \tau) + \vec A\X{L}(t) \, i} \\
  &&{} \times \wp^{(m)}_{\mathrm L, \vec p\X{st}(t, \tau) +
    \vec A\X{L}(t - \tau) \, a} \; . \nonumber
\end{eqnarray}
The factor~$\sqrt{\tfrac{(-2 \pi \imag)^3}{\tau^3}}$ in Eq.~(\ref{eq:dipcompst})
stems from the integration over~$\vec p$ in saddle-point
approximation [Appendix~\ref{sec:saddlepoint}].

The dipole components without x~rays~$\mathfrak d^{(m)}(t)$
for~$m \in \mathbb M_1$ are obtained from the case~$i=a$ and $j=+$
in Eq.~(\ref{eq:dipcompst})
by replacing~$(-1)^{\delta_{i\,c}} \> U^{(m)}_{ij} \, w^{(m)}_j \>
\euler^{-\imag \, \delta_{i\,c} \, \omega\X{X} \, t} \to 1$ and
$\lambda^{(m)}_{\pm} \to - \delta - \imag \, \Gamma_a$
in Eq.~(\ref{eq:actionstat}).

The resulting equation becomes the same as Eq.~(13) in
Ref.~\onlinecite{Lewenstein:HH-94}---by setting~$\epsilon = 0$ and
suppressing the $\mathrm{c.c.}$~summand there---and further replacing~$a^*(t)
\to -1$ and $a(t - \tau) \to 1$.
Additionally, Eq.~(\ref{eq:atomicTerms}) needs to be changed,
before it is inserted into Eq.~(\ref{eq:dipcompst}), by
replacing~$E\X{L}(t) \to E \cos(t - \tau)$,
$\wp^{(m) \, *}_{\mathrm H, \vec p\X{st}(t, \tau) + \vec A\X{L}(t) \, a} \to
d^{\,*}_x(p\X{st}(t, \tau) - A_x(t))$, $\wp^{(m)}_{\mathrm L, \vec p\X{st}
(t, \tau) + \vec A\X{L}(t - \tau) \, a} \to d_x(p\X{st}(t, \tau)
- A_x(t - \tau))$ and adapting~$\vec p\X{st}(t, \tau)$ from
Eq.~(\ref{eq:pstat}) as for Eq.~(14) from Ref.~\onlinecite{Lewenstein:HH-94}.
Finally, Eq.~(\ref{eq:actionstat}) is adapted as for Eq.~(9)
in Ref.~\onlinecite{Lewenstein:HH-94} and I let~$m = 0$
in Eq.~(\ref{eq:dipcompst}).

The field-dipole product~(\ref{eq:atomicTerms}) is periodic in time with the
optical-laser period~$T\X{L} = \tfrac{2 \pi}{\omega\X{L}}$ and
can thus be expanded into a Fourier
series~\cite{Lewenstein:HH-94,Arfken:MM-05} for~$i \in \{a, c\}$ yielding
\begin{equation}
  \label{eq:FourSeries}
  \mathfrak A^{(m)}_i(t, \tau) = \Sum_{M = -\infty}^{\infty}
    \tilde {\mathfrak A}^{(m)}_{M, i}(\tau) \, \euler^{-\imag \,
    (2 \, M + \delta_{i\,a}) \, \omega\X{L} \, t} \; ,
\end{equation}
with the Fourier coefficients
\begin{equation}
  \label{eq:FourAtomic}
  \tilde {\mathfrak A}^{(m)}_{M, i}(\tau) = \dfrac{1}{T\X{L}} \Int_0^{T\X{L}}
    \mathfrak A^{(m)}_i(t, \tau) \; \euler^{\imag \, (2 \, M
    + \delta_{i\,a}) \, \omega\X{L} \, t} \differential t \; .
\end{equation}
The~$\tilde {\mathfrak A}^{(m)}_{M, i}(\tau)$ for~$i \in \{a, c\}$ reflect
the symmetry properties of the involved atomic states:
the Fourier coefficients for $M$~odd (even)---restricted to an integration
range from $0$ to~$T\X{L}/2$---can be examined by substituting~$t \to t
+ \tfrac{T\X{L}}{2}$.
Using Eqs.~(\ref{eq:lasfield}), (\ref{eq:lasvector}), (\ref{eq:pstat}),
and (\ref{eq:atomicTerms}) with this replacement,
I find that the kinetic momentum in the argument of the dipole matrix
elements in Eq.~(\ref{eq:atomicTerms}) is negated.
To investigate the consequences of the substitution~$\vec k \to -\vec k$
for the atomic dipole matrix element~$\wp^{(m)}_{\mathrm j, \vec k \, i}$
for~$i \in \{ a, c\}$, I express it with Eq.~(\ref{eq:freeelmom}) and
the spatial atomic orbital~$\bracket{\vec r}{i ; m}$ as follows
\begin{equation}
  \label{eq:dipMom}
  \wp^{(m)}_{\mathrm j, \vec k \, i} = \dfrac{1}{(2 \pi)^{3/2}}
    \Int_{\mathbb R^3} \euler^{-\imag \, \vec k \mul \vec r} \;
    \vec e\X{j} \mul \vec r \; \bracket{\vec r}{i ; m} \differential^3 r \; ,
\end{equation}
for~$j \in \{\mathrm L, \mathrm H\}$.
The central symmetry of the atomic potential implies that the orbitals
are parity eigenstates~\cite{Merzbacher:QM-98} and a
spatial inversion~$\vec r \to -\vec r$ yields~$\wp^{(m)}_{\mathrm j,
-\vec k \, i} = \pm \wp^{(m)}_{\mathrm j, \vec k \, i}$ depending on
the parity of~$\ket{i ; m}$.
Going back to the analysis of Eq.~(\ref{eq:FourAtomic}), I find that
for~$i = a$, the even Fourier coefficients vanish.
As $a \to c$~is an electric dipole transition, the one-electron
states~$\ket{a ; m}$ and $\ket{c ; m}$ have opposite
parity~\cite{Rose:ET-57}.
Therefore, for~$i = c$, the odd Fourier coefficients are zero.
In Appendix~\ref{sec:BCtrans}, Eq.~(\ref{eq:dipMom}) is decomposed into
a sum over radial integrals times spherical harmonics times Clebsch-Gordan
coefficients.

\subsection{High-order harmonic spectrum}
\label{sec:HHGspectrum}
\subsubsection{Harmonic photon emission rate}

The HH~spectrum for low ground state depopulation due to ionization by
the optical laser and the x~rays is calculated by setting~$\Gamma_0 = 0$.
In this case, $a(t) \approx 1$ for all times~$t$ and Eq.~(\ref{eq:groundepl})
can be neglected altogether.
\emph{Without} the $\omega\X{X}$-dependent term, the dipole components
[Eq.~(\ref{eq:dipcompst})] are periodic with period~$T\X{L}$ because, in
general, $\omega\X{X}$~is not a harmonic of~$\omega\X{L}$.
Consequently, $\euler^{\imag \, \delta_{i\,c} \, \omega\X{X} \, t} \>
\mathfrak D^{(m)}_{ij}(t)$ can be expanded into a Fourier series in analogy
to Eqs.~(\ref{eq:FourSeries}) and (\ref{eq:FourAtomic}).
The even (odd) Fourier coefficients of~$\euler^{\imag \, \delta_{i\,c} \,
\omega\X{X} \, t} \> \mathfrak D^{(m)}_{ij}(t)$ are present when the even (odd)
Fourier coefficients of the Fourier series expansion of the field-dipole
product~(\ref{eq:FourSeries}) are nonzero.

I simplify~$\euler^{\imag \, \delta_{i\,c} \, \omega\X{X} \, t} \>
\mathfrak D^{(m)}_{ij}(t)$ [Eq.~(\ref{eq:dipcompst})]
by expanding the exponential of the quasiclassical action by
decomposing the $t$-dependent terms in~$\euler^{-\imag \, S^{(m)}_{\mathrm{st},
j}(t, \tau)}$ with the Jacobi-Anger expansion
\begin{equation}
  \label{eq:JacobiAnger}
  \euler^{\imag \, z \, \cos \theta} = \Sum^{\infty}_{M = -\infty}
    \imag^M \, J_{M}(z) \; \euler^{-\imag \, M \, \theta} \; ,
\end{equation}
involving the Bessel functions~$J_M(z)$~\cite{Arfken:MM-05}
by letting~$\theta = \omega\X{L} \, (2 \, t - \tau)$ and
$z = \tfrac{U\X{P}}{\omega\X{L}} \> C(\tau)$.
The expression for the Fourier coefficients of the dipole
components~(\ref{eq:dipcompst}), with the dependence on~$\omega\X{X}$ removed
and an extra factor~$2 \, \pi$ introduced, reads
\begin{eqnarray}
  \label{eq:dipfreqgen}
  \tilde{\mathfrak D}^{(m)}_{ij, 2K + \delta_{i\,a}} &=&
    \dfrac{2 \, \pi}{T\X{L}} \Int_{0}^{T\X{L}} \euler^{\imag \, \delta_{i\,c} \,
    \omega\X{X} \, t^{\,\prime}} \> \mathfrak D^{(m)}_{ij}(t^{\,\prime})
    \nonumber \\
  &&{} \times \euler^{\imag \, (2 \, K + \delta_{i\,a}) \,
    \omega\X{L} \, t^{\,\prime}} \differential t^{\,\prime}\\
  &=& -2 \, \pi \, \imag \> (-1)^{\delta_{i\,c}} \> U^{(m)}_{ij} \, w^{(m)}_j
    \Int_0^{\infty} \sqrt{\tfrac{(-2 \pi \imag)^3}{\tau^3}} \nonumber \\
  &&{} \times \euler^{-\imag \, F^{(m)}_{0,j}(\tau)}
    \Sum_{M = -\infty}^{\infty} \tilde{\mathfrak A}^{(m)}_{K-M, i}(\tau)
    \> \imag^M \nonumber \\
  &&{} \times J_M \Bigl(\dfrac{U\X{P}}{\omega\X{L}}
    \> C(\tau) \Bigr) \; \euler^{\imag \, M \, \omega\X{L} \, \tau}
    \differential \tau \; , \nonumber
\end{eqnarray}
where the integration over~$t^{\,\prime}$ leads to~$\delta_{K - M \, N}$,
I translate the sum by~$M \to K - M$, and define
\begin{eqnarray}
  \label{eq:lewF0}
  F^{(m)}_{0,j}(\tau) &=& \dfrac{\lambda^{(m)}_j}{2} \, \tau
    + F^{\,\prime}_0(\tau) \nonumber \\
    &=& \Bigl(\dfrac{\lambda^{(m)}_j}{2} + \Cal I\X{P} + U\X{P} \Bigr)
      \, \tau \\
    &&{} - 2 \, \dfrac{U\X{P}}{\omega\X{L}^2 \tau} \, \bigl(1 -
      \cos (\omega\X{L} \, \tau) \bigr) \; . \nonumber
\end{eqnarray}
Inspecting Eqs.~(\ref{eq:dipfreqgen}) and (\ref{eq:lewF0}), I find that there
is a $\euler^{-\imag \, \tfrac{\lambda^{(m)}_j}{2} \, \tau}$~dependence
of the integrand in Eq.~(\ref{eq:dipfreqgen}).
The eigenvalues of the Rabi matrix~(\ref{eq:RabiEigval}) have a
negative imaginary summand~$-\dfrac{\imag}{2} \, (\Gamma_a + \Gamma_c)$
which causes~$\euler^{-\imag \, \tfrac{\lambda^{(m)}_j}{2} \, \tau}$
to acquire an exponentially decaying factor.
In other words, the longer the excursion time of an electron in the continuum,
the stronger is the suppression of its contribution to the HH~spectrum
due to destruction of the system by ionization and decay.
This reduces the importance of long trajectories versus
short trajectories further in addition to the reduction due to the increased
spreading of the wave packet~(\ref{eq:wavepacket}) for long trajectories in the
continuum~\cite{Corkum:PP-93,Lewenstein:HH-94}.
Along the same line, multiple returns of the continuum electron to the nucleus
are disadvantaged.

The expression for~$\tilde{\mathfrak d}^{(m)}_{2K + 1}$
for~$m \in \mathbb M_1$ is obtained from the case~$i = a$ and
$j=+$ in Eq.~(\ref{eq:dipfreqgen}) by letting~$\lambda^{(0)}_0 = -\delta
- \imag \, \Gamma_a$ in Eq.~(\ref{eq:lewF0}) and
changing~$(-1)^{\delta_{i\,c}} \> U^{(m)}_{ij} \> w^{(m)}_j \to 1$ and
$F^{(m)}_{0,j}(\tau) \to F^{(m)}_{0,0}(\tau)$.

From the expression for~$\tilde{\mathfrak d}^{(m)}_{2K + 1}$, Eq.~(18) of
Ref.~\cite{Lewenstein:HH-94} follows by the
replacements~$-2 \, \pi \, \imag \to \imag$,
$\tilde{\mathfrak A}^{(m)}_{K-M, i}(\tau) \to b_{K - M}(\tau)$
and $\omega\X{L} \to 1$---where the
modification~$\omega\X{L} \to 1$ is made to Eq.~(\ref{eq:FourSeries}) and
$C(\tau)$~is adapted as before---and
truncating the~$\Sum_{M = -\infty}^{\infty}$ to~$\Sum_{M = 0}^{\infty}$ and
expressing the factor from the saddle-point approximation
with the infinitesimal~$\epsilon$ set to zero.
After removing the~$\omega\X{L}$ from~$F^{(m)}_{0,j}(\tau)$, the
replacement~$\tfrac{\lambda^{(m)}_j}{2} + \Cal I\X{P} \to I\X{P}$
in~$F^{(m)}_{0,j}(\tau)$ turns it into~$F_0(\tau)$ after Eq.~(18) in
Ref.~\onlinecite{Lewenstein:HH-94}.

The dipole moment~$\Cal D(t)$ [Eq.~(\ref{eq:diptransmat})] is obtained
from the Fourier coefficients of the dipole components~(\ref{eq:dipfreqgen}),
with the $\omega\X{X}$-dependence removed, as follows:
\begin{equation}
  \label{eq:FourDip}
  \Cal D(t) = \tfrac{1}{2 \, \pi} \!\!\! \Sum_{\atopa{\scriptstyle i \in
    \{a, c\}}{\scriptstyle j \in \{+, -\}}} \!\! \Sum_{K = -\infty}^{\infty}
    \check{\mathfrak D}_{ij, 2K + \delta_{i\,a}} \> \euler^{-\imag
    \, [ (2 \, K + \delta_{i\,a}) \, \omega\X{L} + \delta_{i\,c} \,
    \omega\X{X} ] \, t} \; ,
\end{equation}
with the compact notation for the dipole components
\begin{equation}
  \label{eq:FourComp}
  \check{\mathfrak D}_{ij, 2K + \delta_{i\,a}} = \dfrac{\delta_{i\,a}}{2}
    \Sum_{m \in \mathbb M_1} \tilde{\mathfrak d}^{(m)}_{2K + 1} +
    \Sum_{m \in \mathbb M_2} \tilde{\mathfrak D}^{(m)}_{ij, 2K +
    \delta_{i\,a}} \; ,
\end{equation}
where the prefactor~$\tfrac{\delta_{i\,a}}{2}$ ensures that only
for valence recombination, \ie, $i=a$, the $\tilde{\mathfrak d}^{(m)}_{2K + 1}$
make a contribution and the sum over~$j$ in Eq.~(\ref{eq:FourDip})
removes the prefactor~$\tfrac{1}{2}$.
The Fourier coefficients for negative~$K$ correspond to the inverse process
in which the HH~photon is emitted before the required number of optical laser
photons is absorbed to satisfy energy conservation~\cite{Kuchiev:QT-99}.
The energies of the emitted HH~photons are certainly the same as
in the regular process.
The inverse process is strongly suppressed because the responsible
Fourier coefficients have a highly oscillatory integrand and thus
are very small;
hence, they may be omitted in Eq.~(\ref{eq:FourDip}) as in
Ref.~\onlinecite{Lewenstein:HH-94}.
I express~$\Cal D(t)$ in frequency domain by a Fourier transformation
\begin{eqnarray}
  \label{eq:DipOmFin}
  \tilde{\Cal D}(\omega) &=& \Int_{-\infty}^{\infty} \Cal D(t)
    \> \euler^{\imag \, \omega \, t} \differential t \nonumber \\
    &=& \Sum_{\atopa{\scriptstyle i \in \{a, c\}}
    {\scriptstyle j \in \{+, -\}}} \Sum_{K = -\infty}^{\infty}
      \check{\mathfrak D}_{ij, 2K + \delta_{i\,a}} \\
    &&{} \times \delta\bigl( (2 \, K + \delta_{i\,a}) \, \omega\X{L}
      + \delta_{i\,c} \, \omega\X{X} - \omega \bigr) \; , \nonumber
\end{eqnarray}
where I use the representation
\begin{equation}
  \label{eq:diracFour}
  \Int^{\infty}_{-\infty} \euler^{\imag \, \omega \, t}
    \differential t = 2 \, \pi \, \delta(\omega) \; ,
\end{equation}
of the Dirac $\delta$~distribution~\cite{Arfken:MM-05}.

The HH~spectrum is given by the frequency-resolved and solid-angle-dependent
rate of HH~photon emission~\cite{Lewenstein:HH-94,Kuchiev:QT-99,%
Milosevic:SI-01} where I consider only HH~emission along the $x$~axis with
polarization vector~$\vec e\X{H} = \vec e_z$.
The derivation in Appendix~\ref{sec:nodepletion} leads to
\begin{equation}
  \label{eq:HHGspecNoDepl}
  \dfrac{\partial^2 \Gamma\X{H}(\omega)}{\partial \omega \,
    \partial \Omega} = 8 \, \pi^2 \, \omega \, \varrho(\omega) \,
    W(\omega) \; ,
\end{equation}
with the solid angle~$\Omega$ into which the HH~photons are emitted.
The density of free-photon states in the photon energy
interval~$[\omega ; \omega + \!\differential
\omega]$~\cite{Scully:QO-97,Meystre:QO-99}, for a fixed
polarization~$\vec e_z$ and propagation along~$\vec e_x$, reads
\begin{equation}
  \label{eq:densfreephot}
  \varrho(\omega) = \dfrac{\omega^2}{(2\pi)^3 \, c^3} \; ,
\end{equation}
with the speed of light~$c$ in vacuum, and
\begin{subeqnarray}
  \label{eq:HHGstrength}
  W(\omega) &=& W_1(\omega) \nonumber \\
  &&{} + \Sum_{H=1}^{\infty} \delta_{(2\,H+1) \; \omega\X{L} \; \omega\X{X}}
    \> W_{2,H}(\omega) \\
  W_1(\omega) &=& \Sum_{K=-\infty}^{\infty} \Sum_{\atopa{\scriptstyle i \in
    \{a, c\}}{\scriptstyle j, j' \in \{+, -\}}} \bigl[ \check{\mathfrak
    D}^*_{ij , 2K + \delta_{i\,a}} \> \check{\mathfrak D}  _{ij', 2K
    + \delta_{i\,a}} \bigr] \nonumber \\
  &&{} \times \delta\bigl( (2 \, K + \delta_{i\,a}) \, \omega\X{L}
    + \delta_{i\,c} \, \omega\X{X} - \omega \bigr) \\
  W_{2,H}(\omega) &=& \Sum_{K=-\infty}^{\infty} \Sum_{j, j' \in \{+, -\}}
    \bigl[ \check{\mathfrak D}^*_{aj , 2K + 1} \>
           \check{\mathfrak D}  _{cj', 2 \, (K-H)} + \mathrm{c.c.} \bigr]
    \nonumber \\
  &&{} \times \delta \bigl( (2 \, K + 1) \, \omega\X{L} - \omega \bigr) \; .
\end{subeqnarray}
With~``$\mathrm{c.c.}$'' I denote the complex conjugate of the preceding
summand.
The HH~spectrum consist, via Eq.~(\ref{eq:DipOmFin}), of discrete lines
where the coefficients in front of the Dirac $\delta$~distributions represent
the strength of the respective HH~emission line.
The width of the lines, however, is zero, despite the widths of the
singly-excited states $\ket{\Phi_{\vec k}^{(m_1)}}$,
$\ket{\Phi_{\vec k \, c}^{(m_2)}}$, and $\ket{\Phi_{\vec k \, a}^{(m_2)}}$
for~$m_1 \in \mathbb M_1$ and $m_2 \in \mathbb M_2$, in
Eq.~(\ref{eq:wavepacket}) with destruction rates~$\Gamma_a$ and $\Gamma_c$,
respectively.
The widths~$\Gamma_a$ and $\Gamma_c$ enter Eq.~(\ref{eq:dipcompst}) only
via~$U^{(m_2)}_{ij}$, $w^{(m_2)}_j$, and~$\lambda^{(m_2)}_j$ which do not
depend on~$t$ and thus cause only a change of the strength of the lines.

There are several contributions to the HH~rate in
Eq.~(\ref{eq:HHGspecNoDepl}) which I analyze in the following.
If $\omega\X{X} = (2 \, H+1) \, \omega\X{L}$~holds for an~$H \in
\mathbb N$, then I obtain~$W(\omega) = W_1(\omega) + W_{2,H}(\omega)$
from Eq.~(\ref{eq:HHGstrength}).
The HH~lines due to the recombination with the core hole align with higher
orders of the HH~emission from the recombination with a valence hole,
if the latter plateau is sufficiently extended.
Then this leads to interference between the light from both processes
mediated by~$W_{2,H}(\omega)$.
Otherwise, if~$\omega\X{X}$ is \emph{not} an odd harmonic of~$\omega\X{L}$,
then Eq.~(\ref{eq:HHGstrength}) reduces to~$W(\omega) = W_1(\omega)$
and the lines in the HH~spectrum belong to two distinct groups:
first, $i = a$, HH~light from the recombination with the valence
vacancy gives rise to lines at the energies~$(2 \, K + 1) \, \omega\X{L}$
for~$K \in \mathbb Z$.
The x~rays induce interference effects in the recombination step via the two
\xray~dressed states with~$j \in \{+, -\}$.
If the x~rays are sufficiently intense, the core electron in the
two-level system may even undergo Rabi flopping~\cite{Rabi:OP-36,Rabi:SQ-37,%
Rabi:MN-38,Cohen:QM-77,Scully:QO-97,Meystre:QO-99,Rohringer:RA-08,%
Rohringer:PN-08,Kanter:MA-11,Rohringer:SD-12,Cavaletto:RF-12,Adams:QO-13}
prior recombination of the continuum electron.
In the bare-states picture, this means that there are multiple pathways for
the recombination with the valence hole: once, direct recombination and,
alternatively, recombination after an even number of Rabi cycles
of the second electron.
Second, $i = c$, HH~light due to recombination with the \xray~dressed core
vacancy leads to lines at energies~$2 \, K \, \omega\X{L} + \omega\X{X}$
for~$K \in \mathbb Z$.
In other words, the energy of the HH~photons from a recombination with a
core hole are shifted by the photon energy of the x~rays~$\omega\X{X}$
to higher energies.
In the bare-states picture, there are multiple pathways: once, recombination
with the core hole after excitation of a core electron into the valence
vacancy and, alternatively, recombination after an odd number of
Rabi cycles of the second electron.

The HH~spectrum~(\ref{eq:HHGspecNoDepl}) gives the emission rate of
HH~photons into a small solid angle centered on~$\vec e_x$ with
specific energy.
However, the experimental observable of HHG is the harmonic photon number
spectrum~(HPNS)~\cite{Diestler:HG-08}.
The HPNS from a single atom is the probability to observe a HH~photon
in a solid angle~$\Omega$ with specific energy~$\omega$ after
irradiation of the atom with a light pulse.
It is obtained along the $x$~axis from Eq.~(\ref{eq:HHGspecNoDepl}) by
multiplying with the pulse duration~$T\X{P}$ of the optical laser
and the x~rays~\footnote{%
The multiplication with the duration~$T\X{P}$ corresponds to
determining the HPNS from optical-laser and \xray~pulses with constant field
strength starting at~$t = 0$ and stopping at~$t = T\X{P}$ where turn-on
effects neglected.}
for which the HPNS is recorded.
The photon-energy-resolved and solid-angle-resolved HPNS is given by the
probability density
\begin{equation}
  \label{eq:HHGspecNoDeplHPNS}
  \dfrac{\partial^2 P\X{H}(\omega)}{\partial \omega \,
    \partial \Omega} = T\X{P} \; \dfrac{\partial^2 \Gamma\X{H}(\omega)}
    {\partial \omega \, \partial \Omega} \; .
\end{equation}
The integration~$\Int_{\Delta \Omega} \; \Int_{\omega\X{H} - \Delta
\omega\X{H}}^{\omega\X{H} + \Delta \omega\X{H}} \differential \omega
\differential \Omega$ of~$\dfrac{\partial^2 \Gamma\X{H}(\omega)}
{\partial \omega \, \partial \Omega}$ over the solid
angle~$\Delta \Omega$ and the photon energy range~$[\omega\X{H} - \Delta
\omega\X{H}; \omega\X{H} + \Delta \omega\X{H}]$ yields the
rate for the atom to emit a HH~photon per unit of time in
the course of the interaction with the light pulses.

In Fig.~35.2 of Ref.~\onlinecite{Kohler:EC-12}, we display \XUV/\xray~boosted
HH~spectra for \XUV/\xray~light tuned to the Kr$^+$~$3d \to 4d$ and
Ne$^+$~$1s \to 2p$~transitions in the parent ions.
Two plateaux are visible from valence- and core-hole
recombinations that are separated by the photon energy of the \XUV/\xray~light.
The amplitude of the plateaux depends on the intensity of the \XUV/\xray~light.
By adjusting the intensity suitably, a similar amplitude of both
plateaux is achieved.

\subsubsection{Harmonic photon number spectrum}
\label{sec:depletion}

When the influence on the HH~spectrum from ground state depopulation
due to ionization by the optical laser and the x~rays becomes appreciable,
one needs to account for it by letting~$\Gamma_0 > 0$.
This, however, breaks the periodicity in time imposed by assuming a cw~optical
laser and cw~x~rays such that one needs to assume finite optical-laser
and \xray~pulses with a constant field strength starting at~$t = 0$
and ending at~$t = T\X{P}$.
In doing so, the \xray~dressed states of Eq.~(\ref{eq:decoupledEOM})
become an approximation for the nonperiodic system.

To determine the time-dependent ground-state amplitude, I need to
solve Eq.~(\ref{eq:groundepl}) where I drop the second term on
the right-hand side and include its influence---the destruction of
the system by tunnel ionization~\cite{Perelomov:I1-66,Perelomov:I2-66,%
Perelomov:I3-67,Popov:I4-67,Ammosov:TI-86,Delone:MP-00,Yudin:NA-01} via
the optical laser---in~$\Gamma_0$.
For a constant ionization rate~$\Gamma_0 > 0$ [Eq.~(\ref{eq:phendecave})]
beginning at~$t = 0$ and ending at~$t = T\X{P}$, the ground-state amplitude is
\begin{equation}
  \label{eq:xraydepl}
  a(t) = \theta(-t) + \euler^{-\tfrac{\Gamma_0}{2} \, t} \> \theta(t) \,
   \theta(T\X{P} - t) + \euler^{-\tfrac{\Gamma_0}{2} \, T\X{P}} \>
   \theta(t - T\X{P}) \; ,
\end{equation}
with the Heaviside step function~$\theta$ and~$\theta(0) =
\tfrac{1}{2}$~\cite{Arfken:MM-05}.
Clearly, $a(t)$~is not periodic in time.

Due to the lack of temporal periodicity of the dipole components---with
the $\omega\X{X}$-dependence removed---when $a(t)$ from
Eq.~(\ref{eq:xraydepl}) is inserted into Eq.~(\ref{eq:dipcompst}),
the HH~spectrum cannot be expanded into a Fourier series any longer, \ie,
a HH~photon emission rate [compare Eq.~(\ref{eq:HHGspecNoDepl})] cannot
be defined anymore.
I need to calculate the Fourier transform of~$\Cal D(t)$
[Eq.~(\ref{eq:diptransmat})]
instead~\cite{Lewenstein:HH-94,Kuchiev:QT-99,Milosevic:SI-01}
in order to determine the HPNS [compare Eq.~(\ref{eq:HHGspecNoDeplHPNS})]
directly.
As a Fourier transform is linear~\cite{Arfken:MM-05}, it is applied
componentwise.
The dipole components~(\ref{eq:dipcompst}) in frequency domain
for~$m \in \mathbb M_2$ read
\begin{eqnarray}
  \label{eq:dipcontinuous}
  \tilde{\mathfrak D}^{(m)}_{ij}(\omega) &=& \Int_{0}^{T\X{P}}
    \mathfrak D^{(m)}_{ij}(t) \> \euler^{\imag \, \omega \, t}
    \differential t \\
  &=& -2 \, \pi \, \imag \> (-1)^{\delta_{i\,c}} \, U^{(m)}_{ij} \, w^{(m)}_j
    \Int_0^{\infty} \sqrt{\tfrac{(-2 \pi \imag)^3}{\tau^3}} \nonumber \\
  &&{} \times \euler^{-\imag
    \, F^{(m)}_{0,j}(\tau)} \!\!\! \Sum_{N = -\infty}^{\infty} \imag^N
    J_N\bigl(\tfrac{U\X{P}}{\omega\X{L}} \> C(\tau) \bigr) \;
    \euler^{\imag \, N \, \omega\X{L} \, \tau} \nonumber \\
  &&{} \times \Sum_{M = -\infty}^{\infty} \tilde {\mathfrak
    A}^{(m)}_{M - N, i}(\tau) \, h_{M,i}(\omega, \tau)
    \differential \tau \; , \nonumber
\end{eqnarray}
where I expand the exponential of the quasiclassical action at the
stationary point in analogy to the derivation of Eq.~(\ref{eq:dipfreqgen}),
translate the sum over~$M$ by~$M \to M - N$, and define
the HH~line shape by
\begin{eqnarray}
  \label{eq:line}
  h_{M,i}(\omega, \tau) &=& \dfrac{1}{2 \pi} \Int_0^{T\X{P}}
    \euler^{-\imag \, [ (2 \, M + \delta_{i\,a}) \, \omega\X{L}
    + \delta_{i\,c} \, \omega\X{X} - \omega ] \, t} \nonumber \\
    &&\qquad{} \times a^*(t) \, a(t-\tau) \differential t \; .
\end{eqnarray}
The time integration in Eq.~(\ref{eq:dipcontinuous}) extends only from~$0$
to~$T\X{P}$ as the optical laser and \xray~pulses lie
in this time interval.
The equation has a straightforward interpretation:
comparing it with the result without ground-state
depletion~(\ref{eq:DipOmFin}), I see that the Dirac $\delta$~distribution
HH~line shapes in the latter equation are replaced with finite-width
line shapes~(\ref{eq:line}) in the former.
Yet Eq.~(\ref{eq:line}) still depends on~$\tau$ which is integrated over
in Eq.~(\ref{eq:dipcontinuous}) in contrast to Eq.~(\ref{eq:DipOmFin}).
Ground-state depletion thus causes a finite width of the HH~lines.

Using expression~(\ref{eq:xraydepl}) for the ground-state depletion,
from Eq.~(\ref{eq:line}), I find the HH~line shape~\cite{SuppData} to be
\begin{eqnarray}
  \label{eq:lineConst}
  &&{}h_{M,i}(\omega, \tau) \approx \dfrac{\euler^{\tfrac{\Gamma_0}{2} \, \tau}}
    {2 \pi} \\
  &&\qquad{} \times \dfrac{\euler^{-(\Gamma_0 + \imag \, (\tilde\omega_{M,i}
    - \omega)) \, \tau}
    - \euler^{-(\Gamma_0 + \imag \, (\tilde\omega_{M,i}
    - \omega)) \, T\X{P}}}{\Gamma_0 + \imag \, ( \tilde\omega_{M,i}
    - \omega)} \; . \nonumber
\end{eqnarray}
The absolute value of the denominator of~$h_{M,i}(\omega, \tau)$
is minimal at~$\omega = \tilde\omega_{M,i}$ for
\begin{equation}
  \label{eq:linepeak}
  \tilde\omega_{M,i} = (2 \, M + \delta_{i\,a}) \, \omega\X{L}
    + \delta_{i\,c} \, \omega\X{X} \; .
\end{equation}
In other words, the HH~peaks with ground-state depletion are
centered on the positions of
the harmonics without ground-state depletion;
for~$i = c$, the harmonics are shifted by~$\omega\X{X}$ with respect to the
harmonics for~$i = a$.
The Eqs.~(\ref{eq:dipcontinuous}) and (\ref{eq:lineConst}) depend
on the excursion time~$\tau$ of the electron in the continuum
which distributes the electron's contribution to HH~emission
over all harmonic peaks.
Yet the dependence on~$\tilde\omega_{M,i} - \omega$
ensures that only a sizable contribution is made for the peak
around~$\tilde \omega_{M,i}$ to which the return energy of the
continuum electron corresponds.
For~$T\X{P} \to \infty$, the~$4 \, \pi \, \Gamma_0 \; |h_{M,i}(\omega, 0)|^2$
is a Lorentzian of a FWHM of~$2 \, \Gamma_0$.
In Ref.~\onlinecite{Buth:NL-11}, Eq.~(6), the factor~$2 \pi$ was not
included in the definition of the HH~line shape~(\ref{eq:lineConst})
and integration started at~$0$ instead of~$\tau$.

Next I turn to the dipole components~$\tilde{\mathfrak d}^{(m)}(\omega)$
for~$m \in \mathbb M_1$.
They follow from the case~$i = a$ and $j = +$ in
Eq.~(\ref{eq:dipcontinuous}) by the replacements~$(-1)^{\delta_{i\,c}} \,
U^{(m)}_{ij} \, w^{(m)}_j \to 1$, and
$F^{(m)}_{0,j}(\tau) \to F^{(0)}_{0,0}(\tau)$.
Additionally, $\lambda^{(0)}_0 = -\delta - \imag \, \Gamma_a$ holds and
the HH~line shape~$h_{M,i}(\omega, \tau)$ is given by
Eq.~(\ref{eq:lineConst}).

From the Fourier transform of the dipole components, I calculate
the HPNS.
For this, I introduce the compact notation [see also Eq.~(\ref{eq:FourDip})]:
\begin{equation}
  \tilde{\Cal D}(\omega) = \Sum_{\atopa{\scriptstyle i \in \{a, c\}}
    {\scriptstyle j \in \{+, -\}}} \check{\mathfrak D}_{ij}(\omega) \; ,
\end{equation}
with [see also Eq.~(\ref{eq:FourComp})]
\begin{equation}
  \label{eq:compounddipcomp}
  \check{\mathfrak D}_{ij}(\omega) = \dfrac{\delta_{i\,a}}{2}
    \Sum_{m \in \mathbb M_1} \tilde{\mathfrak d}^{(m)}(\omega) +
    \Sum_{m \in \mathbb M_2} \tilde{\mathfrak D}_{ij}(\omega) \; .
\end{equation}
Then the HH~photon-energy-resolved and solid-angle-resolved HPNS for a
single atom---the probability density of emitting a HH~photon
with specified energy along the $x$~axis---is derived in
Appendix~\ref{sec:gsdepletion} and is given by
\begin{equation}
  \label{eq:HHGspecXray}
  \dfrac{\partial^2 P\X{H}(\omega)}{\partial \omega \,
    \partial \Omega} = 4 \, \pi \, \omega \, \varrho(\omega)
    \> |\tilde{\Cal D}(\omega)|^2 \; ,
\end{equation}
similar to Eq.~(\ref{eq:HHGspecNoDepl}).

An emission rate of HH~photons can be obtained from the
HPNS~(\ref{eq:HHGspecXray}), if its temporal evolution is
known via\break $\dfrac{\partial}{\partial t} \dfrac{\partial^2
P^{\,\prime}\X{H}(\omega, t)}{\partial \omega \, \partial \Omega}$.
This instantaneous HH~emission rate (not optical-laser-cycle averaged),
however, presently cannot be determined experimentally and thus is of
litte use in contrast to the rate for HH~emission without ground-state
depletion~(\ref{eq:HHGspecNoDepl}).
However, an average HH~emission rate for harmonic~$H$ can be determined
from the HPNS~(\ref{eq:HHGspecXray}) via
\begin{equation}
  \label{eq:averatehpns}
  \dfrac{\partial^2 \bar\Gamma\X{H}(\omega)}{\partial \omega \,
    \partial \Omega} = \dfrac{1}{T\X{P}} \dfrac{\partial^2 P\X{H}(\omega)}
    {\partial \omega \, \partial \Omega} \; .
\end{equation}
An integration similar to the one after Eq.~(\ref{eq:HHGspecNoDeplHPNS})
may be used to determine the probability for an atom to emit a HH~photon in
the course of the interaction with the light pulses of duration~$T\X{P}$.
Dividing by~$T\X{P}$ then gives the average rate of HH~photon emission.
This is the inverse expression to Eq.~(\ref{eq:HHGspecNoDeplHPNS})
where the HPNS is determined from the HH~rate.
In Sect.~V of Ref.~\onlinecite{Lewenstein:HH-94}, an approximation for
the average HH~emission rate is constructed
from~$|\tilde{\Cal D}(\tilde\omega_{M,i})|^2$ for
nonvanishing ground-state depletion by multiplying its value at the position
of a HH~peak~$\omega = \tilde\omega_{M,i}$ with a factor that accounts
for the area of the peak.

The Fig.~2 of Ref.~\onlinecite{Buth:NL-11} shows \XUV~enhanced HHG for the
Kr$^+$~$3d \to 4d$~transition in the parent ion with ground
state depletion and Fig.~35.2 of Ref.~\onlinecite{Kohler:EC-12}
shows a comparison between \XUV/\xray~boosting with and without
ground-state depletion for the transitions~Kr$^+$~$3d \to 4d$ and
Ne$^+$~$1s \to 2p$.
Two \XUV/\xray~intensities are investigated: for the lower intensity the
plateaux due to core-hole recombination have low amplitude whereas
for the higher intensity roughly the same amplitude is found for
valence- and core-hole recombination.
For low \XUV/\xray~intensity there is a very good agreement between the
HH~spectra with treating ground-state depletion and without treating it.
This agreement, however, becomes less good for high
\XUV/\xray~intensity due to increased destruction of the system by
\XUV/\xray-based ionization.

\subsubsection{Cutoff law}

The cutoff of the HH~spectrum with resonant excitation by x~rays
is changed dramatically with respect to the optical laser-only case.
First, inspecting Eq.~(\ref{eq:diptransmat}), I see that the
terms for the recombination of the continuum electron with a
core hole are shifted by~$\omega\X{X}$ towards higher energy
compared with the terms for the recombination with a
valence hole.
Second, the phenomenological semiclassical cutoff law of an optical laser-only
HH~spectrum~\cite{Krause:HO-92,Schafer:AT-93,Corkum:PP-93,Lewenstein:HH-94}
is given by
\begin{equation}
  \label{eq:cutoff}
  \omega\X{cut} = 3.17 \, U\X{P} + I\X{P} \; ,
\end{equation}
with the valence ionization potential of the atom~$I\X{P}$.
\Xray~dressing leads, however, to a different situation;
namely, I observe that Eq.~(\ref{eq:diptransmat}) contains two terms
with~$j \in \{-, +\}$ for valence- and core-hole recombination.
Associated with the \xray-dressed states are the
energies~$\tfrac{\lambda^{(m)}_j}{2} + \Cal I\X{P}$ [Eq.~(\ref{eq:quasiIP})].
Consequently, there are two contributions shifted for each recombination
spectrum with respect to each other.
Inserting them for~$I\X{P}$ in Eq.~(\ref{eq:cutoff}), I see that the two terms
of valence- and core-hole recombination have different cutoffs
due to~$\lambda^{(m)}_{\pm}$;
the larger of the two terms determines the cutoff of the valence-
and core-hole recombination HH~spectra.

\section{High-order harmonic generation with arbitrarily-shaped \xray~pulses}
\label{sec:theoryGen}

The formalism of the previous Sect.~\ref{sec:theoryCW} relies
on an \xray-dressed-states picture [Eq.~(\ref{eq:decoupledEOM})] that
is only sensible for constant-amplitude x~rays [Eq.~(\ref{eq:honeX})].
For arbitrarily-shaped \xray~pulses, this is no longer practicable.
In this section, I pursue a derivation of \xray-boosted HHG that
circumvents the \xray-dressed-states picture.
This generalization is motivated by the fact that highly-intense x~rays
are presently only available from free electron
lasers~\cite{LCLS:CDR-02,Altarelli:TDR-06,Emma:FL-10} that frequently operate
by the SASE principle~\cite{Kondratenko:GC-79,Bonifacio:CI-84,Saldin:PF-00}
and consequently have short temporal coherence with a rapidly fluctuating
pulse envelope.
Those expressions in this section which have counterparts in
Sect.~\ref{sec:theoryCW} are noted with the same symbols but
with a prime attached.

\subsection{Hamiltonian and equations of motion}
\label{sec:HamWavEOM}

I assume an optical laser which produces pulses
of a constant amplitude starting at~$t = 0$ and stopping at~$t = T\X{P}$.
So the optical laser is basically approximated as cw~light and the
field-dipole product~(\ref{eq:atomicTerms}) is expanded into a Fourier
series.
Conversely, the \xray~pulse is chosen to be arbitrarily-shaped inside
the time interval~$[0 ; T\X{P}]$ and zero otherwise;
it is written as
\begin{equation}
  \label{eq:EXarb}
  E^{\,\prime}\X{X}(t) = \dfrac{E^{\,\prime}\X{0X}(t)}{2} \, \bigl[
    \euler^{\imag \, (\omega\X{X} \, t + \varphi\X{X}(t))}
    + \euler^{-\imag \, (\omega\X{X} \, t + \varphi\X{X}(t))} \bigr] \; ,
\end{equation}
where $E^{\,\prime}\X{0X}(t)$~is the time-dependent amplitude of the pulse
and $\varphi\X{X}(t)$~is its time-dependent
phase~\cite{Diels:UL-06}.
In order to account for arbitrarily-shaped \xray~pulses in~$\hat h\X{X}$
[Eq.~(\ref{eq:honeX})], I need to replace~$E\X{X}(t)$ by~$E^{\,\prime}\X{X}(t)$
therein.
The $N/2$-electron Hamiltonian for the interaction with the
x~rays~$\hat H'\X{X}$ has the same form as Eq.~(\ref{eq:htwoX})
with~$\hat h\X{X}$ replaced by~$\hat h^{\prime}\X{X}$.

The newly arising time-dependent phase of the \xray~field~$\varphi\X{X}(t)$
[Eq.~(\ref{eq:EXarb})] needs to be taken into account in the ansatz
for the $N/2$-electron wave packet~(\ref{eq:wavepacket});
it is modified by multiplying the prefactor of~$\ket{\Phi_{\vec k \, a}^{(m)}}$
by~$\euler^{- \imag \, \varphi\X{X}(t)}$ for~$m \in \mathbb M_2$.
With this modified ansatz, I obtain new EOMs which are very similar to the ones
obtained previously apart from the phenomenological destruction
rates~$\Gamma^{\,\prime}_0(t)$, $\Gamma^{\,\prime}_a(t)$, and
$\Gamma^{\,\prime}_c(t)$ [Eq.~(\ref{eq:phendec})]
whose time dependence is now accounted for.
Further, I introduce the temporal destruction exponents of the
system~\cite{Buth:KE-13} by the optical laser, the x~rays, and
decay processes up to time~$t$ via
\begin{equation}
  \label{eq:destruct}
  \digamma_i(t) = \theta(t) \Int_0^t \Gamma^{\,\prime}_i(t^{\,\prime})
    \differential t^{\,\prime} \; ,
\end{equation}
for~$i \in \{0, a, c\}$.
First, the new EOM for the ground-state amplitude is formally identical
to Eq.~(\ref{eq:groundepl}).
Second, also the new EOMs with~$m \in \mathbb M_2$ for the amplitude of
a continuum electron with a valence hole in the parent ion are formally
identical to Eq.~(\ref{eq:revalence}).
So are the valence-only EOMs with~$m \in \mathbb M_1$.
Third, the new EOMs for the amplitude of a continuum electron with
a core hole in the parent ion, however, differs from Eq.~(\ref{eq:recore})
in the first term on the right hand side
that needs to be augmented by the summand~$-2 \, \dot\varphi\X{X}(t) =
-2 \, \tfrac{\differential \varphi\X{X}(t)} {\differential t}$;
I use a dot over symbols to denote a time derivative.
The new EOMs~(\ref{eq:revalence}) and (\ref{eq:recore}) are
recast in terms of matrix equations for the combined
amplitudes~$\vec b^{\,\prime(m)}(\vec k, t) \equiv \binom{b'^{(m)}_a(\vec k, t)}
{b'^{(m)}_c(\vec k, t)}$ [compare with Eq.~(\ref{eq:contelamp})]
which are formally identical to Eq.~(\ref{eq:coupledAmps}) and read
\begin{eqnarray}
  \label{eq:coupledAmpsArb}
  \dfrac{\partial}{\partial t} \, \vec b^{\,\prime\,(m)}(\vec k, t)
    &=& -\dfrac{\imag}{2} \, \bigl( \mat R^{\,\prime\,(m)} + (\vec k^{\,2}
    + 2 \, \Cal I\X{P}) \, \unitmatrix \bigr) \,
    \vec b^{\,\prime\,(m)}(\vec k, t) \nonumber \\
  &&{} + E\X{L}(t) \, \dfrac{\partial}{\partial k_z}
    \, \vec b^{\,\prime\,(m)}(\vec k, t) \\
  &&{} - \imag \, E\X{L}(t) \, \wp^{(m)}_{\mathrm L,
    \vec k \, a} \, \binom{1}{0} \, a'_0(t) \; , \nonumber
\end{eqnarray}
for~$m \in \mathbb M_2$ and $\vec k \in \mathbb R^3$ where the
instantaneous Rabi matrix [compare with Eq.~(\ref{eq:RabiMat})] is
\begin{eqnarray}
  \label{eq:RabiMatGen}
  \mat R^{\,\prime\,(m)}(t) &=& - \delta \, \mat \sigma_z
    + R^{\,\prime\,(m)}\X{0X}(t) \, \mat \sigma_x \\
  &&{} - \imag \, \diag\bigl( \Gamma^{\,\prime}_a(t), \Gamma^{\,\prime}_c(t)
    - 2 \, \imag \, \dot \varphi\X{X}(t) \bigr) \; , \nonumber
\end{eqnarray}
with the Pauli matrices~$\mat \sigma_x$ and $\mat
\sigma_z$~\cite{Cohen:QM-77,Merzbacher:QM-98}.

\subsection{Transformation of the equations of motion}
\label{sec:integrationEOM}

The solution of the EOMs for the one-electron amplitudes
with~$m \in \mathbb M_1$, can be determined, in straight analogy to
Sect.~\ref{sec:EOMs}.
In order to solve the system of EOMs [new Eq.~(\ref{eq:groundepl})
and Eqs.~(\ref{eq:coupledAmpsArb})], I need to realize that, in contrast to
the EOMs for cw~x~rays, the Rabi matrix~$\mat R^{\,\prime\,(m)}(t)$
[Eq.~(\ref{eq:RabiMatGen})] is time dependent.
Hence a decoupling of the two recombination amplitudes by going to an
\xray~dressed-states picture, as in Eq.~(\ref{eq:decoupledAmps}), is
no longer possible because the eigenvector matrix
of~$\mat R^{\,\prime\,(m)}(t)$ does not commute with the time derivative.
However, I can approximately decouple the EOMs [Eq.~(\ref{eq:coupledAmpsArb})]
as follows.
I remove the diagonal elements of~$\mat R^{\,\prime\,(m)}(t)$
for~$m \in \mathbb M_2$ with the substitution
\begin{eqnarray}
  \label{eq:timesub}
  b^{\,\prime\,(m)}_a(\vec k, t) &=& b^{\,\prime\prime\,(m)}_a(\vec k, t)
    \; \euler^{\tfrac{\imag}{2} \, \delta \, t - \tfrac{\digamma_a(t)}{2}} \\
  b^{\,\prime\,(m)}_c(\vec k, t) &=& b^{\,\prime\prime\,(m)}_c(\vec k, t)
    \; \euler^{-\tfrac{\imag}{2} \, \delta \, t + \imag \, \varphi\X{X}(t)
    - \tfrac{\digamma_c(t)}{2}} \; . \nonumber
\end{eqnarray}
I let~$\vec b^{\,\prime\prime\,(m)}(\vec k, t) \equiv
\binom{b^{\,\prime\prime\,(m)}_a(\vec k, t)}
{b^{\,\prime\prime\,(m)}_c(\vec k, t)}$ and introduce the temporal
destruction exponents from Eq.~(\ref{eq:destruct}).
Inserting the ansatz~(\ref{eq:timesub}) into Eq.~(\ref{eq:coupledAmpsArb})
leads to
\begin{eqnarray}
  \label{eq:coupledAmpsNoDiagGen}
  \dfrac{\partial}{\partial t} \, \vec b^{\,\prime\prime\,(m)}(\vec k, t)
  &=& \Bigl( -\dfrac{\imag}{2} \, R^{\,\prime\,(m)}\X{0X}(t) \,
    \mat \Sigma_x(t) \nonumber \\
  &&\qquad{} - \imag \, \Bigl(\dfrac{\vec k^{\,2}}{2} + \Cal I\X{P} \Bigr) \,
    \unitmatrix \Bigr) \, \vec b^{\,\prime\prime\,(m)}(\vec k, t) \nonumber \\
  &&{} + E\X{L}(t) \, \dfrac{\partial}{\partial k_z} \,
    \vec b^{\,\prime\prime\,(m)}(\vec k, t) \\
  &&{} - \imag \, E\X{L}(t) \, \wp^{(m)}_{\mathrm L, \vec k \, a}
    \, \binom{\euler^{-\tfrac{\imag}{2} \, \delta \, t +
    \tfrac{\digamma_a(t)}{2}}}{0} \, a'(t) \; , \nonumber
\end{eqnarray}
where I employ the matrix
\begin{equation}
  \label{eq:offdiag}
  \mat \Sigma_x(t) = \left(
    \begin{matrix}
      0 & \euler^{\Delta(t)} \\
      \euler^{-\Delta(t)} & 0
  \end{matrix}
  \right) \; ,
\end{equation}
with the time- but not momentum-dependent exponent
\begin{equation}
  \label{eq:diffphase}
  \Delta(t) \equiv -\imag \, \delta \, t + \imag \, \varphi\X{X}(t)
    + \dfrac{1}{2} \, \bigl( \digamma_a(t) - \digamma_c(t) \bigr) \; .
\end{equation}
A diagonalization of~$\mat \Sigma_x(t)$ [Eq.~(\ref{eq:offdiag})] leads
to~\cite{SuppData} time-independent eigenvalues
\begin{equation}
  \label{eq:eigenvalARB}
  \mat \Lambda' \equiv \diag(\lambda'_+, \lambda'_-) = \mat \sigma_z \; ,
\end{equation}
and~\cite{SuppData} time-dependent eigenvectors
\begin{equation}
  \label{eq:eigenvecARB}
  \mat U^{\,\prime}(t) = \left( \begin{matrix}
    \euler^{\Delta(t)} & -\euler^{\Delta(t)} \\
    1 & 1
  \end{matrix} \right) \; .
\end{equation}
I transform the EOMs~(\ref{eq:coupledAmpsNoDiagGen}) to the
eigenbasis~(\ref{eq:eigenvecARB}) of~$\mat \Sigma_x(t)$
[Eq.~(\ref{eq:offdiag})]---in analogy to the \xray-dressed-state basis used
before [Eq.~(\ref{eq:decoupledEOM})]---yielding the new amplitudes
[compare with Eq.~(\ref{eq:decoupledAmps})]:
\begin{equation}
  \label{eq:decoupledAmpsArb}
  \vec {\mathfrak b}^{\,\prime\,(m)}(\vec k, t) \equiv \binom{{\mathfrak
    b}^{\,\prime\,(m)}_+(\vec k, t)}{{\mathfrak b}^{\,\prime\,(m)}_-(\vec k, t)}
    = \mat U^{\,\prime\,-1}(t) \, \vec b^{\,\prime\prime\,(m)}(\vec k, t) \; ,
\end{equation}
and the new EOMs
\begin{eqnarray}
  \label{eq:eigenSigmax}
  \dfrac{\partial}{\partial t} \, \vec {\mathfrak b}^{\,\prime\,(m)}(\vec k, t)
    &=& \Bigl( \dot {\mat U}^{\prime\,-1}(t) \; \mat U'(t)
    - \dfrac{\imag}{2} \, R^{\,\prime\,(m)}\X{0X}(t) \, \mat \Lambda'
    \nonumber \\
  &&\quad{} - \imag \, \Bigl( \dfrac{\vec k^{\,2}}{2} + \Cal I\X{P} \Bigr) \,
    \unitmatrix \Bigr) \, \vec {\mathfrak b}^{\,\prime\,(m)}(\vec k, t) \\
  &&{} + E\X{L}(t) \, \dfrac{\partial}{\partial k_z} \,
    \vec {\mathfrak b}^{\,\prime\,(m)}(\vec k, t) \nonumber \\
  &&{} - \imag \, E\X{L}(t) \, \wp^{(m)}_{\mathrm L, \vec k \, a}
    \, \vec w^{\,\prime}(t) \, a'(t) \; . \nonumber
\end{eqnarray}
I added~$\dot{\mat{U}}^{\,\prime\,-1}(t) \> \mat U(t) \> \vec {\mathfrak
b}^{\,\prime}(\vec k, t)$ on both sides of Eq.~(\ref{eq:eigenSigmax})
which is the transformation rest that reads using Eq.~(\ref{eq:diffphase}):
\begin{equation}
  \label{eq:trafrest}
  \dot{\mat U}^{\,\prime\,-1}(t)\; \mat U^{\,\prime}(t) =
    \dfrac{\dot\Delta(t)}{2} \> (-\unitmatrix + \mat \sigma_x) \; .
\end{equation}
The valence ionization fraction in the eigenbasis~(\ref{eq:eigenvecARB}) becomes
\begin{eqnarray}
  \label{eq:valfracARB}
  \vec w^{\,\prime}(t) &\equiv& \binom{w^{\,\prime}_+(t)}{w^{\,\prime}_-(t)}
  = \mat U^{\,\prime\,-1}(t) \, \binom{\euler^{-\tfrac{\imag}{2}
    \, \delta \, t + \tfrac{\digamma_a(t)}{2}}}{0} \\
  &=& \tfrac{1}{2} \> \euler^{\tfrac{\imag}{2} \, \delta \, t - \imag
    \, \varphi\X{X}(t) + \tfrac{\digamma_c(t)}{2}} \binom{1}{-1} \; , \nonumber
\end{eqnarray}
[compare with Eq.~(\ref{eq:valionfrac})].

\subsection{Iterative solution of the equations of motion}

Due to the fact that $\dot{\textbf{\textit{U}}}^{\,\prime\,-1}(t) \;
\mat U^{\,\prime}(t)$ [Eq.~(\ref{eq:trafrest})] is not diagonal and
time dependent, I cannot simplify the EOMs further without making
additional approximations.
Therefore, I derive an iterative solution of the EOMs in what follows.

\subsubsection{Zeroth-iteration amplitudes}
\label{sec:zerothsolution}

In order to decouple the two differential equations~(\ref{eq:eigenSigmax}),
I need to approximate the transformation rest~(\ref{eq:trafrest})
by its diagonal elements, \ie, $\dot{\textbf{\textit{U}}}'\,^{-1}(t)\;
\mat U'(t) \approx -\tfrac{\dot\Delta(t)}{2} \> \unitmatrix$.
I refer to this as zeroth iteration and attach a superscript~``$^{(m,0)}$'' to
the involved coefficients.

As in Eq.~(\ref{eq:betasubstitution}), I make the
substitution~$\vec k^{\,\prime} = \vec k - \vec A\X{L}(t) + \vec
A\X{L}(t^{\,\prime})$ at time~$t^{\,\prime}$ giving
\begin{eqnarray}
  \dfrac{\differential}{\differential t^{\,\prime}} \,
    \vec{\mathfrak b}^{\,\prime\,(m,0)}
    (\vec k^{\,\prime}, t^{\,\prime}) &=& -\dfrac{\imag}{2} \, \bigl[
    R^{\,\prime\,(m)}\X{0X}(t) \, \mat \Lambda' + 2 \, \bigl( \Cal I\X{P}
    - \imag \, \dot\Delta(t) \bigr) \, \unitmatrix \nonumber \\
  &&\qquad{} + \bigl(\vec k - \vec A\X{L}(t) + \vec A\X{L}(t^{\,\prime})
    \bigr)^2 \, \unitmatrix \bigr] \nonumber \\
  &&\qquad{} \times \vec{\mathfrak b}^{\,\prime\,(m,0)} (\vec k^{\,\prime},
    t^{\,\prime}) \\
  &&{} - \imag \, E\X{L}(t^{\,\prime}) \, \wp^{(m)}_{\mathrm L, \vec k
    - \vec A\X{L}(t) + \vec A\X{L}(t^{\,\prime}) \, a} \nonumber \\
  &&\qquad{} \times \vec w^{\,\prime\,(m)} \, a'(t^{\,\prime}) \; . \nonumber
\end{eqnarray}
This equation can be integrated exactly~\cite{Arfken:MM-05},
analogously to Eq.~(\ref{eq:betasolution}), yielding
\begin{eqnarray}
  \label{eq:solutionArb}
  {\mathfrak b}'^{\,(m,0)}_{\pm}(\vec k, t) &=& -\imag \, \Int_0^t
    E\X{L}(t^{\,\prime}) \, \wp^{(m)}_{\mathrm L, \vec k - \vec A\X{L}(t)
    + \vec A\X{L}(t^{\,\prime}) \, a} \\
  &&{} \times \euler^{-\imag \, S^{\,\prime\,(m)}_{\pm}(\vec k
    - \vec A\X{L}(t), t, t^{\,\prime})} \> w'_{\pm}(t^{\,\prime}) \,
    a'(t^{\,\prime}) \differential t^{\,\prime} \; . \nonumber
\end{eqnarray}
The quasiclassical action [compare with Eq.~(\ref{eq:action})]---after
introducing the canonical momentum~$\vec p = \vec k - \vec A\X{L}(t)$---is
given by
\begin{eqnarray}
  S^{\,\prime\,(m)}_j(\vec p, t, t^{\,\prime}) &=& \dfrac{1}{2} \,
    \Int_{t^{\,\prime}}^t \bigl(\vec p + \vec A\X{L}(t^{\,\prime\prime})
    \bigr)^2 \differential t^{\,\prime\prime} \nonumber \\
  &&{} + \dfrac{\lambda'_j}{2} \, \bigl( \Theta^{(m)}(t)
    - \Theta^{(m)}(t^{\,\prime}) \bigr) \\
  &&{} - \dfrac{\imag}{2} \, \bigl( \Delta(t) - \Delta(t^{\,\prime}) \bigr)
    + \Cal I\X{P} \, (t - t^{\,\prime}) \; , \nonumber
\end{eqnarray}
where the area~\cite{Meystre:QO-99,Cavaletto:RF-12} of the \xray~pulse is
\begin{equation}
  \label{eq:pulsearea}
  \Theta^{(m)}(t) = \theta(t) \, \Int_0^t R^{\,\prime\,(m)}\X{0X}(t^{\,\prime})
    \differential t^{\,\prime} \; .
\end{equation}

\subsubsection{First-iteration amplitudes}

I obtain the zeroth-iteration solution~(\ref{eq:solutionArb}) of
the coupled EOMs [Eq.~(\ref{eq:coupledAmpsArb})] by taking only the diagonal
elements of the transformation rest~(\ref{eq:trafrest}) into account.
The first-iteration amplitudes~$\vec {\mathfrak b}'^{\,(m,1)}(\vec k, t)$
for~$m \in \mathbb M_2$ are found by inserting the zeroth-iteration
result from Sect.~\ref{sec:zerothsolution} into Eq.~(\ref{eq:eigenSigmax}) to
specify the off-diagonal elements of Eq.~(\ref{eq:trafrest}).
I find with Eqs.~(\ref{eq:eigenSigmax}), (\ref{eq:trafrest}), and
(\ref{eq:solutionArb}) the first-iteration EOMs
\begin{eqnarray}
  \dfrac{\partial}{\partial t} \, \vec {\mathfrak b}^{\,\prime\,(m,1)}
    (\vec k, t)
  &=& \Bigl[ - \imag \, \Bigl( \dfrac{\vec k^{\,2}}{2} + \Cal I\X{P}
    - \dfrac{\imag}{2} \, \dot \Delta(t) \Bigr) \, \unitmatrix \nonumber \\
  && \  {} - \dfrac{\imag}{2} \, R^{\,\prime\,(m)}\X{0X}(t) \, \mat \Lambda'
    \Bigr] \, \vec {\mathfrak b}^{\,\prime\,(m,1)}(\vec k, t) \nonumber \\
  &&{} + E\X{L}(t) \, \dfrac{\partial}{\partial k_z} \,
    \vec {\mathfrak b}^{\,\prime\,(m,1)}(\vec k, t) \\
  &&{} - \imag \, E\X{L}(t) \, \wp_{\mathrm
    L, \vec k \, a} \, \vec w^{\,\prime}(t) \, a'(t) \nonumber \\
  &&{} + \tfrac{1}{2} \, \dot \Delta(t) \> \mat \sigma_x \>
    \vec {\mathfrak b}^{\,\prime\,(m,0)}(\vec k, t) \; . \nonumber
\end{eqnarray}
These equations are still decoupled and can be solved along the lines
leading to Eq.~(\ref{eq:solutionArb}).
By identifying the zeroth-iteration solution~(\ref{eq:solutionArb}), I obtain
\begin{eqnarray}
  \label{eq:firstbeta}
  {\mathfrak b}_{\pm}^{\prime\,(m,1)}(\vec k, t) &=& {\mathfrak
    b}_{\pm}^{\prime\,(m,0)}(\vec k, t) + \dfrac{1}{2} \, \Int_0^t
    \dot \Delta(t^{\,\prime}) \\
  &&{} \times \euler^{-\imag \, S^{\,\prime\,(m)}_{\pm}(\vec k
    - \vec A\X{L}(t), t, t^{\,\prime})} \> {\mathfrak b}_{\mp}^{\prime\,(m,0)}
    (\vec k^{\,\prime}, t^{\,\prime}) \differential t^{\,\prime} \nonumber \\
  &=& {\mathfrak b}_{\pm}^{\prime\,(m,0)}(\vec k, t) + \Delta {\mathfrak
    b}_{\pm}^{\prime\,(m,1)}(\vec k, t) \; . \nonumber
\end{eqnarray}

\subsection{Electric dipole transition matrix element and high-order
harmonic spectrum}
\label{sec:TDdipHHG}

The time-dependent $N/2$-electron dipole transition matrix element
follows from Eqs.~(\ref{eq:Nhalfdipole}) and (\ref{eq:diptransmat}) to
\begin{eqnarray}
  \label{eq:diptransmatArb}
  \Cal D^{\,\prime}(t) &=& \bra{\Psi^{\,\prime}_0,t} \hat D
    \ket{\Psi^{\,\prime}\X{c}, t} \\
  &=& \Sum_{m \in \mathbb M_1} \mathfrak d^{\prime \, (m)}(t)
    + \Sum_{\atopa{\scriptstyle i \in \{a, c\}}{\scriptstyle j
    \in \{+, -\}}} \Sum_{m \in \mathbb M_2} \mathfrak D'^{\,(m)}_{ij}(t)
    \; , \nonumber
\end{eqnarray}
where I denote by~$\ket{\Psi^{\,\prime}_0,t}$~the ground-state part of the new
wave packet~(\ref{eq:wavepacket}) and by~$\ket{\Psi^{\,\prime}\X{c}, t}$~its
continuum part~\cite{Kuchiev:QT-99}.
The dipole components~$\mathfrak D'^{\,(m)}_{ij}(t)$ for~$m \in \mathbb M_2$
are found by transforming~$\vec {\mathfrak b}'^{\,(m)}(\vec k, t)$
[Eq.~(\ref{eq:solutionArb})] back to the bare-state
amplitudes~$\vec b^{\,\prime(m)}(\vec k, t)$ with the inverse of
Eqs.~(\ref{eq:timesub}) and (\ref{eq:decoupledAmpsArb});
they read for~$i \in \{a ; c\}$ and $j \in \{- ; +\}$:
\begin{eqnarray}
  \label{eq:dipcompArb}
  \mathfrak D^{\prime\,(m)}_{ij}(t) &=& (-1)^{\delta_{i\,c}} \,
    U^{\,\prime}_{ij}(t) \; \euler^{\imag \, \phi_i(t)} \, a'^{\,*}(t) \\
  &&{} \times \Int_{\mathbb R^3} \wp^{(m)}_{\mathrm H, i \, \vec k} \;
    {\mathfrak b}^{\prime\,(m)}_j(\vec k,t) \differential^3 k \; , \nonumber
\end{eqnarray}
with the time-dependent phase
\begin{equation}
  \label{eq:dipphase}
  \phi_i(t) = (-1)^{\delta_{i\,c}} \, \dfrac{\delta}{2} \, t
    - \delta_{i\,c} \, \omega\X{X} \, t + \imag \, \dfrac{\digamma_i(t)}{2} \; .
\end{equation}
The zeroth-iteration dipole components~$\mathfrak D^{\prime\,(m,0)}_{ij}(t)$
follow immediately by substituting~${\mathfrak b}^{\prime\,(m,0)}_j(\vec k,t)$
[Eq.~(\ref{eq:solutionArb})] for~${\mathfrak b}^{\prime\,(m)}_j(\vec k,t)$
in Eq.~(\ref{eq:dipcompArb}).
Inserting the first-iteration solution~(\ref{eq:firstbeta}) into
Eq.~(\ref{eq:dipcompArb}), I arrive at the expansion
\begin{equation}
  \label{eq:dipcompArbExp}
  \mathfrak D'^{\,(m,1)}_{ij}(t) = \mathfrak D'^{\,(m,0)}_{ij}(t)
    + \Delta \mathfrak D'^{\,(m,1)}_{ij}(t) \; ,
\end{equation}
for the first-iteration dipole components.

\subsubsection{Zeroth-iteration dipole components}
\label{sec:zerothorder}

In zeroth-iteration, the dipole components~$\mathfrak D'^{\,(m,0)}_{ij}(t)$
[Eq.~(\ref{eq:dipcompArb})] are obtained from the zeroth-iteration
amplitudes~(\ref{eq:solutionArb}).
I assume a constant-amplitude optical
laser pulse which starts at~$t = 0$ and ends at~$t = T\X{P}$.
As the dipole components~(\ref{eq:dipcompArb}) are not periodic in time, they
are Fourier transformed~\cite{Lewenstein:HH-94,Kuchiev:QT-99,Milosevic:SI-01}.
I introduce the canonical momentum~$\vec p = \vec k - \vec A\X{L}(t)$ and
the excursion time~$\tau = t - t^{\,\prime}$, expand the field-dipole product
into a Fourier series [Eqs.~(\ref{eq:atomicTerms}), (\ref{eq:FourSeries}),
(\ref{eq:FourAtomic}), and (\ref{eq:dipMom})] and make the saddle-point
approximation [see Eq.~(\ref{eq:dipcompst}) and Appendix~\ref{sec:saddlepoint}]
which yields
\begin{eqnarray}
  \label{eq:dipcontinuousArb}
  \tilde{\mathfrak D}'^{\,(m,0)}_{ij}(\omega) &=& -2 \, \pi \>
    (-1)^{\delta_{i\,c}} \> \imag \> \Int_0^{\infty}
    \sqrt{\tfrac{(-2 \pi \imag)^3}{\tau^3}} \>
    \euler^{-\imag \, F^{\,\prime}_0(\tau)} \nonumber \\
  &&{} \times \Sum_{N = -\infty}^{\infty} \imag^N J_N \bigl(
    \tfrac{U\X{P}}{\omega\X{L}} \> C(\tau) \bigr) \; \euler^{\imag
    \, N \, \omega\X{L} \, \tau} \\
  &&{} \times \Sum_{M = -\infty}^{\infty} \tilde {\mathfrak
    A}^{(m)}_{M-N,i}(\tau) \, h'^{\,(m,0)}_{M,i,j}(\omega, \tau)
    \differential \tau \; , \nonumber
\end{eqnarray}
where I expand the exponential of the quasiclassical action at the
stationary point, use~$F^{\,\prime}_0(\tau)$ from Eq.~(\ref{eq:lewF0}),
$C(\tau)$~from Eq.~(\ref{eq:funC}), translate the sum over~$M$
by~$M \to M - N$, and define the zeroth-iteration HH~line shape
[compare with Eq.~(\ref{eq:line})]:
\begin{eqnarray}
  \label{eq:lineArb}
  h'^{\,(m,0)}_{M,i,j}(\omega, \tau) &=& \dfrac{1}{2 \, \pi}
    \Int_0^{T\X{P}} \euler^{-\imag \, [ (2 \, M
    + \delta_{i\,a}) \, \omega\X{L} - \omega ] \, t} \>
    \euler^{\imag \, \phi_i(t)} \nonumber \\
  &&{} \times \euler^{-\tfrac{\imag}{2} \, \lambda'_j \,
    (\Theta^{(m)}(t)-\Theta^{(m)}(t-\tau))} \\
  &&{} \times \euler^{-\tfrac{1}{2} \, (\Delta(t) - \Delta(t-\tau))} \>
    U'_{ij}(t) \nonumber \\
  &&{} \times w'_j(t - \tau) \, a'^{\,*}(t) \, a'(t-\tau)
    \differential t \; . \nonumber
\end{eqnarray}
The result [Eqs.~(\ref{eq:dipcontinuousArb}) and (\ref{eq:lineArb})]
has a close resemblance to the dipole components for a constant-amplitude
\xray~pulse with ground-state depletion [Eqs.~(\ref{eq:dipcontinuous}) and
(\ref{eq:line})].
However, here, the eigenvectors~$\mat U^{\,\prime}(t)$ transform
from a different basis to bare states and are time dependent.
Hence $\mat U^{\,\prime}(t)$ and $\vec w^{\,\prime}(t)$~enter
Eq.~(\ref{eq:lineArb}) in contrast to Eq.~(\ref{eq:line}).

To derive a closed-form expression from Eq.~(\ref{eq:lineArb}), I need
to solve the new Eq.~(\ref{eq:groundepl}) where I omit the second term
on the right-hand side of the equation and account for its influence
in~$\Gamma^{\,\prime}_0(t)$.
For an arbitrarily-shaped \xray~pulse and a constant-amplitude optical
laser pulse, where both pulses begin at~$t = 0$ and end at~$t = T\X{P}$,
the solution for the ground-state amplitude reads
\begin{equation}
  \label{eq:grounddeplARB}
  a'(t) = \theta(-t) + \euler^{-\tfrac{\digamma_0(t)}{2}} \>
    \theta(t) \, \theta(T\X{P} - t) + \euler^{-\tfrac{\digamma_0(T\X{P})}{2}}
    \> \theta(t - T\X{P}) \; ,
\end{equation}
similarly to Eq.~(\ref{eq:xraydepl}), however, with~$\Gamma_0$~replaced
by~$\digamma_0(t)$.
This expression is inserted into Eq.~(\ref{eq:lineArb}) to obtain
the line shape for~$m \in \mathbb M_2$;
I find with Eqs.~(\ref{eq:eigenvalARB}), (\ref{eq:eigenvecARB}),
(\ref{eq:valfracARB}), (\ref{eq:pulsearea}), and (\ref{eq:dipphase}) the
zeroth-iteration HH~line shape
\begin{eqnarray}
  \label{eq:linedeplARB}
  h'^{\,(m,0)}_{M,i,j}(\omega, \tau) &\approx& \dfrac{(-1)^{\delta_{i\,c} \,
    \delta_{j\,-}}}{4 \, \pi} \Int_{\tau}^{T\X{P}} \euler^{-\tfrac{\digamma_0(t)
    + \digamma_0(t-\tau)}{2}} \nonumber \\
  &&{} \times \euler^{-\imag \, [ (2 \, M + \delta_{i\,a}) \,
    \omega\X{L} + \delta_{i\,c} \, \omega\X{X} - \omega ] \, t} \nonumber \\
  &&{} \times \euler^{\frac{\imag}{2} \, (\varphi\X{X}(t) -
    \varphi\X{X}(t-\tau))} \> \euler^{-\imag \, \delta_{i\,c} \,
    \varphi\X{X}(t)} \\
  &&{} \times \euler^{\tfrac{-\digamma_a(t) - \digamma_c(t) +
    \digamma_a(t - \tau) + \digamma_c(t - \tau)}{4}} \nonumber \\
  &&{} \times \euler^{-\tfrac{\imag}{2} \, \lambda'_j \, (\Theta^{(m)}(t) -
    \Theta^{(m)}(t - \tau))} \differential t \; . \nonumber
\end{eqnarray}
Disregarding the impact of the other factors in the equation,
the line shape peaks around~$\omega = \tilde\omega_{M,i}$
[Eq.~(\ref{eq:linepeak})].
The other factors differ substantially from the result~(\ref{eq:lineConst}).
The reason for these contributions to occur is the only
partial diagonalization of the instantaneous Rabi
matrix~(\ref{eq:RabiMatGen}).
High-order harmonic emission is suppressed by the
destruction of the ground-state amplitude at the time of tunnel
ionization~$\digamma_0(t - \tau)$ and at the time of
recombination~$\digamma_0(t)$.
Further, the line-shape depends on $\digamma_a(t)$ and $\digamma_c(t)$~which
account for the destruction of the intermediate hole states by the optical
laser, the x~rays, and decay processes.
Inspecting Eq.~(\ref{eq:RabiEigval}), I see that via Eq.~(\ref{eq:destruct})
this term and the term depending on the pulse area~(\ref{eq:pulsearea})
can be understood to result from~$\Int_{t - \tau}^t
\tfrac{\lambda^{(m)}_{\pm}}{2} \differential t$ where
I retain only the leading order in the expansion
of the complex Rabi frequency~$\mu^{(m)} \approx R^{(m)}\X{0X}$
in~$\lambda^{(m)}_{\pm}$ where $R^{(m)}\X{0X}$~is assumed to be large with
respect to all other parameters in~$\mu^{(m)}$.

The dipole components~$\mathfrak d^{\prime \, (m)}(t)$ for~$m \in \mathbb M_1$
follow directly as in Sect.~\ref{sec:HHGtrama} where I, however,
set~$\lambda_0^{\prime \, (0)} = -\delta$ and the HH~line shape is
\begin{eqnarray}
  h'_M(\omega, \tau) &=& \dfrac{1}{2 \, \pi} \Int_0^{T\X{P}}
    \euler^{-\imag \, [ (2 \, M + 1) \, \omega\X{L}
    - \omega ] \, t} \nonumber \\
  &&{} \times \euler^{-\tfrac{\digamma_a(t) - \digamma_a(t-\tau)}{2}}
    \, a^{\,\prime\,*}(t) \, a^{\,\prime}(t-\tau) \differential t \\
  &\approx& \dfrac{1}{2 \, \pi} \Int_{\tau}^{T\X{P}}
    \euler^{-\imag \, [ (2 \, M + 1) \, \omega\X{L}
    - \omega ] \, t} \nonumber \\
  &&{} \times \euler^{-\tfrac{\digamma_0(t) + \digamma_0(t-\tau)}{2}}
    \euler^{-\tfrac{\digamma_a(t) - \digamma_a(t-\tau)}{2}}
    \differential t \nonumber
\end{eqnarray}

Finally, the spectral and solid-angle-dependent probability density of
HH~emission along the $x$~axis follows in zeroth iteration
from Eq.~(\ref{eq:HHGspecXray}).

\subsubsection{First-iteration dipole components}

To obtain first-iteration dipole components, the zeroth-iteration
solution [Eq.~(\ref{eq:solutionArb})] is inserted
into the expression for~$\Delta {\mathfrak b}_{\pm}^{\prime\,(m,1)}(\vec k, t)$
[Eq.~(\ref{eq:firstbeta})], the two quasiclassical actions are united and
treated with the saddle-point method [Appendix~\ref{sec:saddlepoint}].
Defining two excursion times~$\tau' = t - t^{\,\prime}$ and
$\tau'' = t - t^{\,\prime\prime}$ leads, with the expansion of
the field-dipole product~(\ref{eq:FourSeries}), to
the first-iteration correction of the dipole components~(\ref{eq:dipcompArb})
that manifests in a first-iteration correction of the HH~line
shape~(\ref{eq:lineArb}) in Eq.~(\ref{eq:dipcontinuousArb});
the expression for the first-iteration dipole component
corrections~$\Delta \tilde{\mathfrak D}'^{\,(m,1)}_{ij}(\omega)$
[Eq.~(\ref{eq:dipcompArbExp})] are formally the same as in
Eq.~(\ref{eq:dipcontinuousArb}) when letting~$\tau = \tau' + \tau''$,
using the identity~$\Int_0^t \differential \tau' \Int_0^{t - \tau'}
\differential \tau'' = \Int_0^t \differential \tau' \Int_{\tau'}^t
\differential \tau = \Int_0^t \differential \tau \Int_0^{\tau}
\differential \tau'$, and extending the upper bound of the outermost
integral to~$+\infty$.
The first-iteration correction to the HH~line shape for~$m \in \mathbb M_2$ is
\begin{eqnarray}
  \label{eq:lineArbFirst}
  \Delta h'^{\,(m,1)}_{M,i,j}(\omega, \tau) &=& \dfrac{1}{2 \, \pi}
    \Int_0^{T\X{P}} \euler^{-\imag \, [ (2 \, M
    + \delta_{i\,a}) \, \omega\X{L} - \omega ] \, t} \> \euler^{\imag
    \, \phi_i(t)} \nonumber \\
  &&{} \times \euler^{-\tfrac{\imag}{2} \, \lambda'_j \,
    (\Theta^{(m)}(t) + \Theta^{(m)}(t - \tau))} \nonumber \\
  &&{} \times \euler^{-\tfrac{1}{2} \, (\Delta(t) - \Delta(t - \tau))}
    \, U'_{ij}(t) \\
  &&{} \times w'_{-j}(t - \tau) \, a'^{\,*}(t) \, a'(t - \tau)
    \nonumber \\
  &&{} \times \Int_0^{\tau} \dfrac{1}{2} \> \dot \Delta(t - \tau') \;
    \euler^{\imag \, \lambda'_j \, \Theta^{(m)}(t - \tau')}
    \differential \tau' \differential t \; . \nonumber
\end{eqnarray}
The first-iteration correction~(\ref{eq:lineArbFirst}) differs
somewhat from the zeroth-iteration expression~(\ref{eq:lineArb}).
All factors for the integration over~$t$ are the same here as there
apart from~$w'_{-j}(t - \tau)$ here and $w'_j(t - \tau)$ there.
The dependence on~$\dot \Delta(t - \tau')$ [Eq.~(\ref{eq:diffphase})]
accounts for the off-diagonal matrix elements in Eq.~(\ref{eq:trafrest});
the integral over~$\tau'$ represents the accumulated influence of the
off-diagonal elements at time~$\tau$.
This integral is not present in Eq.~(\ref{eq:lineArb}).

I obtain a closed-form expression for~$\Delta h'^{\,(m,1)}_{M,i,j}
(\omega, \tau)$ from Eq.~(\ref{eq:lineArbFirst}) by inserting
Eqs.~(\ref{eq:eigenvalARB}), (\ref{eq:eigenvecARB}),
(\ref{eq:valfracARB}), (\ref{eq:pulsearea}), (\ref{eq:dipphase}),
and (\ref{eq:grounddeplARB}) which yields
\begin{eqnarray}
  \Delta h'^{\,(m,1)}_{M,i,j}(\omega, \tau) &\approx& \dfrac{(-1)^{\delta_{i\,c}
    \, \delta_{j\,-} + 1}}{4 \, \pi} \Int_{\tau}^{T\X{P}}
    \euler^{-\tfrac{\digamma_0(t) + \digamma_0(t-\tau)}{2}} \nonumber \\
  &&{} \times \euler^{-\imag \, [ (2 \, M + \delta_{i\,a}) \,
    \omega\X{L} + \delta_{i\,c} \, \omega\X{X} - \omega ] \, t} \nonumber \\
  &&{} \times \euler^{-\tfrac{\imag}{2} \, \varphi\X{X}(t - \tau)
    + \tfrac{\imag}{2} \, (-1)^{\delta_{i\,c}} \, \varphi\X{X}(t)} \nonumber \\
  &&{} \times \euler^{\tfrac{-\digamma_a(t) - \digamma_c(t) +
    \digamma_a(t - \tau) + \digamma_c(t - \tau)}{4}} \nonumber \\
  &&{} \times \euler^{-\tfrac{\imag}{2} \, \lambda'_j \, [\Theta^{(m)}(t) -
    \Theta^{(m)}(t - \tau)]} \nonumber \\
  &&{} \times \Int_0^{\tau} \tfrac{1}{2} \, \bigl[ -\imag \, \delta
    + \imag \> \dot \varphi\X{X}(t - \tau') \\
  &&\qquad{} + \tfrac{1}{2} \, \bigl( \Gamma^{\,\prime}_a(t - \tau')
    - \Gamma^{\,\prime}_c(t - \tau') \bigr) \bigr] \nonumber \\
  &&\qquad{} \times \euler^{\imag \, \lambda'_j \, \Theta^{(m)}(t - \tau')}
    \differential \tau' \differential t \; . \nonumber
\end{eqnarray}
The first-iteration HH~line shape for~$m \in \mathbb M_2$ reads
\begin{equation}
  h'^{\,(m,1)}_{M,i,j}(\omega, \tau) = h'^{\,(m,0)}_{M,i,j}(\omega, \tau)
    + \Delta h'^{\,(m,1)}_{M,i,j}(\omega, \tau) \;.
\end{equation}
The probability of the HH~emission along the $x$~axis follows by inserting
the sum of the zeroth-iteration dipole components
[Eq.~(\ref{eq:dipcontinuousArb})] and first-iteration
dipole corrections~$\Delta \tilde{\mathfrak D}'^{\,(m,1)}_{ij}(\omega)$
into Eqs.~(\ref{eq:compounddipcomp}) and (\ref{eq:HHGspecXray}).

\section{Conclusion}
\label{sec:conclusion}

I discuss theoretically the impact of \xray~excitation on HHG.
For this purpose, I develop a two-electron model that is based on
the approach of Lewenstein~\etal~\cite{Lewenstein:HH-94} for optical-laser-only
HHG.
In my model, the first electron from the atomic valence is tunnel ionized
and, thereafter, propagates freely in the continuum;
a second electron from the atomic core is driven by intense x~rays
that are tuned to the core-valence resonance in the transient ion.
For ultrahigh \xray~intensities, this electron may even Rabi flop between the
valence and the core state prior recombination of the continuum electron
with the parent ion.
The optical laser eventually reverses its direction and the first electron
may be driven back to the parent ion where it sees a superposition of
valence-hole and core-hole states and recombines with it emitting
HH~radiation that is characteristic of this superposition.
Valence-hole recombination leads to the formation of a first HH~plateau which
is, apart from a slight influence due to \xray~dressing, the same as the
HH~plateau that is obtained from optical-laser-only HHG.
Core-hole recombination causes the formation of a second HH~plateau
that is shifted by the \xray~photon energy to larger HH~energies
with respect to the first plateau.
The expressions in this article are formulated for, first, cw~optical laser
and cw~x~rays, second, constant-amplitude optical-laser and
x~rays with finite pulse durations, and, third, a constant-amplitude
optical laser together with arbitrarily-shaped \xray~pulses.
Yet, for pulsed optical light, one will need to take into account that the
field-dipole product~(\ref{eq:atomicTerms}) can no longer be expanded into
a Fourier series~(\ref{eq:FourSeries}) and the equations discussed here
need to be generalized accordingly.
A benefit of the chosen approach here is that the equations can be treated to
a large degree analytically in contrast to the case when also the
optical laser produces arbitrarily-shaped pulses.

I focus on the HH~spectrum of a single atom.
However, in experiments, a macroscopic sample is used for~HHG.
The copropagation of the optical laser together with the
\xray~radiation from a FEL and HHG through the gas transforms the
HH~spectrum significantly due to interferences effects caused by coherent
emission of light in the sample~\cite{LHuillier:TA-91,Balcou:QP-99,%
Priori:NT-00,Gaarde:MA-08,Popmintchev:BC-12}.
The phase matching shall be examined in future studies specifically under the
objective of the impact of SASE FEL x~rays on the HHG~process;
a simple estimate of the yield of a different \xray-boosted HHG~scheme
in SAE---for which tunnel ionization is replaced by one-\xray-photon ionization
of a core electron---shows that a considerable output can be achieved
from a macroscopic sample~\cite{Buth:KE-13}.

The model of Lewenstein~\etal~\cite{Lewenstein:HH-94} describes excellently
qualitatively HH~spectra but tends to overestimate the HH~yield
significantly by roughly two orders of magnitude~\cite{Gordon:QM-05}.
The inaccuracies of the model can be mitigated by using
the Eikonal-Volkov approximation~(EVA)~\cite{Smirnova:AS-08} and considering
electron correlations~\cite{Santra:TS-06,Gordon:RM-06}.
In the case of a SAE, a numerical integration of the time-dependent
Schr\"odinger equation is a route to obtain an accurate HH~yield.
However, in my case, a numerical integration of a $N/2$-electron
Schr\"odinger equation would be required with substantially increased
complexity~\cite{Schafer:NM-09}.

My prediction of \xray-boosted-HHG offers novel prospects for
nonlinear \xray{} physics that complements the nonlinear \xray-only
processes of sequential and simultaneous absorption of
multiple x~rays~\cite{Doumy:NA-11,Buth:UA-12}.
Namely, the excursion of the tunnel-ionized electron in the HHG~process
determines a time window in which resonant \xray~interactions may occur.
Rabi flopping is a highly nonlinear fundamental process and becomes
experimentally feasible with x~rays for the first time using
FELs~\cite{Rohringer:RA-08,Rohringer:PN-08,Kanter:MA-11,Rohringer:SD-12,%
Cavaletto:RF-12,Adams:QO-13}.
The occurrence of Rabi oscillations of the second electron between
valence and core states prior to recombination of the first electron
may have a much clearer signature if combined with the HHG~scheme
compared with the case if only x~rays are considered.
The manipulation of HHG with x~rays may also facilitate
frequency-resolved optical gating~(FROG)~\cite{DeLong:PR-94,Trebino:FR-02}
with \xray~pulses thus offering the long-sought after pulse characterization
for chaotic SASE FEL x~rays~\cite{Dusterer:FS-11,Meyer:AR-12} but
this issue requires further theoretical research.
The novel scheme also makes single attosecond
\xray~pulses~\cite{Hentschel:AM-01,Sansone:IS-06} and attosecond
\xray~pulse trains~\cite{Paul:TA-01} feasible using the same methods
that are used with conventional~HHG which has been predicted recently
using one-\xray-photon ionization of core electrons for boosting
HHG~\cite{Buth:KE-13}.
Above all, tomographic imaging of core orbitals with HHG comes into
reach~\cite{Itatani:TI-04,Santra:IM-06,Santra:TS-06,Morishita:AR-08,Lin:UD-12}.
Finally, HHG has been used to generate frequency combs in the \XUV{};
potentially, such HHG spectra can be boosted by x~rays in order
to extend frequency-comb-based spectroscopy to the
\xray~regime~\cite{Cingoz:DF-12,Young:PM-12} provided that
one has identically-shaped \xray~pulses which are carefully synchronized
to the optical laser.
This would complement a competing approach of \xray~frequency comb generation
using resonance fluorescence and coherent \xray{} pulse
shaping~\cite{Cavaletto:FC-13,Adams:QO-13,Cavaletto:HF-14}.

\begin{CJK*}{UTF8}{}
\begin{acknowledgments}
I am indebted to Christoph H.~Keitel and Markus C.~Kohler,
for their continuous encouragement, helpful
discussions, and a critical reading of the manuscript.
I am grateful to Karen Z.~Hatsagortsyan, Feng He
({\CJKfamily{gbsn}何峰}), and Robin Santra for helpful discussions.
I was supported by a Marie Curie International Reintegration
Grant within the 7$^{\mathrm{th}}$~European Community Framework Program
(call identifier: FP7-PEOPLE-2010-RG, proposal No.~266551).
\end{acknowledgments}
\end{CJK*}

\appendix
\section{Electronic structure}
\label{sec:elstructmodel}

In the nonrelativistic Hartree-Fock-Slater independent-electron
approximation~\cite{Slater:AS-51,Herman:AS-63,Slater:XA-72,Buth:TX-07},
the electronic structure of an atom is given by the orbital energies
[in~$\hat h\X{A}$, Eq.~(\ref{eq:oneatstruc})] and the spatial atomic orbitals
in spherical polar coordinates~\cite{Arfken:MM-05}---with radius~$r
=||\vec r||$ and solid angle~$\Omega$---expressed
by~$\bracket{\vec r}{i ; m_i} = R_{n_i \, l_i}(r)
\, Y_{l_i \, m_i}(\Omega)$ with the radial part~$R_{n_i \, l_i}(r)$
and the angular part, a spherical harmonic~\cite{Rose:ET-57},
$Y_{l_i \, m_i}(\Omega)$ for~$i \in \{ a, c\}$~\cite{Buth:TX-07}.
In optical-laser and \xray~light~(\ref{eq:hamiltonian}), the spatial
orbitals enter the electric dipole transition matrix elements
in~$\hat h\X{L}$ [Eq.~(\ref{eq:honeL})], $\hat h\X{X}$
[Eq.~(\ref{eq:honeX})], and $\hat D$ [Eq.~(\ref{eq:diptransmat})].
There are three types of dipole transition matrix elements:
first, bound-bound transitions [Sect.~\ref{sec:BBtrans}] between
core and valence states~(\ref{eq:xrayexcite}), second,
bound-continuum transitions [Sect.~\ref{sec:BCtrans}]
between the valence and continuum states~(\ref{eq:tunnelmatel}),
and, third, continuum-continuum transitions [Sect.~\ref{sec:CCtrans}]
between continuum states~(\ref{eq:CCtrans}).

To express the electric dipole transition operator in terms of spherical
polar coordinates~\cite{Arfken:MM-05}, I employ the relation
\begin{equation}
  \label{eq:polposprod}
  \vec e_{\lambda} \mul \vec r = r \; \sqrt{\dfrac{4 \pi}{3}} \;
    Y_{1 \, \lambda}(\Omega) \; ,
\end{equation}
with~$\lambda \in \{-1, 0, 1\}$, $\vec e_0 = \vec e_z$, and
$\vec e_{\pm 1} = \mp \tfrac{1}{\sqrt{2}} \> (\vec e_x \pm \imag \,
\vec e_y)$~\cite{Rose:ET-57,Cavaletto:RF-12}.
The $\vec e_{+1}$~is almost the left-circularly polarization (positive helicity)
vector~$\vec e^{\;(\mathrm L)} = -\vec e_{+1}$ whereas $\vec e_{-1}$~is
equal to the right-circularly polarization (negative helicity)
vector~$\vec e_{-1} = \vec e^{\;(\mathrm R)}$~\cite{Craig:MQ-84}.
In the scalar product~$\vec e_{\lambda} \mul \vec r$,
complex conjugation is on the second factor, here~$\vec r$,
which is real throughout.

\subsection{Bound-bound transitions}
\label{sec:BBtrans}

Electric dipole transition matrix elements between two bound states
are expressed in terms of spatial orbitals as in Eq.~(\ref{eq:dipMom});
they are
\begin{eqnarray}
  \label{eq:totBBmatel}
  &&\bra{c ; m_c} \vec e_{\lambda} \mul \vec r \ket{a ; m_a} \nonumber \\
  &&\hspace{3em} = \Int_{\mathbb R^3} \bracket{c ; m_c}{\vec r} \>
    \vec e_{\lambda} \mul \vec r \; \bracket{\vec r}{a ; m_a}
    \differential^3 r \\
  &&\hspace{3em} = \delta_{m_a \pm \lambda \, m_c} \; \sqrt{\dfrac{4 \pi}{3}} \;
    \threeY{l_a,1,l_c}{m_a,\lambda,m_c} \; \Cal R_{\mathrm{BB}, ca}
    \; , \nonumber
\end{eqnarray}
upon inserting Eq.~(\ref{eq:polposprod}) and the spatial
orbitals~$\bracket{\vec r}{i ; m_i} = R_{n_i \, l_i}(r) \,
Y_{l_i \, m_i}(\Omega)$ with~$i \in \{ a, c\}$.
The angular integral is
\begin{widetext}
\begin{equation}
  \label{eq:threeY}
  \begin{array}{rcl}
    \threeY{l_1,l_2,l_3}{m_1,m_2,m_3} &=& \Int_{4 \, \pi}
      Y^*_{l_3 \, m_3}(\Omega) \, Y_{l_2 \, m_2}(\Omega) \,
      Y_{l_1 \, m_1}(\Omega) \differential \Omega \\
    &=& \sqrt{\dfrac{(2 l_1 + 1) (2 l_2 + 1)}{4 \pi \, (2 l_3 + 1)}} \>
      \cleb{l_1,l_2,l_3}{m_1,m_2,m_3} \> \cleb{l_1,l_2,l_3}{0,0,0} \; ,
  \end{array}
\end{equation}
\end{widetext}
where $\cleb{l_1,l_2,l_3}{m_1,m_2,m_3}$~is a Clebsch-Gordan
coefficient~\cite{Rose:ET-57}.
The integral restricts the accessible angular momenta and magnetic
quantum numbers in the photoexcitation process such that only
transitions occur where~$m_c = m_a \pm \lambda$~holds for the
magnetic and $l_c \in \{ |l_a - 1|, l_a + 1 \}$~for the
angular quantum numbers.
The radial dipole matrix element is
\begin{equation}
  \Cal R_{\mathrm{BB}, ca} = \Int_0^{\infty} R_{n_c \, l_c}(r) \, r^3 \,
    R_{n_a \, l_a}(r) \differential r \; ,
\end{equation}

For~$\vec e\X{X} = \vec e_z = \vec e_0$, the matrix element
in Eq.~(\ref{eq:xrayexcite}) is obtained from Eq.~(\ref{eq:totBBmatel}).

\subsection{Bound-continuum transitions}
\label{sec:BCtrans}

Electric dipole transition matrix elements between bound and continuum
states~\cite{Merzbacher:QM-98,Buth:TA-09} read
\begin{equation}
  \label{eq:XUVdipPlane}
  \bra{\vec k} \vec e_{\lambda} \mul \vec r \ket{a ; m_a}
    = \dfrac{1}{(2\pi)^{3/2}} \Int_{\mathbb R^3} \euler^{-\imag \,
    \vec k \mul \vec r} \; \vec e_{\lambda} \mul \vec r \;
    \bracket{\vec r}{a ; m_a} \differential^3 r \; .
\end{equation}
I use Eq.~(\ref{eq:polposprod}), the spatial atomic valence
orbital~$\bracket{\vec r}{a ; m_a} = R_{n_a \, l_a}(r)
\, Y_{l_a \, m_a}(\Omega_r)$, and the spatial part
of the plane wave~(\ref{eq:freeelmom}).
Expression~(\ref{eq:XUVdipPlane}) is recast employing
the Rayleigh expansion~\cite{Rose:ET-57} of plane waves
\begin{equation}
  \label{eq:sphereexpand}
  \euler^{\imag \, \vec k \mul \vec r} = 4 \pi \Sum_{l = 0}^{\infty}
    \Sum_{m = -l}^l \imag^l \, Y^*_{l \, m}(\Omega_k) \, j_l(k \, r) \,
    Y_{l \, m}(\Omega_r) \; .
\end{equation}
The directions of~$\vec k$ and $\vec r$ are specified by the solid
angles~$\Omega_k$ and $\Omega_r$, respectively, and the magnitudes
by~$k = ||\vec k||$ and $r = ||\vec r||$.
Here, $j_l$~denotes a spherical Bessel function~\cite{Arfken:MM-05}.
Upon inserting Eq.~(\ref{eq:polposprod}) into Eq.~(\ref{eq:XUVdipPlane}),
I arrive at the dipole matrix element~(\ref{eq:XUVdipPlane})
in spherical polar coordinates,
\begin{widetext}
\begin{equation}
  \label{eq:XUVdipexpanded}
  \bra{\vec k} \vec e_{\lambda} \mul \vec r \ket{a ; m_a} = 2 \,
    \sqrt{\tfrac{2}{3}} \Sum_{l \in \{|l_a-1|, l_a+1\}} (-\imag)^l \>
    \threeY{l_a, 1, l}{m_a, \lambda, m_a + \lambda} \> Y_{l \, m_a +
    \lambda}(\Omega_k) \, \Cal R_{\mathrm{BC}, a}^{(l)}(k) \; .
\end{equation}
\end{widetext}
The radial dipole matrix element is
\begin{equation}
  \Cal R_{\mathrm{BC}, a}^{(l)}(k) = \Int_0^{\infty} j_l(k \, r) \,
    r^3 \, R_{n_a \, l_a}(r) \differential r \; .
\end{equation}

From expression~(\ref{eq:XUVdipexpanded}), I realize that the
bound-continuum matrix elements for circularly polarized light,
\ie, for~$\lambda = \pm 1$, are nonzero~\cite{SuppData}.
This implies that emission of circularly polarized HH~photons
may occur also.
This is an artifact of choosing plane waves~(\ref{eq:freeelmom})
as a basis for continuum electrons instead of the part of them with
the appropriate magnetic quantum number~$m_a$.
Letting~$\vec e\X{H} = \vec e_z = \vec e_0$, as it is done in this
article in Eqs.~(\ref{eq:tunnelmatel}) and (\ref{eq:diptransmat}),
resolves this issue because then only the correct part with~$m_a$
of the plane waves is projected onto.

\subsection{Continuum-continuum transitions}
\label{sec:CCtrans}

Electric dipole transition matrix elements between two continuum
states~(\ref{eq:freeelmom}) are given by
\begin{eqnarray}
  \bra{\vec k} \vec e_{\lambda} \mul \vec r \ket{\vec k^{\,\prime}}
    &=& \dfrac{1}{(2 \pi)^{3/2}} \  \vec e_{\lambda} \mul \Int_{\mathbb R^3}
    \euler^{-\imag \, \vec k \mul \vec r} \; \vec r \; \euler^{\imag \,
    \vec k^{\,\prime} \mul \vec r} \differential^3 r \\
    &=& \dfrac{1}{(2 \pi)^{3/2}} \  \vec e_{\lambda} \mul \Int_{\mathbb R^3}
    (-\imag) \; \vec \nabla_{\vec k^{\,\prime}} \; \euler^{\imag \,
    (\vec k^{\,\prime} - \vec k) \mul \vec r} \differential^3 r \; . \nonumber
\end{eqnarray}
Using the representation~(\ref{eq:diracFour}) of the Dirac
$\delta$~distribution~\cite{Arfken:MM-05}, yields
\begin{eqnarray}
  \label{eq:genccplane}
  \bra{\vec k} \vec e_{\lambda} \mul \vec r \ket{\vec k^{\,\prime}}
  &=& \imag \; \vec e_{\lambda} \mul \vec \nabla_{\vec k} \,
    \delta^3(\vec k^{\,\prime} - \vec k) \\
  &=& -\imag \; \vec e_{\lambda} \mul \vec \nabla_{\vec k^{\,\prime}} \,
    \delta^3(\vec k^{\,\prime} - \vec k) \; . \nonumber
\end{eqnarray}

By letting~$\lambda = 0$, \ie, $\vec e_0 = \vec e_z$, and using the last
equality in Eq.~(\ref{eq:genccplane}), I arrive at Eq.~(\ref{eq:CCtrans}).

\section{Saddle-point approximation}
\label{sec:saddlepoint}

The saddle-point approximation or stationary phase
approximation~\cite{Lewenstein:HH-94,Arfken:MM-05,Klaiber:RR-07}
is used frequently in strong-field physics to approximate
integrals with a highly-oscillating integrand of the form
\begin{equation}
  \label{eq:integralgS}
  I = \Int_{\mathbb R^n} f(\vec x) \; \euler^{-\imag \, \Cal S(\vec x)}
    \differential^n x \; ,
\end{equation}
with $n \in \mathbb N$, $\vec x \in \mathbb R^n$ and
functions~$f: \mathbb R^n \to \mathbb C$, and
$\Cal S: \mathbb R^n \to \mathbb R$.
The exponential function is assumed to oscillate rapidly
and the dominant contribution to~$I$ is due to the behavior of the integrand
in the vicinity of a single stationary point~$\vec x_0$ of~$\Cal S(\vec x)$
which is determined via~$\vec \nabla_x \, \Cal S(\vec x) =  \vec 0$.
The function~$f$ shall be continuous and vary comparatively slowly
in a compact space around the stationary point~$\vec x_0$ which is in its
interior.
Outside the compact space, $f$~is sufficiently slowly varying
such that a negligible contribution is made to~$I$ which
does not change noticeably, if I let~$f(\vec x) \approx f(\vec x_0)$ there.
The following derivation is inspired by the account of Klaiber~\footnote{%
The derivation of Ref.~\onlinecite{Klaiber:RR-07} is restricted
to a positive definite Hessian in Eq.~(\ref{eq:TaylorS})
as the square root of the Hessian matrix is needed which is not required
here~\cite{Golub:MC-96}.}.

I expand~$\Cal S$ around~$\vec x_0$ in terms of a Taylor series
\begin{equation}
  \label{eq:TaylorS}
  \Cal S(\vec x) = \Cal S(\vec x_0) + \dfrac{1}{2!} \Sum_{i,j = 1}^n
    \Cal H_{\Cal S}(\vec x_0)_{ij} \,
    \Delta x_i \, \Delta x_j + \ldots \; ,
\end{equation}
where~$\vec x = \vec x_0 + \Delta \vec x$ and $\Delta \vec x =
(\Delta x_1, \ldots, \Delta x_n)\transpose$~\cite{Arfken:MM-05}
and the series needs to converge for~$\vec x \in \mathbb R^n$.
There is no first-order term~$\tfrac{1}{1!} \, \vec \nabla_x \>
\Cal S(\vec x_0) \mul \Delta \vec x$
in Eq.~(\ref{eq:TaylorS}), as~$\vec \nabla_x \, \Cal S(\vec x_0) = \vec 0$.
The matrix elements of the Hessian of~$\Cal S$ in~$\vec x_0$
are~$\Cal H_{\Cal S}(\vec x_0)_{ij} = \tfrac{\partial^2
\Cal S(\vec x_0)} {\partial x_i \, \partial x_j}$
for~$i,j \in \{1, \ldots, n\}$.
The Hessian matrix has to be regular, \ie, its
determinant~$\det \mat{\Cal H}_{\Cal S}(\vec x_0) \neq 0$
and thus its eigenvalues are nonzero.
With the expansion~(\ref{eq:TaylorS}), truncated after the second summand,
I approximate the integral~$I$ [Eq.~(\ref{eq:integralgS})] by
\begin{eqnarray}
  \label{eq:approxInt}
  I \approx I\X{st} &=& f(\vec x_0) \; \euler^{-\imag \, \Cal S(\vec x_0)} \\
  &&{} \times \Int_{\mathbb R^n}
    \euler^{-\tfrac{\imag}{2} \Delta \vec x\transpose \,
    \mat{\Cal H}_{\Cal S}(\vec x_0) \, \Delta \vec x} \differential^n x
    \; . \nonumber
\end{eqnarray}
It is sufficient, to assume that $f$~is continuous and slowly varying
around~$\vec x_0$.
However, if~$f$ can be expanded into a Taylor series around~$\vec x_0$
which converges in~$\vec x \in \mathbb R^n$,
then, in~$I\X{st}$, the series expansion of~$f$ is understood to be
truncated after the zeroth-order term, \ie,
$f(\vec x) = f(\vec x_0) + \ldots$.
Together with Eq.~(\ref{eq:TaylorS}), this provides a prescription
to obtain higher-order corrections to~$I\X{st}$ in Eq.~(\ref{eq:approxInt})
and to determine the error of the truncations from the remainder terms
of the respective Taylor expansions~\cite{Arfken:MM-05}.

The Hessian matrix~$\mat{\Cal H}_{\Cal S}(\vec x_0)$ in
Eq.~(\ref{eq:approxInt}) is real-symmetric and thus diagonalizable
with an orthogonal matrix~$\mat{\Cal O}(\vec x_0)$
where~$\mat{\Cal O}(\vec x_0)^{-1} = \mat{\Cal O}(\vec x_0)
\transpose$~\cite{Arfken:MM-05}.
This provides the decomposition~$\mat{\Cal H}(\vec x_0)
= \mat{\Cal O}(\vec x_0) \, \mat{\Cal E}(\vec x_0) \,
\mat{\Cal O}(\vec x_0)\transpose$ where $\mat{\Cal E}(\vec x_0)$~is
the diagonal matrix of the real eigenvalues of~$\mat{\Cal H}(\vec x_0)$.
To simplify the remaining integral in Eq.~(\ref{eq:approxInt}),
I transform to new coordinates~$\vec y$ with the
diffeomorphism~$\Phi_{\vec x_0}: \mathbb R^n \to \mathbb R^n$,
$\vec y \mapsto \mat{\Cal O}(\vec x_0)
\> \vec y + \vec x_0 = \vec x$, an affine transformation
for which $\mat{\Cal O}(\vec x_0) \equiv \mat J_{\Phi}(\vec x_0)$~is
the Jacobian matrix~\cite{Arfken:MM-05}.
This yields
\begin{equation}
  I\X{st} = f(\vec x_0) \; \euler^{-\imag \,
    \Cal S(\vec x_0)} \Int_{\mathbb R^n} \euler^{-\tfrac{\imag}{2} \,
    \vec y\transpose \, \mat{\Cal E} \, \vec y} \differential^n y \; ,
\end{equation}
where the absolute value of the Jacobian determinant~$\det
\mat J_{\Phi}(\vec y_0)$ is unity as~$\mat{\Cal O}(\vec x_0)$
is an orthogonal matrix~\cite{Arfken:MM-05}.
The remaining integral is known~\cite{Arfken:MM-05} and found
from~$\Int_{-\infty}^{+\infty} \euler^{-\tfrac{\imag}{2} \, \epsilon \, y^2}
\differential y = \sqrt{\tfrac{-2 \, \pi \, \imag}
{\epsilon}}$ for~$\epsilon \in \mathbb R$, $\epsilon \neq 0$.
The eigenvalues are multiplied in~$\det \mat{\Cal H}(\vec x_0) = \det \bigl(
\mat{\Cal O}(\vec x_0) \, \mat{\Cal E}(\vec x_0) \,
\mat{\Cal O}(\vec x_0)\transpose \bigr) = \det \mat{\Cal O}(\vec x_0) \,
\det \mat{\Cal E}(\vec x_0) \, \det \mat{\Cal O}(\vec x_0)\transpose
= \Prod_{i=1}^n \Cal E(\vec x_0)_{ii}$
because $\det \mat{\Cal O}(\vec x_0) = \det \mat{\Cal O}(\vec x_0)\transpose$
and $|\det \mat{\Cal O}(\vec x_0)| = 1$.
Then I arrive at the integral~$I$ [Eq.~(\ref{eq:integralgS})] in
saddle-point approximation
\begin{equation}
  I\X{st} = \sqrt{(-2 \, \pi \, \imag)^n} \; f(\vec x_0) \;
    \euler^{-\imag \, \Cal S(\vec x_0)} \; \sqrt{\bigl( \det
    \mat{\Cal H}(\vec x_0) \bigr)^{-1}} \; .
\end{equation}

I use the saddle-point approximation in Eqs.~(\ref{eq:dipcompst}),
(\ref{eq:dipcontinuousArb}), and (\ref{eq:lineArbFirst}).

\section{Harmonic photon emission}
\label{sec:HPNS}
\subsection{Quantum electrodynamic formulation}

The recombination of a continuum electron with a hole in the parent ion
leads to the emission of a HH~photon.
This is fluorescence which cannot be treated with semiclassical light fields;
instead, quantum electrodynamics~(QED) is required~\cite{Craig:MQ-84,%
Scully:QO-97,Meystre:QO-99,Milosevic:SI-01,Pukhov:TS-03,Buth:TX-07,%
Diestler:HG-08}.
The energy of a photon in the $\vec q \,$th~mode is~$\omega_{\vec q}$
with~$\vec q \in \mathbb V$ where $\mathbb V$~is the set of allowed
modes in the volume~$V$ that is used to quantize the electromagnetic field.
The photon vacuum state is~$\ket{0}$ and the photon number state with
one HH~photon in mode~$\vec q \in \mathbb V$ is denoted by~$\ket{1_{\vec q}}$,
Then the annihilation~$\hat a^{\vphantom{\dagger}}_{\vec q}$ and
creation~$\hat a^{\dagger}_{\vec q}$ operators for photons in
mode~$\vec q \in \mathbb V$ can be expressed
by~$\hat a^{\vphantom{\dagger}}_{\vec q} = \ket{0} \bra{1_{\vec q}}$
and $\hat a^{\dagger}_{\vec q} = \ket{1_{\vec q}} \bra{0}$, respectively.
\emph{Nota bene}, I restrict the description to, at maximum,
a single HH~photon, \ie, multi-HH-photon states are omitted.
This is an excellent approximation as multi-HH-photon emission
is highly improbable.
The Hamiltonian of the free HH~photon field is
\begin{equation}
  \label{eq:Hem}
  \hat H\X{EM} = \Sum_{\vec q \in \mathbb V} \omega_{\vec q} \;
    \unitop\X{el}^{N/2} \otimes \hat a^{\dagger}_{\vec q} \>
    \hat a^{\vphantom{\dagger}}_{\vec q}
  = \Sum_{\vec q \in \mathbb V} \omega_{\vec q} \;
    \unitop\X{el}^{N/2} \otimes \ket{1_{\vec q}} \bra{1_{\vec q}} \; ,
\end{equation}
where I set the energy of the photon vacuum to
zero~\cite{Craig:MQ-84,Buth:TX-07}.

The electronic structure after recombination is represented by the
no-decay $N/2$-electron atomic electronic structure Hamiltonian from
Eq.~(\ref{eq:twoatstruc}) times the projector on the number
states with one HH~photon which is
\begin{equation}
  \label{eq:He}
  H\X{E} = \Bigl[ \Sum_{i=1}^{N/2} \unitop\X{el}^{i-1} \otimes \hat h\X{A}
    \otimes \unitop\X{el}^{N/2-i} \Bigr] \otimes \Sum_{\vec q \in \mathbb V}
    \ket{1_{\vec q}} \bra{1_{\vec q}} \; .
\end{equation}
I do not include destruction of the $N/2$-electron ground state with one
HH~photon as the amplitude of HH~photon emission is desired.

The Hamiltonian for HH~fluorescence~$\hat H\X{H}$ is constructed
based on Eq.~(\ref{eq:Nhalfdipole}) by replacing in~$\hat d_1$ the polarization
vector~$\vec e\X{H}$ with the operator for the quantized electric
field of the HH~light~\cite{Craig:MQ-84} which is represented in
electric dipole approximation by
\begin{equation}
  \label{eq:quantHH}
  \hat{\vec E}\X{H} = -\imag \Sum_{\vec q \in \mathbb V} \sqrt{\dfrac{2
    \, \pi \, \omega_{\vec q}}{V}} \; \bigl[ \vec e\X{H}^{\,*} \otimes
    \hat a^{\dagger}_{\vec q} - \vec e\X{H}^{\vphantom{\,*}} \otimes \hat
    a^{\vphantom{\dagger}}_{\vec q} \bigr] \; ,
\end{equation}
with the polarization vector~$\vec e\X{H} = \vec e_z$
and propagation of the HH~photons along the $x$~axis.
The Hamiltonian~$\hat H\X{H}$ is expressed in terms of the basis states
from Sect.~\ref{sec:wavefunction} [Eq.~(\ref{eq:fullground})] via
\begin{widetext}
\begin{eqnarray}
  \label{eq:Hh}
  \hat H\X{H} &=&  -\imag \, a^{\,\prime\,*}(t) \Sum_{\vec q \in \mathbb V}
    \sqrt{\tfrac{2 \, \pi \, \omega_{\vec q}}{V}} \; \Bigl[
    \Sum_{m \in \mathbb M_1} \  \Int_{\mathbb R^3} \ket{\Phi_0} \otimes
    \ket{1_{\vec q}} \bra{a ; m} \vec e^{\>*}\X{H} \mul \vec r \ket{\vec k}
    \bra{\Phi_{\vec k}^{(m)}} \otimes \bra{0} \differential^3 k \\
  &&{} + \Sum_{m \in \mathbb M_2} \ \Int_{\mathbb R^3} \bigl(
    \ket{\Phi_0} \otimes \ket{1_{\vec q}} \bra{a ; m} \vec e^{\>*}\X{H} \mul
    \vec r \ket{\vec k} \bra{\Phi_{\vec k \, c}^{(m)}} \otimes \bra{0}
    - \ket{\Phi_0} \otimes \ket{1_{\vec q}} \bra{c ; m}
    \vec e^{\>*}\X{H} \mul \vec r \ket{\vec k} \bra{\Phi_{\vec k \, a}^{(m)}}
    \otimes \bra{0} \bigr) \differential^3 k \Bigr] + \mathrm{h.c.} \; .
    \nonumber
\end{eqnarray}
\end{widetext}
Ground-state depletion is accounted for by the
factor~$a^{\,\prime\,*}(t)$~\footnote{%
All coupling terms that involve a HH~photon with continuum states, the
``continuum-continuum terms'' are negligible.}.

The QED~Hamiltonian for the description of the HHG~process reads
\begin{equation}
  \label{eq:HQED}
  \hat H\X{QED} = \hat H \otimes \ket{0} \bra{0} + \hat H\X{EM}
    + \hat H\X{E} + \hat H\X{H} \; ,
\end{equation}
with the total $N/2$-electron Hamiltonian with semiclassical optical-laser
and \xray~fields~$\hat H$ from Eq.~(\ref{eq:hamiltonian}) here augmented
by the projector on the vacuum photon state~$\ket{0} \bra{0}$ of
the HH~field, and Eqs.~(\ref{eq:Hem}), (\ref{eq:Hh}), and (\ref{eq:He}).

To determine the amplitude of fluorescence, I make a wave packet ansatz
similar to Eq.~(\ref{eq:wavepacket}) which, however, explicitly includes the
photon number states of the HH~field.
Similar to Ref.~\onlinecite{Molmer:MC-96}, I have
\begin{equation}
  \label{eq:wavepacketQED}
  \ket{\Psi^{\,\prime\prime}, t} = \Sum_{\vec q \in \mathbb V}
    c^{\,\prime}_{\vec q}(t) \, \euler^{-\imag \, E_0 \, t}
    \ket{\Phi_0} \otimes \ket{1_{\vec q}} + \ket{\Psi^{\,\prime}, t}
    \otimes \ket{0} \; ,
\end{equation}
where $\ket{\Psi^{\,\prime}, t}$~is taken from Eq.~(\ref{eq:wavepacket}),
generalized to arbitrary \xray~pulses [Sect.~\ref{sec:HamWavEOM}],
and extended to encompass the vacuum photon state.
The first term on the right-hand side of~(\ref{eq:wavepacketQED})
is needed to describe recombination, in which a continuum electron fills a
vacancy in the valence or the core, returning the atom to its electronic
ground state~$\ket{\Phi_0}$ where the excess energy is emitted in terms
of a HH~photon in state~$\ket{1_{\vec q}}$ with mode~$\vec q \in \mathbb V$.

The wave packet~(\ref{eq:wavepacketQED}) is inserted into the time-dependent
Schr\"odinger equation~(\ref{eq:tdschroedi}) with the
QED~Hamiltonian~(\ref{eq:HQED}) and projected
onto the state~$\bra{\Phi_0} \otimes \bra{0}$ to obtain the same
EOM~(\ref{eq:groundepl}) for ground-state depletion---also true
for arbitrary \xray~pulses---as was obtained with the semiclassical
Hamiltonian~(\ref{eq:hamiltonian}).

Projecting for~$\vec k \in \mathbb R^3$ onto~$\bra{\Phi_{\vec k}^{(m_1)}}
\otimes \bra{0}$ with~$m_1 \in \mathbb M_1$ and
onto~$\bra{\Phi_{\vec k \, c}^{(m_2)}} \otimes \bra{0}$ and
$\bra{\Phi_{\vec k \, a}^{(m_2)}} \otimes \bra{0}$ with~$m_2 \in \mathbb M_2$,
respectively, yields EOMs for recombination with a valence hole
and a core hole.
These EOMs are similar to those in Eqs.~(\ref{eq:revalence})
and (\ref{eq:recore}), however, I need to allow for arbitrary \xray~pulses
[Sect.~\ref{sec:HamWavEOM}] and the first term on the right-hand side
of~(\ref{eq:wavepacketQED}) which leads to an extra term on the right-hand
side of these EOMs.
For valence-hole recombination I need to add~${} + \Sum_{\vec q \in \mathbb V}
\sqrt{\tfrac{2 \, \pi \, \omega_{\vec q}}{V}} \bra{\vec k}
\vec e\X{H} \mul \vec r \ket{a ; m} \, a'(t) \, c^{\,\prime}_{\vec q}(t)$
with~$m \in \mathbb M_a$ and for core-hole
recombination~${} - \Sum_{\vec q \in \mathbb V}
\sqrt{\tfrac{2 \, \pi \, \omega_{\vec q}}{V}} \bra{\vec k}
\vec e\X{H} \mul \vec r \ket{c ; m_2} \, \euler^{\imag \, \omega\X{X} \, t
+ \imag \, \varphi\X{X}(t)} \, a'(t) \, c^{\,\prime}_{\vec q}(t)$
with~$m_2 \in \mathbb M_2$.
The extra summands account for changes in the continuum
amplitudes~$b^{\,\prime\,(m)}_a(\vec k, t)$ and $b^{\,\prime\,(m_2)}_c(\vec k,
t)$, respectively, because of recombination with HH~emission.
Yet as I consider only HH~photons with~$\vec q \in \mathbb V$,
the two terms do not comprise the entire change in the continuum
amplitudes, \ie, due to emission of HH~photons into
mode~$\vec q \in \bigl( \mathbb R^3 \setminus \{ \vec 0 \} \bigr)
\setminus \mathbb V$ and suitable polarization~\cite{Craig:MQ-84}.
As HHG is inefficient, the contributions of these summands is tiny
and can be neglected in excellent approximation.

Projecting onto~$\bra{\Phi_0} \otimes \bra{1_{\vec q}}$ leads to the
EOMs for the amplitude to find a fluorescence photon with mode~$\vec q \in
\mathbb V$ [compare with Eq.~(\ref{eq:groundepl})] reading
\begin{equation}
  \label{eq:groundeplQED}
  \dfrac{\differential}{\differential t} \, c^{\,\prime}_{\vec q}(t)
    = - \imag \, \omega^{\vphantom{\,\prime}}_{\vec q} \,
    c^{\,\prime}_{\vec q}(t) - \sqrt{\dfrac{2 \, \pi \, \omega_{\vec q}}{V}}
    \; \Cal D^{\,\prime}(t) \; ,
\end{equation}
with the appropriately-adapted time-dependent $N/2$-electron dipole
transition matrix element~$\Cal D^{\,\prime}(t)$ from
Eq.~(\ref{eq:diptransmat}).
Then Eq.~(\ref{eq:groundeplQED}) can be integrated directly~\cite{Arfken:MM-05}
yielding
\begin{equation}
  \label{eq:groundeplQEDamp}
  c^{\,\prime}_{\vec q}(t) = - \sqrt{\dfrac{2 \, \pi \, \omega_{\vec q}}{V}}
   \; \euler^{-\imag \, \omega_{\vec q} \, t}
   \Int_{-t}^t \euler^{\imag \, \omega_{\vec q} \, t^{\,\prime}}
    \; \Cal D^{\,\prime}(t^{\,\prime}) \differential t^{\,\prime} \; .
\end{equation}
For vanishing ground-state depletion, a HH~emission rate is derived in
Sect.~\ref{sec:nodepletion};
otherwise only a HPNS~\cite{Diestler:HG-08} can be obtained in
Sect.~\ref{sec:gsdepletion}.

\subsection{Harmonic photon emission rate}
\label{sec:nodepletion}

I assume cw~optical-laser and \xray~light and no ground-state
depletion, \ie, $a'(t) = 1$ and $\Gamma^{\,\prime}_0(t) = 0$ for all~$t$.
Then $\Cal D^{\,\prime}(t^{\,\prime})$ can be replaced by a Fourier
series~(\ref{eq:FourDip}).
I carry out the time integration in Eq.~(\ref{eq:groundeplQEDamp}) using
\begin{equation}
  \Int_{-t}^t \euler^{-\imag \, \omega \, t^{\,\prime}} \differential
    t^{\,\prime} = 2 \, t \, \sinc(\omega \, t) \; ,
\end{equation}
with the sinus cardinalis~\cite{SuppData} for~$z \in \mathbb C$ which
is defined by
\begin{equation}
  \label{eq:sinc}
  \sinc z = \begin{cases}
    \dfrac{\sin z}{z} & ; z \neq 0 \\
    1 & ; z = 0 \; .
   \end{cases}
\end{equation}
The amplitude of fluorescence photons follows to
\begin{eqnarray}
  \label{eq:amplfluores}
  c^{\,\prime}_{\vec q}(t) &=& - \sqrt{\dfrac{2 \, \pi \, \omega_{\vec q}}{V}}
    \; \euler^{-\imag \, \omega_{\vec q} \, t} \; (2 \, t) \!\!\!
    \Sum_{j \in \{+, -\}} \Sum_{K=-\infty}^{\infty} \\
  &&{} \times \bigl[ \tilde{\mathfrak D}_{aj, 2K + 1} \> \sinc \bigl(
    (2 \, K + 1) \, \omega\X{L} \, t - \omega_{\vec q} \, t \bigr) \nonumber \\
  &&\ \,\quad {} + \tilde{\mathfrak D}_{cj, 2K} \> \sinc\bigl(2 \, K \,
    \omega\X{L} \, t + \omega\X{X} \, t - \omega_{\vec q} \, t
    \bigr) \bigr] \; . \nonumber
\end{eqnarray}
The probability to find a HH~photon at time~$t$ in
mode~$\vec q \in \mathbb V$ is given by~$|c'_{\vec q}(t)|^2$.
From this, the rate of HH~emission into mode~$\vec q$ follows
to~$\Gamma_{\vec q} = \Lim_{t \to \infty} \tfrac{|c_{\vec q}(t)|^2}{2 \, t}$.
I transform the relation for~$\Gamma_{\vec q}$, after inserting
Eq.~(\ref{eq:amplfluores}), with the representation
\begin{equation}
  \label{eq:diracsinc}
  \Lim_{t \to \infty} \bigl( t \; \sinc\!^2 (\omega \, t) \bigr)
    = \pi \, \delta(\omega) \; ,
\end{equation}
of the Dirac $\delta$~distribution~\cite{Merzbacher:QM-98}.
The cross terms in~$\Gamma_{\vec q}$ for harmonics with different
arguments in the sinus cardinalis functions vanish~\cite{SuppData} due
to
\begin{equation}
  \Lim_{t \to \infty} \bigl( t \; \sinc(\omega \, t) \;
    \sinc \bigr((\omega + \omega') \, t \bigl) \bigr) = 0
\end{equation}
for~$\omega' \neq 0$~\footnote{%
I determine the cases~$\omega = 0$ and $\omega + \omega' = 0$ \emph{not}
by inserting~$0$ into the sinus cardinalis~(\ref{eq:sinc})---which
would give~$\sinc \, 0 = 1$---but by taking the limits~$\omega \to 0$
and $\omega + \omega' \to 0$, respectively, instead after the
limit~$t \to \infty$ has been taken.}.

The total rate of HH~photon emission is~$\Sum_{\vec q \in \mathbb V}
\Gamma_{\vec q}$.
As free-photon states are spaced densely, the sum over~$\vec q \in \mathbb V$
can be replaced by an integral~$\Int_0^{\infty} \differential \omega$ over
$V$~times the density of free-photon states~$\varrho(\omega)$
[Eq.~(\ref{eq:densfreephot})].
No angular integration is performed as the propagation direction of the
HH~photons is fixed along the $x$~axis.
Then the integrand represents the angular-resolved (for the chosen angles)
spectral rate of HH~photon emission where the extra factor of~2 in
Eq.~(\ref{eq:HHGspecNoDepl}) accounts for the two possible spin states
of electrons per spatial orbital.

\subsection{Harmonic photon number spectrum}
\label{sec:gsdepletion}

The situation of HH~emission changes as soon as ground-state depletion
is considered by letting~$\Gamma^{\,\prime}_0(t) > 0$ leading to a
decreasing ground-state amplitude~$a^{\,\prime\,*}(t)$ with
time~$0 \leq t \leq T\X{P}$ in Eq.~(\ref{eq:Hh}) for optical-laser
and \xray~pulses restricted to this range.
In this case, HH~emission is not periodic anymore;
instead of a HH~emission rate, I, therefore, calculate the
HPNS~\cite{Diestler:HG-08}, \ie, the probability
to emit a HH~photon into mode~$\vec q \in \mathbb V$ via
\begin{equation}
  \label{eq:Capprox}
  \Lim_{t \to \infty} |c_{\vec q}(t)|^2 = \dfrac{2 \, \pi \,
    \omega_{\vec q}}{V} \; |\tilde{\Cal D}^{\,\prime}(\omega_{\vec q})|^2 \; ,
\end{equation}
where the time integral in Eq.~(\ref{eq:groundeplQEDamp})
over~$\Cal D^{\,\prime}(t^{\,\prime})$ becomes the Fourier
transform~$\tilde{\Cal D}^{\,\prime}(\omega_{\vec q})$ in the
limit~$t \to \infty$.
The HPNS for a single atom follows by replacing the sum over
modes~$\vec q$ by an integral over~$V \, \varrho(\omega)$
[Eq.~(\ref{eq:densfreephot})] yielding Eq.~(\ref{eq:HHGspecXray})
where the extra factor of~2 stems from the two spin states.


\begin{thebibliography}{158}%
\makeatletter
\providecommand \@ifxundefined [1]{%
 \@ifx{#1\undefined}
}%
\providecommand \@ifnum [1]{%
 \ifnum #1\expandafter \@firstoftwo
 \else \expandafter \@secondoftwo
 \fi
}%
\providecommand \@ifx [1]{%
 \ifx #1\expandafter \@firstoftwo
 \else \expandafter \@secondoftwo
 \fi
}%
\providecommand \natexlab [1]{#1}%
\providecommand \enquote  [1]{``#1''}%
\providecommand \bibnamefont  [1]{#1}%
\providecommand \bibfnamefont [1]{#1}%
\providecommand \citenamefont [1]{#1}%
\providecommand \href@noop [0]{\@secondoftwo}%
\providecommand \href [0]{\begingroup \@sanitize@url \@href}%
\providecommand \@href[1]{\@@startlink{#1}\@@href}%
\providecommand \@@href[1]{\endgroup#1\@@endlink}%
\providecommand \@sanitize@url [0]{\catcode `\\12\catcode `\$12\catcode
  `\&12\catcode `\#12\catcode `\^12\catcode `\_12\catcode `\%12\relax}%
\providecommand \@@startlink[1]{}%
\providecommand \@@endlink[0]{}%
\providecommand \url  [0]{\begingroup\@sanitize@url \@url }%
\providecommand \@url [1]{\endgroup\@href {#1}{\urlprefix }}%
\providecommand \urlprefix  [0]{URL }%
\providecommand \Eprint [0]{\href }%
\providecommand \doibase [0]{http://dx.doi.org/}%
\providecommand \selectlanguage [0]{\@gobble}%
\providecommand \bibinfo  [0]{\@secondoftwo}%
\providecommand \bibfield  [0]{\@secondoftwo}%
\providecommand \translation [1]{[#1]}%
\providecommand \BibitemOpen [0]{}%
\providecommand \bibitemStop [0]{}%
\providecommand \bibitemNoStop [0]{.\EOS\space}%
\providecommand \EOS [0]{\spacefactor3000\relax}%
\providecommand \BibitemShut  [1]{\csname bibitem#1\endcsname}%
\let\auto@bib@innerbib\@empty
%</preamble>
\bibitem [{\citenamefont {McPherson}\ \emph {et~al.}(1987)\citenamefont
  {McPherson}, \citenamefont {Gibson}, \citenamefont {Jara}, \citenamefont
  {Johann}, \citenamefont {Luk}, \citenamefont {McIntyre}, \citenamefont
  {Boyer},\ and\ \citenamefont {Rhodes}}]{McPherson:MP-87}%
  \BibitemOpen
  \bibfield  {author} {\bibinfo {author} {\bibfnamefont {A.}~\bibnamefont
  {McPherson}}, \bibinfo {author} {\bibfnamefont {G.}~\bibnamefont {Gibson}},
  \bibinfo {author} {\bibfnamefont {H.}~\bibnamefont {Jara}}, \bibinfo {author}
  {\bibfnamefont {U.}~\bibnamefont {Johann}}, \bibinfo {author} {\bibfnamefont
  {T.~S.}\ \bibnamefont {Luk}}, \bibinfo {author} {\bibfnamefont {I.~A.}\
  \bibnamefont {McIntyre}}, \bibinfo {author} {\bibfnamefont {K.}~\bibnamefont
  {Boyer}}, \ and\ \bibinfo {author} {\bibfnamefont {C.~K.}\ \bibnamefont
  {Rhodes}},\ }\bibfield  {title} {\enquote {\bibinfo {title} {Studies of
  multiphoton production of vacuum-ultraviolet radiation in the rare gases},}\
  }\href {\doibase 10.1364/JOSAB.4.000595} {\bibfield  {journal} {\bibinfo
  {journal} {J. Opt. Soc. Am. B}\ }\textbf {\bibinfo {volume} {4}},\ \bibinfo
  {pages} {595--601} (\bibinfo {year} {1987})}\BibitemShut {NoStop}%
\bibitem [{\citenamefont {Ferray}\ \emph {et~al.}(1988)\citenamefont {Ferray},
  \citenamefont {L'Huillier}, \citenamefont {Li}, \citenamefont {Lompre},
  \citenamefont {Mainfray},\ and\ \citenamefont {Manus}}]{Ferray:MH-88}%
  \BibitemOpen
  \bibfield  {author} {\bibinfo {author} {\bibfnamefont {M.}~\bibnamefont
  {Ferray}}, \bibinfo {author} {\bibfnamefont {A.}~\bibnamefont {L'Huillier}},
  \bibinfo {author} {\bibfnamefont {X.~F.}\ \bibnamefont {Li}}, \bibinfo
  {author} {\bibfnamefont {L.~A.}\ \bibnamefont {Lompre}}, \bibinfo {author}
  {\bibfnamefont {G.}~\bibnamefont {Mainfray}}, \ and\ \bibinfo {author}
  {\bibfnamefont {C.}~\bibnamefont {Manus}},\ }\bibfield  {title} {\enquote
  {\bibinfo {title} {Multiple-harmonic conversion of 1064$\,$nm radiation in
  rare gases},}\ }\href {\doibase 10.1088/0953-4075/21/3/001} {\bibfield
  {journal} {\bibinfo  {journal} {J. Phys. B}\ }\textbf {\bibinfo {volume}
  {21}},\ \bibinfo {pages} {L31--L35} (\bibinfo {year} {1988})}\BibitemShut
  {NoStop}%
\bibitem [{\citenamefont {Agostini}\ and\ \citenamefont
  {DiMauro}(2004)}]{Agostini:PA-04}%
  \BibitemOpen
  \bibfield  {author} {\bibinfo {author} {\bibfnamefont {Pierre}\ \bibnamefont
  {Agostini}}\ and\ \bibinfo {author} {\bibfnamefont {Louis~F.}\ \bibnamefont
  {DiMauro}},\ }\bibfield  {title} {\enquote {\bibinfo {title} {The physics of
  attosecond light pulses},}\ }\href {\doibase 10.1088/0034-4885/67/6/R01}
  {\bibfield  {journal} {\bibinfo  {journal} {Rep. Prog. Phys.}\ }\textbf
  {\bibinfo {volume} {67}},\ \bibinfo {pages} {813--855} (\bibinfo {year}
  {2004})}\BibitemShut {NoStop}%
\bibitem [{\citenamefont {Scrinzi}\ \emph {et~al.}(2006)\citenamefont
  {Scrinzi}, \citenamefont {Ivanov}, \citenamefont {Kienberger},\ and\
  \citenamefont {Villeneuve}}]{Scrinzi:AP-06}%
  \BibitemOpen
  \bibfield  {author} {\bibinfo {author} {\bibfnamefont {A.}~\bibnamefont
  {Scrinzi}}, \bibinfo {author} {\bibfnamefont {M.~Yu}\ \bibnamefont {Ivanov}},
  \bibinfo {author} {\bibfnamefont {R.}~\bibnamefont {Kienberger}}, \ and\
  \bibinfo {author} {\bibfnamefont {D.~M.}\ \bibnamefont {Villeneuve}},\
  }\bibfield  {title} {\enquote {\bibinfo {title} {Attosecond physics},}\
  }\href {\doibase 10.1088/0953-4075/39/1/R01} {\bibfield  {journal} {\bibinfo
  {journal} {J. Phys. B}\ }\textbf {\bibinfo {volume} {39}},\ \bibinfo {pages}
  {R1--R37} (\bibinfo {year} {2006})}\BibitemShut {NoStop}%
\bibitem [{\citenamefont {Bucksbaum}(2007)}]{Bucksbaum:FA-07}%
  \BibitemOpen
  \bibfield  {author} {\bibinfo {author} {\bibfnamefont {Philip~H.}\
  \bibnamefont {Bucksbaum}},\ }\bibfield  {title} {\enquote {\bibinfo {title}
  {The future of attosecond spectroscopy},}\ }\href {\doibase
  10.1126/science.1142135} {\bibfield  {journal} {\bibinfo  {journal}
  {Science}\ }\textbf {\bibinfo {volume} {317}},\ \bibinfo {pages} {766--769}
  (\bibinfo {year} {2007})}\BibitemShut {NoStop}%
\bibitem [{\citenamefont {Krausz}\ and\ \citenamefont
  {Ivanov}(2009)}]{Krausz:AP-09}%
  \BibitemOpen
  \bibfield  {author} {\bibinfo {author} {\bibfnamefont {Ferenc}\ \bibnamefont
  {Krausz}}\ and\ \bibinfo {author} {\bibfnamefont {Misha}\ \bibnamefont
  {Ivanov}},\ }\bibfield  {title} {\enquote {\bibinfo {title} {Attosecond
  physics},}\ }\href {\doibase 10.1103/RevModPhys.81.163} {\bibfield  {journal}
  {\bibinfo  {journal} {Rev. Mod. Phys.}\ }\textbf {\bibinfo {volume} {81}},\
  \bibinfo {pages} {163--234} (\bibinfo {year} {2009})}\BibitemShut {NoStop}%
\bibitem [{\citenamefont {Kohler}\ \emph
  {et~al.}(2012{\natexlab{a}})\citenamefont {Kohler}, \citenamefont {Pfeifer},
  \citenamefont {Hatsagortsyan},\ and\ \citenamefont {Keitel}}]{Kohler:FA-12}%
  \BibitemOpen
  \bibfield  {author} {\bibinfo {author} {\bibfnamefont {Markus~C.}\
  \bibnamefont {Kohler}}, \bibinfo {author} {\bibfnamefont {Thomas}\
  \bibnamefont {Pfeifer}}, \bibinfo {author} {\bibfnamefont {Karen~Z.}\
  \bibnamefont {Hatsagortsyan}}, \ and\ \bibinfo {author} {\bibfnamefont
  {Christoph~H.}\ \bibnamefont {Keitel}},\ }\bibfield  {title} {\enquote
  {\bibinfo {title} {Frontiers of atomic high-harmonic generation},}\ }\href
  {\doibase 10.1016/B978-0-12-396482-3.00004-1} {\bibfield  {journal} {\bibinfo
   {journal} {Adv. At. Mol. Opt. Phys.}\ }\textbf {\bibinfo {volume} {61}},\
  \bibinfo {pages} {159--208} (\bibinfo {year} {2012}{\natexlab{a}})},\ \Eprint
  {http://arxiv.org/abs/arXiv:1201.5094} {arXiv:1201.5094} \BibitemShut
  {NoStop}%
\bibitem [{\citenamefont {Kuchiev}(1987)}]{Kuchiev:AA-87}%
  \BibitemOpen
  \bibfield  {author} {\bibinfo {author} {\bibfnamefont {M.~Yu.}\ \bibnamefont
  {Kuchiev}},\ }\bibfield  {title} {\enquote {\bibinfo {title} {Atomic
  antenna},}\ }\href@noop {} {\bibfield  {journal} {\bibinfo  {journal} {Pis'ma
  Zh. Eksp. Teor. Fiz.}\ }\textbf {\bibinfo {volume} {45}},\ \bibinfo {pages}
  {319--332} (\bibinfo {year} {1987})},\ \bibinfo {note} {{JETP} Lett.
  \textbf{45}, 404--406 (1987)}\BibitemShut {NoStop}%
\bibitem [{\citenamefont {Krause}\ \emph {et~al.}(1992)\citenamefont {Krause},
  \citenamefont {Schafer},\ and\ \citenamefont {Kulander}}]{Krause:HO-92}%
  \BibitemOpen
  \bibfield  {author} {\bibinfo {author} {\bibfnamefont {Jeffrey~L.}\
  \bibnamefont {Krause}}, \bibinfo {author} {\bibfnamefont {Kenneth~J.}\
  \bibnamefont {Schafer}}, \ and\ \bibinfo {author} {\bibfnamefont
  {Kenneth~C.}\ \bibnamefont {Kulander}},\ }\bibfield  {title} {\enquote
  {\bibinfo {title} {High-order harmonic generation from atoms and ions in the
  high intensity regime},}\ }\href {\doibase 10.1103/PhysRevLett.68.3535}
  {\bibfield  {journal} {\bibinfo  {journal} {Phys. Rev. Lett.}\ }\textbf
  {\bibinfo {volume} {68}},\ \bibinfo {pages} {3535--3538} (\bibinfo {year}
  {1992})}\BibitemShut {NoStop}%
\bibitem [{\citenamefont {Schafer}\ \emph {et~al.}(1993)\citenamefont
  {Schafer}, \citenamefont {Yang}, \citenamefont {DiMauro},\ and\ \citenamefont
  {Kulander}}]{Schafer:AT-93}%
  \BibitemOpen
  \bibfield  {author} {\bibinfo {author} {\bibfnamefont {K.~J.}\ \bibnamefont
  {Schafer}}, \bibinfo {author} {\bibfnamefont {Baorui}\ \bibnamefont {Yang}},
  \bibinfo {author} {\bibfnamefont {L.~F.}\ \bibnamefont {DiMauro}}, \ and\
  \bibinfo {author} {\bibfnamefont {K.~C.}\ \bibnamefont {Kulander}},\
  }\bibfield  {title} {\enquote {\bibinfo {title} {Above threshold ionization
  beyond the high harmonic cutoff},}\ }\href {\doibase
  10.1103/PhysRevLett.70.1599} {\bibfield  {journal} {\bibinfo  {journal}
  {Phys. Rev. Lett.}\ }\textbf {\bibinfo {volume} {70}},\ \bibinfo {pages}
  {1599--1602} (\bibinfo {year} {1993})}\BibitemShut {NoStop}%
\bibitem [{\citenamefont {Corkum}(1993)}]{Corkum:PP-93}%
  \BibitemOpen
  \bibfield  {author} {\bibinfo {author} {\bibfnamefont {Paul~B.}\ \bibnamefont
  {Corkum}},\ }\bibfield  {title} {\enquote {\bibinfo {title} {Plasma
  perspective on strong-field multiphoton ionization},}\ }\href {\doibase
  10.1103/PhysRevLett.71.1994} {\bibfield  {journal} {\bibinfo  {journal}
  {Phys. Rev. Lett.}\ }\textbf {\bibinfo {volume} {71}},\ \bibinfo {pages}
  {1994--1997} (\bibinfo {year} {1993})}\BibitemShut {NoStop}%
\bibitem [{\citenamefont {Lewenstein}\ \emph {et~al.}(1994)\citenamefont
  {Lewenstein}, \citenamefont {Balcou}, \citenamefont {Ivanov}, \citenamefont
  {L'Huillier},\ and\ \citenamefont {Corkum}}]{Lewenstein:HH-94}%
  \BibitemOpen
  \bibfield  {author} {\bibinfo {author} {\bibfnamefont {M.}~\bibnamefont
  {Lewenstein}}, \bibinfo {author} {\bibfnamefont {Ph.}\ \bibnamefont
  {Balcou}}, \bibinfo {author} {\bibfnamefont {M.~Yu.}\ \bibnamefont {Ivanov}},
  \bibinfo {author} {\bibfnamefont {Anne}\ \bibnamefont {L'Huillier}}, \ and\
  \bibinfo {author} {\bibfnamefont {Paul~B.}\ \bibnamefont {Corkum}},\
  }\bibfield  {title} {\enquote {\bibinfo {title} {Theory of high-harmonic
  generation by low-frequency laser fields},}\ }\href {\doibase
  10.1103/PhysRevA.49.2117} {\bibfield  {journal} {\bibinfo  {journal} {Phys.
  Rev. A}\ }\textbf {\bibinfo {volume} {49}},\ \bibinfo {pages} {2117--2132}
  (\bibinfo {year} {1994})}\BibitemShut {NoStop}%
\bibitem [{\citenamefont {Becker}\ \emph {et~al.}(1994)\citenamefont {Becker},
  \citenamefont {Long},\ and\ \citenamefont {McIver}}]{Becker:MH-94}%
  \BibitemOpen
  \bibfield  {author} {\bibinfo {author} {\bibfnamefont {W.}~\bibnamefont
  {Becker}}, \bibinfo {author} {\bibfnamefont {S.}~\bibnamefont {Long}}, \ and\
  \bibinfo {author} {\bibfnamefont {J.~K.}\ \bibnamefont {McIver}},\ }\bibfield
   {title} {\enquote {\bibinfo {title} {Modeling harmonic generation by a
  zero-range potential},}\ }\href {\doibase 10.1103/PhysRevA.50.1540}
  {\bibfield  {journal} {\bibinfo  {journal} {Phys. Rev. A}\ }\textbf {\bibinfo
  {volume} {50}},\ \bibinfo {pages} {1540--1560} (\bibinfo {year}
  {1994})}\BibitemShut {NoStop}%
\bibitem [{\citenamefont {Kuchiev}\ and\ \citenamefont
  {Ostrovsky}(1999)}]{Kuchiev:QT-99}%
  \BibitemOpen
  \bibfield  {author} {\bibinfo {author} {\bibfnamefont {M.~Yu.}\ \bibnamefont
  {Kuchiev}}\ and\ \bibinfo {author} {\bibfnamefont {V.~N.}\ \bibnamefont
  {Ostrovsky}},\ }\bibfield  {title} {\enquote {\bibinfo {title} {Quantum
  theory of high harmonic generation as a three-step process},}\ }\href
  {\doibase 10.1103/PhysRevA.60.3111} {\bibfield  {journal} {\bibinfo
  {journal} {Phys. Rev. A}\ }\textbf {\bibinfo {volume} {60}},\ \bibinfo
  {pages} {3111--3124} (\bibinfo {year} {1999})}\BibitemShut {NoStop}%
\bibitem [{\citenamefont {Milo{\v{s}}evi{\'c}}(2001)}]{Milosevic:SI-01}%
  \BibitemOpen
  \bibfield  {author} {\bibinfo {author} {\bibfnamefont {Dejan~B.}\
  \bibnamefont {Milo{\v{s}}evi{\'c}}},\ }\bibfield  {title} {\enquote {\bibinfo
  {title} {A semi-classical model for high-harmonic generation},}\ }in\
  \href@noop {} {\emph {\bibinfo {booktitle} {Super-Intense Laser-Atom
  Physics}}},\ \bibinfo {series} {NATO Science Series~II: Mathematics, Physics
  and Chemistry}, Vol.~\bibinfo {volume} {12},\ \bibinfo {editor} {edited by\
  \bibinfo {editor} {\bibfnamefont {Bernard}\ \bibnamefont {Piraux}}\ and\
  \bibinfo {editor} {\bibfnamefont {Kazimierz}\ \bibnamefont
  {Rz{\c{a}}{\.z}ewski}}}\ (\bibinfo  {publisher} {Kluwer Academic
  Publishers},\ \bibinfo {address} {Dordrecht, Boston, London},\ \bibinfo
  {year} {2001})\ pp.\ \bibinfo {pages} {229--238}\BibitemShut {NoStop}%
\bibitem [{\citenamefont {Becker}\ \emph {et~al.}(2002)\citenamefont {Becker},
  \citenamefont {Grasbon}, \citenamefont {Kopold}, \citenamefont
  {Milo{\v{s}}evi{\'c}}, \citenamefont {Paulus},\ and\ \citenamefont
  {Walther}}]{Becker:AT-02}%
  \BibitemOpen
  \bibfield  {author} {\bibinfo {author} {\bibfnamefont {W.}~\bibnamefont
  {Becker}}, \bibinfo {author} {\bibfnamefont {F.}~\bibnamefont {Grasbon}},
  \bibinfo {author} {\bibfnamefont {R.}~\bibnamefont {Kopold}}, \bibinfo
  {author} {\bibfnamefont {D.~B.}\ \bibnamefont {Milo{\v{s}}evi{\'c}}},
  \bibinfo {author} {\bibfnamefont {G.~G.}\ \bibnamefont {Paulus}}, \ and\
  \bibinfo {author} {\bibfnamefont {H.}~\bibnamefont {Walther}},\ }\bibfield
  {title} {\enquote {\bibinfo {title} {Above-threshold ionization: {From}
  classical features to quantum effects},}\ }\href {\doibase
  10.1016/S1049-250X(02)80006-4} {\bibfield  {journal} {\bibinfo  {journal}
  {Adv. At. Mol. Opt. Phys.}\ }\textbf {\bibinfo {volume} {48}},\ \bibinfo
  {pages} {35--98} (\bibinfo {year} {2002})}\BibitemShut {NoStop}%
\bibitem [{\citenamefont {Schafer}(2009)}]{Schafer:NM-09}%
  \BibitemOpen
  \bibfield  {author} {\bibinfo {author} {\bibfnamefont {Kenneth~J.}\
  \bibnamefont {Schafer}},\ }\bibfield  {title} {\enquote {\bibinfo {title}
  {Numerical methods in strong field physics},}\ }in\ \href {\doibase
  10.1007/978-0-387-34755-4_6} {\emph {\bibinfo {booktitle} {Strong Field Laser
  Physics}}},\ \bibinfo {series} {Springer Series in Optical Sciences}, Vol.\
  \bibinfo {volume} {134},\ \bibinfo {editor} {edited by\ \bibinfo {editor}
  {\bibfnamefont {Thomas}\ \bibnamefont {Brabec}}}\ (\bibinfo  {publisher}
  {Springer},\ \bibinfo {address} {New York},\ \bibinfo {year} {2009})\ pp.\
  \bibinfo {pages} {111--145}\BibitemShut {NoStop}%
\bibitem [{\citenamefont {L'Huillier}\ \emph {et~al.}(1991)\citenamefont
  {L'Huillier}, \citenamefont {Schafer},\ and\ \citenamefont
  {Kulander}}]{LHuillier:TA-91}%
  \BibitemOpen
  \bibfield  {author} {\bibinfo {author} {\bibfnamefont {A.}~\bibnamefont
  {L'Huillier}}, \bibinfo {author} {\bibfnamefont {K.~J.}\ \bibnamefont
  {Schafer}}, \ and\ \bibinfo {author} {\bibfnamefont {K.~C.}\ \bibnamefont
  {Kulander}},\ }\bibfield  {title} {\enquote {\bibinfo {title} {Theoretical
  aspects of intense field harmonic generation},}\ }\href {\doibase
  10.1088/0953-4075/24/15/004} {\bibfield  {journal} {\bibinfo  {journal} {J.
  Phys. B}\ }\textbf {\bibinfo {volume} {24}},\ \bibinfo {pages} {3315--3341}
  (\bibinfo {year} {1991})}\BibitemShut {NoStop}%
\bibitem [{\citenamefont {Balcou}\ \emph {et~al.}(1999)\citenamefont {Balcou},
  \citenamefont {Dederichs}, \citenamefont {Gaarde},\ and\ \citenamefont
  {L'Huillier}}]{Balcou:QP-99}%
  \BibitemOpen
  \bibfield  {author} {\bibinfo {author} {\bibfnamefont {Philippe}\
  \bibnamefont {Balcou}}, \bibinfo {author} {\bibfnamefont {Anne~S.}\
  \bibnamefont {Dederichs}}, \bibinfo {author} {\bibfnamefont {Mette~B.}\
  \bibnamefont {Gaarde}}, \ and\ \bibinfo {author} {\bibfnamefont {Anne}\
  \bibnamefont {L'Huillier}},\ }\bibfield  {title} {\enquote {\bibinfo {title}
  {Quantum-path analysis and phase matching of high-order harmonic generation
  and high-order frequency mixing processes in strong laser fields},}\ }\href
  {\doibase 10.1088/0953-4075/32/12/315} {\bibfield  {journal} {\bibinfo
  {journal} {J. Phys. B}\ }\textbf {\bibinfo {volume} {32}},\ \bibinfo {pages}
  {2973--2989} (\bibinfo {year} {1999})}\BibitemShut {NoStop}%
\bibitem [{\citenamefont {Priori}\ \emph {et~al.}(2000)\citenamefont {Priori},
  \citenamefont {Cerullo}, \citenamefont {Nisoli}, \citenamefont {Stagira},
  \citenamefont {De~Silvestri}, \citenamefont {Villoresi}, \citenamefont
  {Poletto}, \citenamefont {Ceccherini}, \citenamefont {Altucci}, \citenamefont
  {Bruzzese},\ and\ \citenamefont {de~Lisio}}]{Priori:NT-00}%
  \BibitemOpen
  \bibfield  {author} {\bibinfo {author} {\bibfnamefont {E.}~\bibnamefont
  {Priori}}, \bibinfo {author} {\bibfnamefont {G.}~\bibnamefont {Cerullo}},
  \bibinfo {author} {\bibfnamefont {M.}~\bibnamefont {Nisoli}}, \bibinfo
  {author} {\bibfnamefont {S.}~\bibnamefont {Stagira}}, \bibinfo {author}
  {\bibfnamefont {S.}~\bibnamefont {De~Silvestri}}, \bibinfo {author}
  {\bibfnamefont {P.}~\bibnamefont {Villoresi}}, \bibinfo {author}
  {\bibfnamefont {L.}~\bibnamefont {Poletto}}, \bibinfo {author} {\bibfnamefont
  {P.}~\bibnamefont {Ceccherini}}, \bibinfo {author} {\bibfnamefont
  {C.}~\bibnamefont {Altucci}}, \bibinfo {author} {\bibfnamefont
  {R.}~\bibnamefont {Bruzzese}}, \ and\ \bibinfo {author} {\bibfnamefont
  {C.}~\bibnamefont {de~Lisio}},\ }\bibfield  {title} {\enquote {\bibinfo
  {title} {Nonadiabatic three-dimensional model of high-order harmonic
  generation in the few-optical-cycle regime},}\ }\href {\doibase
  10.1103/PhysRevA.61.063801} {\bibfield  {journal} {\bibinfo  {journal} {Phys.
  Rev. A}\ }\textbf {\bibinfo {volume} {61}},\ \bibinfo {pages} {063801}
  (\bibinfo {year} {2000})}\BibitemShut {NoStop}%
\bibitem [{\citenamefont {Gaarde}\ \emph {et~al.}(2008)\citenamefont {Gaarde},
  \citenamefont {Tate},\ and\ \citenamefont {Schafer}}]{Gaarde:MA-08}%
  \BibitemOpen
  \bibfield  {author} {\bibinfo {author} {\bibfnamefont {Mette~B.}\
  \bibnamefont {Gaarde}}, \bibinfo {author} {\bibfnamefont {Jennifer~L.}\
  \bibnamefont {Tate}}, \ and\ \bibinfo {author} {\bibfnamefont {Kenneth~J.}\
  \bibnamefont {Schafer}},\ }\bibfield  {title} {\enquote {\bibinfo {title}
  {Macroscopic aspects of attosecond pulse generation},}\ }\href {\doibase
  10.1088/0953-4075/41/13/132001} {\bibfield  {journal} {\bibinfo  {journal}
  {J. Phys. B}\ }\textbf {\bibinfo {volume} {41}},\ \bibinfo {pages} {132001}
  (\bibinfo {year} {2008})}\BibitemShut {NoStop}%
\bibitem [{\citenamefont {Popmintchev}\ \emph {et~al.}(2012)\citenamefont
  {Popmintchev}, \citenamefont {Chen}, \citenamefont {Popmintchev},
  \citenamefont {Arpin}, \citenamefont {Brown}, \citenamefont
  {Ali{\v{s}}auskas}, \citenamefont {Andriukaitis}, \citenamefont
  {Bal{\v{c}}iunas}, \citenamefont {M{\"u}cke}, \citenamefont {Pugzlys},
  \citenamefont {Baltu{\v{s}}ka}, \citenamefont {Shim}, \citenamefont
  {Schrauth}, \citenamefont {Gaeta}, \citenamefont {Hern{\'a}ndez-Garc{\'i}a},
  \citenamefont {Plaja}, \citenamefont {Becker}, \citenamefont {Jaron-Becker},
  \citenamefont {Murnane},\ and\ \citenamefont {Kapteyn}}]{Popmintchev:BC-12}%
  \BibitemOpen
  \bibfield  {author} {\bibinfo {author} {\bibfnamefont {Tenio}\ \bibnamefont
  {Popmintchev}}, \bibinfo {author} {\bibfnamefont {Ming-Chang}\ \bibnamefont
  {Chen}}, \bibinfo {author} {\bibfnamefont {Dimitar}\ \bibnamefont
  {Popmintchev}}, \bibinfo {author} {\bibfnamefont {Paul}\ \bibnamefont
  {Arpin}}, \bibinfo {author} {\bibfnamefont {Susannah}\ \bibnamefont {Brown}},
  \bibinfo {author} {\bibfnamefont {Skirmantas}\ \bibnamefont
  {Ali{\v{s}}auskas}}, \bibinfo {author} {\bibfnamefont {Giedrius}\
  \bibnamefont {Andriukaitis}}, \bibinfo {author} {\bibfnamefont {Tadas}\
  \bibnamefont {Bal{\v{c}}iunas}}, \bibinfo {author} {\bibfnamefont
  {Oliver~D.}\ \bibnamefont {M{\"u}cke}}, \bibinfo {author} {\bibfnamefont
  {Audrius}\ \bibnamefont {Pugzlys}}, \bibinfo {author} {\bibfnamefont
  {Andrius}\ \bibnamefont {Baltu{\v{s}}ka}}, \bibinfo {author} {\bibfnamefont
  {Bonggu}\ \bibnamefont {Shim}}, \bibinfo {author} {\bibfnamefont {Samuel~E.}\
  \bibnamefont {Schrauth}}, \bibinfo {author} {\bibfnamefont {Alexander}\
  \bibnamefont {Gaeta}}, \bibinfo {author} {\bibfnamefont {Carlos}\
  \bibnamefont {Hern{\'a}ndez-Garc{\'i}a}}, \bibinfo {author} {\bibfnamefont
  {Luis}\ \bibnamefont {Plaja}}, \bibinfo {author} {\bibfnamefont {Andreas}\
  \bibnamefont {Becker}}, \bibinfo {author} {\bibfnamefont {Agnieszka}\
  \bibnamefont {Jaron-Becker}}, \bibinfo {author} {\bibfnamefont {Margaret~M.}\
  \bibnamefont {Murnane}}, \ and\ \bibinfo {author} {\bibfnamefont {Henry~C.}\
  \bibnamefont {Kapteyn}},\ }\bibfield  {title} {\enquote {\bibinfo {title}
  {Bright coherent ultrahigh harmonics in the {keV} x-ray regime from
  mid-infrared femtosecond lasers},}\ }\href {\doibase 10.1126/science.1218497}
  {\bibfield  {journal} {\bibinfo  {journal} {Science}\ }\textbf {\bibinfo
  {volume} {336}},\ \bibinfo {pages} {1287--1291} (\bibinfo {year}
  {2012})}\BibitemShut {NoStop}%
\bibitem [{\citenamefont {Swoboda}\ \emph {et~al.}(2010)\citenamefont
  {Swoboda}, \citenamefont {Fordell}, \citenamefont {Kl{\"u}nder},
  \citenamefont {Dahlstr{\"o}m}, \citenamefont {Miranda}, \citenamefont {Buth},
  \citenamefont {Schafer}, \citenamefont {Mauritsson}, \citenamefont
  {L'Huillier},\ and\ \citenamefont {Gisselbrecht}}]{Swoboda:PM-10}%
  \BibitemOpen
  \bibfield  {author} {\bibinfo {author} {\bibfnamefont {Marko}\ \bibnamefont
  {Swoboda}}, \bibinfo {author} {\bibfnamefont {Thomas}\ \bibnamefont
  {Fordell}}, \bibinfo {author} {\bibfnamefont {Kathrin}\ \bibnamefont
  {Kl{\"u}nder}}, \bibinfo {author} {\bibfnamefont {J.~Marcus}\ \bibnamefont
  {Dahlstr{\"o}m}}, \bibinfo {author} {\bibfnamefont {Miguel}\ \bibnamefont
  {Miranda}}, \bibinfo {author} {\bibfnamefont {Christian}\ \bibnamefont
  {Buth}}, \bibinfo {author} {\bibfnamefont {Kenneth~J.}\ \bibnamefont
  {Schafer}}, \bibinfo {author} {\bibfnamefont {Johan}\ \bibnamefont
  {Mauritsson}}, \bibinfo {author} {\bibfnamefont {Anne}\ \bibnamefont
  {L'Huillier}}, \ and\ \bibinfo {author} {\bibfnamefont {Mathieu}\
  \bibnamefont {Gisselbrecht}},\ }\bibfield  {title} {\enquote {\bibinfo
  {title} {Phase measurement of resonant two-photon ionization in helium},}\
  }\href {\doibase 10.1103/PhysRevLett.104.103003} {\bibfield  {journal}
  {\bibinfo  {journal} {Phys. Rev. Lett.}\ }\textbf {\bibinfo {volume} {104}},\
  \bibinfo {pages} {103003} (\bibinfo {year} {2010})},\ \Eprint
  {http://arxiv.org/abs/1002.2550} {arXiv:1002.2550} \BibitemShut {NoStop}%
\bibitem [{\citenamefont {Itatani}\ \emph {et~al.}(2004)\citenamefont
  {Itatani}, \citenamefont {Levesque}, \citenamefont {Zeidler}, \citenamefont
  {Niikura}, \citenamefont {P{\'e}pin}, \citenamefont {Kieffer}, \citenamefont
  {Corkum},\ and\ \citenamefont {Villeneuve}}]{Itatani:TI-04}%
  \BibitemOpen
  \bibfield  {author} {\bibinfo {author} {\bibfnamefont {J.}~\bibnamefont
  {Itatani}}, \bibinfo {author} {\bibfnamefont {J.}~\bibnamefont {Levesque}},
  \bibinfo {author} {\bibfnamefont {D.}~\bibnamefont {Zeidler}}, \bibinfo
  {author} {\bibfnamefont {Hiromichi}\ \bibnamefont {Niikura}}, \bibinfo
  {author} {\bibfnamefont {H.}~\bibnamefont {P{\'e}pin}}, \bibinfo {author}
  {\bibfnamefont {J.~C.}\ \bibnamefont {Kieffer}}, \bibinfo {author}
  {\bibfnamefont {P.~B.}\ \bibnamefont {Corkum}}, \ and\ \bibinfo {author}
  {\bibfnamefont {D.~M.}\ \bibnamefont {Villeneuve}},\ }\bibfield  {title}
  {\enquote {\bibinfo {title} {Tomographic imaging of molecular orbitals},}\
  }\href {\doibase 10.1038/nature03183} {\bibfield  {journal} {\bibinfo
  {journal} {Nature}\ }\textbf {\bibinfo {volume} {432}},\ \bibinfo {pages}
  {867--871} (\bibinfo {year} {2004})}\BibitemShut {NoStop}%
\bibitem [{\citenamefont {Santra}(2006)}]{Santra:IM-06}%
  \BibitemOpen
  \bibfield  {author} {\bibinfo {author} {\bibfnamefont {Robin}\ \bibnamefont
  {Santra}},\ }\bibfield  {title} {\enquote {\bibinfo {title} {Imaging
  molecular orbitals using photoionization},}\ }\href {\doibase
  10.1016/j.chemphys.2006.07.008} {\bibfield  {journal} {\bibinfo  {journal}
  {Chem. Phys.}\ }\textbf {\bibinfo {volume} {329}},\ \bibinfo {pages}
  {357--364} (\bibinfo {year} {2006})}\BibitemShut {NoStop}%
\bibitem [{\citenamefont {Morishita}\ \emph {et~al.}(2008)\citenamefont
  {Morishita}, \citenamefont {Le}, \citenamefont {Chen},\ and\ \citenamefont
  {Lin}}]{Morishita:AR-08}%
  \BibitemOpen
  \bibfield  {author} {\bibinfo {author} {\bibfnamefont {Toru}\ \bibnamefont
  {Morishita}}, \bibinfo {author} {\bibfnamefont {Anh-Thu}\ \bibnamefont {Le}},
  \bibinfo {author} {\bibfnamefont {Zhangjin}\ \bibnamefont {Chen}}, \ and\
  \bibinfo {author} {\bibfnamefont {C.~D.}\ \bibnamefont {Lin}},\ }\bibfield
  {title} {\enquote {\bibinfo {title} {Accurate retrieval of structural
  information from laser-induced photoelectron and high-order harmonic spectra
  by few-cycle laser pulses},}\ }\href {\doibase
  10.1103/PhysRevLett.100.013903} {\bibfield  {journal} {\bibinfo  {journal}
  {Phys. Rev. Lett.}\ }\textbf {\bibinfo {volume} {100}},\ \bibinfo {pages}
  {013903} (\bibinfo {year} {2008})}\BibitemShut {NoStop}%
\bibitem [{\citenamefont {Lin}\ and\ \citenamefont {Xu}(2012)}]{Lin:UD-12}%
  \BibitemOpen
  \bibfield  {author} {\bibinfo {author} {\bibfnamefont {C.~D.}\ \bibnamefont
  {Lin}}\ and\ \bibinfo {author} {\bibfnamefont {Junliang}\ \bibnamefont
  {Xu}},\ }\bibfield  {title} {\enquote {\bibinfo {title} {Imaging ultrafast
  dynamics of molecules with laser-induced electron diffraction},}\ }\href
  {\doibase 10.1039/C2CP41606A} {\bibfield  {journal} {\bibinfo  {journal}
  {Phys. Chem. Chem. Phys.}\ }\textbf {\bibinfo {volume} {14}},\ \bibinfo
  {pages} {13133--13145} (\bibinfo {year} {2012})}\BibitemShut {NoStop}%
\bibitem [{\citenamefont {Santra}\ and\ \citenamefont
  {Gordon}(2006)}]{Santra:TS-06}%
  \BibitemOpen
  \bibfield  {author} {\bibinfo {author} {\bibfnamefont {Robin}\ \bibnamefont
  {Santra}}\ and\ \bibinfo {author} {\bibfnamefont {Ariel}\ \bibnamefont
  {Gordon}},\ }\bibfield  {title} {\enquote {\bibinfo {title} {Three-step model
  for high-harmonic generation in many-electron systems},}\ }\href {\doibase
  10.1103/PhysRevLett.96.073906} {\bibfield  {journal} {\bibinfo  {journal}
  {Phys. Rev. Lett.}\ }\textbf {\bibinfo {volume} {96}},\ \bibinfo {pages}
  {073906} (\bibinfo {year} {2006})}\BibitemShut {NoStop}%
\bibitem [{\citenamefont {Patchkovskii}\ \emph {et~al.}(2007)\citenamefont
  {Patchkovskii}, \citenamefont {Zhao}, \citenamefont {Brabec},\ and\
  \citenamefont {Villeneuve}}]{Patchkovskii:OT-07}%
  \BibitemOpen
  \bibfield  {author} {\bibinfo {author} {\bibfnamefont {Serguei}\ \bibnamefont
  {Patchkovskii}}, \bibinfo {author} {\bibfnamefont {Zengxiu}\ \bibnamefont
  {Zhao}}, \bibinfo {author} {\bibfnamefont {Thomas}\ \bibnamefont {Brabec}}, \
  and\ \bibinfo {author} {\bibfnamefont {David~M.}\ \bibnamefont
  {Villeneuve}},\ }\bibfield  {title} {\enquote {\bibinfo {title} {High
  harmonic generation and molecular orbital tomography in multielectron
  systems},}\ }\href {\doibase 10.1063/1.2711809} {\bibfield  {journal}
  {\bibinfo  {journal} {J. Chem. Phys.}\ }\textbf {\bibinfo {volume} {126}},\
  \bibinfo {pages} {114306} (\bibinfo {year} {2007})}\BibitemShut {NoStop}%
\bibitem [{\citenamefont {Kling}\ \emph {et~al.}(2006)\citenamefont {Kling},
  \citenamefont {Siedschlag}, \citenamefont {Verhoef}, \citenamefont {Khan},
  \citenamefont {Schultze}, \citenamefont {Uphues}, \citenamefont {Ni},
  \citenamefont {Uiberacker}, \citenamefont {Drescher}, \citenamefont
  {Krausz},\ and\ \citenamefont {Vrakking}}]{Kling:CE-06}%
  \BibitemOpen
  \bibfield  {author} {\bibinfo {author} {\bibfnamefont {M.~F.}\ \bibnamefont
  {Kling}}, \bibinfo {author} {\bibfnamefont {Ch.}\ \bibnamefont {Siedschlag}},
  \bibinfo {author} {\bibfnamefont {A.~J.}\ \bibnamefont {Verhoef}}, \bibinfo
  {author} {\bibfnamefont {J.~I.}\ \bibnamefont {Khan}}, \bibinfo {author}
  {\bibfnamefont {M.}~\bibnamefont {Schultze}}, \bibinfo {author}
  {\bibfnamefont {Th.}\ \bibnamefont {Uphues}}, \bibinfo {author}
  {\bibfnamefont {Y.}~\bibnamefont {Ni}}, \bibinfo {author} {\bibfnamefont
  {M.}~\bibnamefont {Uiberacker}}, \bibinfo {author} {\bibfnamefont
  {M.}~\bibnamefont {Drescher}}, \bibinfo {author} {\bibfnamefont
  {F.}~\bibnamefont {Krausz}}, \ and\ \bibinfo {author} {\bibfnamefont
  {M.~J.~J.}\ \bibnamefont {Vrakking}},\ }\bibfield  {title} {\enquote
  {\bibinfo {title} {Control of electron localization in molecular
  dissociation},}\ }\href {\doibase 10.1126/science.1126259} {\bibfield
  {journal} {\bibinfo  {journal} {Science}\ }\textbf {\bibinfo {volume}
  {312}},\ \bibinfo {pages} {246--248} (\bibinfo {year} {2006})}\BibitemShut
  {NoStop}%
\bibitem [{\citenamefont {Cavalieri}\ \emph {et~al.}(2007)\citenamefont
  {Cavalieri}, \citenamefont {M{\"u}ller}, \citenamefont {Uphues},
  \citenamefont {Yakovlev}, \citenamefont {Baltuka}, \citenamefont {Horvath},
  \citenamefont {Schmidt}, \citenamefont {Holzwarth}, \citenamefont {Drescher},
  \citenamefont {Kleineberg}, \citenamefont {Echenique}, \citenamefont
  {Kienberger}, \citenamefont {Krausz},\ and\ \citenamefont
  {Heinzmann}}]{Cavalieri:AS-07}%
  \BibitemOpen
  \bibfield  {author} {\bibinfo {author} {\bibfnamefont {A.~L.}\ \bibnamefont
  {Cavalieri}}, \bibinfo {author} {\bibfnamefont {N.}~\bibnamefont
  {M{\"u}ller}}, \bibinfo {author} {\bibfnamefont {Th.}\ \bibnamefont
  {Uphues}}, \bibinfo {author} {\bibfnamefont {V.~S.}\ \bibnamefont
  {Yakovlev}}, \bibinfo {author} {\bibfnamefont {A.}~\bibnamefont {Baltuka}},
  \bibinfo {author} {\bibfnamefont {B.}~\bibnamefont {Horvath}}, \bibinfo
  {author} {\bibfnamefont {L.}~\bibnamefont {Schmidt}, \bibfnamefont
  {B.~Bl{\"u}mel}}, \bibinfo {author} {\bibfnamefont {S.}~\bibnamefont
  {Holzwarth}, \bibfnamefont {R.~Hendel}}, \bibinfo {author} {\bibfnamefont
  {M.}~\bibnamefont {Drescher}}, \bibinfo {author} {\bibfnamefont
  {U.}~\bibnamefont {Kleineberg}}, \bibinfo {author} {\bibfnamefont {P.~M.}\
  \bibnamefont {Echenique}}, \bibinfo {author} {\bibfnamefont {R.}~\bibnamefont
  {Kienberger}}, \bibinfo {author} {\bibfnamefont {F.}~\bibnamefont {Krausz}},
  \ and\ \bibinfo {author} {\bibfnamefont {U.}~\bibnamefont {Heinzmann}},\
  }\bibfield  {title} {\enquote {\bibinfo {title} {Attosecond spectroscopy in
  condensed matter},}\ }\href {\doibase 10.1038/nature06229} {\bibfield
  {journal} {\bibinfo  {journal} {Nature}\ }\textbf {\bibinfo {volume} {449}},\
  \bibinfo {pages} {1029--1032} (\bibinfo {year} {2007})}\BibitemShut {NoStop}%
\bibitem [{\citenamefont {Kulander}\ and\ \citenamefont
  {Rescigno}(1991)}]{Kulander:EP-91}%
  \BibitemOpen
  \bibfield  {author} {\bibinfo {author} {\bibfnamefont {K.~C.}\ \bibnamefont
  {Kulander}}\ and\ \bibinfo {author} {\bibfnamefont {T.~N.}\ \bibnamefont
  {Rescigno}},\ }\bibfield  {title} {\enquote {\bibinfo {title} {Effective
  potentials for time-dependent calculations of multiphoton processes in
  atoms},}\ }\href {\doibase 10.1016/0010-4655(91)90273-N} {\bibfield
  {journal} {\bibinfo  {journal} {Comput. Phys. Commun.}\ }\textbf {\bibinfo
  {volume} {63}},\ \bibinfo {pages} {523--528} (\bibinfo {year}
  {1991})}\BibitemShut {NoStop}%
\bibitem [{\citenamefont {Kulander}\ \emph {et~al.}(1993)\citenamefont
  {Kulander}, \citenamefont {Schafer},\ and\ \citenamefont
  {Krause}}]{Kulander:SI-93}%
  \BibitemOpen
  \bibfield  {author} {\bibinfo {author} {\bibfnamefont {K.~C.}\ \bibnamefont
  {Kulander}}, \bibinfo {author} {\bibfnamefont {K.~J.}\ \bibnamefont
  {Schafer}}, \ and\ \bibinfo {author} {\bibfnamefont {J.~L.}\ \bibnamefont
  {Krause}},\ }\bibfield  {title} {\enquote {\bibinfo {title} {Dynamics of
  short-pulse excitation, ionization and harmonic conversion},}\ }in\
  \href@noop {} {\emph {\bibinfo {booktitle} {Super-Intense Laser-Atom
  Physics}}},\ \bibinfo {series} {NATO Advanced Study Institute Series B:
  Physics}, Vol.\ \bibinfo {volume} {316},\ \bibinfo {editor} {edited by\
  \bibinfo {editor} {\bibfnamefont {Bernard}\ \bibnamefont {Piraux}}, \bibinfo
  {editor} {\bibfnamefont {Anne}\ \bibnamefont {L'Huillier}}, \ and\ \bibinfo
  {editor} {\bibfnamefont {Kazimierz}\ \bibnamefont
  {{Rz{\c{a}}{\.{z}}ewski}}}}\ (\bibinfo  {publisher} {Plenum Press},\ \bibinfo
  {address} {New York},\ \bibinfo {year} {1993})\ pp.\ \bibinfo {pages}
  {95--110}\BibitemShut {NoStop}%
\bibitem [{\citenamefont {Pukhov}\ \emph {et~al.}(2003)\citenamefont {Pukhov},
  \citenamefont {Gordienko},\ and\ \citenamefont {Baeva}}]{Pukhov:TS-03}%
  \BibitemOpen
  \bibfield  {author} {\bibinfo {author} {\bibfnamefont {A.}~\bibnamefont
  {Pukhov}}, \bibinfo {author} {\bibfnamefont {S.}~\bibnamefont {Gordienko}}, \
  and\ \bibinfo {author} {\bibfnamefont {T.}~\bibnamefont {Baeva}},\ }\bibfield
   {title} {\enquote {\bibinfo {title} {Temporal structure of attosecond pulses
  from intense laser-atom interactions},}\ }\href {\doibase
  10.1103/PhysRevLett.91.173002} {\bibfield  {journal} {\bibinfo  {journal}
  {Phys. Rev. Lett.}\ }\textbf {\bibinfo {volume} {91}},\ \bibinfo {pages}
  {173002} (\bibinfo {year} {2003})}\BibitemShut {NoStop}%
\bibitem [{\citenamefont {Paul}\ \emph {et~al.}(2001)\citenamefont {Paul},
  \citenamefont {Toma}, \citenamefont {Breger}, \citenamefont {Mullot},
  \citenamefont {Aug{\'e}}, \citenamefont {Balcou}, \citenamefont {Muller},\
  and\ \citenamefont {Agostini}}]{Paul:TA-01}%
  \BibitemOpen
  \bibfield  {author} {\bibinfo {author} {\bibfnamefont {P.~M.}\ \bibnamefont
  {Paul}}, \bibinfo {author} {\bibfnamefont {E.~S.}\ \bibnamefont {Toma}},
  \bibinfo {author} {\bibfnamefont {P.}~\bibnamefont {Breger}}, \bibinfo
  {author} {\bibfnamefont {G.}~\bibnamefont {Mullot}}, \bibinfo {author}
  {\bibfnamefont {F.}~\bibnamefont {Aug{\'e}}}, \bibinfo {author}
  {\bibfnamefont {Ph.}\ \bibnamefont {Balcou}}, \bibinfo {author}
  {\bibfnamefont {H.~G.}\ \bibnamefont {Muller}}, \ and\ \bibinfo {author}
  {\bibfnamefont {P.}~\bibnamefont {Agostini}},\ }\bibfield  {title} {\enquote
  {\bibinfo {title} {Observation of a train of attosecond pulses from high
  harmonic generation},}\ }\href {\doibase 10.1126/science.1059413} {\bibfield
  {journal} {\bibinfo  {journal} {Science}\ }\textbf {\bibinfo {volume}
  {292}},\ \bibinfo {pages} {1689--1692} (\bibinfo {year} {2001})}\BibitemShut
  {NoStop}%
\bibitem [{\citenamefont {Goulielmakis}\ \emph {et~al.}(2008)\citenamefont
  {Goulielmakis}, \citenamefont {Schultze}, \citenamefont {Hofstetter},
  \citenamefont {Yakovlev}, \citenamefont {Gagnon}, \citenamefont {Uiberacker},
  \citenamefont {Aquila}, \citenamefont {Gullikson}, \citenamefont {Attwood},
  \citenamefont {Kienberger}, \citenamefont {Krausz},\ and\ \citenamefont
  {Kleineberg}}]{Goulielmakis:SC-08}%
  \BibitemOpen
  \bibfield  {author} {\bibinfo {author} {\bibfnamefont {E.}~\bibnamefont
  {Goulielmakis}}, \bibinfo {author} {\bibfnamefont {M.}~\bibnamefont
  {Schultze}}, \bibinfo {author} {\bibfnamefont {M.}~\bibnamefont
  {Hofstetter}}, \bibinfo {author} {\bibfnamefont {V.~S.}\ \bibnamefont
  {Yakovlev}}, \bibinfo {author} {\bibfnamefont {J.}~\bibnamefont {Gagnon}},
  \bibinfo {author} {\bibfnamefont {M.}~\bibnamefont {Uiberacker}}, \bibinfo
  {author} {\bibfnamefont {A.~L.}\ \bibnamefont {Aquila}}, \bibinfo {author}
  {\bibfnamefont {E.~M.}\ \bibnamefont {Gullikson}}, \bibinfo {author}
  {\bibfnamefont {D.~T.}\ \bibnamefont {Attwood}}, \bibinfo {author}
  {\bibfnamefont {R.}~\bibnamefont {Kienberger}}, \bibinfo {author}
  {\bibfnamefont {F.}~\bibnamefont {Krausz}}, \ and\ \bibinfo {author}
  {\bibfnamefont {U.}~\bibnamefont {Kleineberg}},\ }\bibfield  {title}
  {\enquote {\bibinfo {title} {Single-cycle nonlinear optics},}\ }\href
  {\doibase 10.1126/science.1157846} {\bibfield  {journal} {\bibinfo  {journal}
  {Science}\ }\textbf {\bibinfo {volume} {320}},\ \bibinfo {pages} {1614--1617}
  (\bibinfo {year} {2008})}\BibitemShut {NoStop}%
\bibitem [{\citenamefont {Chipperfield}\ \emph {et~al.}(2009)\citenamefont
  {Chipperfield}, \citenamefont {Robinson}, \citenamefont {Tisch},\ and\
  \citenamefont {Marangos}}]{Chipperfield:IW-09}%
  \BibitemOpen
  \bibfield  {author} {\bibinfo {author} {\bibfnamefont {L.~E.}\ \bibnamefont
  {Chipperfield}}, \bibinfo {author} {\bibfnamefont {J.~S.}\ \bibnamefont
  {Robinson}}, \bibinfo {author} {\bibfnamefont {J.~W.~G.}\ \bibnamefont
  {Tisch}}, \ and\ \bibinfo {author} {\bibfnamefont {J.~P.}\ \bibnamefont
  {Marangos}},\ }\bibfield  {title} {\enquote {\bibinfo {title} {Ideal waveform
  to generate the maximum possible electron recollision energy for any given
  oscillation period},}\ }\href {\doibase 10.1103/PhysRevLett.102.063003}
  {\bibfield  {journal} {\bibinfo  {journal} {Phys. Rev. Lett.}\ }\textbf
  {\bibinfo {volume} {102}},\ \bibinfo {pages} {063003} (\bibinfo {year}
  {2009})}\BibitemShut {NoStop}%
\bibitem [{\citenamefont {Watson}\ \emph {et~al.}(1996)\citenamefont {Watson},
  \citenamefont {Sanpera}, \citenamefont {Chen},\ and\ \citenamefont
  {Burnett}}]{Watson:CS-96}%
  \BibitemOpen
  \bibfield  {author} {\bibinfo {author} {\bibfnamefont {J.~B.}\ \bibnamefont
  {Watson}}, \bibinfo {author} {\bibfnamefont {A.}~\bibnamefont {Sanpera}},
  \bibinfo {author} {\bibfnamefont {X.}~\bibnamefont {Chen}}, \ and\ \bibinfo
  {author} {\bibfnamefont {K.}~\bibnamefont {Burnett}},\ }\bibfield  {title}
  {\enquote {\bibinfo {title} {Harmonic generation from a coherent
  superposition of states},}\ }\href {\doibase 10.1103/PhysRevA.53.R1962}
  {\bibfield  {journal} {\bibinfo  {journal} {Phys. Rev. A}\ }\textbf {\bibinfo
  {volume} {53}},\ \bibinfo {pages} {R1962--R1965} (\bibinfo {year}
  {1996})}\BibitemShut {NoStop}%
\bibitem [{\citenamefont {Sanpera}\ \emph {et~al.}(1996)\citenamefont
  {Sanpera}, \citenamefont {Watson}, \citenamefont {Lewenstein},\ and\
  \citenamefont {Burnett}}]{Sanpera:HG-96}%
  \BibitemOpen
  \bibfield  {author} {\bibinfo {author} {\bibfnamefont {A.}~\bibnamefont
  {Sanpera}}, \bibinfo {author} {\bibfnamefont {J.~B.}\ \bibnamefont {Watson}},
  \bibinfo {author} {\bibfnamefont {M.}~\bibnamefont {Lewenstein}}, \ and\
  \bibinfo {author} {\bibfnamefont {K.}~\bibnamefont {Burnett}},\ }\bibfield
  {title} {\enquote {\bibinfo {title} {Harmonic-generation control},}\ }\href
  {\doibase 10.1103/PhysRevA.54.4320} {\bibfield  {journal} {\bibinfo
  {journal} {Phys. Rev. A}\ }\textbf {\bibinfo {volume} {54}},\ \bibinfo
  {pages} {4320--4326} (\bibinfo {year} {1996})}\BibitemShut {NoStop}%
\bibitem [{\citenamefont {Milosevic}(2006)}]{Milosevic:TA-06}%
  \BibitemOpen
  \bibfield  {author} {\bibinfo {author} {\bibfnamefont {Dejan~B.}\
  \bibnamefont {Milosevic}},\ }\bibfield  {title} {\enquote {\bibinfo {title}
  {Theoretical analysis of high-order harmonic generation from a coherent
  superposition of states},}\ }\href {\doibase 10.1364/JOSAB.23.000308}
  {\bibfield  {journal} {\bibinfo  {journal} {J. Opt. Soc. Am. B}\ }\textbf
  {\bibinfo {volume} {23}},\ \bibinfo {pages} {308--317} (\bibinfo {year}
  {2006})}\BibitemShut {NoStop}%
\bibitem [{\citenamefont {Milo{\v{s}}evi{\'c}}(2007)}]{Milosevic:HE-07}%
  \BibitemOpen
  \bibfield  {author} {\bibinfo {author} {\bibfnamefont {D.~B.}\ \bibnamefont
  {Milo{\v{s}}evi{\'c}}},\ }\bibfield  {title} {\enquote {\bibinfo {title}
  {High-energy stimulated emission from plasma ablation pumped by resonant
  high-order harmonic generation},}\ }\href {\doibase
  10.1088/0953-4075/40/17/005} {\bibfield  {journal} {\bibinfo  {journal} {J.
  Phys. B}\ }\textbf {\bibinfo {volume} {40}},\ \bibinfo {pages} {3367--3376}
  (\bibinfo {year} {2007})}\BibitemShut {NoStop}%
\bibitem [{\citenamefont {Strelkov}(2010)}]{Strelkov:AR-10}%
  \BibitemOpen
  \bibfield  {author} {\bibinfo {author} {\bibfnamefont {V.}~\bibnamefont
  {Strelkov}},\ }\bibfield  {title} {\enquote {\bibinfo {title} {Role of
  autoionizing state in resonant high-order harmonic generation and attosecond
  pulse production},}\ }\href {\doibase 10.1103/PhysRevLett.104.123901}
  {\bibfield  {journal} {\bibinfo  {journal} {Phys. Rev. Lett.}\ }\textbf
  {\bibinfo {volume} {104}},\ \bibinfo {pages} {123901} (\bibinfo {year}
  {2010})}\BibitemShut {NoStop}%
\bibitem [{\citenamefont {Milo\ifmmode \check{s}\else
  \v{s}\fi{}evi\ifmmode~\acute{c}\else \'{c}\fi{}}(2010)}]{Milosevic:RH-10}%
  \BibitemOpen
  \bibfield  {author} {\bibinfo {author} {\bibfnamefont {D.~B.}\ \bibnamefont
  {Milo\ifmmode \check{s}\else \v{s}\fi{}evi\ifmmode~\acute{c}\else
  \'{c}\fi{}}},\ }\bibfield  {title} {\enquote {\bibinfo {title} {Resonant
  high-order harmonic generation from plasma ablation: {Laser} intensity
  dependence of the harmonic intensity and phase},}\ }\href {\doibase
  10.1103/PhysRevA.81.023802} {\bibfield  {journal} {\bibinfo  {journal} {Phys.
  Rev. A}\ }\textbf {\bibinfo {volume} {81}},\ \bibinfo {pages} {023802}
  (\bibinfo {year} {2010})}\BibitemShut {NoStop}%
\bibitem [{\citenamefont {McFarland}\ \emph {et~al.}(2008)\citenamefont
  {McFarland}, \citenamefont {Farrell}, \citenamefont {Bucksbaum},\ and\
  \citenamefont {G{\"u}hr}}]{McFarland:HH-08}%
  \BibitemOpen
  \bibfield  {author} {\bibinfo {author} {\bibfnamefont {Brian~K.}\
  \bibnamefont {McFarland}}, \bibinfo {author} {\bibfnamefont {Joseph~P.}\
  \bibnamefont {Farrell}}, \bibinfo {author} {\bibfnamefont {Philip~H.}\
  \bibnamefont {Bucksbaum}}, \ and\ \bibinfo {author} {\bibfnamefont {Markus}\
  \bibnamefont {G{\"u}hr}},\ }\bibfield  {title} {\enquote {\bibinfo {title}
  {High harmonic generation from multiple orbitals in {N$_2$}},}\ }\href
  {\doibase 10.1126/science.1162780} {\bibfield  {journal} {\bibinfo  {journal}
  {Science}\ }\textbf {\bibinfo {volume} {322}},\ \bibinfo {pages} {1232--1235}
  (\bibinfo {year} {2008})}\BibitemShut {NoStop}%
\bibitem [{\citenamefont {Smirnova}\ \emph {et~al.}(2009)\citenamefont
  {Smirnova}, \citenamefont {Mairesse}, \citenamefont {Patchkovskii},
  \citenamefont {Dudovich}, \citenamefont {Villeneuve}, \citenamefont
  {Corkum},\ and\ \citenamefont {Ivanov}}]{Smirnova:HI-09}%
  \BibitemOpen
  \bibfield  {author} {\bibinfo {author} {\bibfnamefont {Olga}\ \bibnamefont
  {Smirnova}}, \bibinfo {author} {\bibfnamefont {Yann}\ \bibnamefont
  {Mairesse}}, \bibinfo {author} {\bibfnamefont {Serguei}\ \bibnamefont
  {Patchkovskii}}, \bibinfo {author} {\bibfnamefont {Nirit}\ \bibnamefont
  {Dudovich}}, \bibinfo {author} {\bibfnamefont {David}\ \bibnamefont
  {Villeneuve}}, \bibinfo {author} {\bibfnamefont {Paul}\ \bibnamefont
  {Corkum}}, \ and\ \bibinfo {author} {\bibfnamefont {Misha~Yu.}\ \bibnamefont
  {Ivanov}},\ }\bibfield  {title} {\enquote {\bibinfo {title} {High harmonic
  interferometry of multi-electron dynamics in molecules},}\ }\href {\doibase
  10.1038/nature08253} {\bibfield  {journal} {\bibinfo  {journal} {Nature}\
  }\textbf {\bibinfo {volume} {460}},\ \bibinfo {pages} {972--977} (\bibinfo
  {year} {2009})}\BibitemShut {NoStop}%
\bibitem [{\citenamefont {Figueira~de Morisson~Faria}\ and\ \citenamefont
  {Augstein}(2010)}]{Figueira:MH-10}%
  \BibitemOpen
  \bibfield  {author} {\bibinfo {author} {\bibfnamefont {C.}~\bibnamefont
  {Figueira~de Morisson~Faria}}\ and\ \bibinfo {author} {\bibfnamefont {B.~B.}\
  \bibnamefont {Augstein}},\ }\bibfield  {title} {\enquote {\bibinfo {title}
  {Molecular high-order harmonic generation with more than one active orbital:
  {Quantum} interference effects},}\ }\href {\doibase
  10.1103/PhysRevA.81.043409} {\bibfield  {journal} {\bibinfo  {journal} {Phys.
  Rev. A}\ }\textbf {\bibinfo {volume} {81}},\ \bibinfo {pages} {043409}
  (\bibinfo {year} {2010})}\BibitemShut {NoStop}%
\bibitem [{\citenamefont {Kraus}\ \emph {et~al.}(2013)\citenamefont {Kraus},
  \citenamefont {Zhang}, \citenamefont {Gijsbertsen}, \citenamefont {Lucchese},
  \citenamefont {Rohringer},\ and\ \citenamefont {W{\"o}rner}}]{Kraus:HP-13}%
  \BibitemOpen
  \bibfield  {author} {\bibinfo {author} {\bibfnamefont {P.~M.}\ \bibnamefont
  {Kraus}}, \bibinfo {author} {\bibfnamefont {S.~B.}\ \bibnamefont {Zhang}},
  \bibinfo {author} {\bibfnamefont {A.}~\bibnamefont {Gijsbertsen}}, \bibinfo
  {author} {\bibfnamefont {R.~R.}\ \bibnamefont {Lucchese}}, \bibinfo {author}
  {\bibfnamefont {N.}~\bibnamefont {Rohringer}}, \ and\ \bibinfo {author}
  {\bibfnamefont {H.~J.}\ \bibnamefont {W{\"o}rner}},\ }\bibfield  {title}
  {\enquote {\bibinfo {title} {High-harmonic probing of electronic coherence in
  dynamically aligned molecules},}\ }\href {\doibase
  10.1103/PhysRevLett.111.243005} {\bibfield  {journal} {\bibinfo  {journal}
  {Phys. Rev. Lett.}\ }\textbf {\bibinfo {volume} {111}},\ \bibinfo {pages}
  {243005} (\bibinfo {year} {2013})}\BibitemShut {NoStop}%
\bibitem [{\citenamefont {Zanghellini}\ \emph {et~al.}(2006)\citenamefont
  {Zanghellini}, \citenamefont {Jungreuthmayer},\ and\ \citenamefont
  {Brabec}}]{Zanghellini:PS-06}%
  \BibitemOpen
  \bibfield  {author} {\bibinfo {author} {\bibfnamefont {J.}~\bibnamefont
  {Zanghellini}}, \bibinfo {author} {\bibfnamefont {Ch.}\ \bibnamefont
  {Jungreuthmayer}}, \ and\ \bibinfo {author} {\bibfnamefont {T.}~\bibnamefont
  {Brabec}},\ }\bibfield  {title} {\enquote {\bibinfo {title} {Plasmon
  signatures in high harmonic generation},}\ }\href {\doibase
  10.1088/0953-4075/39/3/022} {\bibfield  {journal} {\bibinfo  {journal} {J.
  Phys. B}\ }\textbf {\bibinfo {volume} {39}},\ \bibinfo {pages} {709--728}
  (\bibinfo {year} {2006})}\BibitemShut {NoStop}%
\bibitem [{\citenamefont {Gordon}\ \emph {et~al.}(2006)\citenamefont {Gordon},
  \citenamefont {K{\"a}rtner}, \citenamefont {Rohringer},\ and\ \citenamefont
  {Santra}}]{Gordon:RM-06}%
  \BibitemOpen
  \bibfield  {author} {\bibinfo {author} {\bibfnamefont {Ariel}\ \bibnamefont
  {Gordon}}, \bibinfo {author} {\bibfnamefont {Franz~X.}\ \bibnamefont
  {K{\"a}rtner}}, \bibinfo {author} {\bibfnamefont {Nina}\ \bibnamefont
  {Rohringer}}, \ and\ \bibinfo {author} {\bibfnamefont {Robin}\ \bibnamefont
  {Santra}},\ }\bibfield  {title} {\enquote {\bibinfo {title} {Role of
  many-electron dynamics in high harmonic generation},}\ }\href {\doibase
  10.1103/PhysRevLett.96.223902} {\bibfield  {journal} {\bibinfo  {journal}
  {Phys. Rev. Lett.}\ }\textbf {\bibinfo {volume} {96}},\ \bibinfo {pages}
  {223902} (\bibinfo {year} {2006})}\BibitemShut {NoStop}%
\bibitem [{\citenamefont {Koval}\ \emph {et~al.}(2007)\citenamefont {Koval},
  \citenamefont {Wilken}, \citenamefont {Bauer},\ and\ \citenamefont
  {Keitel}}]{Koval:ND-07}%
  \BibitemOpen
  \bibfield  {author} {\bibinfo {author} {\bibfnamefont {P.}~\bibnamefont
  {Koval}}, \bibinfo {author} {\bibfnamefont {F.}~\bibnamefont {Wilken}},
  \bibinfo {author} {\bibfnamefont {D.}~\bibnamefont {Bauer}}, \ and\ \bibinfo
  {author} {\bibfnamefont {C.~H.}\ \bibnamefont {Keitel}},\ }\bibfield  {title}
  {\enquote {\bibinfo {title} {Nonsequential double recombination in intense
  laser fields},}\ }\href {\doibase 10.1103/PhysRevLett.98.043904} {\bibfield
  {journal} {\bibinfo  {journal} {Phys. Rev. Lett.}\ }\textbf {\bibinfo
  {volume} {98}},\ \bibinfo {pages} {043904} (\bibinfo {year}
  {2007})}\BibitemShut {NoStop}%
\bibitem [{\citenamefont {Ishikawa}(2003)}]{Ishikawa:PE-03}%
  \BibitemOpen
  \bibfield  {author} {\bibinfo {author} {\bibfnamefont {Kenichi}\ \bibnamefont
  {Ishikawa}},\ }\bibfield  {title} {\enquote {\bibinfo {title} {Photoemission
  and ionization of {He$^+$} under simultaneous irradiation of fundamental
  laser and high-order harmonic pulses},}\ }\href {\doibase
  10.1103/PhysRevLett.91.043002} {\bibfield  {journal} {\bibinfo  {journal}
  {Phys. Rev. Lett.}\ }\textbf {\bibinfo {volume} {91}},\ \bibinfo {pages}
  {043002} (\bibinfo {year} {2003})}\BibitemShut {NoStop}%
\bibitem [{\citenamefont {Takahashi}\ \emph {et~al.}(2007)\citenamefont
  {Takahashi}, \citenamefont {Kanai}, \citenamefont {Ishikawa}, \citenamefont
  {Nabekawa},\ and\ \citenamefont {Midorikawa}}]{Takahashi:DE-07}%
  \BibitemOpen
  \bibfield  {author} {\bibinfo {author} {\bibfnamefont {Eiji~J.}\ \bibnamefont
  {Takahashi}}, \bibinfo {author} {\bibfnamefont {Tsuneto}\ \bibnamefont
  {Kanai}}, \bibinfo {author} {\bibfnamefont {Kenichi~L.}\ \bibnamefont
  {Ishikawa}}, \bibinfo {author} {\bibfnamefont {Yasuo}\ \bibnamefont
  {Nabekawa}}, \ and\ \bibinfo {author} {\bibfnamefont {Katsumi}\ \bibnamefont
  {Midorikawa}},\ }\bibfield  {title} {\enquote {\bibinfo {title} {Dramatic
  enhancement of high-order harmonic generation},}\ }\href {\doibase
  10.1103/PhysRevLett.99.053904} {\bibfield  {journal} {\bibinfo  {journal}
  {Phys. Rev. Lett.}\ }\textbf {\bibinfo {volume} {99}},\ \bibinfo {pages}
  {053904} (\bibinfo {year} {2007})}\BibitemShut {NoStop}%
\bibitem [{\citenamefont {Ishikawa}\ \emph {et~al.}(2009)\citenamefont
  {Ishikawa}, \citenamefont {Takahashi},\ and\ \citenamefont
  {Midorikawa}}]{Ishikawa:WD-09}%
  \BibitemOpen
  \bibfield  {author} {\bibinfo {author} {\bibfnamefont {Kenichi~L.}\
  \bibnamefont {Ishikawa}}, \bibinfo {author} {\bibfnamefont {Eiji~J.}\
  \bibnamefont {Takahashi}}, \ and\ \bibinfo {author} {\bibfnamefont {Katsumi}\
  \bibnamefont {Midorikawa}},\ }\bibfield  {title} {\enquote {\bibinfo {title}
  {Wavelength dependence of high-order harmonic generation with independently
  controlled ionization and ponderomotive energy},}\ }\href {\doibase
  10.1103/PhysRevA.80.011807} {\bibfield  {journal} {\bibinfo  {journal} {Phys.
  Rev. A}\ }\textbf {\bibinfo {volume} {80}},\ \bibinfo {pages} {011807}
  (\bibinfo {year} {2009})}\BibitemShut {NoStop}%
\bibitem [{\citenamefont {Popruzhenko}\ \emph {et~al.}(2010)\citenamefont
  {Popruzhenko}, \citenamefont {Zaretsky},\ and\ \citenamefont
  {Becker}}]{Popruzhenko:HO-10}%
  \BibitemOpen
  \bibfield  {author} {\bibinfo {author} {\bibfnamefont {S.~V.}\ \bibnamefont
  {Popruzhenko}}, \bibinfo {author} {\bibfnamefont {D.~F.}\ \bibnamefont
  {Zaretsky}}, \ and\ \bibinfo {author} {\bibfnamefont {W.}~\bibnamefont
  {Becker}},\ }\bibfield  {title} {\enquote {\bibinfo {title} {High-order
  harmonic generation by an intense infrared laser pulse in the presence of a
  weak {UV} pulse},}\ }\href {\doibase 10.1103/PhysRevA.81.063417} {\bibfield
  {journal} {\bibinfo  {journal} {Phys. Rev. A}\ }\textbf {\bibinfo {volume}
  {81}},\ \bibinfo {pages} {063417} (\bibinfo {year} {2010})}\BibitemShut
  {NoStop}%
\bibitem [{\citenamefont {Schafer}\ \emph {et~al.}(2004)\citenamefont
  {Schafer}, \citenamefont {Gaarde}, \citenamefont {Heinrich}, \citenamefont
  {Biegert},\ and\ \citenamefont {Keller}}]{Schafer:SF-04}%
  \BibitemOpen
  \bibfield  {author} {\bibinfo {author} {\bibfnamefont {Kenneth~J.}\
  \bibnamefont {Schafer}}, \bibinfo {author} {\bibfnamefont {Mette~B.}\
  \bibnamefont {Gaarde}}, \bibinfo {author} {\bibfnamefont {Arne}\ \bibnamefont
  {Heinrich}}, \bibinfo {author} {\bibfnamefont {Jens}\ \bibnamefont
  {Biegert}}, \ and\ \bibinfo {author} {\bibfnamefont {Ursula}\ \bibnamefont
  {Keller}},\ }\bibfield  {title} {\enquote {\bibinfo {title} {Strong field
  quantum path control using attosecond pulse trains},}\ }\href {\doibase
  10.1103/PhysRevLett.92.023003} {\bibfield  {journal} {\bibinfo  {journal}
  {Phys. Rev. Lett.}\ }\textbf {\bibinfo {volume} {92}},\ \bibinfo {pages}
  {023003} (\bibinfo {year} {2004})}\BibitemShut {NoStop}%
\bibitem [{\citenamefont {Gaarde}\ \emph {et~al.}(2005)\citenamefont {Gaarde},
  \citenamefont {Schafer}, \citenamefont {Heinrich}, \citenamefont {Biegert},\
  and\ \citenamefont {Keller}}]{Gaarde:LE-05}%
  \BibitemOpen
  \bibfield  {author} {\bibinfo {author} {\bibfnamefont {Mette~B.}\
  \bibnamefont {Gaarde}}, \bibinfo {author} {\bibfnamefont {Kenneth~J.}\
  \bibnamefont {Schafer}}, \bibinfo {author} {\bibfnamefont {Arne}\
  \bibnamefont {Heinrich}}, \bibinfo {author} {\bibfnamefont {Jens}\
  \bibnamefont {Biegert}}, \ and\ \bibinfo {author} {\bibfnamefont {Ursula}\
  \bibnamefont {Keller}},\ }\bibfield  {title} {\enquote {\bibinfo {title}
  {Large enhancement of macroscopic yield in attosecond pulse train-assisted
  harmonic generation},}\ }\href {\doibase 10.1103/PhysRevA.72.013411}
  {\bibfield  {journal} {\bibinfo  {journal} {Phys. Rev. A}\ }\textbf {\bibinfo
  {volume} {72}},\ \bibinfo {pages} {013411} (\bibinfo {year}
  {2005})}\BibitemShut {NoStop}%
\bibitem [{\citenamefont {Figueira~de Morisson~Faria}\ \emph
  {et~al.}(2006)\citenamefont {Figueira~de Morisson~Faria}, \citenamefont
  {Sali{\`e}res}, \citenamefont {Villain},\ and\ \citenamefont
  {Lewenstein}}]{Figueira:CH-06}%
  \BibitemOpen
  \bibfield  {author} {\bibinfo {author} {\bibfnamefont {C.}~\bibnamefont
  {Figueira~de Morisson~Faria}}, \bibinfo {author} {\bibfnamefont
  {P.}~\bibnamefont {Sali{\`e}res}}, \bibinfo {author} {\bibfnamefont
  {P.}~\bibnamefont {Villain}}, \ and\ \bibinfo {author} {\bibfnamefont
  {M.}~\bibnamefont {Lewenstein}},\ }\bibfield  {title} {\enquote {\bibinfo
  {title} {Controlling high-order harmonic generation and above-threshold
  ionization with an attosecond-pulse train},}\ }\href {\doibase
  10.1103/PhysRevA.74.053416} {\bibfield  {journal} {\bibinfo  {journal} {Phys.
  Rev. A}\ }\textbf {\bibinfo {volume} {74}},\ \bibinfo {pages} {053416}
  (\bibinfo {year} {2006})}\BibitemShut {NoStop}%
\bibitem [{\citenamefont {Heinrich}\ \emph {et~al.}(2006)\citenamefont
  {Heinrich}, \citenamefont {Kornelis}, \citenamefont {Anscombe}, \citenamefont
  {Hauri}, \citenamefont {Schlup}, \citenamefont {Biegert},\ and\ \citenamefont
  {Keller}}]{Heinrich:EV-06}%
  \BibitemOpen
  \bibfield  {author} {\bibinfo {author} {\bibfnamefont {A.}~\bibnamefont
  {Heinrich}}, \bibinfo {author} {\bibfnamefont {W.}~\bibnamefont {Kornelis}},
  \bibinfo {author} {\bibfnamefont {M.~P.}\ \bibnamefont {Anscombe}}, \bibinfo
  {author} {\bibfnamefont {C.~P.}\ \bibnamefont {Hauri}}, \bibinfo {author}
  {\bibfnamefont {P.}~\bibnamefont {Schlup}}, \bibinfo {author} {\bibfnamefont
  {J.}~\bibnamefont {Biegert}}, \ and\ \bibinfo {author} {\bibfnamefont
  {U.}~\bibnamefont {Keller}},\ }\bibfield  {title} {\enquote {\bibinfo {title}
  {Enhanced {VUV}-assisted high harmonic generation},}\ }\href {\doibase
  10.1088/0953-4075/39/13/S03} {\bibfield  {journal} {\bibinfo  {journal} {J.
  Phys. B}\ }\textbf {\bibinfo {volume} {39}},\ \bibinfo {pages} {S275--S281}
  (\bibinfo {year} {2006})}\BibitemShut {NoStop}%
\bibitem [{\citenamefont {Biegert}\ \emph {et~al.}(2006)\citenamefont
  {Biegert}, \citenamefont {Heinrich}, \citenamefont {Hauri}, \citenamefont
  {Kornelis}, \citenamefont {Schlup}, \citenamefont {Anscombe}, \citenamefont
  {Gaarde}, \citenamefont {Schafer},\ and\ \citenamefont
  {Keller}}]{Biegert:CH-06}%
  \BibitemOpen
  \bibfield  {author} {\bibinfo {author} {\bibfnamefont {J.}~\bibnamefont
  {Biegert}}, \bibinfo {author} {\bibfnamefont {A.}~\bibnamefont {Heinrich}},
  \bibinfo {author} {\bibfnamefont {C.~P.}\ \bibnamefont {Hauri}}, \bibinfo
  {author} {\bibfnamefont {W.}~\bibnamefont {Kornelis}}, \bibinfo {author}
  {\bibfnamefont {P.}~\bibnamefont {Schlup}}, \bibinfo {author} {\bibfnamefont
  {M.~P.}\ \bibnamefont {Anscombe}}, \bibinfo {author} {\bibfnamefont {M.~B.}\
  \bibnamefont {Gaarde}}, \bibinfo {author} {\bibfnamefont {K.~J.}\
  \bibnamefont {Schafer}}, \ and\ \bibinfo {author} {\bibfnamefont
  {U.}~\bibnamefont {Keller}},\ }\bibfield  {title} {\enquote {\bibinfo {title}
  {Control of high-order harmonic emission using attosecond pulse trains},}\
  }\href {\doibase 10.1080/09500340500167669} {\bibfield  {journal} {\bibinfo
  {journal} {J. Mod. Opt.}\ }\textbf {\bibinfo {volume} {53}},\ \bibinfo
  {pages} {87--96} (\bibinfo {year} {2006})}\BibitemShut {NoStop}%
\bibitem [{\citenamefont {Figueira~de Morisson~Faria}\ and\ \citenamefont
  {Sali{\`e}res}(2007)}]{Figueira:HO-07}%
  \BibitemOpen
  \bibfield  {author} {\bibinfo {author} {\bibfnamefont {C.}~\bibnamefont
  {Figueira~de Morisson~Faria}}\ and\ \bibinfo {author} {\bibfnamefont
  {P.}~\bibnamefont {Sali{\`e}res}},\ }\bibfield  {title} {\enquote {\bibinfo
  {title} {High-order harmonic generation with a strong laser field and an
  attosecond-pulse train: {The Dirac-Delta} comb and monochromatic limits},}\
  }\href {\doibase 10.1134/S1054660X07040147} {\bibfield  {journal} {\bibinfo
  {journal} {Las. Phys.}\ }\textbf {\bibinfo {volume} {17}},\ \bibinfo {pages}
  {390--400} (\bibinfo {year} {2007})}\BibitemShut {NoStop}%
\bibitem [{\citenamefont {Fleischer}\ and\ \citenamefont
  {Moiseyev}(2008)}]{Fleischer:AH-08}%
  \BibitemOpen
  \bibfield  {author} {\bibinfo {author} {\bibfnamefont {Avner}\ \bibnamefont
  {Fleischer}}\ and\ \bibinfo {author} {\bibfnamefont {Nimrod}\ \bibnamefont
  {Moiseyev}},\ }\bibfield  {title} {\enquote {\bibinfo {title} {Amplification
  of high-order harmonics using weak perturbative high-frequency radiation},}\
  }\href {\doibase 10.1103/PhysRevA.77.010102} {\bibfield  {journal} {\bibinfo
  {journal} {Phys. Rev. A}\ }\textbf {\bibinfo {volume} {77}},\ \bibinfo
  {pages} {010102(R)} (\bibinfo {year} {2008})}\BibitemShut {NoStop}%
\bibitem [{\citenamefont {Fleischer}(2008)}]{Fleischer:GH-08}%
  \BibitemOpen
  \bibfield  {author} {\bibinfo {author} {\bibfnamefont {Avner}\ \bibnamefont
  {Fleischer}},\ }\bibfield  {title} {\enquote {\bibinfo {title} {Generation of
  higher-order harmonics upon the addition of high-frequency {XUV} radiation to
  {IR} radiation: {Generalization} of the three-step model},}\ }\href {\doibase
  10.1103/PhysRevA.78.053413} {\bibfield  {journal} {\bibinfo  {journal} {Phys.
  Rev. A}\ }\textbf {\bibinfo {volume} {78}},\ \bibinfo {pages} {053413}
  (\bibinfo {year} {2008})}\BibitemShut {NoStop}%
\bibitem [{\citenamefont {Altarelli}\ \emph {et~al.}(2006)\citenamefont
  {Altarelli}, \citenamefont {Brinkmann}, \citenamefont {Chergui},
  \citenamefont {Decking}, \citenamefont {Dobson}, \citenamefont
  {D{\"u}sterer}, \citenamefont {Gr{\"u}bel}, \citenamefont {Graeff},
  \citenamefont {Graafsma}, \citenamefont {Hajdu}, \citenamefont {Marangos},
  \citenamefont {Pfl{\"u}ger}, \citenamefont {Redlin}, \citenamefont {Riley},
  \citenamefont {Robinson}, \citenamefont {Rossbach}, \citenamefont {Schwarz},
  \citenamefont {Tiedtke}, \citenamefont {Tschentscher}, \citenamefont
  {Vartaniants}, \citenamefont {Wabnitz}, \citenamefont {Weise}, \citenamefont
  {Wichmann}, \citenamefont {Witte}, \citenamefont {Wolf}, \citenamefont
  {Wulff},\ and\ \citenamefont {Yurkov}}]{Altarelli:TDR-06}%
  \BibitemOpen
  \bibinfo {editor} {\bibfnamefont {Massimo}\ \bibnamefont {Altarelli}},
  \bibinfo {editor} {\bibfnamefont {Reinhard}\ \bibnamefont {Brinkmann}},
  \bibinfo {editor} {\bibfnamefont {Majed}\ \bibnamefont {Chergui}}, \bibinfo
  {editor} {\bibfnamefont {Winfried}\ \bibnamefont {Decking}}, \bibinfo
  {editor} {\bibfnamefont {Barry}\ \bibnamefont {Dobson}}, \bibinfo {editor}
  {\bibfnamefont {Stefan}\ \bibnamefont {D{\"u}sterer}}, \bibinfo {editor}
  {\bibfnamefont {Gerhard}\ \bibnamefont {Gr{\"u}bel}}, \bibinfo {editor}
  {\bibfnamefont {Walter}\ \bibnamefont {Graeff}}, \bibinfo {editor}
  {\bibfnamefont {Heinz}\ \bibnamefont {Graafsma}}, \bibinfo {editor}
  {\bibfnamefont {Janos}\ \bibnamefont {Hajdu}}, \bibinfo {editor}
  {\bibfnamefont {Jonathan}\ \bibnamefont {Marangos}}, \bibinfo {editor}
  {\bibfnamefont {Joachim}\ \bibnamefont {Pfl{\"u}ger}}, \bibinfo {editor}
  {\bibfnamefont {Harald}\ \bibnamefont {Redlin}}, \bibinfo {editor}
  {\bibfnamefont {David}\ \bibnamefont {Riley}}, \bibinfo {editor}
  {\bibfnamefont {Ian}\ \bibnamefont {Robinson}}, \bibinfo {editor}
  {\bibfnamefont {J{\"o}rg}\ \bibnamefont {Rossbach}}, \bibinfo {editor}
  {\bibfnamefont {Andreas}\ \bibnamefont {Schwarz}}, \bibinfo {editor}
  {\bibfnamefont {Kai}\ \bibnamefont {Tiedtke}}, \bibinfo {editor}
  {\bibfnamefont {Thomas}\ \bibnamefont {Tschentscher}}, \bibinfo {editor}
  {\bibfnamefont {Ivan}\ \bibnamefont {Vartaniants}}, \bibinfo {editor}
  {\bibfnamefont {Hubertus}\ \bibnamefont {Wabnitz}}, \bibinfo {editor}
  {\bibfnamefont {Hans}\ \bibnamefont {Weise}}, \bibinfo {editor}
  {\bibfnamefont {Riko}\ \bibnamefont {Wichmann}}, \bibinfo {editor}
  {\bibfnamefont {Karl}\ \bibnamefont {Witte}}, \bibinfo {editor}
  {\bibfnamefont {Andreas}\ \bibnamefont {Wolf}}, \bibinfo {editor}
  {\bibfnamefont {Michael}\ \bibnamefont {Wulff}}, \ and\ \bibinfo {editor}
  {\bibfnamefont {Mikhail}\ \bibnamefont {Yurkov}},\ eds.,\ \href@noop {}
  {\emph {\bibinfo {title} {The Technical Design Report of the European
  XFEL}}},\ \bibinfo {number} {DESY 2006-097}\ (\bibinfo  {publisher} {DESY
  XFEL Project Group, Deutsches Elektronen-Synchrotron~(DESY)},\ \bibinfo
  {address} {Notkestra{\ss}e 85, 22607 Hamburg, Germany},\ \bibinfo {year}
  {2006})\BibitemShut {NoStop}%
\bibitem [{\citenamefont {Arthur}\ \emph {et~al.}(2002)\citenamefont {Arthur},
  \citenamefont {Anfinrud}, \citenamefont {Audebert}, \citenamefont {Bane},
  \citenamefont {Ben-Zvi}, \citenamefont {Bharadwaj}, \citenamefont {Bionta},
  \citenamefont {Bolton}, \citenamefont {Borland}, \citenamefont {Bucksbaum},
  \citenamefont {Cauble}, \citenamefont {Clendenin}, \citenamefont
  {Cornacchia}, \citenamefont {Decker}, \citenamefont {Den~Hartog},
  \citenamefont {Dierker}, \citenamefont {Dowell}, \citenamefont {Dungan},
  \citenamefont {Emma}, \citenamefont {Evans}, \citenamefont {Faigel},
  \citenamefont {Falcone}, \citenamefont {Fawley}, \citenamefont {Ferrario},
  \citenamefont {Fisher}, \citenamefont {Freeman}, \citenamefont {Frisch},
  \citenamefont {Galayda}, \citenamefont {Gauthier}, \citenamefont {Gierman},
  \citenamefont {Gluskin}, \citenamefont {Graves}, \citenamefont {Hajdu},
  \citenamefont {Hastings}, \citenamefont {Hodgson}, \citenamefont {Huang},
  \citenamefont {Humphry}, \citenamefont {Ilinski}, \citenamefont {Imre},
  \citenamefont {Jacobsen}, \citenamefont {Kao}, \citenamefont {Kase},
  \citenamefont {Kim}, \citenamefont {Kirby}, \citenamefont {Kirz},
  \citenamefont {Klaisner}, \citenamefont {Krejcik}, \citenamefont {Kulander},
  \citenamefont {Landen}, \citenamefont {Lee}, \citenamefont {Lewis},
  \citenamefont {Limborg}, \citenamefont {Lindau}, \citenamefont {Lumpkin},
  \citenamefont {Materlik}, \citenamefont {Mao}, \citenamefont {Miao},
  \citenamefont {Mochrie}, \citenamefont {Moog}, \citenamefont {Milton},
  \citenamefont {Mulhollan}, \citenamefont {Nelson}, \citenamefont {Nelson},
  \citenamefont {Neutze}, \citenamefont {Ng}, \citenamefont {Nguyen},
  \citenamefont {Nuhn}, \citenamefont {Palmer}, \citenamefont {Paterson},
  \citenamefont {Pellegrini}, \citenamefont {Reiche}, \citenamefont {Renner},
  \citenamefont {Riley}, \citenamefont {Robinson}, \citenamefont {Rokni},
  \citenamefont {Rose}, \citenamefont {Rosenzweig}, \citenamefont {Ruland},
  \citenamefont {Ruocco}, \citenamefont {Saenz}, \citenamefont {Sasaki},
  \citenamefont {Sayre}, \citenamefont {Schmerge}, \citenamefont {Schneider},
  \citenamefont {Schroeder}, \citenamefont {Serafini}, \citenamefont {Sette},
  \citenamefont {Sinha}, \citenamefont {van~der Spoel}, \citenamefont
  {Stephenson}, \citenamefont {Stupakov}, \citenamefont {Sutton}, \citenamefont
  {Sz{\"o}ke}, \citenamefont {Tatchyn}, \citenamefont {Toor}, \citenamefont
  {Trakhtenberg}, \citenamefont {Vasserman}, \citenamefont {Vinokurov},
  \citenamefont {Wang}, \citenamefont {Waltz}, \citenamefont {Wark},
  \citenamefont {Weckert}, \citenamefont {{Wilson-Squire Group}}, \citenamefont
  {Winick}, \citenamefont {Woodley}, \citenamefont {Wootton}, \citenamefont
  {Wulff}, \citenamefont {Xie}, \citenamefont {Yotam}, \citenamefont {Young},\
  and\ \citenamefont {Zewail}}]{LCLS:CDR-02}%
  \BibitemOpen
  \bibfield  {author} {\bibinfo {author} {\bibfnamefont {J.}~\bibnamefont
  {Arthur}}, \bibinfo {author} {\bibfnamefont {P.}~\bibnamefont {Anfinrud}},
  \bibinfo {author} {\bibfnamefont {P.}~\bibnamefont {Audebert}}, \bibinfo
  {author} {\bibfnamefont {K.}~\bibnamefont {Bane}}, \bibinfo {author}
  {\bibfnamefont {I.}~\bibnamefont {Ben-Zvi}}, \bibinfo {author} {\bibfnamefont
  {V.}~\bibnamefont {Bharadwaj}}, \bibinfo {author} {\bibfnamefont
  {R.}~\bibnamefont {Bionta}}, \bibinfo {author} {\bibfnamefont
  {P.}~\bibnamefont {Bolton}}, \bibinfo {author} {\bibfnamefont
  {M.}~\bibnamefont {Borland}}, \bibinfo {author} {\bibfnamefont {P.~H.}\
  \bibnamefont {Bucksbaum}}, \bibinfo {author} {\bibfnamefont {R.~C.}\
  \bibnamefont {Cauble}}, \bibinfo {author} {\bibfnamefont {J.}~\bibnamefont
  {Clendenin}}, \bibinfo {author} {\bibfnamefont {M.}~\bibnamefont
  {Cornacchia}}, \bibinfo {author} {\bibfnamefont {G.}~\bibnamefont {Decker}},
  \bibinfo {author} {\bibfnamefont {P.}~\bibnamefont {Den~Hartog}}, \bibinfo
  {author} {\bibfnamefont {S.}~\bibnamefont {Dierker}}, \bibinfo {author}
  {\bibfnamefont {D.}~\bibnamefont {Dowell}}, \bibinfo {author} {\bibfnamefont
  {D.}~\bibnamefont {Dungan}}, \bibinfo {author} {\bibfnamefont
  {P.}~\bibnamefont {Emma}}, \bibinfo {author} {\bibfnamefont {I.}~\bibnamefont
  {Evans}}, \bibinfo {author} {\bibfnamefont {G.}~\bibnamefont {Faigel}},
  \bibinfo {author} {\bibfnamefont {R.}~\bibnamefont {Falcone}}, \bibinfo
  {author} {\bibfnamefont {W.~M.}\ \bibnamefont {Fawley}}, \bibinfo {author}
  {\bibfnamefont {M.}~\bibnamefont {Ferrario}}, \bibinfo {author}
  {\bibfnamefont {A.~S.}\ \bibnamefont {Fisher}}, \bibinfo {author}
  {\bibfnamefont {R.~R.}\ \bibnamefont {Freeman}}, \bibinfo {author}
  {\bibfnamefont {J.}~\bibnamefont {Frisch}}, \bibinfo {author} {\bibfnamefont
  {J.}~\bibnamefont {Galayda}}, \bibinfo {author} {\bibfnamefont {J.-C.}\
  \bibnamefont {Gauthier}}, \bibinfo {author} {\bibfnamefont {S.}~\bibnamefont
  {Gierman}}, \bibinfo {author} {\bibfnamefont {E.}~\bibnamefont {Gluskin}},
  \bibinfo {author} {\bibfnamefont {W.}~\bibnamefont {Graves}}, \bibinfo
  {author} {\bibfnamefont {J.}~\bibnamefont {Hajdu}}, \bibinfo {author}
  {\bibfnamefont {J.}~\bibnamefont {Hastings}}, \bibinfo {author}
  {\bibfnamefont {K.}~\bibnamefont {Hodgson}}, \bibinfo {author} {\bibfnamefont
  {Z.}~\bibnamefont {Huang}}, \bibinfo {author} {\bibfnamefont
  {R.}~\bibnamefont {Humphry}}, \bibinfo {author} {\bibfnamefont
  {P.}~\bibnamefont {Ilinski}}, \bibinfo {author} {\bibfnamefont
  {D.}~\bibnamefont {Imre}}, \bibinfo {author} {\bibfnamefont {C.}~\bibnamefont
  {Jacobsen}}, \bibinfo {author} {\bibfnamefont {C.-C.}\ \bibnamefont {Kao}},
  \bibinfo {author} {\bibfnamefont {K.~R.}\ \bibnamefont {Kase}}, \bibinfo
  {author} {\bibfnamefont {K.-J.}\ \bibnamefont {Kim}}, \bibinfo {author}
  {\bibfnamefont {R.}~\bibnamefont {Kirby}}, \bibinfo {author} {\bibfnamefont
  {J.}~\bibnamefont {Kirz}}, \bibinfo {author} {\bibfnamefont {L.}~\bibnamefont
  {Klaisner}}, \bibinfo {author} {\bibfnamefont {P.}~\bibnamefont {Krejcik}},
  \bibinfo {author} {\bibfnamefont {K.}~\bibnamefont {Kulander}}, \bibinfo
  {author} {\bibfnamefont {O.~L.}\ \bibnamefont {Landen}}, \bibinfo {author}
  {\bibfnamefont {R.~W.}\ \bibnamefont {Lee}}, \bibinfo {author} {\bibfnamefont
  {C.}~\bibnamefont {Lewis}}, \bibinfo {author} {\bibfnamefont
  {C.}~\bibnamefont {Limborg}}, \bibinfo {author} {\bibfnamefont {E.~I.}\
  \bibnamefont {Lindau}}, \bibinfo {author} {\bibfnamefont {A.}~\bibnamefont
  {Lumpkin}}, \bibinfo {author} {\bibfnamefont {G.}~\bibnamefont {Materlik}},
  \bibinfo {author} {\bibfnamefont {S.}~\bibnamefont {Mao}}, \bibinfo {author}
  {\bibfnamefont {J.}~\bibnamefont {Miao}}, \bibinfo {author} {\bibfnamefont
  {S.}~\bibnamefont {Mochrie}}, \bibinfo {author} {\bibfnamefont
  {E.}~\bibnamefont {Moog}}, \bibinfo {author} {\bibfnamefont {S.}~\bibnamefont
  {Milton}}, \bibinfo {author} {\bibfnamefont {G.}~\bibnamefont {Mulhollan}},
  \bibinfo {author} {\bibfnamefont {K.}~\bibnamefont {Nelson}}, \bibinfo
  {author} {\bibfnamefont {W.~R.}\ \bibnamefont {Nelson}}, \bibinfo {author}
  {\bibfnamefont {R.}~\bibnamefont {Neutze}}, \bibinfo {author} {\bibfnamefont
  {A.}~\bibnamefont {Ng}}, \bibinfo {author} {\bibfnamefont {D.}~\bibnamefont
  {Nguyen}}, \bibinfo {author} {\bibfnamefont {H.-D.}\ \bibnamefont {Nuhn}},
  \bibinfo {author} {\bibfnamefont {D.~T.}\ \bibnamefont {Palmer}}, \bibinfo
  {author} {\bibfnamefont {J.~M.}\ \bibnamefont {Paterson}}, \bibinfo {author}
  {\bibfnamefont {C.}~\bibnamefont {Pellegrini}}, \bibinfo {author}
  {\bibfnamefont {S.}~\bibnamefont {Reiche}}, \bibinfo {author} {\bibfnamefont
  {M.}~\bibnamefont {Renner}}, \bibinfo {author} {\bibfnamefont
  {D.}~\bibnamefont {Riley}}, \bibinfo {author} {\bibfnamefont {C.~V.}\
  \bibnamefont {Robinson}}, \bibinfo {author} {\bibfnamefont {S.~H.}\
  \bibnamefont {Rokni}}, \bibinfo {author} {\bibfnamefont {S.~J.}\ \bibnamefont
  {Rose}}, \bibinfo {author} {\bibfnamefont {J.}~\bibnamefont {Rosenzweig}},
  \bibinfo {author} {\bibfnamefont {R.}~\bibnamefont {Ruland}}, \bibinfo
  {author} {\bibfnamefont {G.}~\bibnamefont {Ruocco}}, \bibinfo {author}
  {\bibfnamefont {D.}~\bibnamefont {Saenz}}, \bibinfo {author} {\bibfnamefont
  {S.}~\bibnamefont {Sasaki}}, \bibinfo {author} {\bibfnamefont
  {D.}~\bibnamefont {Sayre}}, \bibinfo {author} {\bibfnamefont
  {J.}~\bibnamefont {Schmerge}}, \bibinfo {author} {\bibfnamefont
  {D.}~\bibnamefont {Schneider}}, \bibinfo {author} {\bibfnamefont
  {C.}~\bibnamefont {Schroeder}}, \bibinfo {author} {\bibfnamefont
  {L.}~\bibnamefont {Serafini}}, \bibinfo {author} {\bibfnamefont
  {F.}~\bibnamefont {Sette}}, \bibinfo {author} {\bibfnamefont
  {S.}~\bibnamefont {Sinha}}, \bibinfo {author} {\bibfnamefont
  {D.}~\bibnamefont {van~der Spoel}}, \bibinfo {author} {\bibfnamefont
  {B.}~\bibnamefont {Stephenson}}, \bibinfo {author} {\bibfnamefont
  {G.}~\bibnamefont {Stupakov}}, \bibinfo {author} {\bibfnamefont
  {M.}~\bibnamefont {Sutton}}, \bibinfo {author} {\bibfnamefont
  {A.}~\bibnamefont {Sz{\"o}ke}}, \bibinfo {author} {\bibfnamefont
  {R.}~\bibnamefont {Tatchyn}}, \bibinfo {author} {\bibfnamefont
  {A.}~\bibnamefont {Toor}}, \bibinfo {author} {\bibfnamefont {E.}~\bibnamefont
  {Trakhtenberg}}, \bibinfo {author} {\bibfnamefont {I.}~\bibnamefont
  {Vasserman}}, \bibinfo {author} {\bibfnamefont {N.}~\bibnamefont
  {Vinokurov}}, \bibinfo {author} {\bibfnamefont {X.~J.}\ \bibnamefont {Wang}},
  \bibinfo {author} {\bibfnamefont {D.}~\bibnamefont {Waltz}}, \bibinfo
  {author} {\bibfnamefont {J.~S.}\ \bibnamefont {Wark}}, \bibinfo {author}
  {\bibfnamefont {E.}~\bibnamefont {Weckert}}, \bibinfo {author} {\bibnamefont
  {{Wilson-Squire Group}}}, \bibinfo {author} {\bibfnamefont {H.}~\bibnamefont
  {Winick}}, \bibinfo {author} {\bibfnamefont {M.}~\bibnamefont {Woodley}},
  \bibinfo {author} {\bibfnamefont {A.}~\bibnamefont {Wootton}}, \bibinfo
  {author} {\bibfnamefont {M.}~\bibnamefont {Wulff}}, \bibinfo {author}
  {\bibfnamefont {M.}~\bibnamefont {Xie}}, \bibinfo {author} {\bibfnamefont
  {R.}~\bibnamefont {Yotam}}, \bibinfo {author} {\bibfnamefont
  {L.}~\bibnamefont {Young}}, \ and\ \bibinfo {author} {\bibfnamefont
  {A.}~\bibnamefont {Zewail}},\ }\href
  {http://www-ssrl.slac.stanford.edu/lcls/cdr} {\emph {\bibinfo {title} {{Linac
  Coherent Light Source (LCLS): Conceptual Design Report}}}},\ \bibinfo {type}
  {Tech. Rep.}\ \bibinfo {number} {SLAC-R-593, UC-414}\ (\bibinfo
  {institution} {Stanford Linear Accelerator Center (SLAC)},\ \bibinfo
  {address} {Menlo Park, California, USA},\ \bibinfo {year} {2002})\BibitemShut
  {NoStop}%
\bibitem [{\citenamefont {Emma}\ \emph {et~al.}(2010)\citenamefont {Emma},
  \citenamefont {Akre}, \citenamefont {Arthur}, \citenamefont {Bionta},
  \citenamefont {Bostedt}, \citenamefont {Bozek}, \citenamefont {Brachmann},
  \citenamefont {Bucksbaum}, \citenamefont {Coffee}, \citenamefont {Decker},
  \citenamefont {Ding}, \citenamefont {Dowell}, \citenamefont {Edstrom},
  \citenamefont {Fisher}, \citenamefont {Gilevich}, \citenamefont {Hastings},
  \citenamefont {Hays}, \citenamefont {Hering}, \citenamefont {Huang},
  \citenamefont {Iverson}, \citenamefont {Loos}, \citenamefont {Messerschmidt},
  \citenamefont {Miahnahri}, \citenamefont {Moeller}, \citenamefont {Nuhn},
  \citenamefont {Pile}, \citenamefont {Ratner}, \citenamefont {Rzepiela},
  \citenamefont {Schultz}, \citenamefont {Smith}, \citenamefont {Stefan},
  \citenamefont {Tompkins}, \citenamefont {Turner}, \citenamefont {Welch},
  \citenamefont {White}, \citenamefont {Wu}, \citenamefont {Yocky},\ and\
  \citenamefont {Galayda}}]{Emma:FL-10}%
  \BibitemOpen
  \bibfield  {author} {\bibinfo {author} {\bibfnamefont {P.}~\bibnamefont
  {Emma}}, \bibinfo {author} {\bibfnamefont {R.}~\bibnamefont {Akre}}, \bibinfo
  {author} {\bibfnamefont {J.}~\bibnamefont {Arthur}}, \bibinfo {author}
  {\bibfnamefont {R.}~\bibnamefont {Bionta}}, \bibinfo {author} {\bibfnamefont
  {C.}~\bibnamefont {Bostedt}}, \bibinfo {author} {\bibfnamefont
  {J.}~\bibnamefont {Bozek}}, \bibinfo {author} {\bibfnamefont
  {A.}~\bibnamefont {Brachmann}}, \bibinfo {author} {\bibfnamefont
  {P.}~\bibnamefont {Bucksbaum}}, \bibinfo {author} {\bibfnamefont
  {R.}~\bibnamefont {Coffee}}, \bibinfo {author} {\bibfnamefont {F.-J.}\
  \bibnamefont {Decker}}, \bibinfo {author} {\bibfnamefont {Y.}~\bibnamefont
  {Ding}}, \bibinfo {author} {\bibfnamefont {D.}~\bibnamefont {Dowell}},
  \bibinfo {author} {\bibfnamefont {S.}~\bibnamefont {Edstrom}}, \bibinfo
  {author} {\bibfnamefont {J.}~\bibnamefont {Fisher}, \bibfnamefont
  {A.~Frisch}}, \bibinfo {author} {\bibfnamefont {S.}~\bibnamefont {Gilevich}},
  \bibinfo {author} {\bibfnamefont {J.}~\bibnamefont {Hastings}}, \bibinfo
  {author} {\bibfnamefont {G.}~\bibnamefont {Hays}}, \bibinfo {author}
  {\bibfnamefont {Ph.}\ \bibnamefont {Hering}}, \bibinfo {author}
  {\bibfnamefont {Z.}~\bibnamefont {Huang}}, \bibinfo {author} {\bibfnamefont
  {R.}~\bibnamefont {Iverson}}, \bibinfo {author} {\bibfnamefont
  {H.}~\bibnamefont {Loos}}, \bibinfo {author} {\bibfnamefont {M.}~\bibnamefont
  {Messerschmidt}}, \bibinfo {author} {\bibfnamefont {A.}~\bibnamefont
  {Miahnahri}}, \bibinfo {author} {\bibfnamefont {S.}~\bibnamefont {Moeller}},
  \bibinfo {author} {\bibfnamefont {H.-D.}\ \bibnamefont {Nuhn}}, \bibinfo
  {author} {\bibfnamefont {G.}~\bibnamefont {Pile}}, \bibinfo {author}
  {\bibfnamefont {D.}~\bibnamefont {Ratner}}, \bibinfo {author} {\bibfnamefont
  {J.}~\bibnamefont {Rzepiela}}, \bibinfo {author} {\bibfnamefont
  {D.}~\bibnamefont {Schultz}}, \bibinfo {author} {\bibfnamefont
  {T.}~\bibnamefont {Smith}}, \bibinfo {author} {\bibfnamefont
  {P.}~\bibnamefont {Stefan}}, \bibinfo {author} {\bibfnamefont
  {H.}~\bibnamefont {Tompkins}}, \bibinfo {author} {\bibfnamefont
  {J.}~\bibnamefont {Turner}}, \bibinfo {author} {\bibfnamefont
  {J.}~\bibnamefont {Welch}}, \bibinfo {author} {\bibfnamefont
  {W.}~\bibnamefont {White}}, \bibinfo {author} {\bibfnamefont
  {J.}~\bibnamefont {Wu}}, \bibinfo {author} {\bibfnamefont {G.}~\bibnamefont
  {Yocky}}, \ and\ \bibinfo {author} {\bibfnamefont {J.}~\bibnamefont
  {Galayda}},\ }\bibfield  {title} {\enquote {\bibinfo {title} {First lasing
  and operation of an {\aa}ngstrom-wavelength free-electron laser},}\ }\href
  {\doibase 10.1038/nphoton.2010.176} {\bibfield  {journal} {\bibinfo
  {journal} {Nature Photon.}\ }\textbf {\bibinfo {volume} {4}},\ \bibinfo
  {pages} {641--647} (\bibinfo {year} {2010})}\BibitemShut {NoStop}%
\bibitem [{\citenamefont {Kondratenko}\ and\ \citenamefont
  {Saldin}(1979)}]{Kondratenko:GC-79}%
  \BibitemOpen
  \bibfield  {author} {\bibinfo {author} {\bibfnamefont {A.~M.}\ \bibnamefont
  {Kondratenko}}\ and\ \bibinfo {author} {\bibfnamefont {E.~L.}\ \bibnamefont
  {Saldin}},\ }\bibfield  {title} {\enquote {\bibinfo {title} {Generation of
  coherent radiation by a relativistic-electron beam in an undulator},}\
  }\href@noop {} {\bibfield  {journal} {\bibinfo  {journal} {Dokl. Akad. Nauk
  SSSR}\ }\textbf {\bibinfo {volume} {249}},\ \bibinfo {pages} {843} (\bibinfo
  {year} {1979})},\ \bibinfo {note} {[Sov. Phys. Dokl. \textbf{24}, 986-988
  (1979)]}\BibitemShut {NoStop}%
\bibitem [{\citenamefont {Bonifacio}\ \emph {et~al.}(1984)\citenamefont
  {Bonifacio}, \citenamefont {Pellegrini},\ and\ \citenamefont
  {Narducci}}]{Bonifacio:CI-84}%
  \BibitemOpen
  \bibfield  {author} {\bibinfo {author} {\bibfnamefont {R.}~\bibnamefont
  {Bonifacio}}, \bibinfo {author} {\bibfnamefont {C.}~\bibnamefont
  {Pellegrini}}, \ and\ \bibinfo {author} {\bibfnamefont {L.~M.}\ \bibnamefont
  {Narducci}},\ }\bibfield  {title} {\enquote {\bibinfo {title} {Collective
  instabilities and high-gain regime in a free electron laser},}\ }\href
  {\doibase 10.1016/0030-4018(84)90105-6} {\bibfield  {journal} {\bibinfo
  {journal} {Opt. Commun.}\ }\textbf {\bibinfo {volume} {50}},\ \bibinfo
  {pages} {373--378} (\bibinfo {year} {1984})}\BibitemShut {NoStop}%
\bibitem [{\citenamefont {Saldin}\ \emph {et~al.}(2000)\citenamefont {Saldin},
  \citenamefont {Schneidmiller},\ and\ \citenamefont {Yurkov}}]{Saldin:PF-00}%
  \BibitemOpen
  \bibfield  {author} {\bibinfo {author} {\bibfnamefont {Evgeny~L.}\
  \bibnamefont {Saldin}}, \bibinfo {author} {\bibfnamefont {E.~A.}\
  \bibnamefont {Schneidmiller}}, \ and\ \bibinfo {author} {\bibfnamefont
  {Mikhail~V.}\ \bibnamefont {Yurkov}},\ }\href@noop {} {\emph {\bibinfo
  {title} {The physics of free electron lasers}}}\ (\bibinfo  {publisher}
  {Springer},\ \bibinfo {address} {Berlin, Heidelberg, New York},\ \bibinfo
  {year} {2000})\BibitemShut {NoStop}%
\bibitem [{\citenamefont {Pfeifer}\ \emph {et~al.}(2010)\citenamefont
  {Pfeifer}, \citenamefont {Jiang}, \citenamefont {D{\"u}sterer}, \citenamefont
  {Moshammer},\ and\ \citenamefont {Ullrich}}]{Pfeifer:PC-10}%
  \BibitemOpen
  \bibfield  {author} {\bibinfo {author} {\bibfnamefont {Thomas}\ \bibnamefont
  {Pfeifer}}, \bibinfo {author} {\bibfnamefont {Yuhai}\ \bibnamefont {Jiang}},
  \bibinfo {author} {\bibfnamefont {Stefan}\ \bibnamefont {D{\"u}sterer}},
  \bibinfo {author} {\bibfnamefont {Robert}\ \bibnamefont {Moshammer}}, \ and\
  \bibinfo {author} {\bibfnamefont {Joachim}\ \bibnamefont {Ullrich}},\
  }\bibfield  {title} {\enquote {\bibinfo {title} {Partial-coherence method to
  model experimental {FEL}~pulse statistics},}\ }\href {\doibase
  10.1364/OL.35.003441} {\bibfield  {journal} {\bibinfo  {journal} {Opt.
  Lett.}\ }\textbf {\bibinfo {volume} {35}},\ \bibinfo {pages} {3441--3443}
  (\bibinfo {year} {2010})}\BibitemShut {NoStop}%
\bibitem [{\citenamefont {Jiang}\ \emph {et~al.}(2010)\citenamefont {Jiang},
  \citenamefont {Pfeifer}, \citenamefont {Rudenko}, \citenamefont {Herrwerth},
  \citenamefont {Foucar}, \citenamefont {Kurka}, \citenamefont {K{\"u}hnel},
  \citenamefont {Lezius}, \citenamefont {Kling}, \citenamefont {Liu},
  \citenamefont {Ueda}, \citenamefont {D{\"u}sterer}, \citenamefont {Treusch},
  \citenamefont {Schr{\"o}ter}, \citenamefont {Moshammer},\ and\ \citenamefont
  {Ullrich}}]{Jiang:TC-10}%
  \BibitemOpen
  \bibfield  {author} {\bibinfo {author} {\bibfnamefont {Y.~H.}\ \bibnamefont
  {Jiang}}, \bibinfo {author} {\bibfnamefont {T.}~\bibnamefont {Pfeifer}},
  \bibinfo {author} {\bibfnamefont {A.}~\bibnamefont {Rudenko}}, \bibinfo
  {author} {\bibfnamefont {O.}~\bibnamefont {Herrwerth}}, \bibinfo {author}
  {\bibfnamefont {L.}~\bibnamefont {Foucar}}, \bibinfo {author} {\bibfnamefont
  {M.}~\bibnamefont {Kurka}}, \bibinfo {author} {\bibfnamefont {K.~U.}\
  \bibnamefont {K{\"u}hnel}}, \bibinfo {author} {\bibfnamefont
  {M.}~\bibnamefont {Lezius}}, \bibinfo {author} {\bibfnamefont {M.~F.}\
  \bibnamefont {Kling}}, \bibinfo {author} {\bibfnamefont {X.}~\bibnamefont
  {Liu}}, \bibinfo {author} {\bibfnamefont {K.}~\bibnamefont {Ueda}}, \bibinfo
  {author} {\bibfnamefont {S.}~\bibnamefont {D{\"u}sterer}}, \bibinfo {author}
  {\bibfnamefont {R.}~\bibnamefont {Treusch}}, \bibinfo {author} {\bibfnamefont
  {C.~D.}\ \bibnamefont {Schr{\"o}ter}}, \bibinfo {author} {\bibfnamefont
  {R.}~\bibnamefont {Moshammer}}, \ and\ \bibinfo {author} {\bibfnamefont
  {J.}~\bibnamefont {Ullrich}},\ }\bibfield  {title} {\enquote {\bibinfo
  {title} {Temporal coherence effects in multiple ionization of~{N$_2$} via
  {XUV} pump-probe autocorrelation},}\ }\href {\doibase
  10.1103/PhysRevA.82.041403} {\bibfield  {journal} {\bibinfo  {journal} {Phys.
  Rev. A}\ }\textbf {\bibinfo {volume} {82}},\ \bibinfo {pages} {041403(R)}
  (\bibinfo {year} {2010})}\BibitemShut {NoStop}%
\bibitem [{\citenamefont {Cavaletto}\ \emph {et~al.}(2012)\citenamefont
  {Cavaletto}, \citenamefont {Buth}, \citenamefont {Harman}, \citenamefont
  {Kanter}, \citenamefont {Southworth}, \citenamefont {Young},\ and\
  \citenamefont {Keitel}}]{Cavaletto:RF-12}%
  \BibitemOpen
  \bibfield  {author} {\bibinfo {author} {\bibfnamefont {Stefano~M.}\
  \bibnamefont {Cavaletto}}, \bibinfo {author} {\bibfnamefont {Christian}\
  \bibnamefont {Buth}}, \bibinfo {author} {\bibfnamefont {Zolt{\'a}n}\
  \bibnamefont {Harman}}, \bibinfo {author} {\bibfnamefont {Elliot~P.}\
  \bibnamefont {Kanter}}, \bibinfo {author} {\bibfnamefont {Stephen~H.}\
  \bibnamefont {Southworth}}, \bibinfo {author} {\bibfnamefont {Linda}\
  \bibnamefont {Young}}, \ and\ \bibinfo {author} {\bibfnamefont
  {Christoph~H.}\ \bibnamefont {Keitel}},\ }\bibfield  {title} {\enquote
  {\bibinfo {title} {Resonance fluorescence in ultrafast and intense x-ray free
  electron laser pulses},}\ }\href {\doibase 10.1103/PhysRevA.86.033402}
  {\bibfield  {journal} {\bibinfo  {journal} {Phys. Rev. A}\ }\textbf {\bibinfo
  {volume} {86}},\ \bibinfo {pages} {033402} (\bibinfo {year} {2012})},\
  \Eprint {http://arxiv.org/abs/1205.4918} {arXiv:1205.4918} \BibitemShut
  {NoStop}%
\bibitem [{\citenamefont {Wuilleumier}\ and\ \citenamefont
  {Meyer}(2006)}]{Wuilleumier:PP-06}%
  \BibitemOpen
  \bibfield  {author} {\bibinfo {author} {\bibfnamefont {F.~J.}\ \bibnamefont
  {Wuilleumier}}\ and\ \bibinfo {author} {\bibfnamefont {M.}~\bibnamefont
  {Meyer}},\ }\bibfield  {title} {\enquote {\bibinfo {title} {Pump-probe
  experiments in atoms involving laser and synchrotron radiation: an
  overview},}\ }\href {\doibase 10.1088/0953-4075/39/23/R01} {\bibfield
  {journal} {\bibinfo  {journal} {J. Phys. B}\ }\textbf {\bibinfo {volume}
  {39}},\ \bibinfo {pages} {R425--R477} (\bibinfo {year} {2006})}\BibitemShut
  {NoStop}%
\bibitem [{\citenamefont {Santra}\ \emph {et~al.}(2007)\citenamefont {Santra},
  \citenamefont {Buth}, \citenamefont {Peterson}, \citenamefont {Dunford},
  \citenamefont {Kanter}, \citenamefont {Kr{\"a}ssig}, \citenamefont
  {Southworth},\ and\ \citenamefont {Young}}]{Santra:SF-07}%
  \BibitemOpen
  \bibfield  {author} {\bibinfo {author} {\bibfnamefont {Robin}\ \bibnamefont
  {Santra}}, \bibinfo {author} {\bibfnamefont {Christian}\ \bibnamefont
  {Buth}}, \bibinfo {author} {\bibfnamefont {Emily~R.}\ \bibnamefont
  {Peterson}}, \bibinfo {author} {\bibfnamefont {Robert~W.}\ \bibnamefont
  {Dunford}}, \bibinfo {author} {\bibfnamefont {Elliot~P.}\ \bibnamefont
  {Kanter}}, \bibinfo {author} {\bibfnamefont {Bertold}\ \bibnamefont
  {Kr{\"a}ssig}}, \bibinfo {author} {\bibfnamefont {Stephen~H.}\ \bibnamefont
  {Southworth}}, \ and\ \bibinfo {author} {\bibfnamefont {Linda}\ \bibnamefont
  {Young}},\ }\bibfield  {title} {\enquote {\bibinfo {title} {Strong-field
  control of x-ray absorption},}\ }\href {\doibase
  10.1088/1742-6596/88/1/012052} {\bibfield  {journal} {\bibinfo  {journal} {J.
  Phys.: Conf. Ser.}\ }\textbf {\bibinfo {volume} {88}},\ \bibinfo {pages}
  {012052} (\bibinfo {year} {2007})},\ \Eprint {http://arxiv.org/abs/0712.2556}
  {arXiv:0712.2556} \BibitemShut {NoStop}%
\bibitem [{\citenamefont {Santra}\ \emph {et~al.}(2008)\citenamefont {Santra},
  \citenamefont {Dunford}, \citenamefont {Kanter}, \citenamefont {Kr{\"a}ssig},
  \citenamefont {Southworth},\ and\ \citenamefont {Young}}]{Santra:SF-08}%
  \BibitemOpen
  \bibfield  {author} {\bibinfo {author} {\bibfnamefont {Robin}\ \bibnamefont
  {Santra}}, \bibinfo {author} {\bibfnamefont {Robert~W.}\ \bibnamefont
  {Dunford}}, \bibinfo {author} {\bibfnamefont {Elliot~P.}\ \bibnamefont
  {Kanter}}, \bibinfo {author} {\bibfnamefont {Bertold}\ \bibnamefont
  {Kr{\"a}ssig}}, \bibinfo {author} {\bibfnamefont {Stephen~H.}\ \bibnamefont
  {Southworth}}, \ and\ \bibinfo {author} {\bibfnamefont {Linda}\ \bibnamefont
  {Young}},\ }\bibfield  {title} {\enquote {\bibinfo {title} {Strong-field
  control of x-ray processes},}\ }in\ \href {\doibase
  10.1016/S1049-250X(08)00014-1} {\emph {\bibinfo {booktitle} {Advances in
  Atomic, Molecular, and Optical Physics}}},\ Vol.~\bibinfo {volume} {56},\
  \bibinfo {editor} {edited by\ \bibinfo {editor} {\bibfnamefont {Ennio}\
  \bibnamefont {Arimondo}}, \bibinfo {editor} {\bibfnamefont {Paul~R.}\
  \bibnamefont {Berman}}, \ and\ \bibinfo {editor} {\bibfnamefont {Chun~C.}\
  \bibnamefont {Lin}}}\ (\bibinfo  {publisher} {Academic Press},\ \bibinfo
  {address} {Amsterdam},\ \bibinfo {year} {2008})\ pp.\ \bibinfo {pages}
  {219--259}\BibitemShut {NoStop}%
\bibitem [{\citenamefont {Adams}\ \emph {et~al.}(2013)\citenamefont {Adams},
  \citenamefont {Buth}, \citenamefont {Cavaletto}, \citenamefont {Evers},
  \citenamefont {Harman}, \citenamefont {Keitel}, \citenamefont {P{\'a}lffy},
  \citenamefont {Pic{\'o}n}, \citenamefont {R{\"o}hlsberger}, \citenamefont
  {Rostovtsev},\ and\ \citenamefont {Tamasaku}}]{Adams:QO-13}%
  \BibitemOpen
  \bibfield  {author} {\bibinfo {author} {\bibfnamefont {Bernhard~W.}\
  \bibnamefont {Adams}}, \bibinfo {author} {\bibfnamefont {Christian}\
  \bibnamefont {Buth}}, \bibinfo {author} {\bibfnamefont {Stefano~M.}\
  \bibnamefont {Cavaletto}}, \bibinfo {author} {\bibfnamefont {J{\"o}rg}\
  \bibnamefont {Evers}}, \bibinfo {author} {\bibfnamefont {Zolt{\'a}n}\
  \bibnamefont {Harman}}, \bibinfo {author} {\bibfnamefont {Christoph~H.}\
  \bibnamefont {Keitel}}, \bibinfo {author} {\bibfnamefont {Adriana}\
  \bibnamefont {P{\'a}lffy}}, \bibinfo {author} {\bibfnamefont {Antonio}\
  \bibnamefont {Pic{\'o}n}}, \bibinfo {author} {\bibfnamefont {Ralf}\
  \bibnamefont {R{\"o}hlsberger}}, \bibinfo {author} {\bibfnamefont {Yuri}\
  \bibnamefont {Rostovtsev}}, \ and\ \bibinfo {author} {\bibfnamefont {Kenji}\
  \bibnamefont {Tamasaku}},\ }\bibfield  {title} {\enquote {\bibinfo {title}
  {X-ray quantum optics},}\ }\href {\doibase 10.1080/09500340.2012.752113}
  {\bibfield  {journal} {\bibinfo  {journal} {J. Mod. Opt.}\ }\textbf {\bibinfo
  {volume} {60}},\ \bibinfo {pages} {2--21} (\bibinfo {year}
  {2013})}\BibitemShut {NoStop}%
\bibitem [{\citenamefont {Glover}\ \emph {et~al.}(2012)\citenamefont {Glover},
  \citenamefont {Fritz}, \citenamefont {Cammarata}, \citenamefont {Allison},
  \citenamefont {Coh}, \citenamefont {Feldkamp}, \citenamefont {Lemke},
  \citenamefont {Zhu}, \citenamefont {Feng}, \citenamefont {Coffee},
  \citenamefont {Fuchs}, \citenamefont {Ghimire}, \citenamefont {Chen},
  \citenamefont {Shwartz}, \citenamefont {Reis}, \citenamefont {Harris},\ and\
  \citenamefont {Hastings}}]{Glover:XO-12}%
  \BibitemOpen
  \bibfield  {author} {\bibinfo {author} {\bibfnamefont {T.~E.}\ \bibnamefont
  {Glover}}, \bibinfo {author} {\bibfnamefont {D.~M.}\ \bibnamefont {Fritz}},
  \bibinfo {author} {\bibfnamefont {M.}~\bibnamefont {Cammarata}}, \bibinfo
  {author} {\bibfnamefont {T.~K.}\ \bibnamefont {Allison}}, \bibinfo {author}
  {\bibfnamefont {Sinisa}\ \bibnamefont {Coh}}, \bibinfo {author}
  {\bibfnamefont {J.~M.}\ \bibnamefont {Feldkamp}}, \bibinfo {author}
  {\bibfnamefont {H.}~\bibnamefont {Lemke}}, \bibinfo {author} {\bibfnamefont
  {D.}~\bibnamefont {Zhu}}, \bibinfo {author} {\bibfnamefont {Y.}~\bibnamefont
  {Feng}}, \bibinfo {author} {\bibfnamefont {R.~N.}\ \bibnamefont {Coffee}},
  \bibinfo {author} {\bibfnamefont {M.}~\bibnamefont {Fuchs}}, \bibinfo
  {author} {\bibfnamefont {S.}~\bibnamefont {Ghimire}}, \bibinfo {author}
  {\bibfnamefont {J.}~\bibnamefont {Chen}}, \bibinfo {author} {\bibfnamefont
  {S.}~\bibnamefont {Shwartz}}, \bibinfo {author} {\bibfnamefont {D.~A.}\
  \bibnamefont {Reis}}, \bibinfo {author} {\bibfnamefont {S.~E.}\ \bibnamefont
  {Harris}}, \ and\ \bibinfo {author} {\bibfnamefont {J.~B.}\ \bibnamefont
  {Hastings}},\ }\bibfield  {title} {\enquote {\bibinfo {title} {X-ray and
  optical wave mixing},}\ }\href {\doibase 10.1038/nature11340} {\bibfield
  {journal} {\bibinfo  {journal} {Nature}\ }\textbf {\bibinfo {volume} {488}},\
  \bibinfo {pages} {603--608} (\bibinfo {year} {2012})}\BibitemShut {NoStop}%
\bibitem [{\citenamefont {Freund}\ and\ \citenamefont
  {Levine}(1970)}]{Freund:OM-70}%
  \BibitemOpen
  \bibfield  {author} {\bibinfo {author} {\bibfnamefont {Isaac}\ \bibnamefont
  {Freund}}\ and\ \bibinfo {author} {\bibfnamefont {B.~F.}\ \bibnamefont
  {Levine}},\ }\bibfield  {title} {\enquote {\bibinfo {title} {Optically
  modulated x-ray diffraction},}\ }\href {\doibase 10.1103/PhysRevLett.25.1241}
  {\bibfield  {journal} {\bibinfo  {journal} {Phys. Rev. Lett.}\ }\textbf
  {\bibinfo {volume} {25}},\ \bibinfo {pages} {1241--1245} (\bibinfo {year}
  {1970})}\BibitemShut {NoStop}%
\bibitem [{\citenamefont {Freund}\ and\ \citenamefont
  {Levine}(1971)}]{Freund:OM-71}%
  \BibitemOpen
  \bibfield  {author} {\bibinfo {author} {\bibfnamefont {Isaac}\ \bibnamefont
  {Freund}}\ and\ \bibinfo {author} {\bibfnamefont {B.~F.}\ \bibnamefont
  {Levine}},\ }\bibfield  {title} {\enquote {\bibinfo {title} {Optically
  modulated x-ray diffraction},}\ }\href {\doibase 10.1103/PhysRevLett.26.156}
  {\bibfield  {journal} {\bibinfo  {journal} {Phys. Rev. Lett.}\ }\textbf
  {\bibinfo {volume} {26}},\ \bibinfo {pages} {156--156} (\bibinfo {year}
  {1971})}\BibitemShut {NoStop}%
\bibitem [{\citenamefont {Eisenberger}\ and\ \citenamefont
  {McCall}(1971)}]{Eisenberger:XO-71}%
  \BibitemOpen
  \bibfield  {author} {\bibinfo {author} {\bibfnamefont {P.~M.}\ \bibnamefont
  {Eisenberger}}\ and\ \bibinfo {author} {\bibfnamefont {S.~L.}\ \bibnamefont
  {McCall}},\ }\bibfield  {title} {\enquote {\bibinfo {title} {Mixing of x-ray
  and optical photons},}\ }\href {\doibase 10.1103/PhysRevA.3.1145} {\bibfield
  {journal} {\bibinfo  {journal} {Phys. Rev. A}\ }\textbf {\bibinfo {volume}
  {3}},\ \bibinfo {pages} {1145--1151} (\bibinfo {year} {1971})}\BibitemShut
  {NoStop}%
\bibitem [{\citenamefont {Buth}\ \emph {et~al.}(2011)\citenamefont {Buth},
  \citenamefont {Kohler}, \citenamefont {Ullrich},\ and\ \citenamefont
  {Keitel}}]{Buth:NL-11}%
  \BibitemOpen
  \bibfield  {author} {\bibinfo {author} {\bibfnamefont {Christian}\
  \bibnamefont {Buth}}, \bibinfo {author} {\bibfnamefont {Markus~C.}\
  \bibnamefont {Kohler}}, \bibinfo {author} {\bibfnamefont {Joachim}\
  \bibnamefont {Ullrich}}, \ and\ \bibinfo {author} {\bibfnamefont
  {Christoph~H.}\ \bibnamefont {Keitel}},\ }\bibfield  {title} {\enquote
  {\bibinfo {title} {High-order harmonic generation enhanced by
  \textsc{xuv}~light},}\ }\href {\doibase 10.1364/OL.36.003530} {\bibfield
  {journal} {\bibinfo  {journal} {Opt. Lett.}\ }\textbf {\bibinfo {volume}
  {36}},\ \bibinfo {pages} {3530--3532} (\bibinfo {year} {2011})},\ \Eprint
  {http://arxiv.org/abs/1012.4930} {arXiv:1012.4930} \BibitemShut {NoStop}%
\bibitem [{\citenamefont {Kohler}\ \emph
  {et~al.}(2012{\natexlab{b}})\citenamefont {Kohler}, \citenamefont
  {M{\"u}ller}, \citenamefont {Buth}, \citenamefont {Voitkiv}, \citenamefont
  {Hatsagortsyan}, \citenamefont {Ullrich}, \citenamefont {Pfeifer},\ and\
  \citenamefont {Keitel}}]{Kohler:EC-12}%
  \BibitemOpen
  \bibfield  {author} {\bibinfo {author} {\bibfnamefont {Markus~C.}\
  \bibnamefont {Kohler}}, \bibinfo {author} {\bibfnamefont {Carsten}\
  \bibnamefont {M{\"u}ller}}, \bibinfo {author} {\bibfnamefont {Christian}\
  \bibnamefont {Buth}}, \bibinfo {author} {\bibfnamefont {Alexander~B.}\
  \bibnamefont {Voitkiv}}, \bibinfo {author} {\bibfnamefont {Karen~Z.}\
  \bibnamefont {Hatsagortsyan}}, \bibinfo {author} {\bibfnamefont {Joachim}\
  \bibnamefont {Ullrich}}, \bibinfo {author} {\bibfnamefont {Thomas}\
  \bibnamefont {Pfeifer}}, \ and\ \bibinfo {author} {\bibfnamefont
  {Christoph~H.}\ \bibnamefont {Keitel}},\ }\bibfield  {title} {\enquote
  {\bibinfo {title} {Electron correlation and interference effects in
  strong-field processes},}\ }in\ \href {\doibase 10.1007/978-3-642-28948-4_35}
  {\emph {\bibinfo {booktitle} {Multiphoton Processes and Attosecond
  Physics}}},\ \bibinfo {series} {Springer Proceedings in Physics}, Vol.\
  \bibinfo {volume} {125},\ \bibinfo {editor} {edited by\ \bibinfo {editor}
  {\bibfnamefont {Kaoru}\ \bibnamefont {Yamanouchi}}\ and\ \bibinfo {editor}
  {\bibfnamefont {Katsumi}\ \bibnamefont {Midorikawa}}}\ (\bibinfo  {publisher}
  {Springer},\ \bibinfo {address} {Berlin, Heidelberg},\ \bibinfo {year}
  {2012})\ pp.\ \bibinfo {pages} {209--217},\ \Eprint
  {http://arxiv.org/abs/1111.3555} {arXiv:1111.3555} \BibitemShut {NoStop}%
\bibitem [{\citenamefont {Buth}\ \emph {et~al.}(2013)\citenamefont {Buth},
  \citenamefont {He}, \citenamefont {Ullrich}, \citenamefont {Keitel},\ and\
  \citenamefont {Hatsagortsyan}}]{Buth:KE-13}%
  \BibitemOpen
  \bibfield  {author} {\bibinfo {author} {\bibfnamefont {Christian}\
  \bibnamefont {Buth}}, \bibinfo {author} {\bibfnamefont {Feng}\ \bibnamefont
  {He}}, \bibinfo {author} {\bibfnamefont {Joachim}\ \bibnamefont {Ullrich}},
  \bibinfo {author} {\bibfnamefont {Christoph~H.}\ \bibnamefont {Keitel}}, \
  and\ \bibinfo {author} {\bibfnamefont {Karen~Zaveni}\ \bibnamefont
  {Hatsagortsyan}},\ }\bibfield  {title} {\enquote {\bibinfo {title}
  {Attosecond pulses at kiloelectronvolt photon energies from high-order
  harmonic generation with core electrons},}\ }\href {\doibase
  10.1103/PhysRevA.88.033848} {\bibfield  {journal} {\bibinfo  {journal} {Phys.
  Rev. A}\ }\textbf {\bibinfo {volume} {88}},\ \bibinfo {pages} {033848}
  (\bibinfo {year} {2013})},\ \Eprint {http://arxiv.org/abs/1203.4127}
  {arXiv:1203.4127} \BibitemShut {NoStop}%
\bibitem [{\citenamefont {Milo{\v{s}}evi{\'c}}\ and\ \citenamefont
  {Starace}(2000)}]{Milosevic:CH-00}%
  \BibitemOpen
  \bibfield  {author} {\bibinfo {author} {\bibfnamefont {Dejan~B.}\
  \bibnamefont {Milo{\v{s}}evi{\'c}}}\ and\ \bibinfo {author} {\bibfnamefont
  {A.~F.}\ \bibnamefont {Starace}},\ }\bibfield  {title} {\enquote {\bibinfo
  {title} {Control of high-harmonic generation and laser-assisted x-ray-atom
  scattering with static electric and magnetic fields},}\ }\href@noop {}
  {\bibfield  {journal} {\bibinfo  {journal} {Las. Phys.}\ }\textbf {\bibinfo
  {volume} {10}},\ \bibinfo {pages} {278--293} (\bibinfo {year}
  {2000})}\BibitemShut {NoStop}%
\bibitem [{\citenamefont {Milo{\v{s}}evi{\'c}}\ and\ \citenamefont
  {Ehlotzky}(2003)}]{Milosevic:SR-03}%
  \BibitemOpen
  \bibfield  {author} {\bibinfo {author} {\bibfnamefont {Dejan~B.}\
  \bibnamefont {Milo{\v{s}}evi{\'c}}}\ and\ \bibinfo {author} {\bibfnamefont
  {Fritz}\ \bibnamefont {Ehlotzky}},\ }\bibfield  {title} {\enquote {\bibinfo
  {title} {Scattering and reaction processes in powerful laser fields},}\
  }\href {\doibase 10.1016/S1049-250X(03)80007-1} {\bibfield  {journal}
  {\bibinfo  {journal} {Adv. At. Mol. Opt. Phys.}\ }\textbf {\bibinfo {volume}
  {49}},\ \bibinfo {pages} {373--532} (\bibinfo {year} {2003})}\BibitemShut
  {NoStop}%
\bibitem [{\citenamefont {Buth}(2013)}]{Buth:HO-13}%
  \BibitemOpen
  \bibfield  {author} {\bibinfo {author} {\bibfnamefont {Christian}\
  \bibnamefont {Buth}},\ }\bibfield  {title} {\enquote {\bibinfo {title}
  {High-order harmonic generation with resonant core excitation by ultraintense
  x~rays},}\ }\href {http://arxiv.org/abs/1303.1332} {\  (\bibinfo {year}
  {2013})},\ \Eprint {http://arxiv.org/abs/1303.1332} {arXiv:1303.1332}
  \BibitemShut {NoStop}%
\bibitem [{\citenamefont {Antonov}\ \emph {et~al.}(2013)\citenamefont
  {Antonov}, \citenamefont {Radeonychev},\ and\ \citenamefont
  {Kocharovskaya}}]{Antonov:FS-13}%
  \BibitemOpen
  \bibfield  {author} {\bibinfo {author} {\bibfnamefont {V.~A.}\ \bibnamefont
  {Antonov}}, \bibinfo {author} {\bibfnamefont {Y.~V.}\ \bibnamefont
  {Radeonychev}}, \ and\ \bibinfo {author} {\bibfnamefont {Olga}\ \bibnamefont
  {Kocharovskaya}},\ }\bibfield  {title} {\enquote {\bibinfo {title} {Formation
  of a single attosecond pulse via interaction of resonant radiation with a
  strongly perturbed atomic transition},}\ }\href {\doibase
  10.1103/PhysRevLett.110.213903} {\bibfield  {journal} {\bibinfo  {journal}
  {Phys. Rev. Lett.}\ }\textbf {\bibinfo {volume} {110}},\ \bibinfo {pages}
  {213903} (\bibinfo {year} {2013})}\BibitemShut {NoStop}%
\bibitem [{\citenamefont {Tudorovskaya}\ and\ \citenamefont
  {Lein}(2014)}]{Tudorovskaya:HH-14}%
  \BibitemOpen
  \bibfield  {author} {\bibinfo {author} {\bibfnamefont {M.}~\bibnamefont
  {Tudorovskaya}}\ and\ \bibinfo {author} {\bibfnamefont {M.}~\bibnamefont
  {Lein}},\ }\bibfield  {title} {\enquote {\bibinfo {title} {High-harmonic
  generation with combined infrared and extreme ultraviolet fields},}\ }\href
  {\doibase 10.1080/09500340.2013.854422} {\bibfield  {journal} {\bibinfo
  {journal} {J. Mod. Opt.}\ }\textbf {\bibinfo {volume} {61}},\ \bibinfo
  {pages} {845--850} (\bibinfo {year} {2014})}\BibitemShut {NoStop}%
\bibitem [{\citenamefont {Rabi}(1936)}]{Rabi:OP-36}%
  \BibitemOpen
  \bibfield  {author} {\bibinfo {author} {\bibfnamefont {I.~I.}\ \bibnamefont
  {Rabi}},\ }\bibfield  {title} {\enquote {\bibinfo {title} {On the process of
  space quantization},}\ }\href {\doibase 10.1103/PhysRev.49.324} {\bibfield
  {journal} {\bibinfo  {journal} {Phys. Rev.}\ }\textbf {\bibinfo {volume}
  {49}},\ \bibinfo {pages} {324--328} (\bibinfo {year} {1936})}\BibitemShut
  {NoStop}%
\bibitem [{\citenamefont {Rabi}(1937)}]{Rabi:SQ-37}%
  \BibitemOpen
  \bibfield  {author} {\bibinfo {author} {\bibfnamefont {I.~I.}\ \bibnamefont
  {Rabi}},\ }\bibfield  {title} {\enquote {\bibinfo {title} {Space quantization
  in a gyrating magnetic field},}\ }\href {\doibase 10.1103/PhysRev.51.652}
  {\bibfield  {journal} {\bibinfo  {journal} {Phys. Rev.}\ }\textbf {\bibinfo
  {volume} {51}},\ \bibinfo {pages} {652--654} (\bibinfo {year}
  {1937})}\BibitemShut {NoStop}%
\bibitem [{\citenamefont {Rabi}\ \emph {et~al.}(1938)\citenamefont {Rabi},
  \citenamefont {Zacharias}, \citenamefont {Millman},\ and\ \citenamefont
  {Kusch}}]{Rabi:MN-38}%
  \BibitemOpen
  \bibfield  {author} {\bibinfo {author} {\bibfnamefont {I.~I.}\ \bibnamefont
  {Rabi}}, \bibinfo {author} {\bibfnamefont {J.~R.}\ \bibnamefont {Zacharias}},
  \bibinfo {author} {\bibfnamefont {S.}~\bibnamefont {Millman}}, \ and\
  \bibinfo {author} {\bibfnamefont {P.}~\bibnamefont {Kusch}},\ }\bibfield
  {title} {\enquote {\bibinfo {title} {A new method of measuring nuclear
  magnetic moment},}\ }\href {\doibase 10.1103/PhysRev.53.318} {\bibfield
  {journal} {\bibinfo  {journal} {Phys. Rev.}\ }\textbf {\bibinfo {volume}
  {53}},\ \bibinfo {pages} {318--318} (\bibinfo {year} {1938})}\BibitemShut
  {NoStop}%
\bibitem [{\citenamefont {Cohen-Tannoudji}\ \emph {et~al.}(1977)\citenamefont
  {Cohen-Tannoudji}, \citenamefont {Diu},\ and\ \citenamefont
  {Lalo{\"e}}}]{Cohen:QM-77}%
  \BibitemOpen
  \bibfield  {author} {\bibinfo {author} {\bibfnamefont {Claude}\ \bibnamefont
  {Cohen-Tannoudji}}, \bibinfo {author} {\bibfnamefont {Bernard}\ \bibnamefont
  {Diu}}, \ and\ \bibinfo {author} {\bibfnamefont {Franck}\ \bibnamefont
  {Lalo{\"e}}},\ }\href@noop {} {\emph {\bibinfo {title} {Quantum Mechanics}}}\
  (\bibinfo  {publisher} {John Wiley {\&} Sons},\ \bibinfo {address} {New
  York},\ \bibinfo {year} {1977})\BibitemShut {NoStop}%
\bibitem [{\citenamefont {Scully}\ and\ \citenamefont
  {Zubairy}(1997)}]{Scully:QO-97}%
  \BibitemOpen
  \bibfield  {author} {\bibinfo {author} {\bibfnamefont {Marlan~O.}\
  \bibnamefont {Scully}}\ and\ \bibinfo {author} {\bibfnamefont {M.~Suhail}\
  \bibnamefont {Zubairy}},\ }\href@noop {} {\emph {\bibinfo {title} {Quantum
  Optics}}}\ (\bibinfo  {publisher} {Cambridge University Press},\ \bibinfo
  {address} {Cambridge, New York, Melbourne},\ \bibinfo {year}
  {1997})\BibitemShut {NoStop}%
\bibitem [{\citenamefont {Meystre}\ and\ \citenamefont
  {Sargent~III}(1999)}]{Meystre:QO-99}%
  \BibitemOpen
  \bibfield  {author} {\bibinfo {author} {\bibfnamefont {Pierre}\ \bibnamefont
  {Meystre}}\ and\ \bibinfo {author} {\bibfnamefont {Murray}\ \bibnamefont
  {Sargent~III}},\ }\href@noop {} {\emph {\bibinfo {title} {Elements of Quantum
  Optics}}},\ \bibinfo {edition} {3rd}\ ed.\ (\bibinfo  {publisher}
  {Springer},\ \bibinfo {address} {Berlin},\ \bibinfo {year}
  {1999})\BibitemShut {NoStop}%
\bibitem [{\citenamefont {Als-Nielsen}\ and\ \citenamefont
  {McMorrow}(2001)}]{Als-Nielsen:EM-01}%
  \BibitemOpen
  \bibfield  {author} {\bibinfo {author} {\bibfnamefont {Jens}\ \bibnamefont
  {Als-Nielsen}}\ and\ \bibinfo {author} {\bibfnamefont {Des}\ \bibnamefont
  {McMorrow}},\ }\href@noop {} {\emph {\bibinfo {title} {Elements of Modern
  X-Ray Physics}}}\ (\bibinfo  {publisher} {John Wiley {\&} Sons},\ \bibinfo
  {address} {New York},\ \bibinfo {year} {2001})\BibitemShut {NoStop}%
\bibitem [{\citenamefont {Campbell}\ and\ \citenamefont
  {Papp}(2001)}]{Campbell:WA-01}%
  \BibitemOpen
  \bibfield  {author} {\bibinfo {author} {\bibfnamefont {J.~L.}\ \bibnamefont
  {Campbell}}\ and\ \bibinfo {author} {\bibfnamefont {Tibor}\ \bibnamefont
  {Papp}},\ }\bibfield  {title} {\enquote {\bibinfo {title} {Widths of the
  atomic {K--N7} levels},}\ }\href {\doibase 10.1006/adnd.2000.0848} {\bibfield
   {journal} {\bibinfo  {journal} {At. Data Nucl. Data Tables}\ }\textbf
  {\bibinfo {volume} {77}},\ \bibinfo {pages} {1--56} (\bibinfo {year}
  {2001})}\BibitemShut {NoStop}%
\bibitem [{\citenamefont {Perelomov}\ \emph
  {et~al.}(1966{\natexlab{a}})\citenamefont {Perelomov}, \citenamefont
  {Popov},\ and\ \citenamefont {Terent{\'e}v}}]{Perelomov:I1-66}%
  \BibitemOpen
  \bibfield  {author} {\bibinfo {author} {\bibfnamefont {A.~M.}\ \bibnamefont
  {Perelomov}}, \bibinfo {author} {\bibfnamefont {V.~S.}\ \bibnamefont
  {Popov}}, \ and\ \bibinfo {author} {\bibfnamefont {V.~M.}\ \bibnamefont
  {Terent{\'e}v}},\ }\bibfield  {title} {\enquote {\bibinfo {title} {Ionization
  of atoms in an alternating electrical field},}\ }\href
  {http://www.jetp.ac.ru/cgi-bin/e/index/e/23/5/p924?a=list} {\bibfield
  {journal} {\bibinfo  {journal} {Zh. Exp. Theor. Fiz.}\ }\textbf {\bibinfo
  {volume} {50}},\ \bibinfo {pages} {1393--1409} (\bibinfo {year}
  {1966}{\natexlab{a}})},\ \bibinfo {note} {[Sov. Phys. JETP \textbf{23},
  924--934 (1966)]}\BibitemShut {NoStop}%
\bibitem [{\citenamefont {Perelomov}\ \emph
  {et~al.}(1966{\natexlab{b}})\citenamefont {Perelomov}, \citenamefont
  {Popov},\ and\ \citenamefont {Terent{\'e}v}}]{Perelomov:I2-66}%
  \BibitemOpen
  \bibfield  {author} {\bibinfo {author} {\bibfnamefont {A.~M.}\ \bibnamefont
  {Perelomov}}, \bibinfo {author} {\bibfnamefont {V.~S.}\ \bibnamefont
  {Popov}}, \ and\ \bibinfo {author} {\bibfnamefont {V.~M.}\ \bibnamefont
  {Terent{\'e}v}},\ }\bibfield  {title} {\enquote {\bibinfo {title} {Ionization
  of atoms in an alternating electrical field. {II}},}\ }\href
  {http://www.jetp.ac.ru/cgi-bin/e/index/e/24/1/p207?a=list} {\bibfield
  {journal} {\bibinfo  {journal} {Zh. Exp. Theor. Fiz.}\ }\textbf {\bibinfo
  {volume} {51}},\ \bibinfo {pages} {309--326} (\bibinfo {year}
  {1966}{\natexlab{b}})},\ \bibinfo {note} {[Sov. Phys. JETP \textbf{24},
  207--217 (1967)]}\BibitemShut {NoStop}%
\bibitem [{\citenamefont {Perelomov}\ and\ \citenamefont
  {Popov}(1967)}]{Perelomov:I3-67}%
  \BibitemOpen
  \bibfield  {author} {\bibinfo {author} {\bibfnamefont {A.~M.}\ \bibnamefont
  {Perelomov}}\ and\ \bibinfo {author} {\bibfnamefont {V.~S.}\ \bibnamefont
  {Popov}},\ }\bibfield  {title} {\enquote {\bibinfo {title} {Ionization of
  atoms in an alternating electrical field. {III}},}\ }\href
  {http://www.jetp.ac.ru/cgi-bin/e/index/e/25/2/p336?a=list} {\bibfield
  {journal} {\bibinfo  {journal} {Zh. Exp. Theor. Fiz.}\ }\textbf {\bibinfo
  {volume} {52}},\ \bibinfo {pages} {514--526} (\bibinfo {year} {1967})},\
  \bibinfo {note} {[Sov. Phys. JETP \textbf{25}, 336--343 (1967)]}\BibitemShut
  {NoStop}%
\bibitem [{\citenamefont {Popov}\ \emph {et~al.}(1967)\citenamefont {Popov},
  \citenamefont {Kuznetsov},\ and\ \citenamefont {Perelomov}}]{Popov:I4-67}%
  \BibitemOpen
  \bibfield  {author} {\bibinfo {author} {\bibfnamefont {V.~S.}\ \bibnamefont
  {Popov}}, \bibinfo {author} {\bibfnamefont {V.~P.}\ \bibnamefont
  {Kuznetsov}}, \ and\ \bibinfo {author} {\bibfnamefont {A.~M.}\ \bibnamefont
  {Perelomov}},\ }\bibfield  {title} {\enquote {\bibinfo {title}
  {Quasiclassical approximation for nonstationary problems},}\ }\href
  {http://www.jetp.ac.ru/cgi-bin/e/index/e/26/1/p222?a=list} {\bibfield
  {journal} {\bibinfo  {journal} {Zh. Exp. Theor. Fiz.}\ }\textbf {\bibinfo
  {volume} {53}},\ \bibinfo {pages} {331--347} (\bibinfo {year} {1967})},\
  \bibinfo {note} {[Sov. Phys. JETP \textbf{26}, 222--232 (1968)]}\BibitemShut
  {NoStop}%
\bibitem [{\citenamefont {Ammosov}\ \emph {et~al.}(1986)\citenamefont
  {Ammosov}, \citenamefont {Delone},\ and\ \citenamefont
  {Krainov}}]{Ammosov:TI-86}%
  \BibitemOpen
  \bibfield  {author} {\bibinfo {author} {\bibfnamefont {M.~V.}\ \bibnamefont
  {Ammosov}}, \bibinfo {author} {\bibfnamefont {N.~B.}\ \bibnamefont {Delone}},
  \ and\ \bibinfo {author} {\bibfnamefont {V.~P.}\ \bibnamefont {Krainov}},\
  }\bibfield  {title} {\enquote {\bibinfo {title} {Tunnel ionization of complex
  atoms and of atomic ions in an alternating electromagnetic field},}\ }\href
  {http://www.jetp.ac.ru/cgi-bin/e/index/e/64/6/p1191?a=list} {\bibfield
  {journal} {\bibinfo  {journal} {Zh. Eksp. Teor. Fiz.}\ }\textbf {\bibinfo
  {volume} {91}},\ \bibinfo {pages} {2008--2013} (\bibinfo {year} {1986})},\
  \bibinfo {note} {[Sov. Phys. JETP \textbf{64}, 1191--1194
  (1986)]}\BibitemShut {NoStop}%
\bibitem [{\citenamefont {Delone}\ and\ \citenamefont
  {Krainov}(2000)}]{Delone:MP-00}%
  \BibitemOpen
  \bibfield  {author} {\bibinfo {author} {\bibfnamefont {Nikolai~B.}\
  \bibnamefont {Delone}}\ and\ \bibinfo {author} {\bibfnamefont {Vladimir~P.}\
  \bibnamefont {Krainov}},\ }\href@noop {} {\emph {\bibinfo {title}
  {Multiphoton Processes in Atoms}}},\ \bibinfo {edition} {2nd}\ ed.,\ edited
  by\ \bibinfo {editor} {\bibfnamefont {P.}~\bibnamefont {Lambropoulos}},\
  \bibinfo {series} {Atoms and Plasmas}, Vol.~\bibinfo {volume} {13}\ (\bibinfo
   {publisher} {Springer},\ \bibinfo {address} {Berlin},\ \bibinfo {year}
  {2000})\BibitemShut {NoStop}%
\bibitem [{\citenamefont {Yudin}\ and\ \citenamefont
  {Ivanov}(2001)}]{Yudin:NA-01}%
  \BibitemOpen
  \bibfield  {author} {\bibinfo {author} {\bibfnamefont {Gennady~L.}\
  \bibnamefont {Yudin}}\ and\ \bibinfo {author} {\bibfnamefont {Misha~Yu.}\
  \bibnamefont {Ivanov}},\ }\bibfield  {title} {\enquote {\bibinfo {title}
  {Nonadiabatic tunnel ionization: {Looking} inside a laser cycle},}\ }\href
  {\doibase 10.1103/PhysRevA.64.013409} {\bibfield  {journal} {\bibinfo
  {journal} {Phys. Rev. A}\ }\textbf {\bibinfo {volume} {64}},\ \bibinfo
  {pages} {013409} (\bibinfo {year} {2001})}\BibitemShut {NoStop}%
\bibitem [{\citenamefont {Rohringer}\ and\ \citenamefont
  {Santra}(2008{\natexlab{a}})}]{Rohringer:RA-08}%
  \BibitemOpen
  \bibfield  {author} {\bibinfo {author} {\bibfnamefont {Nina}\ \bibnamefont
  {Rohringer}}\ and\ \bibinfo {author} {\bibfnamefont {Robin}\ \bibnamefont
  {Santra}},\ }\bibfield  {title} {\enquote {\bibinfo {title} {Resonant {Auger}
  effect at high x-ray intensity},}\ }\href {\doibase
  10.1103/PhysRevA.77.053404} {\bibfield  {journal} {\bibinfo  {journal} {Phys.
  Rev. A}\ }\textbf {\bibinfo {volume} {77}},\ \bibinfo {pages} {053404}
  (\bibinfo {year} {2008}{\natexlab{a}})}\BibitemShut {NoStop}%
\bibitem [{\citenamefont {Rohringer}\ and\ \citenamefont
  {Santra}(2008{\natexlab{b}})}]{Rohringer:PN-08}%
  \BibitemOpen
  \bibfield  {author} {\bibinfo {author} {\bibfnamefont {Nina}\ \bibnamefont
  {Rohringer}}\ and\ \bibinfo {author} {\bibfnamefont {Robin}\ \bibnamefont
  {Santra}},\ }\bibfield  {title} {\enquote {\bibinfo {title} {Publisher's
  {Note: Resonant Auger} effect at high x-ray intensity [{Phys. Rev. A} 77,
  053404 (2008)]},}\ }\href {\doibase 10.1103/PhysRevA.77.059903} {\bibfield
  {journal} {\bibinfo  {journal} {Phys. Rev. A}\ }\textbf {\bibinfo {volume}
  {77}},\ \bibinfo {pages} {059903(E)} (\bibinfo {year}
  {2008}{\natexlab{b}})}\BibitemShut {NoStop}%
\bibitem [{\citenamefont {Rohringer}\ and\ \citenamefont
  {Santra}(2012)}]{Rohringer:SD-12}%
  \BibitemOpen
  \bibfield  {author} {\bibinfo {author} {\bibfnamefont {Nina}\ \bibnamefont
  {Rohringer}}\ and\ \bibinfo {author} {\bibfnamefont {Robin}\ \bibnamefont
  {Santra}},\ }\bibfield  {title} {\enquote {\bibinfo {title} {Strongly driven
  resonant {Auger} effect treated by an open-quantum-system approach},}\ }\href
  {\doibase 10.1103/PhysRevA.86.043434} {\bibfield  {journal} {\bibinfo
  {journal} {Phys. Rev. A}\ }\textbf {\bibinfo {volume} {86}},\ \bibinfo
  {pages} {043434} (\bibinfo {year} {2012})}\BibitemShut {NoStop}%
\bibitem [{\citenamefont {Kanter}\ \emph {et~al.}(2011)\citenamefont {Kanter},
  \citenamefont {Kr{\"a}ssig}, \citenamefont {Li}, \citenamefont {March},
  \citenamefont {Ho}, \citenamefont {Rohringer}, \citenamefont {Santra},
  \citenamefont {Southworth}, \citenamefont {DiMauro}, \citenamefont {Doumy},
  \citenamefont {Roedig}, \citenamefont {Berrah}, \citenamefont {Fang},
  \citenamefont {Hoener}, \citenamefont {Bucksbaum}, \citenamefont {Ghimire},
  \citenamefont {Reis}, \citenamefont {Bozek}, \citenamefont {Bostedt},
  \citenamefont {Messerschmidt},\ and\ \citenamefont {Young}}]{Kanter:MA-11}%
  \BibitemOpen
  \bibfield  {author} {\bibinfo {author} {\bibfnamefont {E.~P.}\ \bibnamefont
  {Kanter}}, \bibinfo {author} {\bibfnamefont {B.}~\bibnamefont {Kr{\"a}ssig}},
  \bibinfo {author} {\bibfnamefont {Y.}~\bibnamefont {Li}}, \bibinfo {author}
  {\bibfnamefont {A.~M.}\ \bibnamefont {March}}, \bibinfo {author}
  {\bibfnamefont {P.}~\bibnamefont {Ho}}, \bibinfo {author} {\bibfnamefont
  {N.}~\bibnamefont {Rohringer}}, \bibinfo {author} {\bibfnamefont
  {R.}~\bibnamefont {Santra}}, \bibinfo {author} {\bibfnamefont {S.~H.}\
  \bibnamefont {Southworth}}, \bibinfo {author} {\bibfnamefont {L.~F.}\
  \bibnamefont {DiMauro}}, \bibinfo {author} {\bibfnamefont {G.}~\bibnamefont
  {Doumy}}, \bibinfo {author} {\bibfnamefont {C.~A.}\ \bibnamefont {Roedig}},
  \bibinfo {author} {\bibfnamefont {N.}~\bibnamefont {Berrah}}, \bibinfo
  {author} {\bibfnamefont {L.}~\bibnamefont {Fang}}, \bibinfo {author}
  {\bibfnamefont {M.}~\bibnamefont {Hoener}}, \bibinfo {author} {\bibfnamefont
  {P.~H.}\ \bibnamefont {Bucksbaum}}, \bibinfo {author} {\bibfnamefont
  {S.}~\bibnamefont {Ghimire}}, \bibinfo {author} {\bibfnamefont {D.~A.}\
  \bibnamefont {Reis}}, \bibinfo {author} {\bibfnamefont {J.~D.}\ \bibnamefont
  {Bozek}}, \bibinfo {author} {\bibfnamefont {C.}~\bibnamefont {Bostedt}},
  \bibinfo {author} {\bibfnamefont {M.}~\bibnamefont {Messerschmidt}}, \ and\
  \bibinfo {author} {\bibfnamefont {L.}~\bibnamefont {Young}},\ }\bibfield
  {title} {\enquote {\bibinfo {title} {Unveiling and driving hidden resonances
  with high-fluence, high-intensity x-ray pulses},}\ }\href {\doibase
  10.1103/PhysRevLett.107.233001} {\bibfield  {journal} {\bibinfo  {journal}
  {Phys. Rev. Lett.}\ }\textbf {\bibinfo {volume} {107}},\ \bibinfo {pages}
  {233001} (\bibinfo {year} {2011})}\BibitemShut {NoStop}%
\bibitem [{\citenamefont {Popmintchev}\ \emph {et~al.}(2009)\citenamefont
  {Popmintchev}, \citenamefont {Chen}, \citenamefont {Bahabad}, \citenamefont
  {Gerrity}, \citenamefont {Sidorenko}, \citenamefont {Cohen}, \citenamefont
  {Christov}, \citenamefont {Murnane},\ and\ \citenamefont
  {Kapteyn}}]{Popmintchev:PM-09}%
  \BibitemOpen
  \bibfield  {author} {\bibinfo {author} {\bibfnamefont {Tenio}\ \bibnamefont
  {Popmintchev}}, \bibinfo {author} {\bibfnamefont {Ming-Chang}\ \bibnamefont
  {Chen}}, \bibinfo {author} {\bibfnamefont {Alon}\ \bibnamefont {Bahabad}},
  \bibinfo {author} {\bibfnamefont {Michael}\ \bibnamefont {Gerrity}}, \bibinfo
  {author} {\bibfnamefont {Pavel}\ \bibnamefont {Sidorenko}}, \bibinfo {author}
  {\bibfnamefont {Oren}\ \bibnamefont {Cohen}}, \bibinfo {author}
  {\bibfnamefont {Ivan~P.}\ \bibnamefont {Christov}}, \bibinfo {author}
  {\bibfnamefont {Margaret~M.}\ \bibnamefont {Murnane}}, \ and\ \bibinfo
  {author} {\bibfnamefont {Henry~C.}\ \bibnamefont {Kapteyn}},\ }\bibfield
  {title} {\enquote {\bibinfo {title} {Phase matching of high harmonic
  generation in the soft and hard x-ray regions of the spectrum},}\ }\href
  {\doibase 10.1073/pnas.0903748106} {\bibfield  {journal} {\bibinfo  {journal}
  {Proc. Natl. Acad. Sci. U.S.A.}\ }\textbf {\bibinfo {volume} {106}},\
  \bibinfo {pages} {10516--10521} (\bibinfo {year} {2009})}\BibitemShut
  {NoStop}%
\bibitem [{\citenamefont {Arpin}\ \emph {et~al.}(2009)\citenamefont {Arpin},
  \citenamefont {Popmintchev}, \citenamefont {Wagner}, \citenamefont {Lytle},
  \citenamefont {Cohen}, \citenamefont {Kapteyn},\ and\ \citenamefont
  {Murnane}}]{Arpin:EH-09}%
  \BibitemOpen
  \bibfield  {author} {\bibinfo {author} {\bibfnamefont {P.}~\bibnamefont
  {Arpin}}, \bibinfo {author} {\bibfnamefont {T.}~\bibnamefont {Popmintchev}},
  \bibinfo {author} {\bibfnamefont {N.~L.}\ \bibnamefont {Wagner}}, \bibinfo
  {author} {\bibfnamefont {A.~L.}\ \bibnamefont {Lytle}}, \bibinfo {author}
  {\bibfnamefont {O.}~\bibnamefont {Cohen}}, \bibinfo {author} {\bibfnamefont
  {H.~C.}\ \bibnamefont {Kapteyn}}, \ and\ \bibinfo {author} {\bibfnamefont
  {M.~M.}\ \bibnamefont {Murnane}},\ }\bibfield  {title} {\enquote {\bibinfo
  {title} {Enhanced high harmonic generation from multiply ionized argon above
  $500\,${eV} through laser pulse self-compression},}\ }\href {\doibase
  10.1103/PhysRevLett.103.143901} {\bibfield  {journal} {\bibinfo  {journal}
  {Phys. Rev. Lett.}\ }\textbf {\bibinfo {volume} {103}},\ \bibinfo {pages}
  {143901} (\bibinfo {year} {2009})}\BibitemShut {NoStop}%
\bibitem [{\citenamefont {Chen}\ \emph {et~al.}(2010)\citenamefont {Chen},
  \citenamefont {Arpin}, \citenamefont {Popmintchev}, \citenamefont {Gerrity},
  \citenamefont {Zhang}, \citenamefont {Seaberg}, \citenamefont {Popmintchev},
  \citenamefont {Murnane},\ and\ \citenamefont {Kapteyn}}]{Chen:BC-10}%
  \BibitemOpen
  \bibfield  {author} {\bibinfo {author} {\bibfnamefont {M.-C.}\ \bibnamefont
  {Chen}}, \bibinfo {author} {\bibfnamefont {P.}~\bibnamefont {Arpin}},
  \bibinfo {author} {\bibfnamefont {T.}~\bibnamefont {Popmintchev}}, \bibinfo
  {author} {\bibfnamefont {M.}~\bibnamefont {Gerrity}}, \bibinfo {author}
  {\bibfnamefont {B.}~\bibnamefont {Zhang}}, \bibinfo {author} {\bibfnamefont
  {M.}~\bibnamefont {Seaberg}}, \bibinfo {author} {\bibfnamefont
  {D.}~\bibnamefont {Popmintchev}}, \bibinfo {author} {\bibfnamefont {M.~M.}\
  \bibnamefont {Murnane}}, \ and\ \bibinfo {author} {\bibfnamefont {H.~C.}\
  \bibnamefont {Kapteyn}},\ }\bibfield  {title} {\enquote {\bibinfo {title}
  {Bright, coherent, ultrafast soft x-ray harmonics spanning the water window
  from a tabletop light source},}\ }\href {\doibase
  10.1103/PhysRevLett.105.173901} {\bibfield  {journal} {\bibinfo  {journal}
  {Phys. Rev. Lett.}\ }\textbf {\bibinfo {volume} {105}},\ \bibinfo {pages}
  {173901} (\bibinfo {year} {2010})}\BibitemShut {NoStop}%
\bibitem [{Sup()}]{SuppData}%
  \BibitemOpen
  \href@noop {} {}\bibinfo {note} {See the Electronic Supplementary Material
  for a \textit{Mathematica}~\cite{Mathematica:pgm-V10.1}
  Notebook.}\BibitemShut {Stop}%
\bibitem [{Mat(2015)}]{Mathematica:pgm-V10.1}%
  \BibitemOpen
  \href {http://www.wolfram.com} {\emph {\bibinfo {title}
  {\textit{Mathematica}~10.1}}},\ \bibinfo {organization} {Wolfram Research,
  Inc.},\ \bibinfo {address} {100~Trade Center Drive, Champaign,
  Illinois~61820-7237, USA} (\bibinfo {year} {2015})\BibitemShut {NoStop}%
\bibitem [{\citenamefont {Hartree}(1928)}]{Hartree:WM-28}%
  \BibitemOpen
  \bibfield  {author} {\bibinfo {author} {\bibfnamefont {D.~R.}\ \bibnamefont
  {Hartree}},\ }\bibfield  {title} {\enquote {\bibinfo {title} {The wave
  mechanics of an atom with a non-coulomb central field. {Part}~{I}. {Theory}
  and methods},}\ }\href {\doibase 10.1017/S0305004100011919} {\bibfield
  {journal} {\bibinfo  {journal} {Proc. Camb. Phil. Soc.}\ }\textbf {\bibinfo
  {volume} {24}},\ \bibinfo {pages} {89--110} (\bibinfo {year}
  {1928})}\BibitemShut {NoStop}%
\bibitem [{\citenamefont {Szabo}\ and\ \citenamefont
  {Ostlund}(1989)}]{Szabo:MQC-89}%
  \BibitemOpen
  \bibfield  {author} {\bibinfo {author} {\bibfnamefont {Attila}\ \bibnamefont
  {Szabo}}\ and\ \bibinfo {author} {\bibfnamefont {Neil~S.}\ \bibnamefont
  {Ostlund}},\ }\href@noop {} {\emph {\bibinfo {title} {Modern Quantum
  Chemistry: Introduction to Advanced Electronic Structure Theory}}},\ \bibinfo
  {edition} {{1st, revised}}\ ed.\ (\bibinfo  {publisher} {McGraw-Hill},\
  \bibinfo {address} {New York},\ \bibinfo {year} {1989})\BibitemShut {NoStop}%
\bibitem [{\citenamefont {Slater}(1951)}]{Slater:AS-51}%
  \BibitemOpen
  \bibfield  {author} {\bibinfo {author} {\bibfnamefont {J.~C.}\ \bibnamefont
  {Slater}},\ }\bibfield  {title} {\enquote {\bibinfo {title} {A simplification
  of the {Hartree-Fock} method},}\ }\href {\doibase 10.1103/PhysRev.81.385}
  {\bibfield  {journal} {\bibinfo  {journal} {Phys. Rev.}\ }\textbf {\bibinfo
  {volume} {81}},\ \bibinfo {pages} {385--390} (\bibinfo {year}
  {1951})}\BibitemShut {NoStop}%
\bibitem [{\citenamefont {Slater}\ and\ \citenamefont
  {Johnson}(1972)}]{Slater:XA-72}%
  \BibitemOpen
  \bibfield  {author} {\bibinfo {author} {\bibfnamefont {J.~C.}\ \bibnamefont
  {Slater}}\ and\ \bibinfo {author} {\bibfnamefont {K.~H.}\ \bibnamefont
  {Johnson}},\ }\bibfield  {title} {\enquote {\bibinfo {title}
  {Self-consistent-field $x\alpha$~cluster method for polyatomic molecules and
  solids},}\ }\href {\doibase 10.1103/PhysRevB.5.844} {\bibfield  {journal}
  {\bibinfo  {journal} {Phys. Rev. B}\ }\textbf {\bibinfo {volume} {5}},\
  \bibinfo {pages} {844--853} (\bibinfo {year} {1972})}\BibitemShut {NoStop}%
\bibitem [{\citenamefont {Herman}\ and\ \citenamefont
  {Skillman}(1963)}]{Herman:AS-63}%
  \BibitemOpen
  \bibfield  {author} {\bibinfo {author} {\bibfnamefont {Frank}\ \bibnamefont
  {Herman}}\ and\ \bibinfo {author} {\bibfnamefont {Sherwood}\ \bibnamefont
  {Skillman}},\ }\href@noop {} {\emph {\bibinfo {title} {Atomic Structure
  Calculations}}}\ (\bibinfo  {publisher} {Prentice-Hall},\ \bibinfo {address}
  {Englewood Cliffs, New Jersey},\ \bibinfo {year} {1963})\BibitemShut
  {NoStop}%
\bibitem [{\citenamefont {Buth}\ and\ \citenamefont
  {Santra}(2007)}]{Buth:TX-07}%
  \BibitemOpen
  \bibfield  {author} {\bibinfo {author} {\bibfnamefont {Christian}\
  \bibnamefont {Buth}}\ and\ \bibinfo {author} {\bibfnamefont {Robin}\
  \bibnamefont {Santra}},\ }\bibfield  {title} {\enquote {\bibinfo {title}
  {Theory of x-ray absorption by laser-dressed atoms},}\ }\href {\doibase
  10.1103/PhysRevA.75.033412} {\bibfield  {journal} {\bibinfo  {journal} {Phys.
  Rev. A}\ }\textbf {\bibinfo {volume} {75}},\ \bibinfo {pages} {033412}
  (\bibinfo {year} {2007})},\ \Eprint {http://arxiv.org/abs/physics/0611122}
  {arXiv:physics/0611122} \BibitemShut {NoStop}%
\bibitem [{\citenamefont {Merzbacher}(1998)}]{Merzbacher:QM-98}%
  \BibitemOpen
  \bibfield  {author} {\bibinfo {author} {\bibfnamefont {Eugen}\ \bibnamefont
  {Merzbacher}},\ }\href@noop {} {\emph {\bibinfo {title} {Quantum
  mechanics}}},\ \bibinfo {edition} {3rd}\ ed.\ (\bibinfo  {publisher} {John
  Wiley {\&} Sons},\ \bibinfo {address} {New York},\ \bibinfo {year}
  {1998})\BibitemShut {NoStop}%
\bibitem [{\citenamefont {Young}\ \emph {et~al.}(2006)\citenamefont {Young},
  \citenamefont {Arms}, \citenamefont {Dufresne}, \citenamefont {Dunford},
  \citenamefont {Ederer}, \citenamefont {H{\"o}hr}, \citenamefont {Kanter},
  \citenamefont {Kr{\"a}ssig}, \citenamefont {Landahl}, \citenamefont
  {Peterson}, \citenamefont {Rudati}, \citenamefont {Santra},\ and\
  \citenamefont {Southworth}}]{Young:XR-06}%
  \BibitemOpen
  \bibfield  {author} {\bibinfo {author} {\bibfnamefont {L.}~\bibnamefont
  {Young}}, \bibinfo {author} {\bibfnamefont {D.~A.}\ \bibnamefont {Arms}},
  \bibinfo {author} {\bibfnamefont {E.~M.}\ \bibnamefont {Dufresne}}, \bibinfo
  {author} {\bibfnamefont {R.~W.}\ \bibnamefont {Dunford}}, \bibinfo {author}
  {\bibfnamefont {D.~L.}\ \bibnamefont {Ederer}}, \bibinfo {author}
  {\bibfnamefont {C.}~\bibnamefont {H{\"o}hr}}, \bibinfo {author}
  {\bibfnamefont {E.~P.}\ \bibnamefont {Kanter}}, \bibinfo {author}
  {\bibfnamefont {B.}~\bibnamefont {Kr{\"a}ssig}}, \bibinfo {author}
  {\bibfnamefont {E.~C.}\ \bibnamefont {Landahl}}, \bibinfo {author}
  {\bibfnamefont {E.~R.}\ \bibnamefont {Peterson}}, \bibinfo {author}
  {\bibfnamefont {J.}~\bibnamefont {Rudati}}, \bibinfo {author} {\bibfnamefont
  {R.}~\bibnamefont {Santra}}, \ and\ \bibinfo {author} {\bibfnamefont {S.~H.}\
  \bibnamefont {Southworth}},\ }\bibfield  {title} {\enquote {\bibinfo {title}
  {X-ray microprobe of orbital alignment in strong-field ionized atoms},}\
  }\href {\doibase 10.1103/PhysRevLett.97.083601} {\bibfield  {journal}
  {\bibinfo  {journal} {Phys. Rev. Lett.}\ }\textbf {\bibinfo {volume} {97}},\
  \bibinfo {pages} {083601} (\bibinfo {year} {2006})}\BibitemShut {NoStop}%
\bibitem [{\citenamefont {Santra}\ \emph {et~al.}(2006)\citenamefont {Santra},
  \citenamefont {Dunford},\ and\ \citenamefont {Young}}]{Santra:SO-06}%
  \BibitemOpen
  \bibfield  {author} {\bibinfo {author} {\bibfnamefont {Robin}\ \bibnamefont
  {Santra}}, \bibinfo {author} {\bibfnamefont {Robert~W.}\ \bibnamefont
  {Dunford}}, \ and\ \bibinfo {author} {\bibfnamefont {Linda}\ \bibnamefont
  {Young}},\ }\bibfield  {title} {\enquote {\bibinfo {title} {Spin-orbit effect
  on strong-field ionization of krypton},}\ }\href {\doibase
  10.1103/PhysRevA.74.043403} {\bibfield  {journal} {\bibinfo  {journal} {Phys.
  Rev. A}\ }\textbf {\bibinfo {volume} {74}},\ \bibinfo {pages} {043403}
  (\bibinfo {year} {2006})}\BibitemShut {NoStop}%
\bibitem [{\citenamefont {Rohringer}\ and\ \citenamefont
  {Santra}(2009)}]{Rohringer:MC-09}%
  \BibitemOpen
  \bibfield  {author} {\bibinfo {author} {\bibfnamefont {Nina}\ \bibnamefont
  {Rohringer}}\ and\ \bibinfo {author} {\bibfnamefont {Robin}\ \bibnamefont
  {Santra}},\ }\bibfield  {title} {\enquote {\bibinfo {title} {Multichannel
  coherence in strong-field ionization},}\ }\href {\doibase
  10.1103/PhysRevA.79.053402} {\bibfield  {journal} {\bibinfo  {journal} {Phys.
  Rev. A}\ }\textbf {\bibinfo {volume} {79}},\ \bibinfo {pages} {053402}
  (\bibinfo {year} {2009})}\BibitemShut {NoStop}%
\bibitem [{\citenamefont {Loh}\ \emph {et~al.}(2007)\citenamefont {Loh},
  \citenamefont {Khalil}, \citenamefont {Correa}, \citenamefont {Santra},
  \citenamefont {Buth},\ and\ \citenamefont {Leone}}]{Loh:QS-07}%
  \BibitemOpen
  \bibfield  {author} {\bibinfo {author} {\bibfnamefont {Zhi-Heng}\
  \bibnamefont {Loh}}, \bibinfo {author} {\bibfnamefont {Munira}\ \bibnamefont
  {Khalil}}, \bibinfo {author} {\bibfnamefont {Raoul~E.}\ \bibnamefont
  {Correa}}, \bibinfo {author} {\bibfnamefont {Robin}\ \bibnamefont {Santra}},
  \bibinfo {author} {\bibfnamefont {Christian}\ \bibnamefont {Buth}}, \ and\
  \bibinfo {author} {\bibfnamefont {Stephen~R.}\ \bibnamefont {Leone}},\
  }\bibfield  {title} {\enquote {\bibinfo {title} {Quantum state-resolved
  probing of strong-field-ionized xenon atoms using femtosecond high-order
  harmonic transient absorption spectroscopy},}\ }\href {\doibase
  10.1103/PhysRevLett.98.143601} {\bibfield  {journal} {\bibinfo  {journal}
  {Phys. Rev. Lett.}\ }\textbf {\bibinfo {volume} {98}},\ \bibinfo {pages}
  {143601} (\bibinfo {year} {2007})},\ \Eprint
  {http://arxiv.org/abs/physics/0703149} {arXiv:physics/0703149} \BibitemShut
  {NoStop}%
\bibitem [{\citenamefont {Arfken}\ and\ \citenamefont
  {Weber}(2005)}]{Arfken:MM-05}%
  \BibitemOpen
  \bibfield  {author} {\bibinfo {author} {\bibfnamefont {George~B.}\
  \bibnamefont {Arfken}}\ and\ \bibinfo {author} {\bibfnamefont {Hans~J.}\
  \bibnamefont {Weber}},\ }\href@noop {} {\emph {\bibinfo {title} {Mathematical
  Methods for Physicists}}},\ \bibinfo {edition} {6th}\ ed.\ (\bibinfo
  {publisher} {Elsevier Academic Press},\ \bibinfo {address} {New York},\
  \bibinfo {year} {2005})\BibitemShut {NoStop}%
\bibitem [{\citenamefont {Cederbaum}\ \emph {et~al.}(1980)\citenamefont
  {Cederbaum}, \citenamefont {Domcke},\ and\ \citenamefont
  {Schirmer}}]{Cederbaum:CH-80}%
  \BibitemOpen
  \bibfield  {author} {\bibinfo {author} {\bibfnamefont {Lorenz~S.}\
  \bibnamefont {Cederbaum}}, \bibinfo {author} {\bibfnamefont {W.}~\bibnamefont
  {Domcke}}, \ and\ \bibinfo {author} {\bibfnamefont {Jochen}\ \bibnamefont
  {Schirmer}},\ }\bibfield  {title} {\enquote {\bibinfo {title} {Many-body
  theory of core holes},}\ }\href {\doibase 10.1103/PhysRevA.22.206} {\bibfield
   {journal} {\bibinfo  {journal} {Phys. Rev. A}\ }\textbf {\bibinfo {volume}
  {22}},\ \bibinfo {pages} {206--222} (\bibinfo {year} {1980})}\BibitemShut
  {NoStop}%
\bibitem [{\citenamefont {Angonoa}\ \emph {et~al.}(1987)\citenamefont
  {Angonoa}, \citenamefont {Walter},\ and\ \citenamefont
  {Schirmer}}]{Angonoa:KS-87}%
  \BibitemOpen
  \bibfield  {author} {\bibinfo {author} {\bibfnamefont {G.}~\bibnamefont
  {Angonoa}}, \bibinfo {author} {\bibfnamefont {O.}~\bibnamefont {Walter}}, \
  and\ \bibinfo {author} {\bibfnamefont {Jochen}\ \bibnamefont {Schirmer}},\
  }\bibfield  {title} {\enquote {\bibinfo {title} {Theoretical {K}-shell
  ionization spectra of~{N$_2$} and {CO} by a fourth-order {Green's} function
  method},}\ }\href {\doibase 10.1063/1.453424} {\bibfield  {journal} {\bibinfo
   {journal} {J. Chem. Phys.}\ }\textbf {\bibinfo {volume} {87}},\ \bibinfo
  {pages} {6789--6801} (\bibinfo {year} {1987})}\BibitemShut {NoStop}%
\bibitem [{\citenamefont {Madsen}(2005)}]{Madsen:SF-05}%
  \BibitemOpen
  \bibfield  {author} {\bibinfo {author} {\bibfnamefont {Lars~Bojer}\
  \bibnamefont {Madsen}},\ }\bibfield  {title} {\enquote {\bibinfo {title}
  {Strong-field approximation in laser-assisted dynamics},}\ }\href {\doibase
  10.1119/1.1796791} {\bibfield  {journal} {\bibinfo  {journal} {Am. J. Phys.}\
  }\textbf {\bibinfo {volume} {73}},\ \bibinfo {pages} {57--62} (\bibinfo
  {year} {2005})}\BibitemShut {NoStop}%
\bibitem [{\citenamefont {Buth}\ and\ \citenamefont
  {Schafer}(2009)}]{Buth:TA-09}%
  \BibitemOpen
  \bibfield  {author} {\bibinfo {author} {\bibfnamefont {Christian}\
  \bibnamefont {Buth}}\ and\ \bibinfo {author} {\bibfnamefont {Kenneth~J.}\
  \bibnamefont {Schafer}},\ }\bibfield  {title} {\enquote {\bibinfo {title}
  {Theory of {Auger} decay by laser-dressed atoms},}\ }\href {\doibase
  10.1103/PhysRevA.80.033410} {\bibfield  {journal} {\bibinfo  {journal} {Phys.
  Rev. A}\ }\textbf {\bibinfo {volume} {80}},\ \bibinfo {pages} {033410}
  (\bibinfo {year} {2009})},\ \Eprint {http://arxiv.org/abs/0905.3756}
  {arXiv:0905.3756} \BibitemShut {NoStop}%
\bibitem [{\citenamefont {Cowan}(1981)}]{Cowan:TA-81}%
  \BibitemOpen
  \bibfield  {author} {\bibinfo {author} {\bibfnamefont {Robert~D.}\
  \bibnamefont {Cowan}},\ }\href@noop {} {\emph {\bibinfo {title} {The Theory
  of Atomic Structure and Spectra}}},\ Los Alamos Series in Basic and Applied
  Sciences\ (\bibinfo  {publisher} {University of California Press},\ \bibinfo
  {address} {Berkeley},\ \bibinfo {year} {1981})\BibitemShut {NoStop}%
\bibitem [{LAN()}]{LANL:AP-00}%
  \BibitemOpen
  \href@noop {} {}\bibinfo {note} {Los Alamos National Laboratory, Atomic
  Physics Codes, \href{http://aphysics2.lanl.gov/tempweb/lanl}%
  {http://aphysics2.lanl.gov/tempweb/lanl}}\BibitemShut {NoStop}%
\bibitem [{Note1()}]{Note1}%
  \BibitemOpen
  \bibinfo {note} {Equation~(\ref {eq:lasvector}) ist the only expression of
  this article where the factor~$\protect \mathrm e^{- \eta \protect \tmspace
  +\thinmuskip {.1667em} |t^{\protect \tmspace +\thinmuskip {.1667em}\prime
  }|}$ is required. Thus I took the limit~$\eta \to 0^+$ in all other equations
  and, specifically, the factor is omitted from Eq.~(\ref
  {eq:lasfield}).}\BibitemShut {Stop}%
\bibitem [{Note2()}]{Note2}%
  \BibitemOpen
  \bibinfo {note} {I omit terms for continuum-continuum transitions, i.e.{},
  terms involving one-electron matrix elements of the type~$\left <\right
  .\protect \tmspace -\thinmuskip {.1667em}\protect \mathaccentV
  {vec}17Ek\protect \tmspace -\thinmuskip {.1667em}\left .\right | \protect
  \mathaccentV {hat}05Ed_1 \left |\right .\protect \tmspace -\thinmuskip
  {.1667em}\protect \mathaccentV {vec}17Ek^{\protect \tmspace +\thinmuskip
  {.1667em}\prime }\protect \tmspace -\thinmuskip {.1667em}\left .\right >$
  for~$\protect \mathaccentV {vec}17Ek, \protect \mathaccentV
  {vec}17Ek^{\protect \tmspace +\thinmuskip {.1667em}\prime } \in \protect
  \mathbb R^3$ [Appendix~\ref {sec:CCtrans}] because such terms do not play a
  role for the description of HH~emission in the electric dipole transition
  matrix element in Eq.~(\ref {eq:Nhalfdipole})~\cite
  {Lewenstein:HH-94,Kuchiev:QT-99}.}\BibitemShut {Stop}%
\bibitem [{Note3()}]{Note3}%
  \BibitemOpen
  \bibinfo {note} {As the electron propagates freely in SFA in the continuum
  with kinetic momentum~$\protect \mathaccentV {vec}17Ek$, the canonical
  momentum~$\protect \mathaccentV {vec}17Ep$ at time~$t$ and at
  time~$t^{\protect \tmspace +\thinmuskip {.1667em}\prime }$ are the same~\cite
  {Lewenstein:HH-94}.}\BibitemShut {Stop}%
\bibitem [{Note4()}]{Note4}%
  \BibitemOpen
  \bibinfo {note} {I use only the classical part of~$S^{(m)}_j(\protect
  \mathaccentV {vec}17Ep, t, t^{\protect \tmspace +\thinmuskip {.1667em}\prime
  })$ for the saddle-point approximation, i.e.{}, I let~${\protect \cal
  S}(\protect \mathaccentV {vec}17Ep) = \protect \genfrac {}{}{}1{1}{2} \DOTSI
  \intop \ilimits@ \limits _{t^{\protect \tmspace +\thinmuskip {.1667em}\prime
  }}^t \mathopen {\setbox \z@ \hbox {\frozen@everymath \@emptytoks
  \mathsurround \z@ $\nulldelimiterspace \z@ \left (\vcenter to\@ne \big@size
  {}\right .$}\box \z@ }\protect \mathaccentV {vec}17Ep + \protect \mathaccentV
  {vec}17EA_{\protect \mathrm {L}}(t^{\protect \tmspace +\thinmuskip
  {.1667em}\prime \prime }) \mathclose {\setbox \z@ \hbox {\frozen@everymath
  \@emptytoks \mathsurround \z@ $\nulldelimiterspace \z@ \left )\vcenter to\@ne
  \big@size {}\right .$}\box \z@ }^2 \mskip \medmuskip \protect \mathrm
  dt^{\protect \tmspace +\thinmuskip {.1667em}\prime \prime }$ in Eq.~(\ref
  {eq:integralgS}) and include the remaining factors in~$f(\protect
  \mathaccentV {vec}17Ep)$. For~$S^{(m)}_j(\protect \mathaccentV {vec}17Ep, t,
  t^{\protect \tmspace +\thinmuskip {.1667em}\prime })$, the series~(\ref
  {eq:TaylorS}) actually terminates after the second-order term and thus is
  convergent in the entire~$\protect \mathbb R^3$.}\BibitemShut {Stop}%
\bibitem [{\citenamefont {Rose}(1957)}]{Rose:ET-57}%
  \BibitemOpen
  \bibfield  {author} {\bibinfo {author} {\bibfnamefont {Morris~Edgar}\
  \bibnamefont {Rose}},\ }\href@noop {} {\emph {\bibinfo {title} {Elementary
  Theory of Angular Momentum}}},\ Structure of Matter\ (\bibinfo  {publisher}
  {John Wiley {\&} Sons},\ \bibinfo {address} {New York},\ \bibinfo {year}
  {1957})\BibitemShut {NoStop}%
\bibitem [{\citenamefont {Diestler}(2008)}]{Diestler:HG-08}%
  \BibitemOpen
  \bibfield  {author} {\bibinfo {author} {\bibfnamefont {D.~J.}\ \bibnamefont
  {Diestler}},\ }\bibfield  {title} {\enquote {\bibinfo {title} {Harmonic
  generation: quantum-electrodynamical theory of the harmonic photon-number
  spectrum},}\ }\href {\doibase 10.1103/PhysRevA.78.033814} {\bibfield
  {journal} {\bibinfo  {journal} {Phys. Rev. A}\ }\textbf {\bibinfo {volume}
  {78}},\ \bibinfo {pages} {033814} (\bibinfo {year} {2008})}\BibitemShut
  {NoStop}%
\bibitem [{Note5()}]{Note5}%
  \BibitemOpen
  \bibinfo {note} {The multiplication with the duration~$T_{\protect \mathrm
  {P}}$ corresponds to determining the HPNS from optical-laser and x\hbox
  {-}{}ray~pulses with constant field strength starting at~$t = 0$ and stopping
  at~$t = T_{\protect \mathrm {P}}$ where turn-on effects
  neglected.}\BibitemShut {Stop}%
\bibitem [{\citenamefont {Diels}\ and\ \citenamefont
  {Rudolph}(2006)}]{Diels:UL-06}%
  \BibitemOpen
  \bibfield  {author} {\bibinfo {author} {\bibfnamefont {Jean-Claude}\
  \bibnamefont {Diels}}\ and\ \bibinfo {author} {\bibfnamefont {Wolfgang}\
  \bibnamefont {Rudolph}},\ }\href@noop {} {\emph {\bibinfo {title} {Ultrashort
  Laser Pulse Phenomena}}},\ \bibinfo {edition} {2nd}\ ed.,\ Optics and
  Photonics Series\ (\bibinfo  {publisher} {Academic Press},\ \bibinfo
  {address} {Amsterdam},\ \bibinfo {year} {2006})\BibitemShut {NoStop}%
\bibitem [{\citenamefont {Gordon}\ and\ \citenamefont
  {K{\"a}rtner}(2005)}]{Gordon:QM-05}%
  \BibitemOpen
  \bibfield  {author} {\bibinfo {author} {\bibfnamefont {Ariel}\ \bibnamefont
  {Gordon}}\ and\ \bibinfo {author} {\bibfnamefont {Franz~X.}\ \bibnamefont
  {K{\"a}rtner}},\ }\bibfield  {title} {\enquote {\bibinfo {title}
  {Quantitative modeling of single atom high harmonic generation},}\ }\href
  {\doibase 10.1103/PhysRevLett.95.223901} {\bibfield  {journal} {\bibinfo
  {journal} {Phys. Rev. Lett.}\ }\textbf {\bibinfo {volume} {95}},\ \bibinfo
  {pages} {223901} (\bibinfo {year} {2005})}\BibitemShut {NoStop}%
\bibitem [{\citenamefont {Smirnova}\ \emph {et~al.}(2008)\citenamefont
  {Smirnova}, \citenamefont {Spanner},\ and\ \citenamefont
  {Ivanov}}]{Smirnova:AS-08}%
  \BibitemOpen
  \bibfield  {author} {\bibinfo {author} {\bibfnamefont {Olga}\ \bibnamefont
  {Smirnova}}, \bibinfo {author} {\bibfnamefont {Michael}\ \bibnamefont
  {Spanner}}, \ and\ \bibinfo {author} {\bibfnamefont {Misha}\ \bibnamefont
  {Ivanov}},\ }\bibfield  {title} {\enquote {\bibinfo {title} {Analytical
  solutions for strong field-driven atomic and molecular one- and two-electron
  continua and applications to strong-field problems},}\ }\href {\doibase
  10.1103/PhysRevA.77.033407} {\bibfield  {journal} {\bibinfo  {journal} {Phys.
  Rev. A}\ }\textbf {\bibinfo {volume} {77}},\ \bibinfo {pages} {033407}
  (\bibinfo {year} {2008})}\BibitemShut {NoStop}%
\bibitem [{\citenamefont {Doumy}\ \emph {et~al.}(2011)\citenamefont {Doumy},
  \citenamefont {Roedig}, \citenamefont {Son}, \citenamefont {Blaga},
  \citenamefont {DiChiara}, \citenamefont {Santra}, \citenamefont {Berrah},
  \citenamefont {Bostedt}, \citenamefont {Bozek}, \citenamefont {Bucksbaum},
  \citenamefont {Cryan}, \citenamefont {Fang}, \citenamefont {Ghimire},
  \citenamefont {Glownia}, \citenamefont {Hoener}, \citenamefont {Kanter},
  \citenamefont {Kr{\"a}ssig}, \citenamefont {Kuebel}, \citenamefont
  {Messerschmidt}, \citenamefont {Paulus}, \citenamefont {Reis}, \citenamefont
  {Rohringer}, \citenamefont {Young}, \citenamefont {Agostini},\ and\
  \citenamefont {DiMauro}}]{Doumy:NA-11}%
  \BibitemOpen
  \bibfield  {author} {\bibinfo {author} {\bibfnamefont {G.}~\bibnamefont
  {Doumy}}, \bibinfo {author} {\bibfnamefont {C.}~\bibnamefont {Roedig}},
  \bibinfo {author} {\bibfnamefont {S.-K.}\ \bibnamefont {Son}}, \bibinfo
  {author} {\bibfnamefont {C.~I.}\ \bibnamefont {Blaga}}, \bibinfo {author}
  {\bibfnamefont {A.~D.}\ \bibnamefont {DiChiara}}, \bibinfo {author}
  {\bibfnamefont {R.}~\bibnamefont {Santra}}, \bibinfo {author} {\bibfnamefont
  {N.}~\bibnamefont {Berrah}}, \bibinfo {author} {\bibfnamefont
  {C.}~\bibnamefont {Bostedt}}, \bibinfo {author} {\bibfnamefont {J.~D.}\
  \bibnamefont {Bozek}}, \bibinfo {author} {\bibfnamefont {P.~H.}\ \bibnamefont
  {Bucksbaum}}, \bibinfo {author} {\bibfnamefont {J.~P.}\ \bibnamefont
  {Cryan}}, \bibinfo {author} {\bibfnamefont {L.}~\bibnamefont {Fang}},
  \bibinfo {author} {\bibfnamefont {S.}~\bibnamefont {Ghimire}}, \bibinfo
  {author} {\bibfnamefont {J.~M.}\ \bibnamefont {Glownia}}, \bibinfo {author}
  {\bibfnamefont {M.}~\bibnamefont {Hoener}}, \bibinfo {author} {\bibfnamefont
  {E.~P.}\ \bibnamefont {Kanter}}, \bibinfo {author} {\bibfnamefont
  {B.}~\bibnamefont {Kr{\"a}ssig}}, \bibinfo {author} {\bibfnamefont
  {M.}~\bibnamefont {Kuebel}}, \bibinfo {author} {\bibfnamefont
  {M.}~\bibnamefont {Messerschmidt}}, \bibinfo {author} {\bibfnamefont {G.~G.}\
  \bibnamefont {Paulus}}, \bibinfo {author} {\bibfnamefont {D.~A.}\
  \bibnamefont {Reis}}, \bibinfo {author} {\bibfnamefont {N.}~\bibnamefont
  {Rohringer}}, \bibinfo {author} {\bibfnamefont {L.}~\bibnamefont {Young}},
  \bibinfo {author} {\bibfnamefont {P.}~\bibnamefont {Agostini}}, \ and\
  \bibinfo {author} {\bibfnamefont {L.~F.}\ \bibnamefont {DiMauro}},\
  }\bibfield  {title} {\enquote {\bibinfo {title} {Nonlinear atomic response to
  intense ultrashort x~rays},}\ }\href {\doibase
  10.1103/PhysRevLett.106.083002} {\bibfield  {journal} {\bibinfo  {journal}
  {Phys. Rev. Lett.}\ }\textbf {\bibinfo {volume} {106}},\ \bibinfo {pages}
  {083002} (\bibinfo {year} {2011})}\BibitemShut {NoStop}%
\bibitem [{\citenamefont {Buth}\ \emph {et~al.}(2012)\citenamefont {Buth},
  \citenamefont {Liu}, \citenamefont {Chen}, \citenamefont {Cryan},
  \citenamefont {Fang}, \citenamefont {Glownia}, \citenamefont {Hoener},
  \citenamefont {Coffee},\ and\ \citenamefont {Berrah}}]{Buth:UA-12}%
  \BibitemOpen
  \bibfield  {author} {\bibinfo {author} {\bibfnamefont {Christian}\
  \bibnamefont {Buth}}, \bibinfo {author} {\bibfnamefont {Ji-Cai}\ \bibnamefont
  {Liu}}, \bibinfo {author} {\bibfnamefont {Mau~Hsiung}\ \bibnamefont {Chen}},
  \bibinfo {author} {\bibfnamefont {James~P.}\ \bibnamefont {Cryan}}, \bibinfo
  {author} {\bibfnamefont {Li}~\bibnamefont {Fang}}, \bibinfo {author}
  {\bibfnamefont {James~M.}\ \bibnamefont {Glownia}}, \bibinfo {author}
  {\bibfnamefont {Matthias}\ \bibnamefont {Hoener}}, \bibinfo {author}
  {\bibfnamefont {Ryan~N.}\ \bibnamefont {Coffee}}, \ and\ \bibinfo {author}
  {\bibfnamefont {Nora}\ \bibnamefont {Berrah}},\ }\bibfield  {title} {\enquote
  {\bibinfo {title} {Ultrafast absorption of intense x~rays by nitrogen
  molecules},}\ }\href {\doibase 10.1063/1.4722756} {\bibfield  {journal}
  {\bibinfo  {journal} {J. Chem. Phys.}\ }\textbf {\bibinfo {volume} {136}},\
  \bibinfo {pages} {214310} (\bibinfo {year} {2012})},\ \Eprint
  {http://arxiv.org/abs/1201.1896} {arXiv:1201.1896} \BibitemShut {NoStop}%
\bibitem [{\citenamefont {DeLong}\ \emph {et~al.}(1994)\citenamefont {DeLong},
  \citenamefont {Fittinghoff}, \citenamefont {Trebino}, \citenamefont
  {Kohler},\ and\ \citenamefont {Wilson}}]{DeLong:PR-94}%
  \BibitemOpen
  \bibfield  {author} {\bibinfo {author} {\bibfnamefont {Kenneth~W.}\
  \bibnamefont {DeLong}}, \bibinfo {author} {\bibfnamefont {David~N.}\
  \bibnamefont {Fittinghoff}}, \bibinfo {author} {\bibfnamefont {Rick}\
  \bibnamefont {Trebino}}, \bibinfo {author} {\bibfnamefont {Bern}\
  \bibnamefont {Kohler}}, \ and\ \bibinfo {author} {\bibfnamefont {Kent}\
  \bibnamefont {Wilson}},\ }\bibfield  {title} {\enquote {\bibinfo {title}
  {Pulse retrieval in frequency-resolved optical gating based on the method of
  generalized projections},}\ }\href {\doibase 10.1364/OL.19.002152} {\bibfield
   {journal} {\bibinfo  {journal} {Opt. Lett.}\ }\textbf {\bibinfo {volume}
  {19}},\ \bibinfo {pages} {2152--2154} (\bibinfo {year} {1994})}\BibitemShut
  {NoStop}%
\bibitem [{\citenamefont {Trebino}(2000)}]{Trebino:FR-02}%
  \BibitemOpen
  \bibfield  {author} {\bibinfo {author} {\bibfnamefont {Rick}\ \bibnamefont
  {Trebino}},\ }\href@noop {} {\emph {\bibinfo {title} {Frequency-Resolved
  Optical Gating: The Measurement of Ultrashort Laser Pulses}}}\ (\bibinfo
  {publisher} {Kluwer Academic Publishers},\ \bibinfo {address} {Boston,
  Dordrecht, London},\ \bibinfo {year} {2000})\BibitemShut {NoStop}%
\bibitem [{\citenamefont {D{\"u}sterer}\ \emph {et~al.}(2011)\citenamefont
  {D{\"u}sterer}, \citenamefont {Radcliffe}, \citenamefont {Bostedt},
  \citenamefont {Bozek}, \citenamefont {Cavalieri}, \citenamefont {Coffee},
  \citenamefont {Costello}, \citenamefont {Cubaynes}, \citenamefont {DiMauro},
  \citenamefont {Ding}, \citenamefont {Doumy}, \citenamefont {Gr{\"u}ner},
  \citenamefont {Helml}, \citenamefont {Schweinberger}, \citenamefont
  {Kienberger}, \citenamefont {Maier}, \citenamefont {Messerschmidt},
  \citenamefont {Richardson}, \citenamefont {Roedig}, \citenamefont
  {Tschentscher},\ and\ \citenamefont {Meyer}}]{Dusterer:FS-11}%
  \BibitemOpen
  \bibfield  {author} {\bibinfo {author} {\bibfnamefont {S.}~\bibnamefont
  {D{\"u}sterer}}, \bibinfo {author} {\bibfnamefont {P.}~\bibnamefont
  {Radcliffe}}, \bibinfo {author} {\bibfnamefont {C.}~\bibnamefont {Bostedt}},
  \bibinfo {author} {\bibfnamefont {J.}~\bibnamefont {Bozek}}, \bibinfo
  {author} {\bibfnamefont {A.~L.}\ \bibnamefont {Cavalieri}}, \bibinfo {author}
  {\bibfnamefont {R.}~\bibnamefont {Coffee}}, \bibinfo {author} {\bibfnamefont
  {J.~T.}\ \bibnamefont {Costello}}, \bibinfo {author} {\bibfnamefont
  {D.}~\bibnamefont {Cubaynes}}, \bibinfo {author} {\bibfnamefont {L.~F.}\
  \bibnamefont {DiMauro}}, \bibinfo {author} {\bibfnamefont {Y.}~\bibnamefont
  {Ding}}, \bibinfo {author} {\bibfnamefont {G.}~\bibnamefont {Doumy}},
  \bibinfo {author} {\bibfnamefont {F.}~\bibnamefont {Gr{\"u}ner}}, \bibinfo
  {author} {\bibfnamefont {W.}~\bibnamefont {Helml}}, \bibinfo {author}
  {\bibfnamefont {W.}~\bibnamefont {Schweinberger}}, \bibinfo {author}
  {\bibfnamefont {R.}~\bibnamefont {Kienberger}}, \bibinfo {author}
  {\bibfnamefont {A.~R.}\ \bibnamefont {Maier}}, \bibinfo {author}
  {\bibfnamefont {M.}~\bibnamefont {Messerschmidt}}, \bibinfo {author}
  {\bibfnamefont {V.}~\bibnamefont {Richardson}}, \bibinfo {author}
  {\bibfnamefont {C.}~\bibnamefont {Roedig}}, \bibinfo {author} {\bibfnamefont
  {T.}~\bibnamefont {Tschentscher}}, \ and\ \bibinfo {author} {\bibfnamefont
  {M.}~\bibnamefont {Meyer}},\ }\bibfield  {title} {\enquote {\bibinfo {title}
  {Femtosecond x-ray pulse length characterization at the {Linac Coherent Light
  Source} free-electron laser},}\ }\href {\doibase
  10.1088/1367-2630/13/9/093024} {\bibfield  {journal} {\bibinfo  {journal}
  {New J. Phys.}\ }\textbf {\bibinfo {volume} {13}},\ \bibinfo {pages} {093024}
  (\bibinfo {year} {2011})}\BibitemShut {NoStop}%
\bibitem [{\citenamefont {Meyer}\ \emph {et~al.}(2012)\citenamefont {Meyer},
  \citenamefont {Radcliffe}, \citenamefont {Tschentscher}, \citenamefont
  {Costello}, \citenamefont {Cavalieri}, \citenamefont {Grguras}, \citenamefont
  {Maier}, \citenamefont {Kienberger}, \citenamefont {Bozek}, \citenamefont
  {Bostedt}, \citenamefont {Schorb}, \citenamefont {Coffee}, \citenamefont
  {Messerschmidt}, \citenamefont {Roedig}, \citenamefont {Sistrunk},
  \citenamefont {DiMauro}, \citenamefont {Doumy}, \citenamefont {Ueda},
  \citenamefont {Wada}, \citenamefont {D{\"u}sterer}, \citenamefont
  {Kazansky},\ and\ \citenamefont {Kabachnik}}]{Meyer:AR-12}%
  \BibitemOpen
  \bibfield  {author} {\bibinfo {author} {\bibfnamefont {M.}~\bibnamefont
  {Meyer}}, \bibinfo {author} {\bibfnamefont {P.}~\bibnamefont {Radcliffe}},
  \bibinfo {author} {\bibfnamefont {T.}~\bibnamefont {Tschentscher}}, \bibinfo
  {author} {\bibfnamefont {J.~T.}\ \bibnamefont {Costello}}, \bibinfo {author}
  {\bibfnamefont {A.~L.}\ \bibnamefont {Cavalieri}}, \bibinfo {author}
  {\bibfnamefont {I.}~\bibnamefont {Grguras}}, \bibinfo {author} {\bibfnamefont
  {A.~R.}\ \bibnamefont {Maier}}, \bibinfo {author} {\bibfnamefont
  {R.}~\bibnamefont {Kienberger}}, \bibinfo {author} {\bibfnamefont
  {J.}~\bibnamefont {Bozek}}, \bibinfo {author} {\bibfnamefont
  {C.}~\bibnamefont {Bostedt}}, \bibinfo {author} {\bibfnamefont
  {S.}~\bibnamefont {Schorb}}, \bibinfo {author} {\bibfnamefont
  {R.}~\bibnamefont {Coffee}}, \bibinfo {author} {\bibfnamefont
  {M.}~\bibnamefont {Messerschmidt}}, \bibinfo {author} {\bibfnamefont
  {C.}~\bibnamefont {Roedig}}, \bibinfo {author} {\bibfnamefont
  {E.}~\bibnamefont {Sistrunk}}, \bibinfo {author} {\bibfnamefont {L.~F.}\
  \bibnamefont {DiMauro}}, \bibinfo {author} {\bibfnamefont {G.}~\bibnamefont
  {Doumy}}, \bibinfo {author} {\bibfnamefont {K.}~\bibnamefont {Ueda}},
  \bibinfo {author} {\bibfnamefont {S.}~\bibnamefont {Wada}}, \bibinfo {author}
  {\bibfnamefont {S.}~\bibnamefont {D{\"u}sterer}}, \bibinfo {author}
  {\bibfnamefont {A.~K.}\ \bibnamefont {Kazansky}}, \ and\ \bibinfo {author}
  {\bibfnamefont {N.~M.}\ \bibnamefont {Kabachnik}},\ }\bibfield  {title}
  {\enquote {\bibinfo {title} {Angle-resolved electron spectroscopy of
  laser-assisted {Auger} decay induced by a few-femtosecond x-ray pulse},}\
  }\href {\doibase 10.1103/PhysRevLett.108.063007} {\bibfield  {journal}
  {\bibinfo  {journal} {Phys. Rev. Lett.}\ }\textbf {\bibinfo {volume} {108}},\
  \bibinfo {pages} {063007} (\bibinfo {year} {2012})}\BibitemShut {NoStop}%
\bibitem [{\citenamefont {Hentschel}\ \emph {et~al.}(2001)\citenamefont
  {Hentschel}, \citenamefont {Kienberger}, \citenamefont {Spielmann},
  \citenamefont {Reider}, \citenamefont {Milosevic}, \citenamefont {Brabec},
  \citenamefont {Corkum}, \citenamefont {Heinzmann}, \citenamefont {Drescher},\
  and\ \citenamefont {Krausz}}]{Hentschel:AM-01}%
  \BibitemOpen
  \bibfield  {author} {\bibinfo {author} {\bibfnamefont {M.}~\bibnamefont
  {Hentschel}}, \bibinfo {author} {\bibfnamefont {R.}~\bibnamefont
  {Kienberger}}, \bibinfo {author} {\bibfnamefont {Ch.}\ \bibnamefont
  {Spielmann}}, \bibinfo {author} {\bibfnamefont {G.~A.}\ \bibnamefont
  {Reider}}, \bibinfo {author} {\bibfnamefont {N.}~\bibnamefont {Milosevic}},
  \bibinfo {author} {\bibfnamefont {T.}~\bibnamefont {Brabec}}, \bibinfo
  {author} {\bibfnamefont {P.}~\bibnamefont {Corkum}}, \bibinfo {author}
  {\bibfnamefont {U.}~\bibnamefont {Heinzmann}}, \bibinfo {author}
  {\bibfnamefont {M.}~\bibnamefont {Drescher}}, \ and\ \bibinfo {author}
  {\bibfnamefont {F.}~\bibnamefont {Krausz}},\ }\bibfield  {title} {\enquote
  {\bibinfo {title} {Attosecond metrology},}\ }\href {\doibase
  10.1038/35107000} {\bibfield  {journal} {\bibinfo  {journal} {Nature}\
  }\textbf {\bibinfo {volume} {414}},\ \bibinfo {pages} {509--513} (\bibinfo
  {year} {2001})}\BibitemShut {NoStop}%
\bibitem [{\citenamefont {Sansone}\ \emph {et~al.}(2006)\citenamefont
  {Sansone}, \citenamefont {Benedetti}, \citenamefont {Calegari}, \citenamefont
  {Vozzi}, \citenamefont {Avaldi}, \citenamefont {Flammini}, \citenamefont
  {Poletto}, \citenamefont {Villoresi}, \citenamefont {Altucci}, \citenamefont
  {Velotta}, \citenamefont {Stagira}, \citenamefont {De~Silvestri},\ and\
  \citenamefont {Nisoli}}]{Sansone:IS-06}%
  \BibitemOpen
  \bibfield  {author} {\bibinfo {author} {\bibfnamefont {G.}~\bibnamefont
  {Sansone}}, \bibinfo {author} {\bibfnamefont {E.}~\bibnamefont {Benedetti}},
  \bibinfo {author} {\bibfnamefont {F.}~\bibnamefont {Calegari}}, \bibinfo
  {author} {\bibfnamefont {C.}~\bibnamefont {Vozzi}}, \bibinfo {author}
  {\bibfnamefont {L.}~\bibnamefont {Avaldi}}, \bibinfo {author} {\bibfnamefont
  {R.}~\bibnamefont {Flammini}}, \bibinfo {author} {\bibfnamefont
  {L.}~\bibnamefont {Poletto}}, \bibinfo {author} {\bibfnamefont
  {P.}~\bibnamefont {Villoresi}}, \bibinfo {author} {\bibfnamefont
  {C.}~\bibnamefont {Altucci}}, \bibinfo {author} {\bibfnamefont
  {R.}~\bibnamefont {Velotta}}, \bibinfo {author} {\bibfnamefont
  {S.}~\bibnamefont {Stagira}}, \bibinfo {author} {\bibfnamefont
  {S.}~\bibnamefont {De~Silvestri}}, \ and\ \bibinfo {author} {\bibfnamefont
  {M.}~\bibnamefont {Nisoli}},\ }\bibfield  {title} {\enquote {\bibinfo {title}
  {Isolated single-cycle attosecond pulses},}\ }\href {\doibase
  10.1126/science.1132838} {\bibfield  {journal} {\bibinfo  {journal}
  {Science}\ }\textbf {\bibinfo {volume} {314}},\ \bibinfo {pages} {443--446}
  (\bibinfo {year} {2006})}\BibitemShut {NoStop}%
\bibitem [{\citenamefont {Cing{\"o}z}\ \emph {et~al.}(2012)\citenamefont
  {Cing{\"o}z}, \citenamefont {Yost}, \citenamefont {Allison}, \citenamefont
  {Ruehl}, \citenamefont {Fermann}, \citenamefont {Hartl},\ and\ \citenamefont
  {Ye}}]{Cingoz:DF-12}%
  \BibitemOpen
  \bibfield  {author} {\bibinfo {author} {\bibfnamefont {Arman}\ \bibnamefont
  {Cing{\"o}z}}, \bibinfo {author} {\bibfnamefont {Dylan~C.}\ \bibnamefont
  {Yost}}, \bibinfo {author} {\bibfnamefont {Thomas~K.}\ \bibnamefont
  {Allison}}, \bibinfo {author} {\bibfnamefont {Axel}\ \bibnamefont {Ruehl}},
  \bibinfo {author} {\bibfnamefont {Martin~E.}\ \bibnamefont {Fermann}},
  \bibinfo {author} {\bibfnamefont {Ingmar}\ \bibnamefont {Hartl}}, \ and\
  \bibinfo {author} {\bibfnamefont {Jun}\ \bibnamefont {Ye}},\ }\bibfield
  {title} {\enquote {\bibinfo {title} {Direct frequency comb spectroscopy in
  the extreme ultraviolet},}\ }\href {\doibase 10.1038/nature10711} {\bibfield
  {journal} {\bibinfo  {journal} {Nature}\ }\textbf {\bibinfo {volume} {482}},\
  \bibinfo {pages} {68--71} (\bibinfo {year} {2012})}\BibitemShut {NoStop}%
\bibitem [{\citenamefont {Young}(2012)}]{Young:PM-12}%
  \BibitemOpen
  \bibfield  {author} {\bibinfo {author} {\bibfnamefont {Linda}\ \bibnamefont
  {Young}},\ }\bibfield  {title} {\enquote {\bibinfo {title} {Precision
  measurement: {A} comb in the extreme ultraviolet},}\ }\href {\doibase
  10.1038/482045a} {\bibfield  {journal} {\bibinfo  {journal} {Nature}\
  }\textbf {\bibinfo {volume} {482}},\ \bibinfo {pages} {45--46} (\bibinfo
  {year} {2012})}\BibitemShut {NoStop}%
\bibitem [{\citenamefont {Cavaletto}\ \emph {et~al.}(2013)\citenamefont
  {Cavaletto}, \citenamefont {Harman}, \citenamefont {Buth},\ and\
  \citenamefont {Keitel}}]{Cavaletto:FC-13}%
  \BibitemOpen
  \bibfield  {author} {\bibinfo {author} {\bibfnamefont {Stefano~M.}\
  \bibnamefont {Cavaletto}}, \bibinfo {author} {\bibfnamefont {Zolt{\'a}n}\
  \bibnamefont {Harman}}, \bibinfo {author} {\bibfnamefont {Christian}\
  \bibnamefont {Buth}}, \ and\ \bibinfo {author} {\bibfnamefont {Christoph~H.}\
  \bibnamefont {Keitel}},\ }\bibfield  {title} {\enquote {\bibinfo {title}
  {X-ray frequency combs from optically controlled resonance fluorescence},}\
  }\href {\doibase 10.1103/PhysRevA.88.063402} {\bibfield  {journal} {\bibinfo
  {journal} {Phys. Rev. A}\ }\textbf {\bibinfo {volume} {88}},\ \bibinfo
  {pages} {063402} (\bibinfo {year} {2013})},\ \Eprint
  {http://arxiv.org/abs/1302.3141} {arXiv:1302.3141} \BibitemShut {NoStop}%
\bibitem [{\citenamefont {Cavaletto}\ \emph {et~al.}(2014)\citenamefont
  {Cavaletto}, \citenamefont {Harman}, \citenamefont {Ott}, \citenamefont
  {Buth}, \citenamefont {Pfeifer},\ and\ \citenamefont
  {Keitel}}]{Cavaletto:HF-14}%
  \BibitemOpen
  \bibfield  {author} {\bibinfo {author} {\bibfnamefont {Stefano~M.}\
  \bibnamefont {Cavaletto}}, \bibinfo {author} {\bibfnamefont {Zolt{\'a}n}\
  \bibnamefont {Harman}}, \bibinfo {author} {\bibfnamefont {Christian}\
  \bibnamefont {Ott}}, \bibinfo {author} {\bibfnamefont {Christian}\
  \bibnamefont {Buth}}, \bibinfo {author} {\bibfnamefont {Thomas}\ \bibnamefont
  {Pfeifer}}, \ and\ \bibinfo {author} {\bibfnamefont {Christoph~H.}\
  \bibnamefont {Keitel}},\ }\bibfield  {title} {\enquote {\bibinfo {title}
  {Broadband high-resolution x-ray frequency combs},}\ }\href {\doibase
  10.1038/NPHOTON.2014.113} {\bibfield  {journal} {\bibinfo  {journal} {Nature
  Photon.}\ }\textbf {\bibinfo {volume} {8}},\ \bibinfo {pages} {520--523}
  (\bibinfo {year} {2014})},\ \Eprint {http://arxiv.org/abs/1402.6652}
  {arXiv:1402.6652} \BibitemShut {NoStop}%
\bibitem [{\citenamefont {Craig}\ and\ \citenamefont
  {Thirunamachandran}(1984)}]{Craig:MQ-84}%
  \BibitemOpen
  \bibfield  {author} {\bibinfo {author} {\bibfnamefont {D.~P.}\ \bibnamefont
  {Craig}}\ and\ \bibinfo {author} {\bibfnamefont {T.}~\bibnamefont
  {Thirunamachandran}},\ }\href@noop {} {\emph {\bibinfo {title} {Molecular
  Quantum Electrodynamics}}}\ (\bibinfo  {publisher} {Academic Press},\
  \bibinfo {address} {London},\ \bibinfo {year} {1984})\BibitemShut {NoStop}%
\bibitem [{\citenamefont {Klaiber}(2007)}]{Klaiber:RR-07}%
  \BibitemOpen
  \bibfield  {author} {\bibinfo {author} {\bibfnamefont {Michael}\ \bibnamefont
  {Klaiber}},\ }\emph {\bibinfo {title} {Relativistische {Rekollisionen} in
  starken {Laserfeldern}}},\ \href@noop {} {\bibinfo {type} {Dissertation}},\
  \bibinfo  {school} {Ruprecht-Karls-Universit{\"a}t Heidelberg}, \bibinfo
  {address} {Fakult{\"a}t f{\"u}r Physik und Astronomie, Heidelberg, Germany}
  (\bibinfo {year} {2007}),\ \bibinfo {note} {uniform Resource Name~(URN):
  \texttt{urn:nbn:de:bsz:16-opus-78423}, HeiDOK~-- Der Heidelberger
  Dokumentenserver: \href{http://www.ub.uni-heidelberg.de/archiv/7842}
  {http://www.ub.uni-heidelberg.de/archiv/7842}}\BibitemShut {NoStop}%
\bibitem [{Note6()}]{Note6}%
  \BibitemOpen
  \bibinfo {note} {The derivation of Ref.~\protect \rev@citealp {Klaiber:RR-07}
  is restricted to a positive definite Hessian in Eq.~(\ref {eq:TaylorS}) as
  the square root of the Hessian matrix is needed which is not required
  here~\cite {Golub:MC-96}.}\BibitemShut {Stop}%
\bibitem [{Note7()}]{Note7}%
  \BibitemOpen
  \bibinfo {note} {All coupling terms that involve a HH~photon with continuum
  states, the ``continuum-continuum terms'' are negligible.}\BibitemShut
  {Stop}%
\bibitem [{\citenamefont {M{\o}lmer}\ and\ \citenamefont
  {Castin}(1996)}]{Molmer:MC-96}%
  \BibitemOpen
  \bibfield  {author} {\bibinfo {author} {\bibfnamefont {Klaus}\ \bibnamefont
  {M{\o}lmer}}\ and\ \bibinfo {author} {\bibfnamefont {Yvan}\ \bibnamefont
  {Castin}},\ }\bibfield  {title} {\enquote {\bibinfo {title} {{Monte Carlo}
  wavefunctions in quantum optics},}\ }\href {\doibase
  10.1088/1355-5111/8/1/007} {\bibfield  {journal} {\bibinfo  {journal}
  {Quantum Semiclass. Opt.}\ }\textbf {\bibinfo {volume} {8}},\ \bibinfo
  {pages} {49--72} (\bibinfo {year} {1996})}\BibitemShut {NoStop}%
\bibitem [{Note8()}]{Note8}%
  \BibitemOpen
  \bibinfo {note} {I determine the cases~$\omega = 0$ and $\omega + \omega ' =
  0$ \protect \emph {not} by inserting~$0$ into the sinus cardinalis~(\ref
  {eq:sinc})---which would give~$\protect \textrm {sinc}\protect \tmspace
  +\thinmuskip {.1667em}\protect \tmspace +\thinmuskip {.1667em} 0 = 1$---but
  by taking the limits~$\omega \to 0$ and $\omega + \omega ' \to 0$,
  respectively, instead after the limit~$t \to \infty $ has been
  taken.}\BibitemShut {Stop}%
\bibitem [{\citenamefont {Golub}\ and\ \citenamefont {van
  Loan}(1996)}]{Golub:MC-96}%
  \BibitemOpen
  \bibfield  {author} {\bibinfo {author} {\bibfnamefont {Gene~H.}\ \bibnamefont
  {Golub}}\ and\ \bibinfo {author} {\bibfnamefont {Charles~F.}\ \bibnamefont
  {van Loan}},\ }\href@noop {} {\emph {\bibinfo {title} {Matrix
  Computations}}},\ \bibinfo {edition} {3rd}\ ed.\ (\bibinfo  {publisher}
  {Johns Hopkins University Press},\ \bibinfo {address} {Baltimore},\ \bibinfo
  {year} {1996})\BibitemShut {NoStop}%
\end{thebibliography}
\end{document}